\documentclass[a4paper,11pt,DIV=15,BCOR=16mm,headinclude=false,footinclude=false,open=right,appendixprefix=true]{scrreprt}
\pdfoutput=1

\makeatletter
\DeclareOldFontCommand{\rm}{\normalfont\rmfamily}{\mathrm}
\DeclareOldFontCommand{\sf}{\normalfont\sffamily}{\mathsf}
\DeclareOldFontCommand{\tt}{\normalfont\ttfamily}{\mathtt}
\DeclareOldFontCommand{\bf}{\normalfont\bfseries}{\mathbf}
\DeclareOldFontCommand{\it}{\normalfont\itshape}{\mathit}
\DeclareOldFontCommand{\sl}{\normalfont\slshape}{\@nomath\sl}
\DeclareOldFontCommand{\sc}{\normalfont\scshape}{\@nomath\sc}
\makeatother

\usepackage[a-1b]{pdfx}
\hypersetup{bookmarksnumbered=true}
\usepackage{lineno}
\usepackage[
    automark,
    headsepline,
    ilines
]{scrpage2}
\clearscrheadfoot
\ohead{\headmark}
\ofoot[\pagemark]{\pagemark}
\automark[section]{chapter}

\usepackage{url}
\usepackage{xspace}
\usepackage{pslatex}
\usepackage{amsmath}
\usepackage{txfonts}
\usepackage{graphicx}
\usepackage{cancel}
\usepackage{amssymb}
\usepackage{tensor}
\usepackage{caption}
\usepackage{subcaption}
 \usepackage{epstopdf}
\usepackage{slashed}
\usepackage{color}
\usepackage{mathtools}
\usepackage{setspace}
\usepackage{soul}
\usepackage[
	ngerman,
	english,
]{babel}

\graphicspath{{figs/}}

\def\MSbar{{\ensuremath{\overline{\text{MS}}}}}
\def\as{{\alpha_s}}
\def\ae{{\alpha_e}}
\def\mt{m_t}

\def\muf{\mu_F}
\def\mur{\mu_R}
\def\nn{\nonumber}

\def\tmin{{\textrm{min}}}
\def\true{{\textrm{true}}}

\def\NLO{{\textrm{NLO}}}
\def\Born{{\textrm{Born}}}
\def\Refs#1{refs.~\cite{#1}}
\def\Ref#1{ref.~\cite{#1}}
\def\Eq#1{eq.~(\ref{#1})}
\def\Fig#1{figure~\ref{#1}}
\def\Lboost#1{\ensuremath{\Lambda^b(#1)}}
\def\Lrotx#1{\ensuremath{\Lambda_x^r(#1)}}
\def\Lroty#1{\ensuremath{\Lambda_y^r(#1)}}

\def\Kallen#1{\lambda\left(#1\right)}
\def\ycut{y_{\text{\scriptsize cut}}}
\def\yij{{y_{ij}}}
\def\ycut{\ensuremath{y_{\textrm{cut}}}}
\def\Lik{\ensuremath{\mathcal{L}}}
\def\ord{\ensuremath{\mathcal{O}}}
\def\MEM{Matrix Element Method\xspace}
\def\ML{Maximum Likelihood\xspace}
\def\SM{Standard Model\xspace}
\def\PSS{phase space slicing\xspace}
\def\IR{infrared\xspace}
\def\UV{ultraviolet\xspace}
\def\incl{{\text{incl.}}}
\def\excl{{\text{excl.}}}
\def\GeV{\ensuremath{~\text{GeV}}\xspace}

\def\ftw{{0.87}}

 \numberwithin{equation}{section}

\begin{document}
\KOMAoptions{twoside = false}
\makeatletter
\begin{titlepage}
\begin{spacing}{1}

\newlength{\parindentbak}
  \setlength{\parindentbak}{\parindent}
\newlength{\parskipbak}
  \setlength{\parskipbak}{\parskip}

\setlength{\parindent}{0pt}
\setlength{\parskip}{\baselineskip}

\thispagestyle{empty}
\vfill
\hfill HU-EP-18/23
\begin{center}
\begin{spacing}{1.5}
{\LARGE \textbf{The \MEM at next-to-leading order QCD using the example of single top-quark production at the LHC}}
\end{spacing}
\large
\vfill
\so{DISSERTATION}
\vfill
zur Erlangung des akademischen Grades

doctor rerum naturalium\\
(Dr. rer. nat.)\\
im Fach Physik\\
Spezialisierung: Theoretische Physik

eingereicht an der\\
Mathematisch-Naturwissenschaftlichen Fakult\"at\\
der Humboldt-Universit\"at zu Berlin

von\\
\textbf{Herrn Dipl. Phys. Till Martini}\\
%(geboren am 14.11.1985 in Bad Ems)
\end{center}

\vfill

\large
Pr\"asidentin der Humboldt-Universit\"at zu Berlin: \\
Prof. Dr.-Ing. Dr. Sabine Kunst

Dekan der Mathematisch-Naturwissenschaftlichen Fakult\"at:\\
Prof. Dr. Elmar Kulke
\end{spacing}

\large
Gutachter/innen:
\begin{enumerate}
\item Prof. Dr. Peter Uwer
\item Prof. Dr. Michael Spannowsky
\item Prof. Dr. Tilman Plehn\\
\end{enumerate}

\begin{spacing}{1}

\textbf{Tag der m\"undlichen Pr\"ufung:} 26.06.2018\\

\setlength{\parindent}{\parindentbak}
\setlength{\parskip}{\parskipbak}
\end{spacing}
\end{titlepage}
\newpage
\thispagestyle{empty}
\mbox{}

\makeatother
\KOMAoptions{twoside}
\begingroup
\selectlanguage{ngerman}
\chapter*{Selbst\"andigkeitserkl\"arung}
\thispagestyle{empty}
Ich erkl\"are, dass ich die Dissertation selbst\"andig und nur unter Verwendung der von mir gem\"a{\ss} Abs. 3 der Promotionsordnung der Mathematisch-Naturwissenschaftlichen Fakult\"at, ver\"offentlicht im Amtlichen Mitteilungsblatt der Humboldt-Universit\"at zu Berlin Nr. 126/2014 am 18.11.2014 angegebenen Hilfsmittel angefertigt habe.

\vspace{5\baselineskip}
\noindent Berlin, den 28.02.2018\hfill Till Martini
\endgroup

\KOMAoptions{DIV=20,BCOR=19mm}
\newpage
\thispagestyle{empty}
\mbox{}
%\chapter*{Abstract}
\newpage
\thispagestyle{empty}
\mbox{}
\thispagestyle{empty}
{\Huge\bf\textsf{Abstract}}
%\vspace{1\baselineskip}\\
\vfill
The task of data analyses in elementary particle physics is to put the Standard Model---its central and commonly accepted theory---to test. A prime example being; by disproving theoretical predictions that are usually only calculable within perturbation theory. However, to draw convincing conclusions, analysis methods are needed which not only offer an unambiguous comparison between data and theory, but which also allow for reliably estimating the related uncertainties.\\
The Matrix Element Method (MEM) is a Maximum Likelihood method whereby theoretically calculated differential cross sections are identified with probability densities to attribute statistical weights to observed events at a collider. With these weights, a cumulative likelihood function is calculated for a data set of events, which can be used in likelihood ratios or for parameter estimation.
The MEM has proven to be beneficial due to its optimal use of the available information and a clean statistical interpretation of the extracted parameters. Since its introduction in the early top-quark analyses at the Tevatron, the MEM has been established as a standard tool for signal searches as well as parameter determinations. \\
Because of the high experimental precision now achievable, a major drawback of the MEM is the limitation of applicable theoretical predictions: In its current formulation, the likelihood can only be consistently defined for cross section predictions that are accurate to the leading perturbative order in the coupling. In perturbative Quantum Field Theory, higher-order corrections not only improve the accuracy of theoretical predictions but also allow for fixing the renormalisation scheme, offering an unambiguous field-theoretical interpretation of the extracted information. \\
In this work the development of the MEM for hadron collider analyses which consistently incorporates corrections of next-to-leading order (NLO) in the coupling of QCD is presented for the first time. An algorithm to calculate event weights in terms of differential cross sections including NLO corrections suited for their application within the MEM is introduced. The ability to calculate these event weights also enables the generation of unweighted events distributed according to the probability density which is given by the differential cross section calculated at NLO accuracy. Treating these events as outcomes of idealised toy experiments facilitates the first simulation of the application of the MEM at NLO accuracy. \\
The presented method is demonstrated for top-quark pairs at a lepton collider and single-top quarks at the LHC. In particular, high-precision studies regarding top-quark mass measurements are relevant for these example processes---in light of both a future linear collider allowing the production of top-quark pairs above the production threshold as well as to resolve current tensions  between mass determinations from events with single top quarks and top-quark pairs at the LHC.\\
As a consistency check, the MEM at NLO accuracy is used for top-quark mass determinations from the generated events. As expected, estimators are obtained which are in agreement with the input values used in the event generation. On the contrary, repeating the mass determinations from the same events, without NLO corrections in the predictions, results in biased estimators. Depending on the specific process, these shifts in the estimators are not accounted for by the attributed estimated theoretical uncertainties. While a bias in the estimator could be removed by a calibration procedure of the MEM at the price of additional systematic uncertainties, the estimation of the theoretical uncertainties of the leading-order analysis remains unreliable.\\
The results presented in this work emphasise that the inclusion of NLO corrections into the MEM is mandatory for reducing the necessary amount of calibration and reliably estimating the theoretical uncertainties, as well as for having a well-defined theoretical interpretation of the extracted parameters. Thus, extending the MEM to NLO accuracy could be crucial not only  for the aforementioned future and current studies of the top-quark mass.
\vfill
\newpage
\thispagestyle{empty}
\mbox{}
%\chapter*{Zusammenfassung}
\newpage
\thispagestyle{empty}
\mbox{}
\begingroup
\selectlanguage{ngerman}
\thispagestyle{empty}
{\Huge\bf\textsf{Zusammenfassung}}
%\vspace{1.5\baselineskip}\\
\vfill
Die Analyse von Daten in der Elementarteilchenphysik dient dem Zweck, das Standardmodell---ihre zentrale und gemeinhin akzeptierte Theorie---auf die Probe zu stellen. Ein wichtiges Beispiel ist das Widerlegen von theoretischen Vorhersagen, die meist nur mithilfe von St\"orungstheorie berechenbar sind. Um jedoch \"uberzeugende Schlussfolgerungen ziehen zu k\"onnen, sind Analysemethoden n\"otig, welche nicht nur einen eindeutigen Vergleich zwischen Daten und Theorie erm\"oglichen, sondern dar\"uberhinaus erlauben die zugeh\"origen Unsicherheiten zuverl\"assig abzusch\"atzen.\\
Die Matrixelement-Methode (MEM) ist eine Maximum-Likelihood-Methode, derzufolge der theoretisch berechnete differentielle Wirkungsquerschnitt mit einer Wahrscheinlichkeitsdichte identifiziert wird um beobachteten Ereignissen an Beschleunigern statistische Gewichte zuzuordnen. Mit diesen Gewichten wird eine kumulative Likelihood-Funktion f\"ur einen Datensatz von Ereignissen berechnet, welche in Likelihood-Quotienten oder zur Parametersch\"atzung genutzt werden kann.
Die MEM hat sich durch die optimale Nutzung der vorhandenen Information und die saubere statistische Interpretation der extrahierten Parameter als vorteilhaft erwiesen. Seit ihrer Einf\"uhrung in den fr\"uhen Top-Quark-Analysen am Tevatron hat sich die MEM als Standardwerkzeug f\"ur Signalsuchen und zur Parameterbestimmung etabliert.\\
Durch die hohe experimentelle Pr\"azision, welche mittlerweile erreichbar ist, ist ein gro{\ss}er Nachteil der Matrix-element-Methode jedoch durch die Einschr\"ankung der verwendbaren theoretische Vorhersagen gegeben: In ihrer derzeitigen Formulierung kann die Likelihood nur f\"ur Vorhersagen von Wirkungsquerschnitten, deren Genauigkeit sich lediglich auf die erste st\"orungstheoretische Ordnung in der Kopplung bel\"auft, konsistent definiert werden. In der perturbativen Quantenfeldtheorie verbessern h\"ohere Ordnungskorrekturen nicht nur die Genauigkeit theoretischer Vorhersagen, sondern erlauben auch die Fixierung des Renormierungsschemas und bieten damit die eindeutige feldtheoretische Interpretation der gewonnen Informationen. \\
In dieser Arbeit wird erstmalig die Entwicklung der MEM unter konsistentem Einbezug der Korrekturen der n\"achstf\"uhrenden Ordnung (NLO) der QCD-Kopplung pr\"asentiert. Es wird ein  Algorithmus zur Berechnung von Ereignisgewichten in Form von differentiellen Wirkungsquerschnitten einschlie{\ss}lich NLO-Korrekturen f\"ur die Anwendung in der Matrixelement-Methode vorgestellt. Die F\"ahigkeit diese Ereignisgewichte zu berechnen erm\"oglicht ebenso die Erzeugung ungewichteter Ereignisse, welche entsprechend der Wahrscheinlichkeitsdichte verteilt sind, die durch den in NLO-Genauigkeit berechneten differentiellen Wirkungsquerschnitt gegeben ist. Die Behandlung dieser Ereignisse als Resultat eines idealisierten Beispielexperiments erm\"oglicht die erste Simulation der Anwendung der Matrixelement-Methode in NLO-Genauigkeit.\\
Der vorgestellte Methode wird am Beispiel von Top-Quark-Paaren an einem Leptonen-Beschleuniger und einzelner Top-Quarks am LHC veranschaulicht. Im Besonderen sind Pr\"azisionsstudien zur Messung der Top-Quark-Masse f\"ur diese Beispielprozesse relevant---sowohl im Hinblick auf einen zuk\"unftigen Linearbeschleuniger, welcher die Produktion von Top-Quark-Paaren jenseits der Produktionsschwelle erlaubt, als auch um aktuelle Spannungen zwischen Massenbestimmungen aus Ereignissen einzelner Top-Quarks und Top-Quark-Paaren am LHC aufzul\"osen.\\
Als Konsistenztest wird die MEM in NLO-Genauigkeit zur Bestimmung der Top-Quark-Masse aus den erzeugten Ereignissen benutzt. Erwartungsgem\"a{\ss} werden Sch\"atzer erhalten, welche mit den  Eingabewerten aus der Ereigniserzeugung \"ubereinstimmen. Die Wiederholung der Massenbestimmung aus denselben Ereignissen, ohne NLO-Korrekturen in den Vorhersagen, f\"uhrt jedoch zu verf\"alschten Sch\"atzern. Abh\"angig vom jeweiligen Prozess werden diese Verschiebungen in den Sch\"atzern nicht durch die zugeordneten theoretischen Unsicherheiten ber\"ucksichtigt. W\"ahrend eine Verf\"alschung des Sch\"atzers durch eine Kalibrierung der MEM zum Preis von zus\"atzlichen systematischen Unsicherheiten behoben werden k\"onnte, bleibt die Absch\"atzung der theoretischen Unsicherheiten in der Analyse in f\"uhrender Ordnung unzuverl\"assig.\\
Die vorgestellten Resultate unterstreichen, dass die Ber\"ucksichtigung von NLO-Korrekturen in der MEM zwingend erforderlich ist, um sowohl den ben\"otigten Umfang an Kalibrierung zu reduzieren und zuverl\"assig die theoretischen Unsicherheiten abzusch\"atzen, als auch um eine wohldefinierte theoretische Interpretation der extrahierten Parameter zu erhalten. Die Erweiterung der MEM auf NLO-Genauigkeit k\"onnte also nicht nur f\"ur die oben genannten zuk\"unftigen und derzeitigen Studien der Top-Quark-Masse entscheidend sein.

\endgroup
\KOMAoptions{DIV=15,BCOR=16mm}
\thispagestyle{empty}

\cleardoublepage
\begingroup
  \renewcommand*{\chapterpagestyle}{empty}
  \pagestyle{empty}
  \tableofcontents
  \clearpage
\endgroup
    \addtocounter{page}{-12}%
\newpage
\thispagestyle{empty}
\mbox{}
\chapter{Introduction}
The Large Hadron Collider (LHC) at CERN creates all-time experimental conditions (see \Ref{Evans:2008zzb}). It represents a unique laboratory for comprehensive tests of the commonly accepted theory of elementary particle physics, the \SM (see e.g. \Refs{Olive:2016xmw,ATLASSM:2018,CMSSM:2018}). At the same time the LHC allows to search for extensions of the theory by `New Physics' beyond the \SM  (see e.g. \Refs{Olive:2016xmw,ATLASSUSY:2018,CMSSUSY:2018,ATLASEXO:2018,CMSEXO:2018}). It is crucial to rule out that the absence of significant New Physics signals reported by the experiments so far is owed to insufficient precision in the analyses. To perform unambigous high precision analyses, a comparison of experimental data and theoretical predictions which match the experimental precision is essential. Due to elaborate analysis methods used by the experiments at the LHC both the statistical and the systematic experimental uncertainties compete with the systematic theoretical uncertainties of the predictions entering the analyses by now  (see \Ref{Olive:2016xmw}).  While in principle many calculations exist which include higher-order corrections, these results are often not readily applicable in the experimental analysis. As precise and accurate theoretical predictions as possible are needed in a form that allows a direct comparison of theory and experiment. 

For signal processes with small production cross sections only a small number of events, which is potentially overwhelmed by background, is expected to be available for analysis. Such situations call for analyses based on powerful efficient methods.
With steadily growing computing power, multi-variate methods have been established as standard tools in many analyses at the LHC. Neuronal networks (see \Ref{McCulloch1943}), boosted decision trees (see \Ref{Friedman:2002:SGB:635939.635941})
and the \MEM (see section~\ref{sec:SmatMEM} and references therein) as specific examples have been proven to be instrumental in many high energy physics analyses. The \MEM is a \ML method which can be used for signal-background discrimination as well as for parameter extraction. The method allows to directly compare measured data to model-dependent theoretical predictions providing a very clean statistical interpretation of the results (cf. section~\ref{sec:MLparest}). 
The \MEM is based on the assumption that the probability to observe a specific event signature in the detectors can be calculated from respective transition matrix elements and transfer functions as a model-dependent likelihood function. The transfer functions model the correspondence between theoretically described and experimentally measured quantities and have to be determined by the experiments through detector simulations. The matrix element describes the transition probability from a given initial state to a certain final state and has to be calculated according to a specific theoretical model within Quantum Field Theory (parameterised by model parameter(s) $\omega$). Determinations of model parameters which are not predicted by the theory can be performed by maximising the overall likelihood for the observation of the whole data set with respect to $\omega$ to obtain an estimator for the model parameter. For the discrimination between signal and background, likelihood ratios for signal+background and the background only hypotheses can be defined. The information content in the data set is optimally used in this likelihood and therefore by the analyses (see \Ref{Cranmer:2006zs}).  Assuming that the underlying theory gives an exact description of the experimentally studied interactions and that unlimited statistics are available, the accuracy of the \MEM is only limited by the accuracy of the theoretical calculation of the matrix elements (and the knowledge of the transfer functions). 

The application of the \MEM in the hadron collider context has been pioneered at the Tevatron in the top-quark mass extraction from only $\ord(100)$ $t\bar{t}$ events (see \Refs{Abbott:1998dn,Abazov:2004cs,Abulencia:2006mi}). It has been established as a standard multi-variate analysis tool both for parameter extraction as well as for (beyond) \SM signal searches  at the Tevatron, LHC and future colliders (see \Refs{Giammanco:2017xyn,FerreiradeLima:2017iwx,Englert:2015dlp,Aad:2015upn,Khachatryan:2015ila,Gritsan:2016hjl,Artoisenet:2013vfa,Englert:2015dlp,Aad:2015upn,Khiem:2015ofa,Gao:2010qx,Bolognesi:2012mm}). For a detailed description of experimental techniques regarding top-quark mass extraction employing the \MEM see \Ref{Fiedler:2010sg}. The automated calculation of the event weights required by the \MEM has been studied within the MadGraph framework in \Ref{Artoisenet:2010cn}. A big drawback of the \MEM is that in the experimental applications it has been limited to a leading-order (LO) formulation so far.

Besides the strength of the \MEM to offer a clean statistical interpretation, it also has a potential weakness: If the theoretical cross section used in the likelihood calculation does not properly describe the underlying probability density, the \MEM is prone to fail to produce reliable results. In this case, parameter extraction with the \MEM no longer provides an unbiased estimator (cf. section~\ref{sec:MLparest}). In principle, the
bias can be accounted for by an additional calibration procedure at the price of introducing additional uncertainties and possibly auxiliary modelling. In light of the increasing precision already achieved or aimed for by current experiments, the inclusion of higher-order QCD corrections in the cross section predictions used in the \MEM is mandatory.

High energetic hadron collisions offer sizeable phase space for extra radiation either contributing to the signal signature as soft/collinear corrections or revealing itself as additional hard jets in the detectors. Restricting the \MEM to a LO formulation prohibits to account for the extra radiation in the likelihood calculation. Events that do not fit the LO picture have either to be discarded from the data set or modified to fit a LO interpretation. This might result in a loss of signal statistics or in possibly distorting the information in the data. Only when including higher-order corrections the non-trivial formation of jets is modelled for the first time as collections of recombined partons, yielding more realistic predictions for jet processes. It is well known that the Born approximation gives only a rough estimate of the cross section and higher-order QCD corrections are sizeable for many processes. In general, taking higher-order corrections into account should reduce the dependence on the unphysical renormalisation and factorisation scales of the fixed order results  providing confidence in the perturbative expansion and improving the accuracy of the theoretical predictions. 
Additionally, for a clear interpretation of model parameters extracted with the \MEM a formulation of the method beyond the LO is crucial: The renormalisation scheme can initially be uniquely defined by performing an next-to-leading-order (NLO) calculation, only then providing distinct parameter definitions within that scheme (e.g. the top-quark mass definition depends on the respective renormalisation scheme, cf. section~\ref{sec:regren}). A consistent inclusion of higher-order corrections is thus mandatory to benefit from the advantages of the \MEM in high precision measurements at the LHC.

NLO corrections are comprised of two contributions (cf. section~\ref{sec:IRdivcanc}): In the virtual contribution loop corrections to the leading-order process have to be considered in the calculation of the transition probabilities which require the evaluation of loop integrals. In general, these integrals exhibit \UV as well as \IR divergences which have to be formally regularised. While the \UV divergences can be absorbed into the definition of the model parameters by the mechanism of renormalisation (cf. section~\ref{sec:regren}), the \IR divergences persist at first. The real contribution describes the emission of additional radiation in the scattering process. Thus, it exhibits a higher parton multiplicity in the final-state than the leading-order process. Unresolved additional radiation has to be integrated out. These phase space integrations over the real matrix elements turn out to be \IR divergent as well. After consistently regularising these \IR divergences they can be cancelled with the respective ones in the virtual correction. In the sum, both contributions yield finite predictions for \IR safe observables according to the KLN theorem (cf. section~\ref{sec:IRdivcanc}).
In the framework of perturbation theory the calculation of partonic cross sections including corrections of next-to-leading order in the strong coupling has been automated for the most part (see e.g. \Refs{Alwall:2014hca,Bevilacqua:2011xh,Gleisberg:2008ta}).

In the context of the \MEM, the required theoretical cross section predictions describing the underlying probability density have to be fully differential in all final-state variables specifying the event signature. However, partons are not the experimentally observed final states but hadrons which are not individually resolvable by the detectors are grouped together to so-called jets according to particular recombination prescriptions (cf. section~\ref{sec:jets}). The theoretical modelling of these jets can be performed by determining their momenta from the partonic momenta according to analogous recombination prescriptions.  The transition probabilities needed for the \MEM thus have to be formulated for the transition from the hadronic initial state to specific configurations of the jet momenta. When considering NLO corrections, the unresolved additional radiation in the real corrections requires a non-trivial mapping from partonic to jet momenta. As stated above, the integration over the unresolved real partonic configurations contributing to a given set of jet of momenta is in general \IR divergent. To ensure the mutual cancellation of these \IR divergences in the real contribution with the respective \IR divergences in the virtual contribution, an unambiguous mapping of the real phase space to jet momenta usable in the virtual matrix elements is needed. Note that in generic NLO calculations, where histograms are filled to describe the distribution of variables, this point-wise cancellation is not necessarily essential because the real and virtual corrections are combined within bins of finite width.

A first study to include the effect of QCD radiation into the \MEM has been conducted in \Ref{Alwall:2010cq}. In \Refs{Soper:2011cr,Soper:2012pb} the radiation profile of QCD jets is compared to the radiation pattern of boosted Higgs bosons and top quarks. The findings from \Refs{Soper:2011cr,Soper:2012pb} are used to combine the information from the Born matrix element and a parton shower in a signal versus background discrimination study of $Z'$ decaying to boosted top quarks in \Ref{Soper:2014rya}. The extension of the \MEM by full NLO corrections including virtual corrections is introduced in \Refs{Campbell:2012cz,Campbell:2012ct,Campbell:2013uha}. In \Ref{Campbell:2013hz} Higgs production with subsequent decay into $H\to Z\gamma$ is studied as a first detailed application. However, the method presented in \Refs{Campbell:2012cz,Campbell:2012ct,Campbell:2013hz} is restricted to the hadronic production of uncoloured objects but a generalisation including the production of coloured particles like for example top quarks is missing. The possibility to include hadronic jet production by mapping NLO and LO jets by a longitudinal boost along the beam axis to account for unbalanced transverse momentum is investigated in \Ref{Campbell:2013uha}.  In \Refs{Martini:2015fsa} we have introduced a complete
algorithm for the systematic inclusion of NLO QCD corrections in the \MEM. The Drell-Yan process and top-quark pair
production in $e^+ e^-$-annihilation are studied as a proof of concept. Inspired by \Ref{Martini:2015fsa}, the authors of \Ref{Baumeister:2016maz} propose general concepts to extend the \MEM beyond the Born approximation by numerically solving a system of non-linear equations or adopting parton-shower matching prescriptions. However, a proof of the feasibility of this approach in practical applications is not given. We have presented the first application of the algorithm introduced in \Ref{Martini:2015fsa} to the hadronic production of jets in \Ref{Martini:2017ydu} with single top-quark production at the LHC as an example process.

In this work the development of the \MEM consistently incorporating full NLO QCD effects as introduced in  \Refs{Martini:2015fsa,Martini:2017ydu} is presented. It is applicable to most of the relevant collider experiments: production of at least $2$ jets at a lepton collider, deep inelastic scattering or hadronic production of electroweak final states and/or jets. The respective final states can have arbitrary masses. The introduced algorithm is process independent and can be automated. As example processes, hadronic Drell-Yan production, top-quark pair production in $e^+e^-$ annihilation and the production of single top quarks at the LHC are considered. The first two are chosen as rather simple exemplary processes to single out and test different aspects of the algorithm individually. The latter represents a state-of-the-art application possibility for the \MEM (see e.g. \Refs{Giammanco:2017xyn,Aad:2015upn}). In these contexts, improvements in the precision of top-quark mass measurements are desirable to not only shed light on some tension between mass determinations from single top-quark production and top-quark pair production at the LHC (see \Ref{Alekhin:2016jjz}) but also to establish mass determinations in the continuum as a competitive alternative to threshold scans at a future linear collider (see e.g. \Ref{Seidel:2013sqa}).

This work is organised as follows. Theoretical prerequisites with a focus on the method of \ML and the calculation of higher-order corrections in perturbation theory are summarised in section~\ref{sec:prerequisites}.\\ In section~\ref{sec:MEMtoNLO} it is shown how the aforementioned unambiguous mapping of the real partonic configurations and the jet phase space can be achieved by modifying the recombination schemes of common jet algorithms. The $2\to1$ clustering typically used in jet algorithms is extended to a $3\to2$ clustering inspired by the dipole subtraction method (see section~\ref{sec:dipsub}) by choosing an additional `spectator parton' as presented in section~\ref{sec:3to2clus}. This allows to satisfy momentum conservation and on-shell conditions for the clustered jets at the same time. Hence, these jet momenta can be used to evaluate the virtual matrix elements, allowing the point-wise cancellation of the \IR divergences in the maximally exclusive differential cross section predictions. Assuming the $4$-momenta of these jets are completely determined, a fully differential jet cross section can be defined at NLO accuracy to be used as an event weight in the \MEM.  The required parameterisations of the real phase space for final- and initial-state clusterings with respective spectators are given in section~\ref{sec:PhaseSpaceparameterisation}.\\
The validation of different aspects of these phase space parameterisations is presented in section~\ref{sec:validphsp}. As a starting point, the issues of coloured initial states and associated initial-state radiation and (massive) coloured final states and associated final-state radiation are isolated. More precisely, the method is developed for Drell-Yan production in hadronic collisions and top-quark pair production in $e^+e^-$ annihilation separately (see sections~\ref{sec:DY} and~\ref{sec:ttprod}). 
Because they are fairly simple, compact analytic results for the NLO calculations are available for both processes making it straightforward to test different aspects of the framework. Combining these aspects paves the way to apply the \MEM at NLO to the production of single top quarks in hadron collisions.
The process of single top-quark production at the LHC and the calculational setup used throughout this work is introduced in section~\ref{sec:sngltp}. Its implementation is thoroughly validated with regards to the phase space parameterisations and the \PSS method.\\ 
In section~\ref{sec:applres} application possibilties of the fully differential cross sections calculated at NLO accuracy are demonstrated.
The differential jet cross sections are used as event weights to generate unweighted events, which are distributed according to the NLO cross section in section~\ref{sec:evgen}. Unweighted top-quark pair events are generated in section~\ref{sec:evgentt} and unweighted single-top quark events are generated in sections~\ref{sec:sgtevgen} and~\ref{sec:21event-definition}.
 To work out a sensible event definition for single top-quark production, the influence of the choice of clustering on distributions of final-state variables is studied in section~\ref{sec:sgtevgen}. Distributions of the event variables calculated at NLO and Born accuracy are also compared in order to estimate the impact of the NLO corrections on the \MEM analysis. In addition, by defining the events in terms of jet variables rather than jet-momenta, it is possible to introduce an inclusive event definition. For the special case of event variables which do not fix the masses of the final-state objects it is even possible to define NLO event weights for jets obtained with a $2\to1$ jet algorithm even without the need to numerically solve a system of non-linear equations as suggested in \Ref{Baumeister:2016maz}. This is demonstrated in section~\ref{sec:21event-definition}.
The distribution of the unweighted events according to the NLO cross section is validated for all cases.\\ 
In section~\ref{sec:MEMNLO} the unweighted event samples generated in section~\ref{sec:evgen} are treated as the outcome of toy experiments. The NLO event weights are used in the \MEM to extract estimators for the top-quark mass from top-quark pair events in section~\ref{sec:mtexttt} and from single top-quark events in sections~\ref{sec:mtextsgt} and~\ref{sec:mtextsgt21}. The consistency of the implementation is validated, also for extended likelihoods. The impact of the NLO corrections on the analyses is studied by using LO predictions in the \MEM and the reliability of theoretical uncertainty estimates inferred from scale variations is commented on. The mass sensitivity of the top-quark mass determination from single top-quark events using the \MEM at NLO accuracy is also examined. With the unweighted event samples obtained with a $2\to1$ jet algorithm the dependence of top-quark mass extractions on the employed clustering prescription is investigated in section~\ref{sec:mtextsgt21}.\\
The conclusion is given in section~\ref{sec:concl}.

\chapter{Prerequisites}\label{sec:prerequisites}
\section{\ML parameter estimation}\label{sec:MLparest}
In this chapter some basic facts about \ML parameter estimation are briefly reviewed which are relevant to lay out the statistical foundation of the \MEM.
The presentation follows the explanations in the book `Statistical data analysis' by Glen Cowan, further details as well as proofs of statements repeated here can be found in \Ref{cowan1998statistical} and references therein.
The method of \ML is a technique to infer information about the underlying probability density from a finite set of measurements $\{x_i\}$ with $i=1,\dots,n$. Let $f(x|\omega)$ be a hypothesis for this probability density with at least one unknown parameter $\omega$. The probability for the $i$-th measurement to lie in $[x_i,x_i+dx_i]$ is predicted to be $f(x_i|\omega)dx_i$. If the measurements in $\{x_i\}$ are independent the joint probability for all $i$ is given by the product $f(x_1|\omega)dx_1f(x_2|\omega)dx_2\dots f(x_{n-1}|\omega)dx_{n-1}f(x_n|\omega)dx_n$ of the individual probabilities. The joint probability for the measured $\{x_i\}$ is expected to be large if the hypothesis $f(x|\omega)$ is the correct underlying probability density and small for incorrect hypotheses. Suppose that the functional form of $f(x|\omega)$ is known but the true value of the model parameter $\omega^{\true}$ is unknown. The joint probability for a fixed set of measured $\{x_i\}$ can be treated as a function of the parameter $\omega$, called the likelihood
\begin{equation}\label{eq:MLlikeli}
\Lik(\omega|\{x_i\})=\prod\limits_{i=1}^{n}f(x_i|\omega).
\end{equation}
The value for $\omega$ that (globally) maximises the likelihood for a given set $\{x_i\}$ is the so-called \ML estimator, denoted by $\widehat{\omega}$. For a probability density $f$ that is differentiable in $\omega$ the estimator is the solution to 
\begin{eqnarray}\label{eq:likelimax}
\nn\left.{\partial \Lik(\omega|\{x_i\})\over \partial \omega}\right|_{\omega=\widehat{\omega}}&=&0,\\
\Lik(\widehat{\omega}|\{x_i\})&=&\sup\limits_{\omega} \Lik(\omega|\{x_i\}).
\end{eqnarray}
Note that the estimator $\widehat{\omega}=\widehat{\omega}(\{x_i\})$ is actually a function of the measured variables $\{x_i\}$ which are distributed according to a probability density and therefore it is itself distributed according to a so-called sampling distribution $g(\widehat{\omega},\omega^{\true})$ with expectation value 
\begin{equation}\label{eq:MLestexp}
E[\widehat{\omega}]=\int\widehat{\omega}\;g(\widehat{\omega},\omega^{\true})\;d\widehat{\omega}=\int\widehat{\omega}(\{x_i\})\;f(x_1|\omega^{\true})\ldots f(x_n|\omega^{\true})\;dx_1\ldots dx_n
\end{equation}
 and variance 
\begin{equation}\label{eq:MLvar}
V[\widehat{\omega}]=E[\widehat{\omega}^2]-E[\widehat{\omega}]^2.
\end{equation}

\subsection{Extended likelihood}\label{sec:MLext}
Since the probability density entering the likelihood calculation in \Eq{eq:MLlikeli} is normalized
\begin{equation}
\int dx\; f(x|\omega) = 1
\end{equation}
the information contained in the total number of measurements $n$ is not made use of in the parameter estimation employing \Eq{eq:MLlikeli}. However, the expected number of measurements according to the theoretical hypothesis defining $f(x|\omega)$ might also depend on the parameter $\omega$. This additional information can be included in the analysis by assuming that the total number of
measurements $n$ is distributed according to a Poisson distribution with the parameter-dependent expectation value $\nu(\omega)$. The likelihood from \Eq{eq:MLlikeli} can be extended as the product of the individual likelihoods for each measurement in the set and the Poisson probability to observe a set of $n$ measurements given the parameter-dependent hypothesis (see e.g. \Ref{Barlow:1990vc})
\begin{equation}\label{eq:MLextlikeli}
\Lik_{\text{ext}}(\omega|\{x_i\})= \frac{\nu(\omega)^n}{n!}e^{-\nu(\omega)}\prod\limits_{i=1}^{n}f(x_i|\omega).
\end{equation}

\subsection{The variance of \ML estimators}\label{sec:MLvar}
The extracted estimator $\widehat{\omega}(\{x_i\})$ for the hypothesis probability density $f(x|\omega)$ from a sample of $n$ measurements $\{x_i\}$ is itself a random number distributed according to the sampling distribution. When repeating the entire experiment of $n$ measurements many times each parameter extraction would yield a different value for $\widehat{\omega}$ spread around $E[\widehat{\omega}]$. The variance $V[\widehat{\omega}]$ is a measure for this spread and therefore for the statistical uncertainty of the estimator. In practical applications it is usually not possible to calculate \Eq{eq:MLvar} analytically but information about the variance can be can be obtained using Monte Carlo methods. Instead of repeating the experiment over and over again, pseudo-experiments (e.g. Monte Carlo simulations) can be used to generate many samples $\{x^{(k),\scriptsize\text{sim}}_i\}$ distributed according to $f(x|\widehat{\omega})$ and the distribution of their \ML estimators $\widehat{\omega}(\{x^{(k),\scriptsize\text{sim}}_i\})$ can be studied.
If the Monte Carlo generation of a large number of experiments is too costly the Rao-Cram\'{e}r-Frechet (RCF) inequality can be used to give a lower bound on an estimator's variance
\begin{equation}\label{eq:RCFbound}
V[\widehat{\omega}]\geq\left(E\left[-{\partial^2\log\Lik\over\partial\omega^2}\right]\right)^{-1}.
\end{equation}
If the equality in \Eq{eq:RCFbound} holds the estimator has minimal variance and is called `efficient'. The calculation of the expectation value of the second derivative of the logarithm of the likelihood in \Eq{eq:RCFbound} requires  integrations over $x_i$ (cf. \Eq{eq:MLestexp}) and is therefore impractical in most applications. For a sufficiently large sample, $V[\widehat{\omega}]$ can be approximated by evaluating the second derivative with the measured data for the value of $\widehat{\omega}$
\begin{equation}\label{eq:apprVar}
V[\widehat{\omega}]\approx\Delta\widehat{\omega}^2=\left(-{\partial^2\log\Lik\over\partial\omega^2}\right)^{-1}\Bigg|_{\omega=\widehat{\omega}}.
\end{equation}
An easy application of \Eq{eq:apprVar} is the `graphical technique' to obtain the variance of the estimator by expanding the negative logarithm of the likelihood (`log-likelihood') in a Taylor series around the value of its minimum
\begin{equation}\label{eq:TaylorLikeli}
-\log\Lik(\omega)=\underbrace{-\log\Lik(\widehat{\omega})}_{\displaystyle=-\log\Lik_{\tmin}}-\underbrace{{\partial \log\Lik\over\partial\omega}\bigg|_{\omega=\widehat{\omega}}}_{\displaystyle=0}(\omega-\widehat{\omega})-{1\over 2}\underbrace{{\partial^2 \log\Lik\over\partial\omega^2}\bigg|_{\omega=\widehat{\omega}}}_{\displaystyle=-\Delta\widehat{\omega}^{-2}}(\omega-\widehat{\omega})^2+\ldots\;.
\end{equation}
Ignoring higher-order terms the log-likelihood can be approximated by a parabola
\begin{equation}\label{eq:apprLikeli}
-\log\Lik(\omega)\approx-\log\Lik_{\tmin}+{(\omega-\widehat{\omega})^2\over2\Delta\widehat{\omega}^{2}}.
\end{equation}
The value for $\Delta\widehat{\omega}$ can then be (graphically) deduced from
\begin{equation}\label{eq:apprLikelisig}
-\log\Lik(\widehat{\omega}\pm\Delta\widehat{\omega})\approx-\log\Lik_{\tmin}+{1\over2}.
\end{equation}
In the large sample limit ($n\rightarrow\infty)$ the likelihood $\Lik$ becomes a Gaussian function with absent higher-order terms in \Eq{eq:TaylorLikeli} and the approximation in \Eq{eq:apprLikeli} becomes an equality.

\subsection{Properties of \ML estimators}\label{sec:MLprop}
An estimator is called `consistent' if it converges to the true value $\omega^{\true}$ for $n\rightarrow \infty$ in the sense of probability
\begin{equation}
\lim\limits_{n\rightarrow\infty}\mathcal{P}(|\widehat{\omega}-\omega^{\true}|>\epsilon)=0\quad\forall\epsilon>0.
\end{equation}
The `bias' of the estimator  $b$ is defined as the difference of the expectation value of $\widehat{\omega}$ and the true value $\omega^{\true}$
\begin{equation}\label{eq:bias}
b=E[\widehat{\omega}]-\omega^{\true}.
\end{equation} 
Estimators with (asymptotically for $n\rightarrow\infty)$ vanishing bias $b=0$ are called `(asymptotically) unbiased'.

An important property of unbiased \ML estimators in the large sample limit ($n\rightarrow\infty)$ is that they exhibit `asymptotic normality', meaning they are distributed according to a Gaussian normal distribution with $\mu=E[\widehat{\omega}]$ and $\sigma^2=V[\widehat{\omega}]$ and are always `efficient' (have minimal variance).

\subsection{The \MEM: A \ML method for scattering theory}\label{sec:SmatMEM}
The `\MEM' belongs to the class of \ML methods and is especially formulated for the application in scattering theory (see \Refs{Kondo:1988yd,Kondo:1991dw}). Under the assumption that an experiment can be described by a probabilistic model, each individual measurement can be assigned a certain theoretical probability for its emergence according to the model. For scattering experiments the cross section $\sigma$ is a measure for this probability. More precisely, for a collision $A+B\rightarrow a_1+a_2+...+a_n$ the differential cross section
\begin{equation}\label{eq:diffcrosssection}
d\sigma={d\sigma\over dR_n(p_1,\ldots,p_n)}dR_n
\end{equation}
is a measure for the probability to observe the $n$-body final state in the infinitesimal phase space region 
\begin{equation}\label{eq:phasespace}
dR_n=(2\pi)^4\delta^4\left(P_A+P_B-\sum\limits_{i=1}^{n} p_i\right)\prod\limits_{j=1}^{n}{d^3p_j\over(2\pi)^32E_j}.
\end{equation}
located at $(p_1,\ldots,p_n)$. 
The calculation of the differential cross section in \Eq{eq:diffcrosssection} under the model assumption depends on associated model parameters $\omega$.
The theoretical prediction for the model-dependent probability density describing the distribution of the observed final states can thus be identified as
\begin{equation}\label{eq:xspdfident}
f\left(p_1,\ldots,p_n|\omega\right)= {1\over \sigma(\omega)}{d\sigma(\omega)\over dR_n(p_1,\ldots, p_n)}.
\end{equation}
In general, the final state can be described by variables collected in the event $\vec{y}$.  The differential cross section can be written in terms of the event variables in $\vec{y}$
\begin{equation}
{d\sigma(\omega)\over dR_n(p_1,\ldots,p_n)}\longrightarrow {d\sigma(\omega)\over d\vec{y}}
\end{equation}
with
\begin{equation}
\int d\vec{y}{1\over \sigma(\omega)}{d\sigma(\omega)\over d\vec{y}}=1.
\end{equation}
Finite detector resolution, imperfect detector efficiency and mismatch between theoretically modelled and actually measured quantities are assumed to be factorisable into universal `transfer functions' $W(\vec{x},\vec{y})$.
Those functions model the probability to observe the (theoretical) quantities $\vec{y}$ as the detector-level events $\vec{x}$
\begin{equation}\label{eq:diffxsdect}
{d\sigma(\omega)\over d\vec{x}}=\int d\vec{y}\;{d\sigma(\omega)\over d\vec{y}}W(\vec{x},\vec{y})
\end{equation}
with
\begin{equation}
\int d\vec{x}\;W(\vec{x},\vec{y}) =1.
\end{equation}
In principle, the variables in $\vec{y}$ may be chosen independently from the variables in $\vec{x}$. For example, when particles escape detection (e.g. neutrinos) both vectors do not even need to have the same dimension. Nevertheless, for this study it may prove beneficial to choose the two sets as closely related as possible. Identifying the two by introducing a $\delta$-function as the transfer function
\begin{equation}\label{eq:trsfdelta}
W(\vec{x},\vec{y})=\delta(\vec{x}-\vec{y})
\end{equation}
would correspond to an ideal detector. Note that different (more inclusive) definitions of $\vec{x}$ can always be obtained by variable transformations (and integrations) of \Eq{eq:diffxsdect}.
In practice, the transfer functions (often modelled by Gaussians or Bi-Gaussians) have to be determined by the experiments using Monte Carlo studies and detector simulations. Additionally, the integration over the transfer functions in \Eq{eq:diffxsdect} looks straightforward but it might turn out to be non-trivial due to the peak structure of the transfer functions requiring sophisticated parameterisations of the phase space. These issues are not considered in this work since experimental analyses based on the \MEM are used to this type of problems. The joint probability for having observed a sample of $N$ independent detector-level events $\{\vec{x}_i\}$ is given by the product of the individual probabilities
\begin{equation}\label{eq:joipro}
\mathcal{P}\left(\{\vec{x}_i\}|\omega\right)=\prod\limits_{i=1}^{N}{1\over \sigma(\omega)}\int d\vec{y}\;{d\sigma(\omega)\over d\vec{y}}W(\vec{x}_i,\vec{y}).
\end{equation}
With the identification in \Eq{eq:xspdfident}, the joint probability can be interpreted as a likelihood for a given set of events (cf. \Eq{eq:MLlikeli})
\begin{equation}\label{eq:likeli}
\mathcal{P}\left(\{\vec{x}_i\}|\omega\right)=\Lik\left(\omega|\{\vec{x}_i\}\right).
\end{equation}
Maximising this likelihood for a given set of events $\{\vec{x}_i\}$ with respect to $\omega$ yields a \ML estimator $\widehat{\omega}$ for the most probable value of the model parameter $\omega$ in order to describe the measured set of events $\{\vec{x}_i\}$ by the model (see \Eq{eq:likelimax}). Since all available information from the event is used for the evaluation of the matrix element in the differential cross section in \Eq{eq:joipro}, $\widehat{\omega}$ is an efficient estimator making optimal use of the information content in the data. Provided that the transfer functions and the differential cross sections in \Eq{eq:joipro} are known with ultimate accuracy and precision and the model used to calculate the scattering amplitudes gives an accurate description of nature, \Eq{eq:likelimax} is expected to yield a consistent, unbiased estimator for the value of $\omega$ realised in the data. If these conditions are not satisfied \ML estimators are prone to have a (unknown) bias. This bias is usually accounted for within the experimental analysis by a calibration of the \MEM: Monte Carlo event generators tuned to the specific experiment are used to generate samples of signal events for different input values $\omega^{\scriptsize\text{input}}$ of the model parameter. The \MEM is used to extract estimators $\widehat{\omega}^{\scriptsize\text{extr.}}$ from the generated event samples to obtain a calibration curve ($\widehat{\omega}^{\scriptsize\text{extr.}}$ vs. $\omega^{\scriptsize\text{input}}$) describing the relation between extracted estimator and input value. This relation can in reverse be used to account for the bias in the actual measurement. Clearly, the calibration procedure comes with its own uncertainties and ambiguities. To increase the accuracy of current measurements it is therefore desirable to limit the amount of calibration to a minimum. 

\section{Cross section calculation in perturbation theory}
Let us assume that the model taken as a basis for the theoretical description of the scattering process provides an accurate representation of nature. In order to minimise a possible bias in the \MEM estimator stemming from improper approximation, the differential cross sections entering \Eq{eq:joipro} have to be calculated with sufficient accuracy. For an interacting quantum field theory it is in general not known how to calculate scattering amplitudes 
\begin{equation}\label{eq:Smatr}
S_{fi}=\left<\phi_{p_1}\phi_{p_2}\ldots\phi_{p_n}\right.\big|\left.\phi_{P_A}\phi_{P_B}\right>
\end{equation}
from an initial state $i$ given by $\left|\phi_{P_A}\phi_{P_B}\right>$ to a final state $f$ given by $\left|\phi_{p_1}\phi_{p_2}\ldots\phi_{p_n}\right>$ exactly. Thus the calculation of \Eq{eq:Smatr} has to be approximated. The idea of `perturbation theory' is briefly sketched and it is illustrated how approximate scattering amplitudes can be expressed by `Feynman diagrams'. See e.g. \Refs{Peskin:1995ev,Boehm:1333727} for a detailed text book derivation of the sketched formalism for actual gauge theories of fermion and vector fields. In perturbation theory the existence of asymptotic states, which approximately can be expressed by free particle states because of their spatial separation suppressing their interactions, is assumed. These asymptotic states are used to describe individual particles, prepared in the distant past, approaching each other, interacting and producing particles leaving the place of interaction to be detected in the detectors as individual particles in the distant future. The scattering amplitudes in \Eq{eq:Smatr} can therefore be approximated by the transition amplitudes of free asymptotic states (described by the Lagrangian $\mathcal{L}_{\text{free}}$) by treating the interactions (described by the Lagrangian $\mathcal{L}_{\text{int}}$) as perturbations to the free theory 
\begin{equation}
\mathcal{L}=\mathcal{L}_{\text{free}}\quad+\quad\underbrace{\mathcal{L}_{\text{int}}}_{\mathclap{\text{perturbation}}}.
\end{equation}
The transition corresponding to \Eq{eq:Smatr} is carried out by the causal, Lorentz invariant, unitary time-evolution operator $U(t,t')$ of the asymptotically free states
from the distant past to the distant future
\begin{equation}\label{eq:Smatr2}
S=\lim\limits_{\substack{t\rightarrow\infty\\ t'\rightarrow-\infty}}U(t,t')=T\Bigg\{\exp\bigg[i\int d^4x\underbrace{\mathcal{L}_{\text{int}}}_{\mathclap{\small\displaystyle\ord(\lambda)}}\bigg]\Bigg\},
\end{equation}
where $T$ denotes time-ordering of products of field operators.
The operator $S$ depends exponentially on $\int d^4x\mathcal{L}_{\text{int}}$ which is proportional to the coupling strength (here schematically denoted by $\lambda$). The scattering amplitudes follow from Green's functions  of the field equations (also called $n$-point correlation functions which are vacuum expectation values of time-ordered products of field operators) according to the Lehmann--Symanzik--Zimmermann reduction.  In the calculation of these Green's functions exponentials involving $\mathcal{L}_{\text{int}}$ can be expanded as a series in the coupling strength of the interaction $ \lambda$. For sufficiently small $ \lambda$ this series can be truncated at $\ord(\lambda^m)$ yielding scattering amplitudes accurate apart from $\ord(\lambda^{m+1})$ corrections. The lowest order of the coupling $m$ necessary to realise the scattering $A+B\rightarrow a_1+\ldots+a_n$ is called the `leading order' (LO) or `Born level'. Higher orders are referred to as `next-to-(next-to-\ldots)leading order' (N(N\ldots)LO).

The linear transformation from the asymptotic initial states to the asymptotic final states given in \Eq{eq:Smatr2} is called `S-matrix' (scattering matrix). It can be written as 
\begin{equation}
S_{fi}=\delta_{fi}+i\mathcal{M}_{fi}(2\pi)^4\delta^4\left(P_A+P_B-\sum\limits_{j=1}^{n}p_j\right)
\end{equation}
with the trivial part $\delta_{fi}$($\neq 0$ only for the same initial and final state) and the matrix elements $i\mathcal{M}_{fi}$ describing the interaction. 
The differential cross section is given by the absolute matrix element squared\footnote{Hence the name `\MEM' for a \ML method with the probability density given in terms of the differential cross section.} and the flux-factor $F=\sqrt{(P_A\cdot P_B)^2-m_A^2m_B^2}$ (with $m_A$ and $m_B$ the masses of $A$ and $B$) which has to be integrated over the phase space $dR_n$
\begin{equation}\label{eq:defdiffxs}
d\sigma={1\over 4F}dR_n\times\left|\mathcal{M}\left(A(P_A)+B(P_B)\rightarrow a_1(p_1)+a_2(p_2)+\ldots+a_n(p_n)\right)\right|^2.
\end{equation}

Matrix elements have the useful property to allow `crossing': The matrix element $\mathcal{M}$ for the process $A(P_A)+B(P_B)\rightarrow a_1(p_1)+a_2(p_2)+\ldots+a_n(p_n)$ is related to the matrix elements $\widetilde{\mathcal{M}}$ with interchanged (crossed) final- and initial-state partons by a reversal of momentum ($p_i\rightarrow -p_i$) and helicity ($h_i\rightarrow -h_i$) of the crossed partons
\begin{eqnarray}\label{eq:crossingsym}
\nn \text{e.g.}&&\widetilde{\mathcal{M}}\left(a^{h_1}(p_1)+B^{h_B}(P_B)\rightarrow A^{h_a}_1(P_A)+a^{h_2}_2(p_2)+\ldots+a^{h_n}_n(p_n)\right)\\
&=&\mathcal{M}\left(A^{-h_A}(-P_A)+B^{h_B}(P_B)\rightarrow a^{-h_1}_1(-p_1)+a^{h_2}_2(p_2)+\ldots+a^{h_n}_n(p_n)\right).
\end{eqnarray}

\begin{figure}[htbp]
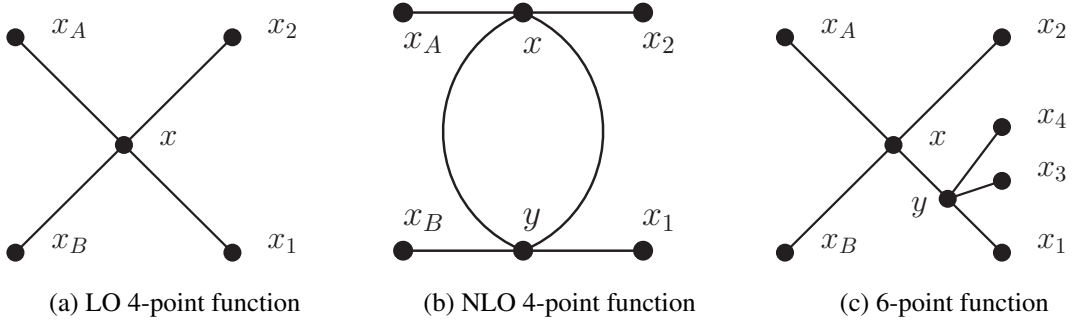

  \begin{center}
    \leavevmode
    \begin{subfigure}[b]{0.32\textwidth}
     \includegraphics[width=1\textwidth]{{{%
          prerequ/phi4lo}}}
     \caption{LO 4-point function}
     \label{fig:sketchlo}
    \end{subfigure}
    \begin{subfigure}[b]{0.32\textwidth}
     \includegraphics[width=1\textwidth]{{{%
          prerequ/phi4vi}}}
     \caption{NLO 4-point function}
     \label{fig:sketchvi}
    \end{subfigure}
    \begin{subfigure}[b]{0.32\textwidth}
     \includegraphics[width=1\textwidth]{{{%
          prerequ/phi4re}}}
     \caption{6-point function}
     \label{fig:sketchre}
    \end{subfigure}
     \caption{Graphical sketch of (a) the connected $4$-point correlation function at leading order $\ord(\lambda)$ and examples of (b) a loop  or (c) $6$-point correlation function with additional external fields at next-to-leading $\ord(\lambda^2)$.}
    \label{fig:sketchlovire}
  \end{center}
\end{figure}
The interaction Lagrangian $\Lik_{\scriptsize\text{int}}$ is not only proportional to the coupling strength but also to the field operators of the respective QFT. Taking higher-order terms of the series expansion in the coupling strength into account adds internal fields to the correlation function which either create additional connections between the $n+2$ external fields at $x_A,x_B,x_1,\ldots,x_n$  (called virtual loops, cf. \Fig{fig:sketchvi}) or are connected to additional external fields (called real radiation, cf. \Fig{fig:sketchre}) as sketched\footnote{All Feynman diagrams shown in this work are produced using the JaxoDraw package (see \Ref{Binosi:2003yf}).} in \Fig{fig:sketchlovire}.
Sketching all possible combinations of fields and their connections like in \Fig{fig:sketchlovire} yields the so-called `Feynman diagrams'. They are an exact representation of the perturbative expansion of $\mathcal{M}$ and each building block can be identified with a mathematical object. The assignments of these objects to the parts of the diagram are called `Feynman rules'. Momentum conservation holds for the external momenta and at every vertex. Undetermined momenta in loops have to be integrated over. Symmetries in the phase space corresponding to certain diagrams have to be accounted for by respective symmetry factors. In principle, knowledge of the Feynman rules for a specific QFT allows to formally calculate the scattering amplitudes by drawing all Feynman diagrams up to the desired order and simply multiplying the respective factors according to the Feynman rules. Feynman rules for gauge theories of fermion and vector fields can be found in the literature (see e.g. \Ref{Peskin:1995ev}).

\subsection{Regularisation and renormalisation}\label{sec:regren}
In practice, including higher-order corrections introduces complications in the amplitude calculation. The integration over the undetermined momenta $p$ of particles running in loops (cf. \Fig{fig:sketchvi}) produces integrals of the form
\begin{equation}\label{eq:loop4d}
I(4,n)=\int{d^4p\over(2\pi)^4}{1\over \left(p^2-\Delta\right)^n}
\end{equation}
which are commonly divergent for $E_p\rightarrow\infty$. These divergences are called \UV divergences. The integral in \Eq{eq:loop4d} might also be divergent for $p^2\rightarrow \Delta$. These divergences are referred to as \IR divergences. For the moment, \IR divergences are ignored in the calculations but will be elaborated on them further in section~\ref{sec:IRdivcanc}. To give a meaning to the amplitude calculation the \UV divergence in the loop integral has to be parameterised as the limit of a so-called `regulator'. The loop integration can then be carried out depending on the regulator. The \UV divergence now corresponds to a divergence in the regulator. 
A Lorentz invariant way of regulating and isolating these divergences is called dimensional regularisation (DREG). The theory (in whole or only certain aspects) is analytically continued to $d$ dimensions
\begin{equation}\label{eq:Lagrangeddim}
\Lik^{d\text{-dim}}=\Lik^{d\text{-dim}}_{\text{free}}+\mur^{\alpha}\Lik^{d\text{-dim}}_{\text{int}}.
\end{equation}
For example, treating internal and external states  as $d$-dimensional objects (their momenta and spin degrees of freedom are continued
from $4$ to $d$ dimensions) corresponds to the so-called `conventional dimensional regularisation' (CDR) scheme (see \Ref{collins1984renormalization}). Considering only the internal states as $d$-dimensional while leaving the external ones $4$-dimensional corresponds to the `'t Hooft-Veltman' (HV) scheme (see \Ref{tHooft:1972tcz}).
The mass scale $\mur$ is introduced to ensure that the coupling strength in $\Lik^{d\text{-dim}}_{\text{int}}$ remains dimensionless also in $d$ dimensions. Depending on the mass dimensions of the fields in $\Lik^{d\text{-dim}}_{\text{int}}$ its exponent $\alpha$ has to be chosen accordingly. 
In $d$ dimensions the respective volume element is given by 
\begin{equation}
{d^dp\over (2\pi)^d}={(2\pi)^{-d}}\;p^{d-1}\;dp\;d\Omega_d.
\end{equation}
The surface of the $d$-dimensional unit sphere is given by
\begin{equation}
\int d\Omega_d = {2\pi^{d/2}\over \Gamma(d/2)}
\end{equation}
where the $\Gamma$-function is defined by
\begin{equation}
\Gamma(x)=\int\limits_0^{\infty}d\tau\;e^{-\tau}\tau^{x-1}.
\end{equation}
Integrals of the form given in \Eq{eq:loop4d} can now be evaluated in $d$ dimensions as
\begin{equation}\label{eq:loopddim}
I(d,n)=\int{d^dp\over(2\pi)^d}{1\over(p^2-\Delta)^n}=i{(-1)^n\over(4\pi)^{d/2}}{\Gamma\left(n-{d/2}\right)\over\Gamma(n)}\Delta^{d/2-n}.
\end{equation}
Setting $d=4-2\varepsilon$ (with $\varepsilon>0$) allows to Laurent expand \Eq{eq:loopddim} around $d=4$.
By using the Laurent expansion of the $\Gamma$-function around $0$
\begin{equation}
\Gamma(x)={1\over x}-\gamma_E+\ord(x)
\end{equation}
(with the Euler-Mascheroni constant $\gamma_E\approx0.5772$), the \UV divergence of the integral in \Eq{eq:loop4d} is expressed as a pole in the dimensional regulator $\varepsilon$.
For example, the result for $n=1$ is
\begin{equation}\label{eq:loopint1}
I(4-2\varepsilon,1)={i\Delta\over 16\pi^2}\mu_R^{-2\varepsilon}\Bigg({1\over\varepsilon}+\underbrace{\ln{4\pi}-\gamma_E+1-\ln{\Delta\over \mur^2}+\ord(\varepsilon)}_{\text{finite for }\varepsilon\rightarrow 0}\Bigg).
\end{equation}
Again, the mass scale $\mur$ is introduced to maintain the right mass dimension in the results.
After isolating the \UV divergences in the amplitude calculations via DREG as poles $1/\varepsilon$ they can be identified with divergent model parameters of the Lagrangian (e.g. masses, couplings or field normalisations in \Eq{eq:Lagrangeddim}) by a procedure called `renormalisation'. The finite parts of the parameters have to be experimentally determined employing a certain renormalisation scheme in order to make predictions within the very same renormalisation scheme. To not lose the predictive power of the model it is mandatory that there is only a fixed number of \UV divergences to be absorbed in a fixed number of model parameters independent of the perturbative order. These models are called renormalisable. A brief sketch how a theory given by the Lagrangian in \Eq{eq:Lagrangeddim} can be renormalised to NLO accuracy by multiplicative renormalisation is given here. The (divergent) parameters of the unrenormalised theory are referred to as `bare quantities' (generic masses, couplings and fields here schematically denoted by $m_{\text{bare}}$, $\lambda_{\text{bare}}$ and $\phi_{\text{bare}}$)
\begin{equation}\label{eq:Lagrangebare}
\Lik^{d\text{-dim}}=\Lik^{d\text{-dim}}_{\text{free}}(m_{\text{bare}},\lambda_{\text{bare}},\phi_{\text{bare}})+\mur^{\alpha}\Lik^{d\text{-dim}}_{\text{int}}(m_{\text{bare}},\lambda_{\text{bare}}, \phi_{\text{bare}}).
\end{equation}
Let us assume the divergences in the bare quantities can be factorised into divergent renormalisation constants $\delta_i$, e.g.
\begin{equation}\label{eq:renconsts}
m^2_{\scriptsize\text{bare}}=Z_mm^2=(1+\delta_m)m^2,\quad
\lambda_{\scriptsize\text{bare}}=Z_{\lambda}\lambda=(1+\delta_{\lambda})\lambda,\quad
\phi_{\scriptsize\text{bare}}=\sqrt{Z_{\phi}}\phi=\sqrt{1+\delta_{\phi}}\phi
\end{equation}
where the quantities without subscript represent the finite (physical) quantities. Inserting \Eq{eq:renconsts} into \Eq{eq:Lagrangebare} (and ignoring terms of $\ord(\delta_i^2)$) allows to separate the bare Lagrangian into a renormalised Lagrangian and a part countering the divergences in the loops (called `counter term' (CT) Lagrangian)
\begin{equation}\label{eq:LagrangerenCT}
\Lik^{d\text{-dim}}=\Lik^{d\text{-dim}}_{\scriptsize\text{ren}}+\Lik^{d\text{-dim}}_{\scriptsize\text{CT}}.
\end{equation}
The renormalised Lagrangian depends only on the (finite) renormalised parameters and yields (divergent) loop amplitudes involving integrals of the form given in  \Eq{eq:loopint1} when calculating virtual corrections. The CT Lagrangian is also a function of the (divergent) renormalisation constants and yields additional Feynman rules with `CT couplings' proportional to the renormalisation constants $\delta_i$ (usually denoted by crosses in the Feynman diagrams).
These Feynman rules generate additional Feynman diagrams that have to be taken into account in the amplitude calculations at the respective perturbative order. In \Fig{fig:renfeynman} this is schematically illustrated for the renormalisation of the so-called self-energy $\Pi$ at $1$-loop order.
\begin{figure}[htbp]
  \begin{center}
     \includegraphics[width=0.87\textwidth]{{{%
          prerequ/renselfen}}}
     \caption{$1$-loop and counter term (CT) Feynman diagrams for  the self-energy.}
    \label{fig:renfeynman}
  \end{center}
\end{figure}
Since they show up in the calculation of the $1$-loop contributions, from the integral in \Eq{eq:loopint1} it can be deduced how to choose the general form of the $\delta_i$ in order to absorb the poles $1/\varepsilon$ and make the $1$-loop Green's functions \UV finite:
\begin{equation}\label{eq:renfacform}
\delta_i\propto-\lambda\left({1\over\varepsilon}+c_i\right).
\end{equation}
The choice for $c_i$ (finite for $\varepsilon\rightarrow0$) is what defines the renormalisation scheme. For $c_i=0$ only the pole is absorbed in $\delta_i$. This is referred to as the `minimal subtraction' (MS) scheme.
It has become common practice to also absorb the summands $c_i=-\gamma_E+\log{4\pi}$ into the counter term thereby
defining the `modified minimal subtraction' ({\MSbar}) scheme. Note that the $\delta_{i}$ are itself of order $\ord(\lambda)$. Thus, the counter term Feynman rules enter only at NLO and beyond. Taking into account all counter terms and the respective Feynman rules makes the scattering amplitudes of the renormalised theory \UV finite up to $1$-loop order. Besides the MS and {\MSbar} scheme other renormalisation conditions can be formulated which offer a more physical interpretation of the renormalised parameters. For example, demanding that the self-energy of a particle with momentum $p$ (cf. \Fig{fig:renfeynman}) and its derivative vanishes for on-shell particles (called the on-shell or pole-mass scheme)
\begin{equation}\label{eq:osren}
\Pi^{\text{ren}}\big|_{p^2=m^2}=0,\quad {d\Pi^{\text{ren}}\over dp^2}\Bigg|_{p^2=m^2}=0
\end{equation}
forces the propagator to have a pole at $p^2=m^2$ with a residue of $1$ allowing the interpretation of $m$ as the `physical mass' of the particle.

\subsubsection{Renormalisation scale (in-)dependence}
The remaining question is how to choose the so-called renormalisation scale $\mur$. Note, that it is introduced to maintain the correct mass dimension of the coupling parameter when performing the calculation of higher-order corrections in $d=4-2\varepsilon$ space-time dimensions. However, after the renormalisation procedure a residual dependence of the fixed-order results on $\mur$ persists even when the limit $\varepsilon\rightarrow0$ is performed. For example, in the massless limit $m=0$ the $1$-loop corrections to the scattering amplitude for the process $P_A+P_B\rightarrow p_1+p_2$ behave like
\begin{equation}
\left.\Gamma^{(4)}\right.^{\text{ren}}_{1\text{-loop}}\propto \lambda^{n+1}\left[\log\left({|t|\over\mur^2}\right)+ \log\left({|u|\over\mur^2}\right)+ \log\left({s\over\mur^2}\right)\right]
\end{equation}
with $s=(P_A+P_B)^2$, $t=(p_1-P_B)^2$ and $u=(p_1+P_A)^2$.
In a perturbative expansion of the correlation functions higher-order corrections $\ord(\lambda^{n+1})$ should be suppressed with respect to the lower orders $\ord(\lambda^{n})$.
To avoid large logarithms in the higher-order corrections the renormalisation scale should be chosen $\mur^2\sim s,|t|,|u|$ (e.g. similar to the typical centre-of-mass energy of the process). Amplitudes for processes with massive particles introduce logarithms $\log(m^2/\mur^2)$ in the higher-order corrections suggesting to choose $\mur^2\sim m^2$. It is clear that the optimal choice of $\mur^2$ is a non-trivial task, especially for processes with multiple intrinsic scales (e.g. different masses, \ldots) or for example differential distributions including extreme kinematic configurations of $s,t,u$. Having chosen a value for the renormalisation scale
$\mur=\mu_0$, the renormalised amplitudes can be used to perform measurements of the coupling and mass within the applied renormalisation scheme (e.g. {MS}) which depend on the chosen renormalisation scale as $\lambda=\lambda^{\scriptsize\text{MS}}(\mu_0)$ or $m=m^{\scriptsize\text{MS}}(\mu_0)$. These measurements can be used to make other theoretical predictions (depending on $\lambda(\mur)$ and $m(\mur)$)  also at different values of the renormalisation scale: Since the introduction of $\mur$ into the calculations is just an artefact of the regularisation procedure, actual physical observables $O_{\text{phys.}}$ must not depend on this scale $\mur$ (at least when all orders of the perturbative expansion are considered)
\begin{equation}
\mur {dO_{\text{phys.}}\over d\mur}=0.
\end{equation}
This demand can be utilised to theoretically predict the evolution of the renormalised parameters as functions of the renormalisation scale, e.g.
\begin{equation}\label{eq:rengroupeq}
{\mur{d \lambda\over d\mur}=\beta(\lambda)}
\end{equation}
with the beta-function $\beta(\lambda)$. Knowledge of the beta-function allows to evolve $\lambda(\mu_0)$ to other values of the renormalisation scale by solving \Eq{eq:rengroupeq}. 

For a fixed-order amplitude calculation the dependence on the renormalisation scale only enters as a residual higher-order effect. Since the scattering amplitude cannot depend on the unphysical scale $\mur$ when all orders of the perturbative expansion are taken into account, the missing higher-order corrections can be estimated by studying the residual dependence on the renormalisation scale. It has become customary to quote the impact of the up- and downwards variation of the scale by a certain factor (e.g. $2$) as a measure for a theoretical uncertainty due to missing higher orders in the perturbative expansion.

\subsection{Quantum Chromodynamics}\label{sec:QCD}
Quarks (up, down, charm, strange, bottom, top) and gluons, the fundamental constituents of hadronic matter, and their interactions are described by the non-Abelian gauge field theory `Quantum Chromodynamics' (QCD) which is locally symmetric under the $\text{SU}(N_c)$ colour group. The degree $N_c$ denotes the number of colours (in QCD: $N_c=3$). A quark of flavour $q=u,d,c,s,b,t$  and mass $m_q$ is a fermion carrying the quantum number colour $a=1,\ldots,N_c$ and is represented by the quark field spinor $\psi_{q,a}$. Quarks are in the fundamental representation of the $\text{SU}(N_c)$ colour group. The massless gluon is a boson carrying the quantum number $C=1,\ldots,N_c^2-1$ and is represented by the vector field $\mathcal{A}_\mu^C$. Gluons are in the adjoint representation of the of the $\text{SU}(N_c)$ colour group. The $N_c^2-1$ generators of the $\text{SU}(N_c)$ colour group are $N_c\times N_c$ matrices $t^C_{ab}$ describing the rotation of the quark's colour in $\text{SU}(N_c)$ colour space by the interaction with gluon. They fulfil the algebra
\begin{equation}\label{eq:colalg}
[t^A,t^B]=if^{ABC}t^C
\end{equation}
with the totally antisymmetric structure constants of the $\text{SU}(N_c)$ group $f^{ABC}$. The explicit values of $f^{ABC}$ are not needed but only the Casimirs (group invariants) $T_F$, $C_F$, $C_A$ show up in the results. They are defined via
\begin{eqnarray}
\nn Tr\left(t^At^B\right)&=&T_F\delta^{AB}\;\text{ with } T_F={1\over2},\\
\nn C_FN_c&=&Tr\left(\sum\limits_At^At^A\right)={N^2_c-1\over2},\\
\sum\limits_{A,B}f^{ABC}f^{ABD}&=&C_A\delta^{CD}\;\text{ with } C_A=N_c.
\end{eqnarray}
The Lagrangian of QCD in $d=4-2\varepsilon$ space-time dimensions (after renormalisation of the fields, masses and coupling) is given by
\begin{eqnarray}\label{eq:QCDLagrangian}
\nn\Lik_{\scriptsize\text{QCD}}&=&\sum\limits_q\bar{\psi}_{q,a}\left(Z_{\psi_q}i\slashed{\partial}\delta_{ab}-Z_{\psi_q}Z
_gZ^{1/2}_{\mathcal{A}}\mur^{\varepsilon}g_s\gamma^{\mu}t^C_{ab}\mathcal{A}^C_{\mu}-Z_{\psi_q}Z_mm_q\delta_{ab}\right)\psi_{q,b}\\
&&-{1\over4}F^A_{\mu\nu}F^{A\;\mu\nu}+{\Lik_{\scriptsize\text{gauge-fix}}}\;\left[+\Lik_{\scriptsize\text{ghost}}\right]
\end{eqnarray}
with the Dirac $\gamma$-matrices $\gamma^{\mu}$. The field strength tensor is given by
\begin{equation}
F^A_{\mu\nu}=Z^{1/2}_{\mathcal{A}}\left(\partial_{\mu}\mathcal{A}^A_{\nu}-\partial_{\nu}\mathcal{A}^A_{\mu}-Z
_gZ^{1/2}_{\mathcal{A}}\mur^{\varepsilon}g_sf^{ABC}\mathcal{A}^B_{\mu}\mathcal{A}^C_{\nu}\right).
\end{equation}
By partial integration the part of $F_{\mu\nu}F^{\mu\nu}$ quadratic in the fields $\mathcal{A}$ can be written as
\begin{eqnarray}
-{1\over4}F^A_{\mu\nu}F^{A\;\mu\nu}={1\over2}Z_{\mathcal{A}}\mathcal{A}^A_{\mu}\left[\partial_{\alpha}\partial^{\alpha}g^{\mu\nu}-\partial^{\mu}\partial^{\nu}\right]\mathcal{A}^A_{\nu}+\ord\left({\mathcal{A}}^3\right)
\end{eqnarray}
where surface terms have been omitted. It turns out that the operator $(\partial_{\alpha}\partial^{\alpha}g^{\mu\nu}-\partial^{\mu}\partial^{\nu})$ is not invertible. In order to define the gluon propagator a gauge fixing term
\begin{equation}\label{eq:gaugefix}
{\Lik_{\scriptsize\text{gauge-fix}}}=-{Z_{\mathcal{A}}\over2\xi}\partial^{\mu}\mathcal{A}^A_{\mu}\partial^{\nu}\mathcal{A}^A_{\nu}={Z_{\mathcal{A}}\over2\xi}\mathcal{A}^A_{\mu}\partial^{\mu}\partial^{\nu}\mathcal{A}^A_{\nu}
\end{equation}
can be added to the Lagrangian yielding the invertible operator $(\partial_{\alpha}\partial^{\alpha}g^{\mu\nu}-(1-1/\xi)\partial^{\mu}\partial^{\nu})$ allowing to define the gluon propagator as its inverse in momentum space. The choice of gauge fixing as in \Eq{eq:gaugefix} is called Lorentz gauge. Note that QCD, being a gauge theory, must be invariant under gauge transformations. Therefore, physical results cannot depend on $\xi$ in the end allowing to freely choose a value for $\xi$ (e.g. `Landau gauge': $\xi=0$, `Feynman gauge': $\xi=1$, \ldots). 

The QCD Feynman rules deduced from the Lagrangian in \Eq{eq:QCDLagrangian} can be found in the literature (see for example \Ref{Peskin:1995ev}).

The contribution of the Faddeev-Popov ghost fields $\Lik_{\text{ghost}}$ which are a mathematical necessity when dealing with the issue of unphysical, longitudinal degrees of freedom of the gluon fields are not discussed. They only couple to gluons and appear in closed loops in the Feynman diagrams. They do not appear in the calculations needed in the context of this work.

\subsubsection{The running coupling of QCD}
\begin{figure}[htbp]
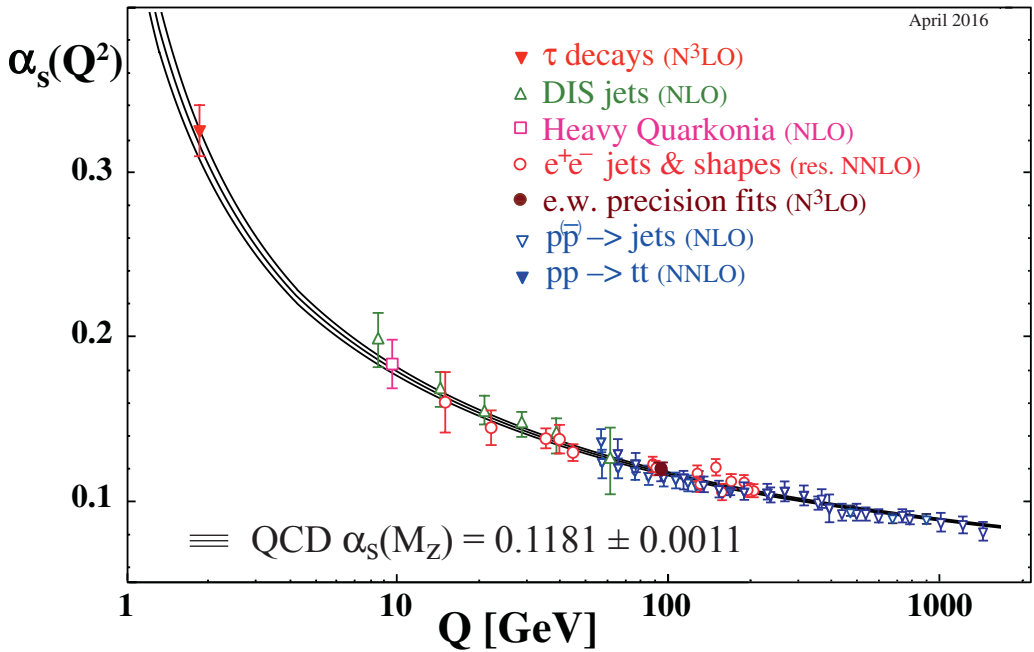

  \begin{center}
     \centering
     \includegraphics[width=0.87\textwidth]{{{%
          prerequ/asq-2015}}}
     \caption{Experimental confirmation of the running coupling of QCD (taken from \Ref{Olive:2016xmw}).}
    \label{fig:runas}
  \end{center}
\end{figure}
Again, by setting $Z_i=1+\delta_i$ the QCD Lagrangian (\Eq{eq:QCDLagrangian}) can be separated in a renormalised Lagrangian and a counter term Lagrangian. Requiring the finiteness of the $1$-loop Green's functions fixes the $\delta_i$ within the respective renormalisation scheme. For example, the $1$-loop coupling counter term in the MS scheme is given by (cf. \Eq{eq:renfacform})
\begin{equation}\label{eq:QCDrenfacg}
\delta_g^{\text{MS}}={\as\over4\pi}{1\over\epsilon}\left[-{11\over6}C_A+{2\over3}n_fT_F\right]=-{1\over\epsilon}{b_0\over2}\as.
\end{equation}
with $\as=g^2_s/(4\pi)$. 
From \Eq{eq:QCDrenfacg} follows
that the $1$-loop evolution of the QCD coupling $\as$ in the MS scheme is given by the solution to the differential equation (cf. \Eq{eq:rengroupeq})
\begin{equation}
\beta(\as)={d\as^{\scriptsize\text{MS}}_{\scriptsize\text{$1$-loop}}\over d\log\mur^2}=-b_0{{\as}^{\scriptsize\text{MS}}_{\mathrlap{\scriptsize\text{$1$-loop}}}}^2\overset{N_c=3}{=}-{1\over12\pi}(33-2n_f){{\as}^{\scriptsize\text{MS}}_{\mathrlap{\scriptsize\text{$1$-loop}}}}^2.
\end{equation}
Note that $b_0>0$ for $N_c=3$ and $n_f<16$ results in the running coupling
\begin{equation}\label{eq:QCDruncoup}
\as^{\scriptsize\text{MS}}_{\scriptsize\text{$1$-loop}}(\mur^2)={12\pi\over(33-2n_f)\log\left(\mur^2/\Lambda^2_{\scriptsize\text{QCD}}\right)}
\end{equation}
which is monotonically decreasing towards $0$ with high scales 
\begin{equation}
\lim\limits_{\mur\rightarrow \infty}\as^{\scriptsize\text{MS}}_{\scriptsize\text{$1$-loop}}(\mur^2)=0.
\end{equation}
This fact, justifying the applicability of perturbation theory to QCD at high energies, is called `asymptotic freedom' and is underpinned experimentally (see \Fig{fig:runas}).
On the other hand, the boundary condition $\Lambda_{\scriptsize\text{QCD}}\approx 200$ GeV corresponds to the scale where the perturbatively deduced running coupling becomes infinity. This means that perturbation theory is not applicable for scales $\mur\approx\Lambda_{\scriptsize\text{QCD}}$. The realm of perturbative QCD is therefore restricted to scales $\mur\gg\Lambda_{\scriptsize\text{QCD}}$. This is a hint at the fact that in the non-perturbative regime QCD forms strongly coupled bound states which have to be described by non-perturbative methods like `Lattice QCD' or even phenomenologically heuristic modelling where the former is not feasible.

\subsection{Infrared divergences: soft and collinear limits}\label{sec:IRdivcanc}
\begin{figure}[htbp]
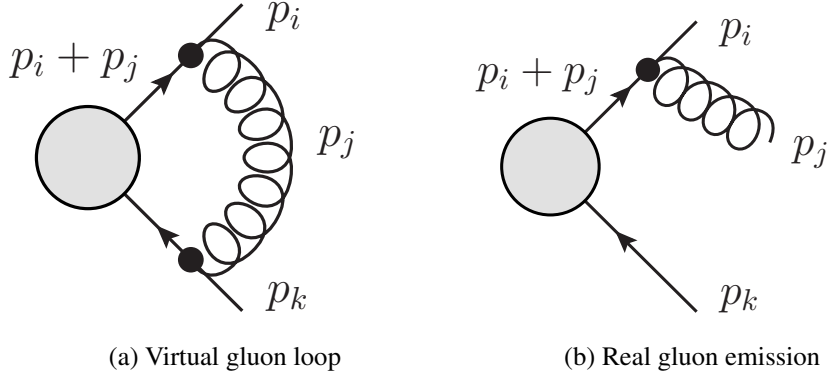

  \begin{center}
    \begin{subfigure}[b]{0.39\textwidth}
     \begin{center}
     \includegraphics[width=1.\textwidth]{{{%
          prerequ/qcdsoftv}}}
     \caption{Virtual gluon loop}
     \label{fig:qcdsoftv}
     \end{center}
    \end{subfigure}
    \begin{subfigure}[b]{0.39\textwidth}
     \begin{center}
     \includegraphics[width=1.\textwidth]{{{%
          prerequ/qcdsoftr}}}
     \caption{Real gluon emission}
     \label{fig:qcdsoftr}
     \end{center}
    \end{subfigure}
     \caption{Exemplary Feynman diagrams for NLO contributions by (a) a virtual gluon loop or (b) real gluon emission.}
    \label{fig:qcdsoft}
  \end{center}
\end{figure}
Both the interference of the virtual amplitude $\mathcal{M}_{n}^{\scriptsize\text{$1$-loop}}$ (e.g. \Fig{fig:qcdsoftv}) with the Born amplitude (without additional gluon) $\mathcal{M}_{n}$ and the squared real amplitude $\mathcal{M}_{n+1}$ (e.g. \Fig{fig:qcdsoftr})
are of the same order in $\as$. They both contribute to the cross section at NLO accuracy
\begin{equation}\label{eq:genericNLOxs}
d\sigma^{\NLO}=d{\sigma^n}\;+\;\delta d\sigma^{\NLO}
\end{equation}
with the cross section in the Born approximation (cf. \Eq{eq:defdiffxs})
\begin{equation}
d{\sigma^n}={1\over 4F}dR_{n}\left|\mathcal{M}_{n}\right|^2
\end{equation}
and the NLO corrections given as the sum of the real $d{\sigma^{n+1}}$ and virtual contributions $d\sigma^{n,\scriptsize\text{$1$-loop}}$
\begin{equation}
\delta d\sigma^{\NLO}=\underbrace{{1\over 4F}dR_{n+1}\left|\mathcal{M}_{n+1}\right|^2}_{\displaystyle=\;d{\sigma^{n+1}}}+\underbrace{{1\over 4F}dR_{n}\;2\Re\left[{\mathcal{M}}^{*}_{n}\mathcal{M}_{n}^{\scriptsize\text{$1$-loop}}\right]}_{\displaystyle=\;d\sigma^{n,\scriptsize\text{$1$-loop}}}.
\end{equation} 
The partonic phase space $dR_n$ is defined in \Eq{eq:phasespace}. Both real and virtual contributions are in general divergent on their own. Brief comments on the nature of these divergences are given and two options how to carry out the mutual cancellation in the sum of both contributions in order to make meaningful predictions from NLO calculations are presented. More details can be found in the literature (see e.g. \Refs{Boehm:1333727,Peskin:1995ev}).

\paragraph{Virtual corrections}
In section~\ref{sec:regren} it has been demonstrated that the \UV divergences in the loop amplitudes can be absorbed into the bare parameters of the Lagrangian by renormalisation. However after renormalisation, when integrating over the momenta of the particles running in loops (e.g. $p_j$ in \Fig{fig:qcdsoftv}), it turns out that the integrand can also have singularities that are not of \UV nature: The integrand of this loop integration is composed of propagators ($\propto 1/(p^2-m^2)$)  which also become singular if the loop momentum $p=p_i+p_j$ goes on shell ($p^2\rightarrow m^2$). If these poles cannot be circumvented by an analytic continuation of the integration path into the complex plain they might give rise to divergences in the amplitude calculation. These divergences stemming from momentum configurations where the loop momentum becomes either soft or collinear to an external particle's momentum are referred to as \IR divergences and cannot be renormalised. Nevertheless, they can be regulated in a similar manner: Employing dimensional regularisation (cf. section~\ref{sec:regren}) allows to analytically calculate the loop integrals in $d=4-2\varepsilon$ (now with $\varepsilon < 0$) dimensions by regulating the \IR divergences as poles in $1/\varepsilon$ and $1/\varepsilon^2$. 

\paragraph{Real corrections}
Integrating over the $n+1$-parton phase space also produces similar \IR divergences in the real contribution:
The propagator for a quark $i$ of mass $m_i$ emitting a massless gluon $j$ appearing for example in \Fig{fig:qcdsoftr} is proportional to
\begin{equation}\label{eq:qprop}
{1\over (p_i+ p_j)^2-m_i^2}={1\over 2p_i\cdot p_j}={1\over2E_iE_j\left(1-\beta_i\cos{\theta_{ij}}\right)}
\end{equation}
with $\theta_{ij}$ the angle between $\vec{p}_i$ and $\vec{p}_j$ and $\beta_i={\sqrt{(E^2_i-m^2_i)/E^2_i}}$ the velocity of the massive quark. These \IR divergences correspond to one massless particle $j$ becoming soft ($E_j\rightarrow 0$) or two massless particles ($ij$) becoming collinear ($p_i\rightarrow zp$ and $p_j\rightarrow (1-z)p$, $z\in\mathbb{R}$). It is obvious that \Eq{eq:qprop} is always singular if the gluon $j$ becomes soft but the collinear divergence only occurs for massless quarks with $\beta_i=1$. Thus, collinear divergences are also referred to as `mass singularities'. Again, these \IR divergences can be regulated by an analytic continuation of the theory to $d=4-2\varepsilon$ (with $\varepsilon < 0$) dimensions where they manifest themselves as poles $1/\varepsilon$ and $1/\varepsilon^2$ in NLO calculations.

 It turns out that when all NLO contributions to `sufficiently inclusive' observables are consistently taken into account the individual \IR divergences from the virtual and the real contributions cancel each other in the sum. This is the statement of the Kinoshita--Lee--Nauenberg (KLN) theorem (see \Refs{Kinoshita:1962ur,Lee:1964is}) ensuring \IR finiteness of physical meaningful observables. In particular, sufficiently inclusive observables $O$ must not be sensitive to the emission of soft or collinear radiation
\begin{eqnarray}\label{eq:IRsafeobs}
O_{n+1}(p_1,\ldots,p_n,p_j=\lambda q)&\xrightarrow{\lambda\rightarrow 0}&O_{n}(p_1,\ldots,p_n),\\
O_{n+1}(p_1,\ldots,p_{n-1},p_i,p_j)&\xrightarrow[p_j\rightarrow (1-z)p_{(ij)}]{p_i\rightarrow zp_{(ij)}}&O_{n}(p_1,\ldots,p_{n-1},p_{(ij)})
\end{eqnarray}
making them `\IR safe'. These \IR safety requirements reflect the fact that arbitrary soft radiation or a pair of collinear particles cannot be individually resolved by the experiments due to finite detector resolutions and thus contributes to a signal signature with a lower multiplicity.
From a practical point of view this cancellation of divergences between different $d$-dimensional integrations with either $n$- or $n+1$-parton kinematics is highly non-trivial. Due to the typical complexity of the individual calculations it is especially preferable to perform as many of the integrations as possible  numerically in an integer number of dimensions. 

 The general behaviour of the real amplitudes in the  soft and collinear limit is illustrated in the following. This knowledge can be exploited to analytically cancel the \IR divergences in the virtual contribution while rendering the real contribution integrable in $d=4$ dimensions at the same time. Two different methods for regulating and cancelling the \IR divergences are presented: `Phase space slicing' (see e.g. \Refs{Giele:1991vf,Giele:1993dj,Brandenburg:1997pu,Harris:2001sx}) and  `dipole subtraction' (see e.g. \Refs{Catani:1996vz,Catani:2002hc}).

\subsubsection{Soft limit}\label{sec:softlimit}
When parameterising the soft parton's momentum as $p_j=\lambda q$  (with an arbitrary massless $4$-momentum $q$ and $\lambda\in\mathbb{R}$), the soft limit corresponds to $\lambda\rightarrow 0$. In this limit the squared $d$-dimensional real matrix element for $n+1$ partons (e.g. \Fig{fig:qcdsoftr} with $n=2$) summed over final-state colours and spins factorises into a divergent prefactor ($\propto 1/\lambda^2$), $d$-dimensional $n$-parton matrix elements (without the soft parton) in colour space $\mathcal{M}^{c_1\ldots c_n}_{n}$, colour operators $T^{c}_{ab}$ mediating colour connection between the $n$-parton matrix elements and the `eikonal factor' $f(p_i,p_k,q,m_i)$ (see e.g. \Ref{Catani:2002hc}) 
\begin{equation}\label{eq:M2soft}
\left|\mathcal{M}_{n+1}\right|^2\xrightarrow{\lambda\rightarrow 0}-{8\pi\mur^{2\varepsilon}\as\over\lambda^2}\sum\limits_{\substack{i\neq j\\{k\neq j,i}}}f(p_i,p_k,q,m_i){\mathcal{M}^{c_1\ldots b_i\ldots b_k\ldots c_n}_{n}}^*T^{c_j}_{b_ic_i}T^{c_j}_{b_kc_k}\mathcal{M}^{c_1\ldots c_i\ldots c_k\ldots c_n}_{n}
\end{equation}
with
\begin{equation}
f(p_i,p_k,q,m_i)={1\over p_i\cdot q}\left[{p_k\cdot p_i\over(p_i+p_k)q}-{m^2_i\over 2p_i\cdot q}\right].
\end{equation}
The colour-charge matrices are given by $T^{c_j}_{c_ic_k}\equiv t^{c_j}_{c_ic_k}$ if $i$ is a final-state quark or an initial-state anti-quark, $T^{c_j}_{c_ic_k}\equiv -t^{c_j}_{c_ic_k}$ if $i$ is a final-state anti-quark or an initial-state quark and $T^{c_j}_{c_ic_k}\equiv if_{c_ic_jc_k}$ if $i$ is a gluon. The eikonal factor $f(p_i,p_k,q,m_i)$ in \Eq{eq:M2soft} also exhibits collinear singularities if the soft momentum $q$ is collinear to $p_i$.
The $d$-dimensional $n+1$ parton phase space in the soft limit factorises into an $n$-parton phase space $dR_{n}$ (without the soft parton $j$) and a phase space factor $dR_{\scriptsize\text{S}}$ for the emission of the soft parton $j$ (see e.g. \Ref{Giele:1991vf})
\begin{equation}\label{eq:phspfacsoft}
\lim\limits_{\lambda\rightarrow 0}dR_{n+1}(p_a,p_b;p_1,\ldots,p_{n-2},p_i,p_k,p_j)\\
=dR_{n}(p_a,p_b;p_1,\ldots,p_{n-2},p_i,p_k)\times dR_{\scriptsize\text{S}}(p_i,p_k,p_j).
\end{equation}
After integrating out the angular orientation of $p_j$ with respect to $p_i$ and $p_k$ the soft phase space factor reads (see e.g. \Ref{Giele:1991vf})
\begin{equation}\label{eq:phspsoft}
dR_{\scriptsize\text{S}}={\pi^{d/2-1}\over 2\Gamma(d/2-1)}s^{1-d/2}_{ik}\;ds_{ij}\;d_{kj}\left[s_{ij}s_{kj}\right]^{d/2-2}
\end{equation}
where $s_{ij}=2p_i\cdot p_j$.

\subsubsection{Collinear limits}\label{sec:collinearlimit}
As can be seen from \Eq{eq:qprop}, the propagator exhibits collinear divergences only for massless quarks. Additional radiation can be collinear either to a final-state parton or to the beam. It turns out that these two cases have to be treated differently in the cancellation of the associated \IR divergences. Thus, the respective limits are studied separately.

\paragraph{Final-state collinear configurations}
By introducing the Sudakov parameterisation of the momenta of two massless final-state partons $i,j$:
\begin{equation}
 p_i=zp_{(ij)}+k_{\perp}-{k^2_{\perp}\over z}{n\over2p_{(ij)}\cdot n},\quad p_j=(1-z)p_{(ij)}-k_{\perp}-{k^2_{\perp}\over(1-z)}{n\over2p_{(ij)}\cdot n}\quad \Rightarrow p_i\cdot p_j={-k^2_{\perp}\over 2z(1-z)}
\end{equation}
the collinear limit corresponds to $k_{\perp}\rightarrow 0$. Here, the collinear direction is denoted by a light-like $4$-vector $p_{(ij)}$ (with $k_{\perp}\cdot p_{(ij)}=0$). The auxiliary light-like vector $n$ specifies how the collinear direction is approached ($k_{\perp}\cdot n=0$).
The variable $z\in[0,1]$ is the momentum fraction contributed by parton $i$ to the momentum of the collinear pair $p_{(ij)}=p_i+p_j$.
The squared $d$-dimensional real matrix element factorises in the collinear limit into a divergent prefactor ($\propto k^{-2}_{\perp}$), $n$-parton matrix elements (with the collinear final-state pair $i,j$ replaced by a single parton $(ij)$ with momentum $p_{(ij)}$) in helicity space $\mathcal{M}^{s_1\ldots s_n}_{n}$ and helicity operators $\hat{P}_{(ij),i}$ introducing spin correlations between the $n$-parton matrix elements (see e.g. \Ref{Catani:1996vz})
\begin{equation}\label{eq:M2colf}
\left|\mathcal{M}_{n+1}\right|^2\xrightarrow{k_{\perp}\rightarrow 0}{4\pi\mur^{2\varepsilon}\as\over2p_i\cdot p_j}{\mathcal{M}^{s_1\ldots s_{(ij)}\ldots s_n}_{n}}^*\hat{P}_{(ij),i}(z,k_{\perp};\varepsilon)\mathcal{M}^{s_1\ldots s'_{(ij)}\ldots s_n}_{n}
\end{equation}
The kernels $\hat{P}_{(ij),i}$ are the $d$-dimensional Altarelli-Parisi splitting functions and they act on the spin indices ($s,s'$ for fermions, $\mu,\nu$ for gluons) as follows (see e.g. \Ref{Catani:1996vz})
\begin{eqnarray}\label{eq:APsplit}
\left<s\big|\hat{P}_{q,q}(z,k_{\perp};\varepsilon)\big|s'\right>&=&\delta_{ss'}C_F\left[{1+z^2\over1-z}-\varepsilon(1-z)\right],\\
\left<s\big|\hat{P}_{q,g}(z,k_{\perp};\varepsilon)\big|s'\right>&=&\delta_{ss'}C_F\left[{1+(1-z)^2\over z}-\varepsilon z\right],\\
\left<\mu\big|\hat{P}_{g,q}(z,k_{\perp};\varepsilon)\big|\nu\right>&=&T_R\left[-g^{\mu\nu}+4z(1-z){k^{\mu}_{\perp}k^{\nu}_{\perp}\over k^{2}_{\perp}}\right],\\
\nn\left<\mu\big|\hat{P}_{g,g}(z,k_{\perp};\varepsilon)\big|\nu\right>&=&2C_A\left[-g^{\mu\nu}\left({z\over 1-z}+{1-z\over z}\right)-2(1-\varepsilon)z(1-z){k^{\mu}_{\perp}k^{\nu}_{\perp}\over k^{2}_{\perp}}\right].\\
\end{eqnarray}
The splitting functions also exhibit soft divergences (encoded in $z\rightarrow 0,1$). Soft and collinear singularities are overlapping in phase space and cannot be separated. Care has to be taken to not overcount them when studying the respective limits.

In the final-state collinear limit the $d$-dimensional $n+1$-parton phase space factorises into an $n$-parton phase space $dR_{n}$ (with the collinear pair $i,j$ replaced by a single parton $(ij)$) and a phase space factor $dR_{\scriptsize\text{Cf}}$ for the emission of the collinear pair $i,j$ from the final state (see e.g. \Ref{Giele:1991vf})
\begin{equation}\label{eq:phspfaccolf}
\lim\limits_{\substack{p_i\rightarrow zp_{(ij)} \\ p_j\rightarrow (1-z)p_{(ij)}}}dR_{n+1}(p_a,p_b;p_1,\ldots,p_{n-1},p_i,p_j)\\
=dR_{n}(p_a,p_b;p_1,\ldots,p_{n-1},p_{(ij)})\times dR_{\scriptsize\text{Cf}}(p_i,p_j).
\end{equation}
After integrating out the azimuthal angle between the plane containing $p_i$ and $p_j$ relative to $p_{(ij)}$ the collinear phase space factor reads (see e.g. \Ref{Giele:1991vf})
\begin{equation}\label{eq:phspcolf}
dR_{\scriptsize\text{Cf}}={\pi^{d/2-1}\over 2\Gamma(d/2-1)}\;ds_{ij}\;d{z}\left[s_{ij}\;z(1-z)\right]^{d/2-2}.
\end{equation}

\paragraph{Initial-state collinear configurations}
If the parton $j$ becomes collinear to an initial-state parton $a$ the following parameterisation can be used to define the collinear limit  $k_{\perp}\rightarrow 0$
\begin{equation}\label{eq:Sudakovinitial}
p_j=(1-x)p_a+k_{\perp}-{k^2_{\perp}\over (1-x)}{n\over2p_a\cdot n},\quad \Rightarrow\; p_j\cdot p_a={-k^2_{\perp}\over 2(1-x)}.
\end{equation}
The squared $d$-dimensional real matrix element factorises in the collinear limit into a divergent prefactor ($\propto k^{-2}_{\perp}$), $n$-parton matrix elements in helicity space $\mathcal{M}^{s_1\ldots s_n}_{n}$ and Altarelli-Parisi splitting functions $\hat{P}_{(aj),i}$ introducing spin correlations between the $n$-parton matrix elements (see e.g. \Ref{Catani:2002hc})
\begin{equation}\label{eq:M2coli}
\left|\mathcal{M}_{n+1}\right|^2\xrightarrow{k_{\perp}\rightarrow 0}{1\over x}{4\pi\mur^{2\varepsilon}\as\over2p_a\cdot p_j}{\mathcal{M}^{s_1\ldots s_{(aj)}\ldots s_n}_{n}}^*\hat{P}_{(aj),a}(x,k_{\perp};\varepsilon)\mathcal{M}^{s_1\ldots s'_{(aj)}\ldots s_n}_{n}.
\end{equation}
In \Eq{eq:M2coli} the $n$-parton matrix elements are obtained by omitting the final-state parton $j$ and replacing the initial-state parton $a$ by an initial-state parton $(aj)$ with momentum fraction $xp_a$ with $x\in[0,1]$.

In the collinear limit the $d$-dimensional $n+1$ parton phase space factorises into an $n$ parton phase space $dR_{n}$ (without the final-state parton $j$ and the initial-state parton $(aj)$) and a phase space factor $dR_{\scriptsize\text{Ci}}$ for the emission of the parton $j$ collinear to the initial-state parton $a$ (see e.g. \Ref{Giele:1993dj})
\begin{equation}\label{eq:phspfaccoli}
\lim\limits_{\substack{p_a \rightarrow xp_a\\ p_j \rightarrow (1-x)p_{a}}}dR_{n+1}(p_a,p_b;p_1,\ldots,p_{n},p_j)\\
=dR_{n}(xp_a,p_b;p_1,\ldots,p_{n})\times dR_{\scriptsize\text{Ci}}(p_a,p_j).
\end{equation}
After integrating over the $(d-3)$ azimuthal angles relative to the
direction of $p_a$ the collinear phase space factor reads (see e.g. \Ref{Giele:1993dj})
\begin{equation}\label{eq:phspcoli}
dR_{\scriptsize\text{Ci}}={\pi^{d/2-1}\over 2\Gamma(d/2-1)}\;d|s_{aj}|\;dx\;x\left[|s_{aj}|\;(1-x)\right]^{d/2-2}.
\end{equation}
Note that both the Born matrix elements in \Eq{eq:M2coli} and the Born phase space in \Eq{eq:phspfaccoli} correspond to a scattering of two partons $a$ and $b$ with a reduced centre-of-mass energy squared of $xs=2xp_a\cdot p_b$. These initial-state collinear divergences can therefore not be cancelled by the virtual contribution $d\sigma^{n,\scriptsize\text{$1$-loop}}$ for the respective scattering with the centre-of-mass energy squared of $s$ (except at the endpoints $x=1$).

\subsection{Parton model and collinear factorisation}\label{sec:factorisation}

\begin{figure}[htbp]
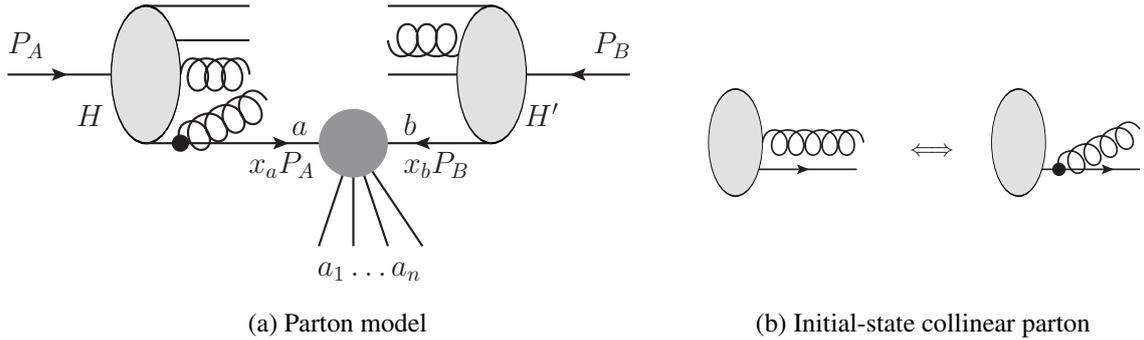

  \begin{center}
    \leavevmode
    \begin{subfigure}[b]{0.59\textwidth}
     \includegraphics[width=1\textwidth]{{{%
          prerequ/hardpart}}}
     \caption{Parton model}
     \label{fig:partmod}
    \end{subfigure}
    \begin{subfigure}[b]{0.39\textwidth}
     \raisebox{.7\height}{\includegraphics[width=1\textwidth]{{{%
          prerequ/collhard}}}}
     \caption{Initial-state collinear parton}
     \label{fig:inicol}
    \end{subfigure}
     \caption{Sketch of the parton model (a) and ambiguous assignment of collinear partons in the initial state (b).}
    \label{fig:partonmodel}
  \end{center}
\end{figure}

Though coloured partons are the fundamental fields of QCD they are not the asymptotic particle states in the experimental spectrum which consists of colour singlet bound states of partons (hadrons) instead. This is because the QCD coupling is only small at large energy scales (short distances) but increases towards smaller energy scales (long distances) (see \Fig{fig:runas}). The phenomenon that coloured partons cannot be isolated but form hadrons as asymptotic particle states is called `colour confinement' and is subject to current research and scientific discussion (see e.g. ref. \cite{Greensite:2011zz}). A consequence of confinement is that perturbative QCD is not applicable at large distances and non-perturbative methods must be employed to connect the short-distance partonic interactions to the long-distance bound states that are preparable and detectable by experiments. As a first approximation, the prepared initial states (e.g. protons in a hadron collider) can be described within the naive parton model (see \Ref{Feynman:1969wa}):
Neglecting the mass of the hadron, a fast moving hadron $H$ can be thought of as a bundle of partons flying longitudinal to the hadron's direction each carrying a fraction of the hadronic momentum (see \Fig{fig:partmod}). This allows to relate the collision of two incoming hadrons $H,H'$ with momenta $P_{a/b}={\sqrt{s}/2}\;(1,0,0,\pm1)$ and the center-of-mass energy squared $s$
\begin{equation}
H(P_a)+H'(P_b)\rightarrow a_1(p_1)+\ldots+a_n(p_n)
\end{equation}
to an incoherent sum of partonic reactions 
\begin{equation}\label{eq:hardpartrea}
a(p_a)+b(p_b)\rightarrow a_1(p_1)+\ldots+a_n(p_n)
\end{equation}
with the momentum fractions 
\begin{equation}\label{eq:partonmom}
p_a=x_aP_a,\quad p_b=x_bP_b.
\end{equation}
Provided large momentum transfers, the latter are calculable within perturbative QCD.
The partonic cross sections $d\sigma_{ab}$ describing the reactions in \Eq{eq:hardpartrea} have to be convoluted with non-perturbative parton distribution functions (PDFs)
\begin{equation}\label{eq:hadpart}
d\sigma_{HH'}(P_a,P_b)=\sum\limits_{a,b}dx_a\;dx_b\;{f^{H}_{a}(x_a)\;f^{H'}_{b}(x_b)}d\sigma_{ab}(x_aP_a,x_bP_b)
\end{equation}
to obtain the hadronic cross section $d\sigma_{HH'}$.
Qualitatively, the PDF $f^{H}_{a}(x_a)$ describes the probability density for finding a parton $a$ in the hadron $H$ with a certain longitudinal momentum fraction in $[x_a,x_a+dx_a]$. When $H$ is a proton the superscript $H$ is usually suppressed.
Note that when all $n$ final-state $4$-momenta of the reaction in \Eq{eq:hardpartrea} are determined the values of the momentum fractions $x_a,x_b$ can be inferred from momentum conservation assuming leading-order kinematics to be
\begin{equation}
x_a=\frac{1}{\sqrt{s}}\sum\limits_{i=1}^{n}\left(E_i+p^z_i\right), \quad x_b=\frac{1}{\sqrt{s}}\sum\limits_{i=1}^{n}\left(E_i-p^z_i\right).
\end{equation}
When including higher-order corrections, additional parton emission from the initial state is allowed. Partons that are radiated strictly collinear to initial-state partons (cf. $k_\perp\rightarrow 0$ in \Eq{eq:Sudakovinitial}) are indistinguishable from partons that are part of the incoming hadron (see \Fig{fig:inicol}) and can also be assigned to the hadronic initial state. Consequently, they have to be accounted for in the PDF definition. This allows to absorb the uncancelled initial-state collinear singularities appearing in the real amplitude calculation (cf. \Eq{eq:M2coli}) into the PDF definition by a procedure called `collinear factorisation': The (singular) PDFs ${f^{H}_{a}}(x_a)$ are interpreted as the sum of finite `renormalised' PDFs ${f^{H}_{a}}(x_a,\muf)$ and the collinear initial-state counter-term PDFs $f^{H,\scriptsize\text{CT}}_{a}(x_a,\muf)$
\begin{equation}\label{eq:pdfren}
{f^{H}_{a}}(x_a)={f^{H}_{a}}(x_a,\muf)+f^{H,\scriptsize\text{CT}}_{a}(x_a,\muf)
\end{equation}
After plugging \Eq{eq:pdfren} into \Eq{eq:hadpart}, the hadronic cross section accurate to NLO reads
\begin{equation}\label{eq:hadpartren}
d\sigma^{\NLO}_{HH'}(P_a,P_b)=\sum\limits_{a,b}dx_a\;dx_b\;{f^{H}_{a}(x_a,\muf)\;f^{H'}_{b}(x_b,\muf)}\;d\sigma^{\NLO}_{ab}(x_aP_a,x_bP_b)\\\;+\;d\sigma^{\scriptsize\text{CT}}_{HH'}(P_A,P_B,\muf).
\end{equation}
Neglecting higher orders in $\as$, the hadronic initial-state collinear counter term at NLO is obtained by a convolution of the (counter-term) PDFs and the partonic Born level cross section
\begin{equation}\label{eq:hadcollCT}
d\sigma^{\scriptsize\text{CT}}_{HH'}(P_A,P_B,\muf)=\left(f^{H,\scriptsize\text{CT}}_{a,\MSbar}(x_a,\muf)f^{H'}_{b}(x_b,\muf)+f^{H}_{a}(x_a,\muf)f^{H',\scriptsize\text{CT}}_{b,\MSbar}(x_b,\muf)\right)\\
\; d\sigma^n_{ab}(x_aP_A,x_bP_B)\;dx_adx_b.
\end{equation}
In order to cancel the initial-state collinear divergences in the real amplitude, the form of the initial-state collinear counter terms can be deduced from the universal factorisations of the $d$-dimensional real matrix element squared (see \Eq{eq:M2coli}) and the phase space (see \Eq{eq:phspfaccoli}) in the initial-state collinear limits.
In a certain factorisation scheme, e.g. the $\MSbar$ scheme, they are consequently given by (see e.g. \Ref{Boehm:1333727}) 
\begin{equation}\label{eq:pdfCT}
 f^{H,\scriptsize\text{CT}}_{a,\MSbar}(x_a,\muf)=-{\as\over 2\pi}\left({1\over\varepsilon}+\gamma_E+\log\left({\muf^2\over4\pi\mur^2}\right)\right)\sum\limits_{(aj)}\int\limits_{x_a}^{1}{dx\over x}\tilde{P}_{(aj),a}\left({x_a\over x}\right){f^{H}_{(aj)}}(x,\muf)+\;\ord(\as^2).
\end{equation}
The sum over $(aj)$ runs over all types of partons that are considered as part of the hadron.

The functions $\tilde{P}_{(aj),a}$ are the $4$-dimensional, regularised (finite for $x\rightarrow 1$), spin-averaged Altarelli-Parisi Splitting probabilities.
On $1$-loop level they are scheme independent and are given by (see e.g. \Ref{Peskin:1995ev})
\begin{subequations}
\begin{eqnarray}
\tilde{P}_{q,q'}(x)&=&\delta_{qq'}C_F\left({1+x^2\over(1-x)_+}+{3\over 2}\delta(1-x)\right),\\
\tilde{P}_{q,g}(x)&=&C_F\left({1+(1-x)^2\over x}\right),\\
\tilde{P}_{g,q}(x)&=&T_R\left(x^2+(1-x)^2\right),\\
\nn\tilde{P}_{g,g}(x)&=&2C_A\left({x\over(1-x)_+}+(1-x)\left(x+{1\over x}\right)\right)+{11C_A-4n_fT_F\over 6}\delta(1-x).\\
\end{eqnarray}
\end{subequations}
The `$+$-distribution' is defined by its action on a generic test function $g(x)$
\begin{equation}
\int_{x_a}^{1}dx{g(x)\over (1-x)_+}=\int_{x_a}^{1}dx{g(x)-g(1)\over (1-x)}+g(1)\log(1-x_a).
\end{equation}
The `factorisation scale' $\muf$ controls the assignment of collinear partons to the hadron or to the hard process. 
The scale $\muf$ can a priori be chosen freely but \Eq{eq:pdfCT} suggests not to choose it very different to the renormalisation scale $\mur$ (typical energy scale of the process) to avoid the appearance of large logarithms spoiling the perturbative expansion.

The sum over all gluon and quark PDFs (denoted by $(aj)$) in \Eq{eq:pdfCT} suggests that there is a non-vanishing probability to find a parton of type $(aj)$ in a parton of type $a$ at NLO accuracy.
This is called the` QCD-improved parton model'. It allows to make finite predictions for \IR safe observables (see \Eq{eq:IRsafeobs}) that fulfil the property of factorisability of initial-state collinear singularities 
\begin{equation}\label{eq:collfactable}
O_{n+1}(p_a,p_b;p_1,\ldots,p_{n},p_j)\xrightarrow{p_j\rightarrow (1-x)p_a}O_{n}(xp_a,p_b;p_1,\ldots,p_{n}).
\end{equation}
Requiring that the hadronic cross section (see \Eq{eq:hadpartren}) is independent of the unphysical scale $\muf$ (at least when all perturbative orders are considered)
\begin{equation}\label{eq:hadxsfacdep}
\muf^2{d\over d\muf^2}\sigma_{HH'}=0
\end{equation}
 yields the Dokshitzer-Gribov-Lipatov-Altarelli-Parisi (DGLAP) evolution of the PDFs at NLO accuracy (see \Refs{Altarelli:1977zs,Dokshitzer:1977sg,Gribov:1972ri})
\begin{equation}\label{eq:DGLAP}
\muf^2{d\over d\muf^2}f^{H}_{a}(x_a,\muf)=\sum\limits_{(aj)}{\as\over 2\pi}\int\limits_{x}^{1}{dy\over y}\tilde{P}_{(aj),a}(y)f^{H}_{(aj)}\left({x\over y},\muf\right)+\ord(\as^2)
\end{equation}
analytically predicting the $\muf$-dependence of the PDFs. Because any factorisation scale dependence in \Eq{eq:hadxsfacdep} can only be a higher-order effect, missing higher-order corrections can be estimated by studying the residual dependence on the factorisation scale. Analogously to the variation of the renormalisation scale, it has become customary to quote the impact of the up- and downwards variation of $\muf$ by a certain factor (e.g. $2$) as a systematic uncertainty contributing to the theoretical uncertainty due to missing higher orders in the perturbative expansion.

Because they describe the non-perturbative distributions of partons in the hadron, PDFs cannot be calculated perturbatively. Instead, they are modelled by using parameterisations of the form
\begin{equation}\label{eq:PDFpara}
x_af^{H}_{a}(x_a,\mu_0)=Ax_a^\alpha(1-x_a)^\beta(1+\gamma x_a+\eta x_a^2+\kappa x_a^3) 
\end{equation}
in theoretical predictions which are fitted to experimental data. The initial scale $\mu_0$ is set to values around $2-4\GeV$ and the evolution to different values of the scale $\mu_0\rightarrow\muf$ is carried out by (numerically) solving the DGLAP equations. Depending on which parton flavours are assumed to be part of the proton different schemes are defined: For example, assuming the parton content of the proton does only include up-, down-, strange- and charm-quark contributions constitutes the `$4$-flavour scheme'. While also including the bottom-quark PDF is referred to as the `$5$-flavour scheme'. Note that due to its high mass the top quark is usually not considered to be part of the proton for energies achievable in current experiments. Diverse proton `PDF sets' which are tables of the fit parameters (e.g. $\alpha,\beta,\gamma,\eta,\kappa$ for \Eq{eq:PDFpara}) for each quark flavour $a$ together with the respective evolution (according to \Eq{eq:DGLAP}) are provided by various groups and are conveniently available via the LHAPDF library (see \Ref{Whalley:2005nh}).

\subsection{Phase space slicing}\label{sec:PSS}
Regulating the loop integrals in the virtual corrections by dimensional regularisation (with $d=4-2\varepsilon$) allows to extract the \IR divergences as poles in $\varepsilon$. Consequently, the integration of the real matrix element over the real phase space has to be performed in $d=4-2\varepsilon$ dimensions as well in order to ensure a consistent mutual cancellation of the respective \IR divergences. However, due to the common complexity of the matrix elements, carrying out the phase space integration analytically in $d$ dimensions is in general not feasible, especially with generic phase space cuts. Instead, one has to rely on numerical phase space integration techniques wherever possible which however require an integer number of integration dimensions. Within the \PSS method (see \Refs{Giele:1991vf,Giele:1993dj}) the $d$-dimensional integration is split into a singular part necessitating dimensional regularisation and the finite complement which can be safely integrated in $d=4$. In the singular part of the integration the integrand is approximated according to its singular limits. This allows to perform the integration of the simplified integrand over the singular regions analytically in $d$ dimensions yielding factorised poles for $d\rightarrow 4$. For illustration, consider the integration of $x^{-1-\varepsilon}f(x)$ (with $f(x)$ an arbitrary test function regular at $x=0$ and the centre-of-mass energy of the process $s$)
\begin{eqnarray}\label{eq:PSSform}
\nn \overbrace{\int\limits_{0}^{1}dx\;x^{-1-\varepsilon}f(x)}^{\displaystyle\small{\text{divergent for }\varepsilon\rightarrow 0}}&=&\overbrace{\int\limits^{s_{\tmin}/s}_{0}dx\;\underbrace{x^{-1-\varepsilon}f(x)}_{\small{\mathclap{\displaystyle\overset{x\rightarrow0}{\approx} x^{-1-\varepsilon}f(0)}}}}^{\displaystyle\small{\text{divergent for }\varepsilon\rightarrow 0}}\quad+\quad\overbrace{\int\limits_{s_{\tmin}/s}^{1}dx\;x^{-1-\varepsilon}f(x)}^{\displaystyle\small{\text{finite for }\varepsilon\rightarrow 0}}\\
\nn&=&\underbrace{\left(-{1\over\varepsilon}+\log\left({s_{\tmin}\over s}\right)\right)}_{\displaystyle\small{\text{universal factor}}}f(0)\quad+\quad\underbrace{\int\limits_{s_{\tmin}/s}^{1}dx\;{f(x)\over x}}_{\displaystyle\small{\text{$\varepsilon=0$}}}\quad+\quad\ord\left(\varepsilon,{s_{\tmin}\over s}\right).\\
\end{eqnarray}
All remaining integrations can be numerically performed in $d=4$. The singular and finite parts of the integration are separated by introducing a \PSS parameter $s_{\tmin}$. The approximation of the integrand in the singular part introduces a systematic error in the result which scales with $s_{\tmin}$. The divergence of the original integral has been analytically factorised as a pole $\propto 1/\varepsilon$ accompanied by a logarithm of the slicing parameter $\log\left(s_{\tmin}/s\right)$. This factorisation is universal (independent of $f$). The result is accurate up to corrections proportional to $s_{\tmin}$ and $\varepsilon$.

As a start, the focus is on processes without coloured partons in the initial state (e.g. leptonic collisions) and the generalisation to hadronic collisions is given later. The finite hard regions of the $n+1$ parton phase space ($H$) can be defined by requiring that all kinematic invariants $s_{ij}=2p_i\cdot p_j$ are bigger than $s_{\tmin}$ with the singular soft and collinear regions ($S/C$) forming the complement thereof
\begin{equation}\label{eq:phspsep}
dR_{n+1}=dR_{n+1}\bigg[\underbrace{\Theta(s_{\tmin}-\min\limits_{ij}s_{ij})}_{\displaystyle=:\Theta^{\scriptsize\text{S/C}}}+\underbrace{\Theta(\min\limits_{ij}s_{ij}-s_{\tmin})}_{\displaystyle=:\Theta^{\scriptsize\text{H}}}\bigg].
\end{equation}
Note that the exact way of slicing the real phase space is not unique and to avoid overlapping of soft and collinear regions more elaborate prescriptions depending on the respective kinematics (especially masses) may be in order (see \Refs{Brandenburg:1997pu,Cao:2004ky}). One can even introduce different parameters for soft and collinear separation (see \Ref{Harris:2001sx}). As long as the whole real phase space is covered by the chosen slicing and it is consistently carried out in all following steps this does not affect the presented arguments.
The integration over the hard region yielding the \IR finite hard real contribution to the (\IR safe) observable $O$ 
\begin{equation}\label{eq:hard}
{d\sigma}_{n+1}\big|_{\scriptsize\text{H}}O_{n+1}=d{R}_{n+1}\Theta^{\scriptsize\text{H}}\left|\mathcal{M}_{n+1}\right|^2O_{n+1}\bigg|_{d=4}
\end{equation}
can be carried out in $d=4$ dimensions. The numerical integration of \Eq{eq:hard} builds up logarithms $\log\left(s_{\tmin}/s\right)$, $\log^2\left(s_{\tmin}/s\right)$ at the lower phase space boundary $s_{ij}>s_{\tmin}$ due to the pole structure of the real matrix elements (cf. \Eq{eq:M2soft}, \Eq{eq:M2colf} and \Eq{eq:M2coli}) like
\begin{equation}\label{eq:PSSlogbuild}
\int\limits_{s_{\tmin}/s}^{1}dx\;{1\over x}f(x)\propto \log\left(s_{\tmin}/s\right),\qquad \int\limits_{s_{\tmin}/s}^{1}dx\int\limits_{s_{\tmin}/s}^{1}dy\;{1\over xy}g(x,y)\propto \log^2\left(s_{\tmin}/s\right)
\end{equation}
with $f$, $g$ regular for $x=0$, $y=0$.
The $d$-dimensional soft and collinear limits (cf. section~\ref{sec:IRdivcanc}) of the real cross section can be used to approximate the integrand
\begin{equation}\label{eq:scapprox}
d{\sigma^{n+1}}\big|_{\scriptsize\text{S/C}}O_{n+1}=d{R}_{n+1}\Theta^{\scriptsize\text{S/C}}\left|\mathcal{M}_{n+1}\right|^2O_{n+1}=dR_n\times dR_{\scriptsize\text{S/C}}\Theta^{\scriptsize\text{S/C}}\left.\left|\mathcal{M}_{n+1}\right|^2\right|_{\scriptsize\text{S/C}} O_{n+1}\;+\;\ord\left({s_{\tmin}\over s}\right)
\end{equation}
 in the soft and collinear regions which is accurate up to terms proportional to $s_{\tmin}$. The soft or collinear limits of the amplitude squared  $\left.\left|\mathcal{M}_{n+1}\right|^2\right|_{\scriptsize\text{S/C}}$ correspond to the limits in \Eq{eq:M2soft}, \Eq{eq:M2colf} or \Eq{eq:M2coli} and the respective limits of the real phase space are given by \Eq{eq:phspsoft} and \Eq{eq:phspcolf} or \Eq{eq:phspcoli}. The integration of the approximated matrix element squared over the soft or collinear parton momenta can be analytically carried out in $d=4-2\varepsilon$ dimensions regulating the soft and collinear divergences as poles $1/\varepsilon^2$ and $1/\varepsilon$ together with logarithms of the phase space  slicing parameter $s_{\tmin}$ (cf. \Eq{eq:PSSform}).
Note that infrared safety of the observable implies
\begin{equation}
\lim\limits_{s_{\tmin}\rightarrow 0}\Theta^{\scriptsize\text{S/C}}O_{n+1}=O_n.
\end{equation}
In accordance with the KLN theorem, the \IR poles from the soft/collinear part of the real contribution analytically cancel the respective poles in the virtual contribution. The sum of $d{\sigma^{n+1}}\big|_{\scriptsize\text{S/C}}O_{n+1}$ and $d{\sigma^{n}}^{\scriptsize\text{$1$-loop}}O_{n}$ is therefore \IR finite but logarithmically $s_{\tmin}$-dependent (cf. \Eq{eq:PSSform})
\begin{eqnarray}\label{eq:PSSmaster}
\nn\int \delta d\sigma^{\NLO}O&=&\int\limits_{n}\overbrace{d{\sigma^{n}}^{\scriptsize\text{$1$-loop}}O_{n}}^{\mathclap{\displaystyle\small\text{divergent for }d\rightarrow 4}}\quad+\quad\overbrace{\int\limits_{n+1}{d\sigma}_{n+1}O_{n+1}}^{\displaystyle\small{\text{divergent for }d\rightarrow 4}}\\
&=&\lim\limits_{s_{\tmin}\rightarrow 0}\bigg[\underbrace{\int\limits_{n}\overbrace{\big(d{\sigma^{n}}^{\scriptsize\text{$1$-loop}}\quad+d{\sigma^{n+1}}\big|_{\scriptsize\text{S/C}}\big)O_{n}}^{\displaystyle\small{\text{finite for } d\rightarrow 4}}+\int\limits_{n+1}\overbrace{{d\sigma}_{n+1}\big|_{\scriptsize\text{H}}O_{n+1}}^{\displaystyle\small{\text{finite for } d\rightarrow 4}}}_{\displaystyle\small{\text{finite for }s_{\tmin}\rightarrow 0}}\bigg].
\end{eqnarray}
Because these logarithms stem from the the phase space separation (cf. \Eq{eq:PSSlogbuild}) they are cancelled in the sum of the soft and collinear and the hard part for \IR safe observables $O$.

The application of \Eq{eq:PSSmaster} can be generalised to hadronic collisions by crossing coloured partons to the initial state (see \Eq{eq:crossingsym}) and replacing the PDFs by `effective NLO structure functions' $F^H_a(x_a,\muf)$ (see e.g. \Ref{Giele:1993dj}). The hadronic cross section at NLO accuracy therefore reads
\begin{equation}\label{eq:HHcross}
d\sigma^{\NLO}_{HH'}=\sum\limits_{a,b}dx_a\;dx_b\;F^{H}_{a}(x_a,\muf)\;F^{H'}_{b}(x_b,\muf)\left(d\sigma_{ab}(x_a,x_b)+\delta d\sigma^{\NLO}_{ab}(x_a,x_b)\right).
\end{equation}
The effective NLO structure functions are given by the renormalised PDFs and the universal `crossing functions'  $C^H_a(x_a,\muf)$
\begin{equation}\label{eq:effstruct}
F^H_a(x_a,\muf)= f^H_a(x_a,\muf)+\as C^H_a(x_a,\muf).
\end{equation}
Inserting \Eq{eq:effstruct} in \Eq{eq:HHcross} and neglecting terms with higher orders of $\as$ then $\delta d\sigma^{\NLO}_{ab}$ yields
\begin{eqnarray}
\nn d\sigma^{\NLO}_{HH'}&=&\sum\limits_{a,b}dx_a\;dx_b\Big[f^{H}_{a}(x_a,\muf)\;f^{H'}_{b}(x_b,\muf)\left(d\sigma_{ab}(x_a,x_b)+\delta d\sigma^{\NLO}_{ab}(x_a,x_b)\right)\\
\nn&&+\as\left(C^{H}_{a}(x_a,\muf)\;f^{H'}_{b}(x_b,\muf)+f^{H}_{a}(x_a,\muf)\;C^{H'}_{b}(x_b,\muf)\right)d\sigma_{ab}(x_a,x_b)\Big].\\
\end{eqnarray}
The crossing functions account for the absorption of unresolved collinear initial-state radiation (invariant mass of collinear pairs smaller than the slicing parameter $s_{\tmin}$) into the effective structure functions (cf. \Eq{eq:hadcollCT}). They also receive a contribution due to crossing a collinear pair with invariant mass smaller than $s_{\tmin}$ from the final to the initial state. After collinear factorisation (see section~\ref{sec:factorisation}) the crossing functions are finite but depend on the factorisation scheme
\begin{equation}
C^{H,scheme}_{a}(x_a,\muf)={N_c\over 2\pi}\left(A^{H}_{a}(x_a)\log\left(s_{\tmin}\over \muf^2\right)+B^{H,scheme}_{a}(x_a)\right).
\end{equation}
The logarithmic dependence on the slicing parameter $s_{\tmin}$ is cancelled in the sum with hard contribution in $\delta d\sigma^{\NLO}_{ab}$ (with the respective $\log(s_{\tmin}/s)$-dependence stemming from requiring $s_{aj}>s_{\tmin}$ and $s_{bj}>s_{\tmin}$ for all final-state partons $j$).
A detailed derivation of the universal crossing functions and explicit expressions for $A^{H}_{a}(x_a)$ and $B^{H,scheme}_{a}(x_a)$ in the $\MSbar$ scheme can be found in \Ref{Giele:1993dj}.

To summarise, the \PSS method allows to regularise and isolate the \IR divergences in the real contribution via dimensional regularisation: The real phase space is split into hard and soft/collinear regions by introducing a slicing parameter $s_\tmin$. In the soft and collinear regions the $d$-dimensional real cross section is approximated by means of the factorisations given in section~\ref{sec:IRdivcanc} which is accurate up to $\ord(s_\tmin)$ terms. In the factorised limits the integration over the soft or collinear radiation can be carried out analytically yielding the \IR poles of the real contribution which match the respective \IR poles of the virtual corrections and cancel in the sum. The slicing of the real phase space introduces logarithmic dependences on the slicing parameter for both the hard and the soft/collinear parts. These logarithms cancel in the sum of both parts.
On the one hand, the systematic error introduced by the approximation of the real contribution in the soft and collinear limits scales with the \PSS parameter $s_{\tmin}$. On the other hand, the potentially large logarithms $\log(s_{\tmin}/s)$ may induce a loss of significant digits when cancelling in the sum of the $n$- and $n+1$-parton contribution (see \Eq{eq:PSSmaster}) blowing up the statistical uncertainty from the numerical integration. The value for $s_{\tmin}$ therefore has to be chosen as a compromise between systematic error and statistical uncertainty. For more details on \PSS with different slicing prescriptions see e.g. \Refs{Giele:1991vf,Giele:1993dj,Brandenburg:1997pu,Cao:2004ky,Harris:2001sx}.

\subsection{Dipole subtraction}\label{sec:dipsub}
The factorisation of the real contribution in the soft limit shows a dipole structure between the pair of emitted soft parton $j$ and emitter $i$  and an additional spectator $k$ that accounts for the colour correlations (see \Eq{eq:M2soft}). The prefactor $(p_i\cdot q)^{-1}$ also signals the presence of a collinear singularity. In fact, in the collinear limit the factorisation of the real contribution (see \Eq{eq:M2coli}) also exhibits a dipole structure which is hidden in the azimuthal dependence of the spin correlations introduced by the splitting functions in \Eq{eq:APsplit} (see \Ref{Catani:1996vz}). On the one hand, exploiting these dipole structures with respect to colour and spin allows to formulate exact factorisation formulae of the real $n+1$ parton phase space into an $n$-parton phase space and a dipole phase space accounting for the emission of soft or collinear radiation.
On the other hand, the dipole factorisations (see \Eq{eq:M2soft} and \Eq{eq:M2coli}) also suggest how to construct an approximation of the real contribution to match the soft and collinear limits (see section~\ref{sec:IRdivcanc}).
Both fulfil exact momentum conservation also away from the soft and collinear limits while approaching these limits smoothly.
With the exact phase space factorisations the approximated real contribution can be integrated over the soft and collinear radiation analytically in $d=4-2\varepsilon$ dimensions yielding the \IR divergences of the real corrections as poles in $\varepsilon$ in a universal way.
Therefore, they can be used both as local counter terms rendering the integration of the real contribution finite in $d=4$ dimensions and to cancel the respective \IR poles in the virtual contribution after integration. For illustration, consider the integration of $x^{-1-\varepsilon}f(x)$ (with $f(x)$ an arbitrary test function regular at $x=0$ and $x^{-1-\varepsilon}f(x)\overset{x\rightarrow0}{\approx}x^{-1-\varepsilon}f(0)$)
\begin{eqnarray}
\nn \overbrace{\int\limits_{0}^{1}dx\;x^{-1-\varepsilon}f(x)}^{\displaystyle\small{\text{divergent for }}\varepsilon\rightarrow 0}&=&\int\limits_{0}^{1}dx\;x^{-1-\varepsilon}f(0)\quad+\quad\int\limits_{0}^{1}dx\;\left(x^{-1-\varepsilon}f(x)-x^{-1-\varepsilon}f(0)\right)\\
\nn&=&\underbrace{-{1\over\varepsilon}}_{\mathclap{\displaystyle\small{\text{universal factor}}}}f(0)\quad+\quad\underbrace{\int\limits_{0}^{1}dx\;{f(x)-f(0)\over x}}_{\displaystyle\small{\text{$\varepsilon=0$}}}\quad+\quad\ord\left(\varepsilon\right) \\
\end{eqnarray}
yielding an integral which can be numerically integrated in $d=4$ dimensions. The divergence of the original integral has been analytically extracted as a pole $\propto 1/\varepsilon$. This factorisation is universal (independent of $f$). In contrast to the \PSS method (see section~\ref{sec:PSS}), the result is accurate because the integrand is not approximated but a counter term is added and subtracted to regulate the \IR divergences. As indicated above, the counter terms for the real cross section $ d\sigma^{n+1,A}$ can be constructed according to its dipole structure (with `emitter' $ij$ and `spectator' $k$) by `dipole operators' $\mathcal{D}_{ij,k}$ acting on Born matrix elements in colour and helicity space ${\mathcal{M}_{n}}^{\ldots c_{(ij)}\ldots c_k\ldots,\;\ldots s_{(ij)}\ldots s_k\ldots}$
\begin{equation}\label{eq:dipoles}
 d\sigma^{n+1,A}=dR_{n+1}\sum\limits_{j\in\{\text{\scriptsize f}\}}\sum\limits_{\substack{i\in\{\text{\scriptsize i, f}\}\\ i\neq j}}\sum\limits_{\substack{k\in\{\text{\scriptsize i, f}\}\\ k\neq i,j}}\underbrace{{{\mathcal{M}}^*_{n}}^{\ldots c_{(ij)}\ldots c_k\ldots,\;\ldots s_{(ij)}\ldots s_k\ldots}\mathcal{D}_{ij,k}{{\mathcal{M}}_{n}}^{\ldots c_{(ij)}\ldots c_k\ldots,\;\ldots s_{(ij)}\ldots s_k\ldots}}_{\displaystyle\small{\equiv D_{ij,k}}}.
\end{equation}
With coloured partons in the initial state ($\{\text{i}\}$) and in the final state ($\{\text{f}\}$) four different types of dipole operators (acting in colour space (indices $c_i$) and spin space (indices $s_i$)) are needed: $\mathcal{D}_{ij,k}$ (final-state emitter $i\in\{\text{f}\}$ with final-state spectator $k\in\{\text{f}\}$), $\mathcal{D}_{ij,a}$ (final-state emitter $i\in\{\text{f}\}$ with initial-state spectator $a\in\{\text{i}\}$), $\mathcal{D}_{aj,k}$ (initial-state emitter $a\in\{\text{i}\}$ with final-state spectator $k\in\{\text{f}\}$) and $\mathcal{D}_{aj,b}$ (initial-state emitter $a\in\{\text{i}\}$ with initial-state spectator $b\in\{\text{i}\}$)
Because of the correlations between emitter and spectator only dipoles with
\begin{equation}
{{\mathcal{M}}^*_{n}}^{\ldots c_{(ij)}\ldots c_k\ldots,\;\ldots s_{(ij)}\ldots s_k\ldots}\mathcal{D}_{ij,k}{{\mathcal{M}}_{n}}^{\ldots c_{(ij)}\ldots c_k\ldots,\;\ldots s_{(ij)}\ldots s_k\ldots}\neq 0
\end{equation}
can contribute to cancel \IR divergences occurring in these limits.
Each dipole encodes a unique factorisation of the real phase space ($i,j,k\mapsto (ij),k$)
\begin{equation}\label{eq:dipphspff}
dR_{n+1}=dR_{n}(p_1,\ldots,p_{(ij)},\ldots,p_k,\ldots,p_n)dR_{ij,k}(p_i,p_j,p_k)
\end{equation}
and approximation of the real matrix element ($|\mathcal{M}_{n+1}|^2\rightarrow D_{ij,k}$) to approach the respective singular limits.
For more details on dipole subtraction and derivation of the actual form of the dipoles $D_{ij,k}$ in \Eq{eq:dipoles} for massless and massive partons see e.g. \Refs{Catani:1996vz,Catani:2002hc}).

\subsection{Jets}\label{sec:jets}
The process of coloured partons forming colour-singlet bound states at large distances is called `hadronisation'. In order for hadronisation to take place, the energy scale of the partons from the hard process has to decrease (see \Fig{fig:runas}). This happens through cascades of colour radiation dominated by low energy and/or collinear partons decreasing the emitting partons' energies towards the hadronisation scale to finally form detectable baryons and mesons. Because soft and collinear emissions are dominant, the radiation cascade does not significantly change the direction of the energy flow. Therefore, hard partons manifest themselves in experiments as clusters of energy depositions in the hadronic calorimeters which can be grouped to so-called jets with the help of jet finding algorithms. For a detailed discussion of QCD and jets see e.g. ref. \cite{Sterman:2004pd}.

The mapping of partonic momenta to jet momenta is given by the jet algorithm. Because of potential features like e.g. \IR safety, close match to QCD radiation cascades and longitudinal boost invariance sequential jet algorithms have prevailed in the hadron collider context. They are specified by a (boost invariant) resolution criterion and a clustering prescription. The resolution criterion $d_{ij}$ is a distance measure for the  particles' momenta in phase space. It is used to identify the pair $k,l$ of particles with the smallest distance according to the measure $d_{kl}=\min\;{d_{ij}}$ as being unresolved. The unresolved pair is then clustered according to the clustering prescription and replaced by a proto-jet with momentum $J_{(kl)}$: $p_k,p_l\mapsto J_{(kl)}$. This procedure is repeated (including the proto-jets) until a desired number of resolved proto-jets is obtained (exclusive formulation) or until all $d_{ij}$ are bigger than some minimal distance $d_{\scriptsize\mbox{cut}}$ (inclusive formulation). The remaining proto-jets are called jets. For hadron collider processes the following resolution criterion can be used to define a class of sequential, \IR safe, boost invariant jet algorithms
\begin{equation}\label{eq:jetsrescrit}
d_{ij}=\min\left({p^{\perp}_{i}}^{\alpha},{p^{\perp}_{j}}^{\alpha}\right)\frac{\Delta R^2_{ij}}{R^2},\quad \Delta R^2_{ij}=\left(y_i-y_j\right)^2+\left(\phi_i-\phi_j\right)^2, \quad d_{iB}={p^{\perp}_{i}}^{\alpha}.
\end{equation}
Here, $p_i^{\perp}$, $y_i$ and $\phi_i$ are the transverse momentum, the rapidity and the azimuthal angle of particle $i$. The $d_{ij}$ measures the distance between two final state particles $i$ and $j$ in terms of the rapidity and azimuthal angle inside a `jet cone' characterised by the radius parameter $R$. The $d_{iB}$ measures the distance of particle $i$ to the beam. Setting the exponent $\alpha=2$, $\alpha=-2$ or $\alpha=0$ is called `$kt$-algorithm', `anti-$kt$-algorithm' or `Cambridge-Aachen-algorithm'. Note that the $kt$-algorithm clusters soft particles first while the anti-$kt$-algorithm starts with the hardest particles. The clustering prescription specifies how to combine unresolved particles to (proto-)jets. 
\begin{figure}[htbp]
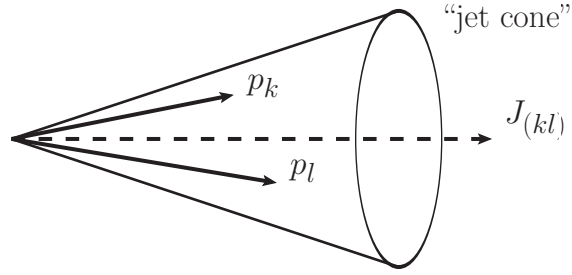

  \begin{center}
    \leavevmode
     \includegraphics[width=0.49\textwidth]{{{%
          prerequ/jetcone}}}
     \caption{Recombination of partonic momenta $p_k$ and $p_l$ to form jet momentum $J_{(kl)}.$}
    \label{fig:jetcone}
  \end{center}
\end{figure}
For the commonly used $2\to1$ clustering algorithms different schemes have become customary. A widely used scheme is the `E-scheme' where the $4$-momenta of the clustered particles are added to form the (proto-)jet momentum $J_{(kl)}=p_k+p_l$ (cf. \Fig{fig:jetcone}). This satisfies $4$-momentum conservation but produces off-shell jets: $J^2_{(kl)}=p^2_k+p^2_l+2p_k\cdot p_l$. On-shell jets can be obtained with the `P-scheme' where the $3$-momenta of the clustered particles are added $\vec{J}_{(kl)}=\vec{p}_k+\vec{p}_l$ while the energy of the (proto-)jet follows from the on-shell condition $J^0_{(kl)}=\sqrt{\vec{J}^2_{(kl)}+m^2_{J_{(kl)}}}$. Note that satisfying both on-shell condition and $4$-momentum conservation is not possible with  $2\to1$ clustering prescriptions.
Based on the resolution criterion and the recombination scheme, a $2\to1$ clustering algorithm can be defined as follows:
\begin{enumerate}
\item[1.] Pick final-state partons $k$ and $l$ or final-state parton $k$ and beam $B$ with minimal $d_{kl}=\min\limits_{i,j}\;\left({d_{ij}},d_{iB}\right)$ or $d_{kB}=\min\limits_{i,j}\;\left({d_{ij}},d_{iB}\right)$.
\item[2a] If $d_{kl}$ is minimal cluster $k,l$ according to the recombination scheme.
\item[2b] If $d_{kB}$ is minimal call $k$ a jet and remove it from the list of particles and proto-jets.
\item[3.] Repeat the procedure (including the proto-jets) until the desired number of resolved jets is obtained or all $d_{ij},d_{iB}$ are bigger than $d_{\scriptsize\mbox{cut}}$.
\end{enumerate}
After clustering, only the jets that pass the experimental cuts (on $p^{\perp}, \eta, \ldots$) are kept in the list of jets.
In order to make meaningful higher-order perturbative predictions for observables of jets obtained from the jet algorithm (`jet observables'), these jet observables have to fulfil the requirements of \IR safety (see \Eq{eq:IRsafeobs}) and factorisability of initial-state collinear singularities (see \Eq{eq:collfactable}).
See e.g. \Refs{Salam:2009jx,Cacciari:2011ma,Sapeta:2015gee} for more details on different jet algorithms, resolution criteria, recombination schemes and jet phenomenology.

In theory, the production of $n$ jets at a hadron collider is thus sketched by
\begin{equation}\label{eq:hadjets}
H(P_A)+H'(P_B)\rightarrow j_1(J_1)+\ldots+j_n(J_n)
\end{equation}
For a consistent comparison of experimental data and theoretical predictions the same jet finding algorithm, used by the experiment to define the measured jets of hadrons, has to be used to define the jets of partons in perturbative calculations.
The respective cross section for the transition depicted in \Eq{eq:hadjets} factorises in a convolution of the partonic cross section for the production of $m$ partons $d\sigma^m_{ab}$ with the PDFs and functions $F^m_{J_1,\ldots,J_n}(p_1,...,p_m)$ which incorporate the mapping of $m$ partonic momenta $p_1,\ldots,p_m$ to the momenta $J_1,\ldots,J_n$ of $n\leq m$ resolved jets. It reads schematically
\begin{equation}\label{eq:xshadjet}
d\sigma_{HH'\rightarrow nj}= dx_a\;dx_b\;f^{H}_a(x_a,\muf)\;f^{H'}_b(x_b,\muf)\; dR_m\; {d\sigma^m_{ab}(x_aP_A,x_bP_B;\muf)\over dR_m(p_1,...,p_m)}F^m_{J_1,\ldots,J_n}(p_1,...,p_m).
\end{equation}
In the Born approximation ($m=n$) the momenta of the jets $J_i$ are identified with the underlying partons' momenta $p_i$
\begin{equation}
J_i=p_i.
\end{equation}

When including the next-to-leading order in the perturbative series in $\alpha_s$, this naive identification is no longer justified because of additional real radiation which can contribute to the jets ($m=n+1$).
The jet momenta are in general non-trivial functions of the partonic
momenta
\begin{equation}\label{eq:realmap}
  J_i=J_i(p_1,\ldots,p_{n+1}),\quad i=1,\ldots,n.
\end{equation} 
The mapping might also depend on the phase space region.

It is clear that the higher the number of partons $m$ considered in the partonic cross section $d\sigma^m_{ab}$ in \Eq{eq:xshadjet}, the better the approximation of the radiation cascades that led to the $n$ observed jets. It is therefore crucial to include higher-order corrections to the cross section calculations in order to more realistically model jet formation and additional jet activity. Additionally, one can combine fixed-order parton level calculations with `parton showers' which model the QCD radiation cascades by collinear parton splitting or soft parton emission followed by non-perturbatively modelled hadronisation and decays (see e.g. ref. \cite{Buckley:2011ms} for a review on parton showers).

\chapter{Developing the \MEM at NLO accuracy}\label{sec:MEMtoNLO}
The extension of the \MEM beyond the Born approximation requires the inclusion of NLO corrections into the theoretical predictions used to calculate the likelihoods. By interpreting differential cross sections as probability densities describing distributions of the produced final state, cross section predictions which are fully differential in all observed final-state variables are needed in the \MEM. At NLO accuracy, individually \IR divergent virtual and real corrections have to be taken into account which only together yield finite predictions for \IR safe observables. In principle, employing \IR safe jet algorithms to define the jet momenta ensures \IR safety of the jet variables. At the same time, jet algorithms introduce non-trivial interrelations between the momenta of unresolved partonic configurations and the momenta of the resolved jets. In the following, it is illustrated how that prevents the naive evaluation of the fully differential NLO cross section for a given set of jet variables in general. A possible solution is presented by modifying the recombination step of conventional jet algorithms resulting in an algorithm allowing to consistently define fully differential cross sections at NLO accuracy to be used in the \MEM. The presented algorithm has been published in \Refs{Martini:2015fsa,Martini:2017ydu}.

As a first step, the event definition describing the observed final states $\vec{x}$ has to be specified in order to define the differential cross section or event weight $d\sigma/d\vec{x}$ needed for the \MEM (see \Eq{eq:diffxsdect}). Experimentally, coloured final states manifest themselves in the detectors as clusters of energy depositions which can be grouped by jet algorithms defining them as experimentally observed jets (see section \ref{sec:jets}). A description of the observed jets within perturbation theory may be obtained through the merging of collinear or soft partons in higher-order calculations according to the same clustering procedure. The momenta of these theoretically modelled jets are called `jet momenta' in what follows to distinguish them from the partonic momenta. To be as general as possible, an event $\vec{x}$ is defined to be the collection of jet momenta of the observed final state for the moment. A threefold motivation to use the jet momenta in the differential cross section can be given:
\begin{enumerate}
\item Being fully differential in the jet momenta, all relevant information contained in the
  cross section for the production of the observed final state is kept, allowing to
 subsequently change or reduce the set of variables by transformations or integrations.
\item Equating the momenta of the observed jets with the jet momenta of the theory jets by identifying the transfer functions with
  delta-functions may provide a reasonable first approximation (cf. \Eq{eq:trsfdelta}).
\item The ability to calculate fully differential event weights for `jet events' including
  higher-order corrections may prove to be of versatile use on its own.
\end{enumerate}

The extension of the \MEM to include higher-order corrections is non-trivial in general. To point out
the origin of the difficulties, the general structure of NLO corrections is inspected. For a set of $r$ variables $\{x_r\}$ describing $n$- as well as $n+1$-jet final states, the generic differential cross section including NLO corrections reads
\begin{eqnarray}\label{eq:diffjetxs}
\nn{d\sigma^{\NLO}\over dx_1\ldots dx_r}\quad=&\displaystyle\int dR_{n}(\vec{y}){d\sigma^{\text{BV}}\over dy_1\ldots dy_s}&F^{n}_{J_1,\ldots,J_n}(\vec{y})\;\delta\left(\vec{x}-\vec{x}(J_1,\ldots,J_n)\right)\\
\nn&+\displaystyle\int dR_{n+1}(\vec{z}){d\sigma^{\text{R}}\over dz_1\ldots dz_t}&\bigg[F^{n+1}_{J_1,\ldots,J_n}(\vec{z})\;\delta\left(\vec{x}-\vec{x}(J_1,\ldots,J_n)\right)\\
&&+F^{n+1}_{J_1,\ldots,J_{n+1}}(\vec{z})\;\delta\left(\vec{x}-\vec{x}(J_1,\ldots,J_{n+1})\right)\bigg].
\end{eqnarray}
The Born and virtual (real) contributions to the NLO differential cross section are denoted by $d\sigma^{\text{BV}}$ ($d\sigma^{\text{R}}$). They have to be integrated over the $n$- ($n+1$-) parton phase space parameterised by the partonic variables collected in $\vec{y}$ ($\vec{z}$). 
Real corrections appear as additional contributions in higher-order corrections describing additional parton emission. Depending on the jet algorithm the extra radiation can either be clustered into a jet or manifest itself in the detectors as an additional resolved jet. The resolution criterion of the jet algorithm is used in $F^{n}_{J_1,\ldots,J_n}$, $F^{n+1}_{J_1,\ldots,J_n}$ and $F^{n+1}_{J_1,\ldots,J_{n+1}}$  to determine whether the $n$ or $n+1$ partons are resolved as $n$ or $n+1$ individual jets with momenta $J_1,\ldots,J_n$ or $J_1,\ldots,J_{n+1}$. Any resolved jet's momentum has to pass certain phase space cuts which are also incorporated in $F^{n}_{J_1,\ldots,J_n}$, $F^{n+1}_{J_1,\ldots,J_n}$ and $F^{n+1}_{J_1,\ldots,J_{n+1}}$. Note that these cuts might be required already to render the Born cross section finite if the LO predictions contain soft and/or collinear singularities (e.g. for $pp\rightarrow ttj$). 
 As long as \IR safe jet algorithms are used to obtain the jet momenta, distributions of  \IR safe jet variables can be calculated with \Eq{eq:diffjetxs}.
In the contribution due to $F^{n+1}_{J_1,\ldots,J_n}$ the resolved jets are modelled in a non-trivial way by clustering unresolved partonic configurations. Additional jet activity is described by the contribution due to $F^{n+1}_{J_1,\ldots,J_{n+1}}$.  Both of these contributions appear for the first time when considering NLO corrections in the perturbative calculation. In the Born approximation (and also the virtual corrections) no recombination can occur and the jet momenta and partonic momenta (parameterised by $\vec{y}$) are identified
\begin{equation}\label{eq:LOmap}
J_i=J_i(p_i)=p_i,\quad i=1,\ldots,n.
\end{equation}
This mapping of $n$ partonic momenta to $n$ jet momenta is implicitly given by the functional dependence of the jet momenta on the partonic momenta (or equivalently $\vec{y}$) in
\begin{equation}
F^{n}_{J_1,\ldots,J_n}(\vec{y})\sim\prod\limits_{i=1}^{m}\Theta_i(J_1(p_1),\ldots,J_n(p_n)).
\end{equation}
The functions $\Theta_i$ are used to determine if the $n$ jet momenta correspond to a resolved $n$-jet final state depending on the resolution criterion and the phase space cuts.
In case of the mapping of $n+1$ partons to $n+1$ jets, $F^{n+1}_{J_1,\ldots,J_{n+1}}$ has an analogous form because on its own it can be regarded as a Born level contribution with a higher final-state multiplicity. In the contribution from the real corrections where $n+1$ partons are recombined to form $n$ jets the resulting jet momenta are in general non-trivial functions of the partonic momenta 
\begin{equation}\label{eq:realmap}
J_i=J_i(p_1,\ldots,p_{n+1}),\quad i=1,\ldots,n.
\end{equation} 
Again, the mapping of $n+1$ partonic momenta to $n$ jet momenta is implicitly given by the functional dependence of the jet momenta on the real partonic momenta (or equivalently $\vec{z}$) in
\begin{equation}
F^{n+1}_{J_1,\ldots,J_n}(\vec{z})\sim\prod\limits_{i=1}^{m}\Theta_i(J_1(p_1,\ldots,p_{n+1}),\ldots,J_n(p_1,\ldots,p_{n+1})).
\end{equation} 
Note that because of several soft and/or collinear configurations the specific mapping depends on the respective phase space region. The clustered jets' momenta also have to pass the phase space cuts and the resolution criterion (encoded in $\Theta_i$) to correspond to a resolved  $n$-jet final state.
Using the functions $\vec{x}(J_1,\ldots,J_n)$ and $\vec{x}(J_1,\ldots,J_{n+1})$ to relate the $r$ variables of the differential cross section (see \Eq{eq:diffjetxs}) to the $n$ and $n+1$ jet momenta makes the differential cross section in \Eq{eq:diffjetxs} differential in `jet variables' rather than partonic variables.

\section{Introducing $3\to2$ jet clustering prescriptions}\label{sec:3to2clus}
For the application in the \MEM event weights or fully differential cross sections for the observed final-states are needed.
When trying to apply \Eq{eq:diffjetxs} to calculate fully differential jet cross sections various problems become obvious:
\begin{enumerate}
\item Because of soft and collinear singularities in the real and virtual amplitudes, the first two contributions are separately 
  ill-defined on their own. Only by summing the two a finite result can be obtained according to the KLN theorem when a mutual cancellation of the singularities is achieved (cf. section~\ref{sec:IRdivcanc}). 
\item In contrast to standard Monte Carlo calculations of histogrammed distributions the event weights are maximally exclusive in the sense that they are differential in all variables describing a specific measured final state. The $\delta$-functions fixing the clustered jet momenta according to the measured jet variables prevent an
  efficient numerical integration over the unresolved real phase space. 
\item In the $n$-parton contribution the $n$ jet momenta are identified with the $n$ partonic momenta satisfying the $n$-parton
  kinematics: They respect the on-shell condition as well as $4$-momentum conservation. For jet momenta obtained
  by a recombination of the $n+1$ partons this is in general not the case. For instance, defining the momentum of the recombined jet as the sum of the combined partonic momenta violates the on-shell condition away from the soft/collinear limits where
the clustering will create non-vanishing masses even for jets of massless progenitor partons. Because $d\sigma^{BV}$ is only defined for on-shell momenta it cannot be evaluated with these jet momenta which would however be required to achieve a point-wise cancellation of the soft and collinear divergences. Employing a different $2\to 1$ recombination scheme, like e.g. the P-scheme, does not help either because while keeping the jets on shell it violates $4$-momentum conservation (cf. section~\ref{sec:jets}). The common practice of omitting additional radiation which is too close to the beam (e.g. falling below a certain transverse momentum threshold) also violates the overall $4$-momentum balance of the event rendering it inept for the evaluation of $d\sigma^{BV}$.
\end{enumerate}
The last two obstacles can be overcome by modifying the conventional jet algorithms relying on a $2
\to 1$ clustering (see section \ref{sec:jets})
\begin{equation}
p_i,p_j\to J_{(ij)}
\end{equation}
by introducing a $3\to2$ clustering prescription instead
\begin{equation}\label{eq:3to2clus}
p_i,p_j,p_k\to J_j,J_k.
\end{equation}
The additional `spectator particle' $k$ allows to simultaneously satisfy
$4$-momentum conservation
\begin{equation}
p_i+p_j+p_k=J_j+J_k
\end{equation}
 and on-shell conditions
\begin{equation}
J_j^2=p^2_j,\quad J^2_k=p^2_k.
\end{equation}
Actually, jet masses created in perturbative calculations by clustering massless partons with established jet algorithms (relying on a $2\to1$ clustering) have very little to do with the masses observed in the experimentally measured `light' jets\footnote{This statement does not apply to fat jets or highly boosted objects.}. Since experimentally measured jets are composed of hadrons stemming from the hadronisation of the partons their observed masses are rather related to non-perturbative effects. Modifying the clustering to better match the perturbative calculation is at least equally well motivated as the current standard and one may even argue that modifying the clustering in order to fulfil the constraints given above may help to better separate perturbative and non-perturbative physics.

The $3\to2$ clustering prescriptions are inspired by the phase space
parameterisation used in the Catani--Seymour dipole subtraction method (see section \ref{sec:dipsub}). 
According to the resolution criterion of a jet algorithm, the phase space of the $n+1$ partons is formally separated into an `$n+1$-jet region' with no recombination of the $n+1$ partons resulting in $n+1$ jets and an `$n$-jet region' with a recombination of $n+1$ partons forming $n$ jets
\begin{equation}
  dR_{n+1} =  \underbrace{dR_{n+1}\Theta^{\text{$n+1$ jets}}}_{\displaystyle=dR_{n+1}\big|_{\text{$n+1$ jets}}}+ \underbrace{dR_{n+1}\Theta^{\text{$n$ jets}}}_{\displaystyle=dR_{n+1}\big|_{\text{$n$ jets}}}.
\end{equation}
Here, $\Theta^{\text{$n$ jets}}=1$ ($\Theta^{\text{$n+1$ jets}}=1$) if two of the partons (none of the partons) are recombined yielding $n$ ($n+1$) jets and zero otherwise.
The $n$-jet region can now schematically be written as a sum of all possible recombinations of unresolved parton pairs $ij$ forming a single jet
\begin{equation}\label{eq:njetpairs}
 dR_{n+1}\big|_{\text{$n$ jets}} = \sum\limits_{ij} dR_{n+1}\Theta_{ij}
\end{equation}
with $\Theta_{ij}=1$ if partons $i$ and $j$ minimize the resolution criterion and their recombination leads to an $n$-jet final state and $\Theta_{ij}=0$ otherwise. For each unresolved parton pair $ij$ the spectator $k\neq i,j$ can be chosen freely. The choice of the spectator is encoded in the generic functions
\begin{equation}\label{eq:specchoice}
\Theta_k(p_1,\ldots,p_{n+1}),\quad k=1,\ldots,n+1\qquad\text{with} \quad\sum\limits_{k}\Theta_{k}=1.
\end{equation}
The functions $\Theta_k$ may be chosen as simple step-functions or smooth functions.
In accordance with the Catani--Seymour dipole formalism, four qualitatively different mappings depending on the clustered pair $ij$ and the spectator $k$ can be identified (cf. section \ref{sec:dipsub}): final-state clustering with final-state spectator, final-state clustering with initial-state spectator, initial-state clustering with final-state spectator or initial-state clustering with initial-state spectator. Note that parton $i$ is always from the final state.

The Catani--Seymour phase space factorisations allow to factorise the $n$-jet region (from \Eq{eq:njetpairs} together with \Eq{eq:specchoice}) into an $n$-body phase space $dR_n$ times a sum over `dipole phase spaces' $dR_{ij,k}$
\begin{equation}\label{eq:phspfac}
  dR_{n+1}(\vec{z})\big|_{\text{$n$ jets}} =\sum\limits_{\substack{ij\\k\neq i,j}}  dR_{n+1}(\vec{z})\;\Theta_{ij}\Theta_{k}=dR_n(\vec{y}) \sum\limits_{\substack{ij\\k\neq i,j}}dR_{ij,k}\Theta_{ij}\Theta_{k}.
\end{equation}
For each phase space factorisation appearing in the sum the Catani--Seymour dipole formalism introduces a phase space mapping from the respective $n+1$-parton phase space to the corresponding $n$-body phase space
\begin{equation}\label{eq:CSpsmap}
F^{ij,k}:\quad R_{n+1}\rightarrow R_n,\quad (p_1,\ldots,p_i,\ldots,p_j,\ldots,p_k,\ldots,p_{n+1})\mapsto (J_1,\ldots,J_j,\ldots,J_k,\ldots,J_{n}).
\end{equation}
Utilising these mappings as the clustering prescriptions of a jet algorithm allows to identify the $dR_n(\vec{y})$ in \Eq{eq:phspfac} with the phase space of the resulting jets. The particular dipole phase space $dR_{ij,k}$ generates exactly the partonic configurations with momenta $p_i,p_j,p_k$ that would be clustered to form the jet momenta $J_j,J_k$ via the respective clustering $F^{ij,k}$ in \Eq{eq:CSpsmap}.
Factorising the real phase space according to \Eq{eq:phspfac} provides the partonic momenta as the preimage of the jet algorithm based on the Catani--Seymour mappings given in \Eq{eq:CSpsmap}.

\subsection{From $2\to1$ to $3\to2$ jet algorithms}\label{sec:3to2jetalg}
The factorisation in \Eq{eq:phspfac} and the mappings given in \Eq{eq:CSpsmap} guide the way how conventional sequential $2\to1$ jet algorithms as described in section~\ref{sec:jets} can be augmented by the  $3\to2$ clustering prescriptions:
\begin{enumerate}
\item Pick final state partons $i$ and $j$ or final state parton
  $i$ and beam $B$ with minimal $d_{ij}$ or
  $d_{iB}$ to be clustered.
\item 
  \begin{enumerate}
  \item  If $d_{ij}$ is minimal, pick spectator
    parton from final state ($k$) or beam particle ($a$) according to $\Theta_k$ or $\Theta_a$.
  \item If $d_{iB}$ is minimal, pick beam particle
  ($a$) and spectator parton from final state ($k$) or beam ($b$) according to $\Theta_k$ or $\Theta_b$.
  \end{enumerate}
\item Cluster ($ij,k$), ($ij,a$), ($ia,k$) or ($ia,b$) according to the
  respective $3\to2$ phase space mapping introduced in the 
  Catani--Seymour subtraction method (see section~\ref{sec:dipsub} and \Refs{Catani:1996vz,Catani:2002hc}).
\end{enumerate}
In contrast to the definition given in section \ref{sec:jets}, the recombination in step three is modified by the introduction of the additional spectator particle in step two. Instead of the usual $2\to1$-recombination schemes (e.g. E-scheme, P-scheme, \ldots) unresolved partons are clustered according to the $3\to2$ mapping to the `reduced kinematics' introduced by the dipole subtraction method (cf. \Eq{eq:CSpsmap}). For each $3\to2$ clustering a spectator particle has to be chosen. The freedom to choose the spectator particle (as long as the particle itself is not collinear or soft) can be used to reduce the difference between the modified clustering prescription and the conventional $2\to1$ recombination. 
The $n$-body phase space $dR_n(\vec{y})$ in \Eq{eq:phspfac} conforms to the $n$ resolved on-shell jets obtained by the $3\to2$ recombination of the unresolved partonic configurations parameterised by the dipole phase spaces $dR_{ij,k}$. The integration over these dipole phase spaces corresponds to integrating out all $n+1$-parton configurations contributing to the $n$ resolved on-shell jets which are parameterised by $dR_n(\vec{y})$. 
Identifying the $n$-body phase space $dR_n(\vec{y})$ with the one occurring in the $n$-parton contribution solves the second and third issue by factorising the unresolved real contribution due to ${F}^{n+1}_{J_1,\ldots,J_n}$ (defined through a $3\to2$ jet algorithm) in \Eq{eq:diffjetxs} as
\begin{eqnarray}
\label{eq:unres-real-fact}
 && \int dR_{n+1}(\vec{z}) {d\sigma^R\over dz_1\ldots dz_{t}}
  {F}^{n+1}_{J_1,\ldots,J_n}(\vec{z})
  \delta\left(\vec{x} - \vec{x}(J_1,\ldots,J_n)\right)\nn\\
  &=& \int dR_{n}(\vec{y}) { F}^{n}_{J_1,\ldots,J_n}(\vec{y}) \delta\left(\vec{x} - \vec{x}(J_1,\ldots,J_n)\right)
  \underbrace{\sum\limits_{\substack{ij\\k\neq i,j}}\int dR_{ij,k} {d\sigma^R\over dz_1\ldots dz_{t}}}_{\displaystyle\equiv{d\sigma^R\over dy_1\ldots dy_s}}.\nn\\
\end{eqnarray}
For a given set of jet momenta $J_1,\ldots,J_n$ (or equivalently the variables
$\vec{y}=(y_1,\ldots,y_s)$) three additional variables $\Phi=(\phi,z,y)$ parameterising the unresolved phase space are needed to specify the variables $z_1,\ldots,z_t$. After integrating over the unresolved partonic configurations, the real contribution in the $n$-jet region can be combined with the Born and virtual $n$-jet contributions from \Eq{eq:diffjetxs} point by point for each set of jet momenta
\begin{equation}\label{eq:diffnjetxs}
{d\sigma^{\NLO}\over dx_1\ldots dx_r}\Bigg|_{\text{$n$ jets}}=\int dR_{n}(\vec{y})\underbrace{\left({d\sigma^{\text{BV}}\over dy_1\ldots dy_s}+{d\sigma^R\over dy_1\ldots dy_s}\right)F^{n}_{J_1,\ldots,J_n}(\vec{y})}_{\displaystyle\equiv{d\sigma^{\NLO}\over dR_n(J_1,\ldots,J_n)}}\;\delta\left(\vec{x}-\vec{x}(J_1,\ldots,J_n)\right).
\end{equation}

\subsection{Cancellation of \IR divergences}
Note that in general both terms in \Eq{eq:diffnjetxs} are individually divergent due to soft and collinear singularities (cf. the aforementioned first issue). The phase space factorisation given by \Eq{eq:phspfac} is valid in $d$ dimensions (see \Refs{Catani:1996vz,Catani:2002hc}). Therefore, it is straightforward to regularise the \IR divergences via dimensional regularisation. One method to mediate the mutual cancellation of the \IR divergences between the real and the virtual contribution is given by \PSS techniques (see section \ref{sec:PSS}) presenting a straightforward solution to the first issue.

Despite the factorisation of the real phase space being inspired by the Catani--Seymour dipole subtraction (cf. section \ref{sec:dipsub} and \Refs{Catani:1996vz,Catani:2002hc}), the application of this method to subtract the \IR divergences within the formalism presented here is problematic: The contribution of the dipole subtraction term is obtained by evaluating the jet algorithm for the reduced kinematics $\tilde{p}^{ij,k}_1,\ldots,\tilde{p}^{ij,k}_n$ which depend on the respective dipole $ {D}_{ij,k}$.
The term which is added (and subtracted) thus reads
\begin{equation}
  {1\over 2s} \sum_{i\not = j,
    k\not=i,j}F^{n}_{\tilde{J}_1,\ldots,\tilde{J}_n}(\tilde p^{ij,k}_1,\ldots,\tilde p^{ij,k}_n) 
  {D}_{ij,k} 
  dR_{n+1}(p_1,\ldots,p_{n+1}).
\end{equation}
The mapping from the $n+1$ parton momenta $p_1,\ldots,p_{n+1}$ to the reduced kinematics $\tilde{p}^{ij,k}_1,\ldots,\tilde{p}^{ij,k}_n$ which are identified with the jet momenta $\tilde{J}_1,\ldots,\tilde{J}_n$ by $F^{n}_{\tilde{J}_1,\ldots,\tilde{J}_n}$ is encoded in the dipole. The mapping cannot be chosen freely without a mismatch with the analytically integrated contribution
which has to be combined with the respective virtual corrections. Thus, the phase space factorisation matching the clustering of the jet algorithm achieved in \Eq{eq:unres-real-fact} is destroyed. The authors of \Ref{Campbell:2013uha} draw a similar conclusion regarding the application of Catani--Seymour dipole subtraction in combination with this formulation of the \MEM at NLO. 

On the other hand, the \PSS method (cf. section \ref{sec:PSS}) does not introduce any principle problems within the presented formalism. Therefore, \PSS techniques are utilised for the cancellation of infrared divergences in the NLO calculations performed in this work. As an outlook, it might be worthwhile to consider alternative subtraction methods like for example FKS subtraction (see \Refs{Frixione:1995ms,Frixione:1997np}) to overcome numerical challenges inherent in the \PSS method.

\section{Phase space parameterisations}
\label{sec:PhaseSpaceparameterisation}
The $n+1$-parton phase space is parameterised in terms of an $n$-body phase space times an `unresolved' dipole phase space according to the phase space factorisation given in the context of the
Catani--Seymour dipole subtraction method (see \Refs{Catani:1996vz,Catani:2002hc}).
The mapping of the real partonic momenta to the reduced kinematics defines the clustering prescriptions for the augmented
$3\to 2$ jet algorithms. In the augmentation of the jet algorithm the resolution criterion is not modified and the new $3\to 2$ clusterings reproduce the soft and collinear limits (see section \ref{sec:IRdivcanc}). Thus, the augmentation of an \IR safe $2\to 1$ jet algorithm which permits the factorisation of initial-state collinear singularities by the $3\to 2$ clusterings preserves these properties.

Selecting a spectator $k\neq i,j$ (by $\Theta_k$) for each pair of unresolved partons $ij$ singled out by the resolution criterion of the jet algorithm defines a particular mapping to recombine the $n+1$ parton momenta to $n$ jet momenta
\begin{equation}
\left(p_1,\ldots,p_{n+1}\right)\xmapsto{ij,k}\left(J_1,\ldots,J_n\right).
\end{equation}
Factorising the real phase space corresponding to the $3\to2$ clustering prescriptions allows to interpret the respective phase space of unresolved partonic configurations as the preimage of the mapping for each set of $n$ clustered jet momenta: The respective unresolved phase space can be parameterised as a function of a set of $n$ on-shell jet momenta $J_1,\ldots,J_n$ and three additional variables describing the unresolved radiation collected in $\bf{\Phi}$
\begin{equation}
  (J_1,\ldots,J_n; {\bf \Phi})\xmapsto{ij,k} (p_1,\ldots,p_{n+1}).
\end{equation}
serving as an `inversion of the jet clustering'. The required formulae for the four qualitatively different mappings are collected in the following sections. The presented mappings allow a direct integration of the variables $\bf{\Phi}$ or equivalently the unresolved radiation. 

Let us note that some of the deduced formulae share similarities with formulae given in the literature where similar factorisations of the real phase space have been applied in different contexts. The method is briefly compared to existing work while the focus is on the differences.
For instance, the method presented in \Ref{Weinzierl:2001ny} to generate unweighted events of `resolved partons' which follow the NLO cross section shares some features with the method presented here but there are important differences: An artificial resolution parameter is introduced in order to define the intermediate resolved partons (cf. the
\PSS techniques described in section \ref{sec:PSS}). In principle any \IR safe jet algorithm can be applied to the intermediate resolved partons yielding finite results for NLO distributions of jet observables. It is crucial that this artificial cut defining the resolved partons and the physical cut defining jets according to a certain jet algorithm do not interfere. In contrast, the presented approach is directly formulated for given resolved jet momenta defined by the jet algorithm applied in the experimental analysis. Additionally, only final-state singularities at a lepton collider are considered in \Ref{Weinzierl:2001ny} requiring only one phase space mapping. By the generalisation to more mappings, initial-state singularities which appear in the hadron collider context are included in this work as well.\\
The authors of \Refs{Dinsdale:2007mf,Schumann:2007mg} study alternative parton showers based on $3\to2$ dipole splittings. Due to the same foundation, their parameterisations of the phase space are largely analogous to what is used in this work and many of the following results are connected to (and can be cross checked with) formulae given in \Refs{Dinsdale:2007mf,Schumann:2007mg}.\\
In \Ref{Giele:2011tm} the phase space of $n+1$ massless partons is generated by forward branching of configurations of $n$ massless partons. In \Refs{Campbell:2012cz,Campbell:2012ct,Campbell:2013uha,Campbell:2013hz} the method is applied to define two $3\to2$ prescriptions for the clustering of three massless partons to two massless jets. In this work a generalisation of this approach is presented by completing it by two further clustering prescriptions and the extension to massive particles. 

In the following sections, each phase space parameterisation is initially worked out in detail for the massless case since the methodology can be presented in a clearer way due to more concise formulae compared to the massive case. Easier implementation and a better numerical performance might also serve as compelling arguments to rely on the massless instead of the massive algorithm when performing calculations dealing with respective massless particles. 

\subsection{Final-state clustering with final-state spectator}
\subsubsection{Massless particles}\label{sec:ff}
Inspired by \Ref{Catani:1996vz} in combination with the $3\to2$ jet clusterings given in \Eq{eq:CSpsmap}, the phase space of $n+1$ massless partons is parameterised in terms of a phase space of $n$ massless momenta to be identified with the jet momenta $J_1,\ldots,J_n$ and a dipole phase-space measure $dR_{ij,k}$ describing the emission of an additional massless parton $i$ from the final state $j$ with the final-state spectator $k$:
\begin{equation}\label{eq:psff}
  dR_{n+1}\left(p_1,\ldots,p_{n-2},p_j,p_k,p_{i}\right)
  =dR_{n}\left(J_1,\ldots,J_{n-2},J_{j},J_{k}\right)dR_{ij,k}.
\end{equation}
The $n$-body phase space is defined in \Eq{eq:phasespace}. The dipole phase-space measure in four space-time dimensions can be parameterised by the two jet momenta $J_j,$ $J_k$ and three integration variables ${\bf\Phi} = \{\phi,z,y\}$ (cf. ref.~\cite[(5.20)]{Catani:1996vz})
\begin{equation}\label{eq:dippsff}
dR_{ij,k}=\frac{J_j\cdot J_k}{2\left(2\pi\right)^{3}}d\phi\;dz\;dy\left(1-y\right)\Theta\left(\phi\left(2\pi-\phi\right)\right)\Theta\left(z\left(1-z\right)\right)\Theta\left(y\left(1-y\right)\right).
\end{equation}
The $n$ jet momenta are defined by the corresponding clustering prescription of $n+1$ partonic momenta to $n$ jet momenta (see \Eq{eq:CSpsmap})
\begin{equation}\label{eq:psmapff}
(p_1,\ldots,p_i,\ldots,p_j,\ldots,p_k,\ldots,p_{n+1})\mapsto(J_1,\ldots,J_j,\ldots,J_k,\ldots,J_n)
\end{equation}
with
\begin{eqnarray}\label{eq:ffmapmassless}
\nn  J_j &=& p_i+p_j-\frac{y}{1-y}p_k,\\
\nn  J_k &=&\;\frac{1}{1-y}p_k,\\
  J_m &=&\;p_m, \quad\textrm{ for } m\neq j,k.
\end{eqnarray}
Requiring momentum conservation $\sum\limits_{l=1}^n J_l=\sum\limits_{l=1}^{n+1} p_l$ and the respective on-shell conditions $J^2_l=0$ for $l=1,\ldots,n$ leads to (cf. ref.~\cite[(5.4)]{Catani:1996vz})
\begin{equation}
y={p_i\cdot p_j\over p_i\cdot p_j+p_j\cdot p_k+p_i\cdot p_k}.
\end{equation}
In this work however, parameterisations of the unresolved partonic momenta are needed.
In the following, it is shown how the unresolved real phase space can be obtained as the preimage of the clustered jets by inverting the mapping in \Eq{eq:ffmapmassless}:
\begin{equation}\label{eq:psinvmapff}
(J_1,\ldots,J_j,\ldots,J_k,\ldots,J_n;{\bf \Phi})\mapsto(p_1,\ldots,p_i,\ldots,p_j,\ldots,p_k,\ldots,p_{n+1}).
\end{equation}
From \Eq{eq:ffmapmassless} the spectator momentum $p_k$ and the momenta of the remaining partons $p_m$ ($m\neq i,j,k$) are given by the momenta $J_k, J_m$ ($m\neq i,j,k$) and the variable $y$ as
\begin{eqnarray}
\nn  p_m &=& J_m,\\
  p_k &=& \left(1-y\right)J_k.
\end{eqnarray}
Upon inspection of \Eq{eq:ffmapmassless}, the unresolved partonic momenta $p_i$ and $p_j$ can be inferred from the kinematics of a two-body decay with momenta
\begin{equation}
  p_{ij}= p_i+p_j=\;J_j+yJ_k
\end{equation}
and
\begin{equation}
  s_{ij}= \left(J_j+yJ_k\right)^2 = 2y(J_j\cdot J_k).
\end{equation}
The two-body decay is expressed most easily in the rest frame of $p_{ij}$. In this specific frame, rotated such that $J_j$ is aligned with the positive $z$-axis, the respective momentum $J_j'$ is given by (using $s_{ij}=2J_j\cdot p_{ij}$)
\begin{equation}
  \label{eq:JinRestFrame}
  J'_j = \frac{\sqrt{s_{ij}}}{2} (1,0,0,1).
\end{equation}
For a given $J_j$ in the lab frame the momentum $J'_j$ follows by a boost into the rest frame of $p_{ij}$ and two
subsequent rotations for alignment with the $z$-axis
\begin{equation}
  J'_j = \Lrotx{\phi_x} \Lroty{\theta_y}
  \Lboost{\hat p_{ij}}
  J_j.
\end{equation} 
The Lorentz transformations $\Lboost{X}$, $\Lrotx{\phi_x}$ and $\Lroty{\theta_y}$ are listed explicitly for this clustering and are easily transferable to all following clusterings. 
Let $X= (X^0,\vec{X})$ be a $4$-vector with $X^2\not=0$ given in a specific frame. To boost a $4$-momentum $p$ from the rest frame of this $4$-vector $X$ to the specific frame in which $X$ is given, a rotational-free boost defined by
\begin{equation}
  \label{eq:BoostRestframe}
  \Lboost{X} p = \left( {X^0\over \sqrt{X^2}}p^0
    +{(\vec{X}\cdot \vec{p})\over \sqrt{X^2}}, 
    \vec{p} 
    + \left[{(\vec{X}\cdot \vec{p})\over \sqrt{X^2}(X^0+\sqrt{X^2})}
    + {p^0\over \sqrt{X^2}}
    \right] \vec{X}\right)
\end{equation}
can be applied.
Defining the parity transformation of a $4$-vector $X$ as $\hat X = (X^0,-\vec{X})$ the boost from the frame of the given
$X$ to its rest frame is given by $\Lambda^b(\hat{X})$.
For rotations around the $x$- and the $y$-axis the following Lorentz transformations are needed
\begin{equation}
  \Lrotx{\phi} = 
  \left(
    \begin{array}{cccc}
      1 & 0 & 0 & 0 \\
      0 & 1 & 0 & 0 \\
      0 & 0 & \cos(\phi) & \sin(\phi) \\ 
      0 & 0 & -\sin(\phi) & \cos(\phi) \\ 
    \end{array}
  \right),\\ \quad
  \Lroty{\phi} = 
  \left(
    \begin{array}{cccc}
      1 & 0 & 0 & 0 \\
      0 & \cos(\phi)&0 & -\sin(\phi) \\ 
      0 & 0 & 1 & 0 \\
      0 &  \sin(\phi)&0 & \cos(\phi) \\ 
    \end{array}
  \right).
\end{equation}
The product is given by
\begin{equation}
  \Lroty{\theta} \Lrotx{\phi} = 
  \left(\begin{array}{cccc}
    1&0&0&0 \\
    0&\cos{\theta}&\sin{\theta}\sin{\phi}&-\sin{\theta}\cos{\phi}\\
    0&0&\cos{\phi}&\sin{\phi} \\
    0&\sin{\theta}&-\cos{\theta}\sin{\phi}&\cos{\theta}\cos{\phi}\\
  \end{array}
\right).
\end{equation}
In order to align $J_j$ with the positive $z$-axis the rotation angles $\theta_y$ and $\phi_x$ are given by
\begin{eqnarray}
  \label{eq:thetaangle}
  &&\cos(\theta_y)=\frac{J^z_j}{\sqrt{(J^x_j)^{2}+(J^z_j)^{2}}},\quad 
  \sin(\theta_y)=\frac{J^x_j}{\sqrt{(J^x_j)^{2}+(J^z_j)^{2}}},\\
  \label{eq:phiangle}
  &&\cos(\phi_x)=\frac{\sqrt{(J^x_j)^{2}+(J^z_j)^{2}}}{|\vec{J}_j|},\quad 
  \sin(\phi_x)=\frac{-J^y_j}{|\vec{J}_j|}.
\end{eqnarray}
Using the definition from ref.~\cite[(5.6)]{Catani:1996vz}
\begin{equation}
  z = {2 (p_i\cdot J_k)\over 2 (J_j\cdot J_k)},
\end{equation}
the unresolved partonic momenta $p'_i$ and $p'_j$ in this frame can be parameterised in terms of the integration variables ${\bf\Phi} = \{\phi,z,y\}$ in \Eq{eq:dippsff} as
\begin{eqnarray}
  \label{eq:iimassless1}
\nn  p'_i &=& {\sqrt{s_{ij}}\over 2} 
  (1,2\sqrt{z(1-z)}\cos\phi,\;2\sqrt{z(1-z)}\sin\phi,\;2z-1),\\
  p'_j &=& \hat p'_i.
\end{eqnarray}
By inverting the Lorentz transformations the momenta $p_i,p_j$ are obtained from $p'_i,p'_j$ as
\begin{eqnarray}
  \label{eq:iimassless3}
\nn  p_i &=& \Lboost{p_{ij}} \Lroty{-\theta_y} \Lrotx{-\phi_x} p'_i,\\
  \label{eq:iimassless4}
  p_j &=& \Lboost{p_{ij}} \Lroty{-\theta_y} \Lrotx{-\phi_x} p'_j.
\end{eqnarray}
Note that the inverse of $\Lboost{\hat p_{ij}}$ is given by $\Lboost{p_{ij}}$.

\subsubsection{Massive particles}\label{sec:ffm}
The phase space factorisation in terms of a phase space of $n$ jets and the dipole phase space 
measure $dR_{ij,k}$ describing the emission of an additional parton $i$ from the final state $j$ with the final-state spectator $k$ can also be achieved for massive partons and jets (see \Ref{Catani:2002hc}):
\begin{equation}
  dR_{n+1}\left(p_1,\ldots,p_{n-2},p_i,p_k,p_{j}\right)
  = dR_{n}\left(J_1,\ldots,J_{n-2},J_j,J_{k}\right)dR_{ij,k}.
\end{equation}
Again, the phase space of $n$ (massive) momenta is given by \Eq{eq:phasespace} with (some) $m_l\neq 0$. 
In four space-time dimensions the dipole phase-space measure reads (cf. ref.~\cite[(5.11)]{Catani:2002hc})
\begin{eqnarray}\label{eq:dippsffm}
\nn  dR_{ij,k} &=& \frac{Q^2}{4\left(2\pi\right)^{3}}
  \frac{\left(1-\mu^2_i-\mu^2_j-\mu^2_k\right)^2}
  {\sqrt{\Kallen{1,\mu^2_{ij},\mu^2_k}}}
  \Theta\left(1-\mu_i-\mu_j-\mu_k\right)\\
  &\times& d\phi\;dz\;dy\left(1-y\right)
  \Theta\left(\phi\left(2\pi-\phi\right)\right)
  \Theta\left(\left(z-z_-\right)\left(z_+-z\right)\right)
  \Theta\left(\left(y-y_-\right)\left(y_+-y\right)\right),
\end{eqnarray}
where the K\"all\'en function $\lambda$ is defined as
\begin{equation}
\Kallen{x,y,z}=x^2+y^2+z^2 - 2xy - 2xz - 2yz.
\end{equation}
The rescaled masses are given by
\begin{equation}
  \mu_n=\frac{m_n}{\sqrt{Q^2}},\quad
Q=p_i+p_j+p_k=J_j+J_k
\end{equation}
and the clustered jet $J_j$ is required to have the fixed mass 
\begin{equation}
  m_{ij}=\sqrt{J^2_j}.
\end{equation}
The integration boundaries in \Eq{eq:dippsffm} are given in ref.~\cite[(5.13)]{Catani:2002hc} as
\begin{eqnarray}
\nn  y_-&=&\frac{2\mu_i\mu_j}{1-\mu^2_i-\mu^2_j-\mu^2_k},\\
\nn  y_-&=&1-\frac{2\mu_k\left(1-\mu_k\right)}{1-\mu^2_i-\mu^2_j-\mu^2_k},\\
  z_{\pm}&=&\frac{2\mu^2_i+\left(1-\mu^2_i-\mu^2_j-\mu^2_k\right)y}{2\left[\mu^2_i+\mu^2_j+\left(1-\mu^2_i-\mu^2_j-\mu^2_k\right)y\right]}
  \left(1\pm \varv_{ij,i}\varv_{ij,k}\right)
\end{eqnarray}
with the relative velocities between $p_i+p_j$ and $p_i$ or $p_k$ (see ref.~\cite[(5.14)]{Catani:2002hc})
\begin{eqnarray}\label{eq:relvel}
\nn  \varv_{ij,i}&=&\frac{\sqrt{\left(1-\mu^2_i-\mu^2_j-\mu^2_k\right)^2y^2-4\mu^2_i\mu^2_j}}{\left(1-\mu^2_i-\mu^2_j-\mu^2_k\right)y+2\mu^2_i},\\
  \varv_{ij,k}&=&\frac{\sqrt{\left[2\mu^2_k+\left(1-\mu^2_i-\mu^2_j-\mu^2_k\right)\left(1-y\right)\right]^2-4\mu^2_k}}{\left(1-\mu^2_i-\mu^2_j-\mu^2_k\right)\left(1-y\right)}.
\end{eqnarray}

The $n$ jet momenta are defined by the corresponding clustering prescription of $n+1$ massive partonic momenta to $n$ massive jet momenta  (see \Eq{eq:CSpsmap})
\begin{eqnarray}\label{eq:ffmapmassive}
\nn  J_{k} &=& \sqrt{\frac{\Kallen{1,\mu^2_{ij},\mu^2_k}}
   {\Kallen{1,\frac{s_{ij}}{Q^2},\mu^2_k}}}p_k
  +\left(-\sqrt{\frac{\Kallen{1,\mu^2_{ij},\mu^2_k}}
      {\Kallen{1,\frac{s_{ij}}{Q^2},\mu^2_k}}}\frac{2p_k\cdot
  Q}{Q^2}+\mu^2_k-\mu^2_{ij}+1\right)\frac{Q}{2}\;\equiv\; A_k p_k + A_{ij} (p_i+p_j),\\
\nn  J_{j}&=&Q-J_{k},   \\ 
  J_m&=&p_m, \quad (m\neq j,k).
\end{eqnarray}
The clustering fulfils momentum conservation $\sum\limits_{l=1}^n J_l=\sum\limits_{l=1}^{n+1} p_l$
and the on-shell conditions $J^2_j=m^2_{ij},\, J^2_l=m^2_l$ for $l\neq j$.
When inverting this clustering the spectator momentum $p_k$ and the momenta of the remaining partons $p_m$ ($m\neq i,j,k$) can be read off as (using $2 p_k\cdot Q = Q^2 + m_k^2 - s_{ij}$)
\begin{eqnarray}
\nn  p_k&=&\left[J_k-\left(1+\mu^2_k-\mu^2_{ij}\right)
    \frac{Q}{2}\right]\sqrt{
    \frac{\Kallen{1,\frac{s_{ij}}{Q^2},\mu^2_k}}
    {\Kallen{1,\mu^2_{ij},\mu^2_k}}}
  +\left[\left(1-y\right)\left(1-\mu^2_i-\mu^2_j-\mu^2_k\right)
    +2\mu^2_k\right]\frac{Q}{2},\\
  p_m&=&J_m,
\end{eqnarray}
where the following definition is introduced (cf.  ref.~\cite[(5.12)]{Catani:2002hc}):
\begin{equation}
  y  = {s_{ij} - m_i^2 -m_j^2 \over
    Q^2 - m_i^2-m_j^2-m_k^2}.
\end{equation}
As in the massless case the unresolved partonic momenta $p_i$ and $p_j$ can be inferred from the kinematics of a two-body decay with momenta
\begin{equation}
  p_{ij}= p_i+p_j=Q-p_k.
\end{equation} 
Again, it is convenient to parameterise the two-body decay in the rest frame of $p_{ij}$ rotated
such that $Q$ in the respective frame is aligned with the positive $z$-axis
\begin{equation}
  Q'= \Lrotx{\phi_x} \Lroty{\theta_y} \Lboost{\hat p_{ij}} Q.
\end{equation} 
The respective momentum $Q'$ in this specific frame reads (using $2Q\cdot p_{ij} = Q^2 + s_{ij}-m_k^2$)
\begin{equation}
  Q'=\frac{Q^2}{2\sqrt{s_{ij}}}\left(\frac{s_{ij}}{Q^2}+1-\mu^2_k,0,0,
    \sqrt{\Kallen{1,\frac{s_{ij}}{Q^2},\mu^2_k}}\right).
\end{equation}
The angles are similar to the ones defined in \Eq{eq:thetaangle} and \Eq{eq:phiangle}.
In this frame the unresolved momenta $p'_i$ and $p'_j$ can be parameterised in terms of the integration variables ${\bf\Phi} = \{\phi,z,y\}$ in \Eq{eq:dippsffm} as
\begin{eqnarray}
\nn  p'_i&=&\left(\frac{Q^2}{2\sqrt{s_{ij}}}
  \left(\frac{s_{ij}}{Q^2}+\mu^2_i-\mu^2_j\right),
  \left|\vec{p}'_i\right| (\sin\theta'\cos\phi,\;\sin\theta'\sin\phi,
    \;\cos\theta') \right),\\
  p'_j&=&\left( \frac{Q^2}{2\sqrt{s_{ij}}}
  \left(\frac{s_{ij}}{Q^2}+\mu^2_j-\mu^2_i\right), -\vec{p}'_i \right),
\end{eqnarray}
with $  \left|\vec{p}'_i\right|=\sqrt{(p'^0_i)^2-m^2_i} = {1\over 2
  \sqrt{s_{ij}}} \sqrt{\Kallen{s_{ij},m_i^2,m_j^2}}$.
Identifying (see ref.~\cite[(5.12)]{Catani:2002hc})
\begin{equation}
z={p_i\cdot p_k\over p_i\cdot p_k+p_j\cdot p_k}
\end{equation}
and using
\begin{eqnarray}
  s_{ik}&=&\left(p_i+p_k\right)^2=\;Q^2\left[z\left(1-y\right)
    \left(1-\mu^2_i-\mu^2_j-\mu^2_k\right)+\mu^2_i+\mu^2_k\right]
\end{eqnarray}
yields
\begin{equation}
  \cos{\theta'} = {Q^2 (1-y)\*(1-\mu_i^2-\mu_j^2-\mu_k^2)\*
 [ ((1-\mu_i^2-\mu_j^2-\mu_k^2) y 
   + \mu_i^2+\mu_j^2 ) (1-2 z)-\mu_j^2+\mu_i^2]\over
   \sqrt{\Kallen{s_{ij},m_i^2,m_j^2}}
     \sqrt{\Kallen{1,\frac{s_{ij}}{Q^2},\mu^2_k}}}.
\end{equation}
When exchanging $\mu_i^2 \leftrightarrow \mu_j^2, z \to 1-z$ this becomes $\cos(\theta') \to - \cos(\theta')$ as it should be.
The momenta $p_i,p_j$ are obtained again from $p'_i,p'_j$ by an inversion of the Lorentz transformations
\begin{eqnarray} 
\nn  p_i &=& \Lboost{p_{ij}} \Lroty{-\theta_y} \Lrotx{-\phi_x} p'_i,\\
  p_j &=& \Lboost{p_{ij}} \Lroty{-\theta_y} \Lrotx{-\phi_x} p'_j.
\end{eqnarray}
Note that when setting $m_i=m_j=m_k=m_{ij}=0$ in the formulae above the massless case (cf. section \ref{sec:ff}) is recovered.

\subsubsection{Modifications beneficial for \PSS}
When \PSS techniques are used to carry out the mutual cancellation of the \IR singularities, the real phase space is typically split into soft/collinear and hard regions defined through a slicing parameter $s_{\min}$ with regards to the kinematic invariants $s_{lm}=2p_l\cdot p_m$ (see section \ref{sec:PSS}). Since the phase space boundaries of the respective regions are given in terms of the slicing parameter, it proves beneficial for the numerical integration of the hard contribution to express the integration via the kinematic 
invariants $s_{lm}$ which are directly constrained by the slicing parameter $s_{\tmin}$. This way not only the integration boundaries can be explicitly implemented in a straightforward manner but also the logarithms of the slicing parameter which numerically build up due to poles $\propto 1/s_{lm}$ can be made manifest while absorbing the pole into the integration measure by appropriate samplings, e.g. (cf. \Eq{eq:PSSform})
\begin{equation}
\int\limits_{0}^{s}ds_{lm}\;{1\over s_{lm}}\Theta\left(s_{lm}-s_\tmin\right)=-\log{s_{\tmin}\over s}=\int\limits_{s_{\tmin}/s}^{1}dx\;{1\over x}\;\overset{\rho=\log{x}}{=}\int\limits_{\log{s_{\tmin}/s}}^{0}d\rho.
\end{equation}
For the massive final-state clustering with a final-state spectator the dipole phase space $dR_{ij,k}$ from \Eq{eq:dippsffm} can also be parameterised with the invariants
\begin{equation}
  Q^2=\left(J_j+J_k\right)^2, \quad 
  s'_{ij}=2p_i\cdot p_j,\quad
  s_{ij}=s'_{ij}+m^2_i+m^2_j, \quad  
  s'_{ik}=2p_i\cdot p_k
\end{equation}
as
\begin{eqnarray}
\nonumber dR_{ij,k} &=&
  \frac{1}{32\pi^{3}\sqrt{\Kallen{Q^2,J^2_j,m^2_k}}}d\phi\;ds'_{ij}\;
  ds'_{ik}\\&&\times\;\Theta\left(\phi\left(2\pi-\phi\right)\right)
  \Theta\left(\left(s'_{ij}-{s'_{ij}}^{-}\right)
    \left({s'_{ij}}^{+}-s'_{ij}\right)\right)\Theta
  \left(\left(s'_{ik}-{s'_{ik}}^{-}\right)\left({s'_{ik}}^{+}
      -s'_{ik}\right)\right).
\end{eqnarray}
The boundaries for the integration over the invariants are given by
\begin{equation}
{s'_{ij}}^{-}=2m_im_j,\quad {s'_{ij}}^{+}=\left(|Q|-m_k\right)^2-m^2_i-m^2_j,\\
\quad{s'_{ik}}^{\mp}=(Q^2-m^2_k-s_{ij})\frac{(2m_i^2+s'_{ij})(1\mp v_{ij,i}v_{ij,k})}{2s_{ij}}.
\end{equation}
with the relative velocities $v_{ij,i},v_{ij,k}$ defined in \Eq{eq:relvel}.
The parameterisation of the real phase space proceeds as above with
\begin{equation}
 y=\frac{s'_{ij}}{Q^2-m^2_i-m^2_j-m^2_k},\quad 
 z=\frac{s'_{ik}}{Q^2-m^2_k-s_{ij}} .
\end{equation}
The respective modifications for the massless case follow from setting $m_i=m_j=m_k=m_{ij}=0$.

\subsection{Final-state clustering with initial-state spectator}
\subsubsection{Massless particles}\label{sec:fi}
According to \Ref{Catani:1996vz} the phase space of $n+1$ massless partons can be written as the convolution
of the phase space of $n$ massless jets with the dipole phase-space measure $dR_{ij,a}$ for the emission of the additional parton $i$ from the final state $j$ with the initial-state spectator $a$
\begin{equation}
\label{eq:psfi}
dR_{n+1}\left(p_a,p_b; p_1,\ldots,p_{n-1},p_j,p_i\right)
=dR_{n}\left(xp_a,p_b;J_1,\ldots,J_{n-1},J_j\right)dR_{ij,a}.
\end{equation}
The incoming partons' momenta in the lab frame are denoted by the $4$-vectors $p_a$ and $p_b$ defined in \Eq{eq:partonmom}.
In four space-time dimensions the dipole phase-space measure can be parameterised by the jet momentum $J_j$, the initial-state momentum $p_a$ and three integration variables ${\bf\Phi} = \{\phi,z,x\}$  (see ref.~\cite[(5.48)]{Catani:1996vz})
\begin{equation}\label{eq:dippsfi}
dR_{ij,a}=\frac{J_j\cdot p_a}{2(2\pi)^3}d\phi\;dz\;dx\;\Theta\left(\phi\left(2\pi-\phi\right)\right)\Theta\left(z\left(1-z\right)\right)\Theta\left(x\left(1-x\right)\right)
\end{equation}
 including an integration over $x$ leading to a convolution of the phase spaces with the PDFs and the flux factor $1/(2x_ax_bs)$ (cf. \Eq{eq:hadpart})
\begin{eqnarray}\label{eq:ficonv}
\nn& &dx_adx_b\frac{f_{a}\left(x_a\right)f_{b}\left(x_b\right)}{2x_ax_bs}dR_{n+1}\left(p_a,p_b; p_1,\ldots,p_{n-1},p_j,p_i\right)\\
&\overset{x_A=xx_a}{=}&dx_Adx_b\frac{f_{a}\left(\frac{x_A}{x}\right)f_{b}\left(x_b\right)}{2x_Ax_bs}dR_{n}\left(xp_a,p_b;J_1,\ldots,J_{n-1},J_j\right)dR_{ij,a}(\phi,z,x).
\end{eqnarray}
The $n$ jet momenta are defined by the corresponding clustering prescription of $n+1$ partonic momenta to $n$ jet momenta  (see \Eq{eq:CSpsmap})
\begin{equation}\label{eq:psmapfi}
(p_1,\ldots,p_i,\ldots,p_j,\ldots,p_{n+1})\mapsto(J_1,\ldots,J_j,\ldots,J_n)
\end{equation}
with
\begin{eqnarray}\label{eq:fimapmassless}
\nn  J_j&=&\;p_i+p_j-\left(1-x\right)p_a,\\
  J_m&=&\;p_m\quad (m\neq i,j)
\end{eqnarray}
Requiring momentum conservation $\sum\limits_{l=1}^n J_l= x p_a+ p_b$ and the on-shell conditions $J^2_l=0$ for $l=1,\ldots,n$ leads to  (see ref.~\cite[(5.38)]{Catani:1996vz})
\begin{equation}
x={p_i\cdot p_a +p_j\cdot p_a - p_i\cdot p_j \over (p_i+p_j)\cdot p_a}.
\end{equation}
Inversion of this mapping and identifying the unresolved partonic configurations as the preimage of the clustered jets in \Eq{eq:fimapmassless}
 allows to parameterise the real phase space by means of the $n$ jet momenta and three integration variables ${\bf\Phi} = \{\phi,z,x\}$ in \Eq{eq:dippsfi}
\begin{equation}\label{eq:psinvmapfi}
(J_1,\ldots,J_j,\ldots,J_n;{\bf \Phi})\mapsto(p_1,\ldots,p_i,\ldots,p_j,\ldots,p_{n+1}).
\end{equation} 
The momenta of the remaining  partons can be simply read off from \Eq{eq:fimapmassless} as
\begin{equation}
p_m=J_m, \quad (m\neq i,j).
\end{equation}
Because of
\begin{eqnarray}
\nn p_{ij}&=&p_i+p_j=\;J_j+\left(1-x\right)p_a,\\
s_{ij}&=&\left(J_j+\left(1-x\right)p_a\right)^2,
\end{eqnarray}
the unresolved momenta $p_i$ and $p_j$ can be deduced from the kinematics of a two-body decay by steps analogous to 
\Eq{eq:JinRestFrame} to \Eq{eq:iimassless4}.

\subsubsection{Massive particles}\label{sec:fim}
The phase space of $n+1$ massive partons can be expressed as a phase space of $n$ massive jets to be convoluted with the dipole
phase-space measure $dR_{ij,a}$ for the emission of the additional parton $i$ from the final state $j$ with the initial-state spectator $a$ (see \Ref{Catani:2002hc})
\begin{equation}
  dR_{n+1}\left(p_a,p_b; p_1,\ldots,p_{n-1},p_j,p_i\right)
  =dR_{n}\left(xp_a,p_b; J_1,\ldots,J_{n-1},J_j\right)dR_{ij,a}.
\end{equation}
In four space-time dimensions the dipole phase-space measure taken from ref.~\cite[(5.48)]{Catani:2002hc} reads
\begin{equation}\label{eq:dippsfim}
dR_{ij,a}=\frac{J_j\cdot p_a}{2\left(2\pi^3\right)}d\phi\;dz\;dx\;\Theta\left(\phi\left(2\pi-\phi\right)\right)\Theta\left(\left(z-z_-\right)\left(z_+-z\right)\right)\Theta\left(x\left(x_+-x\right)\right).
\end{equation}
 including an integration over $x$ leading to a convolution of the phase spaces with the PDFs and the flux factor analogous to \Eq{eq:ficonv}.
The integration boundaries read (cf. ref.~\cite[(5.47),(5.49)]{Catani:2002hc})
\begin{eqnarray}
\nn x_+&=&1+\mu^2_{ij}-\left(\mu_i+\mu_j\right)^2,\\
z_{\pm}&=&\frac{1-x+\mu^2_{ij}+\mu^2_i-\mu^2_j\pm\sqrt{\left(1-x+\mu^2_{ij}-\mu^2_i-\mu^2_j\right)^2-4\mu^2_i\mu^2_j}}{2\left(1-x+\mu^2_{ij}\right)}.
\end{eqnarray}
The rescaled masses are given by
\begin{equation}
\mu_n=\frac{m_n}{\sqrt{2J_j\cdot p_a}}
\end{equation}
and the clustered jet momentum $J_j$ is required to have the fixed mass
\begin{equation}
m_{ij}=\sqrt{J^2_j}.
\end{equation}
The $n$ jet momenta are defined by the corresponding clustering prescription of $n+1$ partonic momenta to $n$ jet momenta as in \Eq{eq:fimapmassless}
but requiring now the on-shell conditions $J^2_j=m^2_{ij}$ and $J^2_l=m^2_l$ for $l\neq j$ yields (see ref.~\cite[(5.42)]{Catani:2002hc})
\begin{equation}
x={p_i\cdot p_a +p_j\cdot p_a - p_i\cdot p_j +{1\over 2}(m^2_{ij}-m^2_i-m^2_j)\over (p_i+p_j)\cdot p_a}.
\end{equation}
Again, obtaining the unresolved partonic momenta $p_i$ and $p_j$ from the kinematics of a two-body decay with momenta
\begin{equation}
p_{ij}=p_i+p_j=J_j+(1-x)p_a
\end{equation}
is most convenient when starting in the rest frame of $p_{ij}$ with the momentum $J_j$ in this frame aligned with the positive $z$-axis.
The corresponding momentum $J'_j$ in this specific frame is given by (using $2p_{ij}\cdot J_j = s_{ij}+m_{ij}^2$)
\begin{equation}
  J'_j = \frac{1}{2\sqrt{s_{ij}}}\left(s_{ij}+m_{ij}^2,
    0,0,s_{ij}-m_{ij}^2\right).
\end{equation}
The jet momentum $J'_j$ is related to the given jet momentum $J_j$ by a Lorentz boost and two rotations
\begin{equation}
\label{eq:Jtransformation}
J'_j = \Lrotx{\phi_x} \Lroty{\theta_y} \Lboost{\hat p_{ij}}J_j .
\end{equation} 
The partonic momenta $p'_i$ and $p'_j$ can be parameterised in terms of the jet momentum $J_j$ and the integration variables ${\bf\Phi} = \{\phi,z,x\}$ in \Eq{eq:dippsfim} as
\begin{eqnarray}
\nn  p'_i &=& \left(\frac{s_{ij}-m^2_j+m^2_i}{2\sqrt{s_{ij}}},
    |\vec{p}'_i|\left(\sin\theta'\cos\phi,\;\sin\theta'\sin\phi,\;\cos\theta'\right)\right),\\
  p'_j & = & \left(\frac{s_{ij}-m^2_i+m^2_j}{2\sqrt{s_{ij}}},-\vec{p}'_i\right),
\end{eqnarray}
with $\left|\vec{p}'_i\right|=\sqrt{{E'_i}^2-m^2_i}={1\over
  2\sqrt{s_ij}} \sqrt{\Kallen{s_{ij},m_i^2,m_j^2}}$ and
\begin{align}
 \cos{\theta'}
=&\frac{m^2_j-m^2_i - (1-2z)s_{ij}}{
  \sqrt{\Kallen{s_{ij},m_i^2,m_j^2}}
}.
\end{align}
To obtain the unresolved partonic momenta $p_i$ and $p_j$ from $p'_i$ and $p'_j$ the Lorentz transformations in \Eq{eq:Jtransformation} have to be inverted.
Note that when setting $m_i=m_j=0$ in the formulae above the massless case (cf. section \ref{sec:fi}) is recovered.

\subsubsection{Modifications beneficial for \PSS}
Having the application of the \PSS method in mind, it proves again beneficial to perform a transformation of the integration variables in  
section \Eq{eq:dippsfim} in favour of the invariants
\begin{equation}
  Q^2=2xJ_j\cdot p_a, \quad
  s'_{ij}=2p_i\cdot p_j, \quad
 s_{ij}=s'_{ij}+m^2_i+m^2_j, \quad
 s'_{ia}=2p_i\cdot p_a 
\end{equation}
The phase-space measure to be convoluted with the $n$-jet phase space (and the PDFs  and flux factor) now reads
\begin{eqnarray}
\nonumber dR_{ij,a}&=&\frac{Q^2}{32\pi^3(Q^2-J^2_j+s_{ij})^2}
\;d\phi\;ds'_{ij}\;ds'_{ia}\\
&&\times\;\Theta\left(\phi\left(2\pi-\phi\right)\right)\Theta\left(\left(s'_{ia}-{s'_{ia}}^{-}\right)\left({s'_{ia}}^{+} -s'_{ia}\right)\right)\Theta\left(\left(s'_{ij}-{s'_{ij}}^{-}\right)\left({s'_{ij}}^{+} -s'_{ij}\right)\right).
\end{eqnarray}
The boundaries for the integration over the invariants can be written as
\begin{eqnarray}
& z^{\mp}=\frac{s_{ij}+m^2_i-m^2_j\mp\sqrt{(s_{ij}-m^2_i-m^2_j)^2-4m^2_im^2_j}}{2s_{ij}},\quad {s'_{ia}}^{\mp}=z^{\mp}(Q^2-J^2_j+s_{ij}),&\\
&{s'_{ij}}^{-}=2m_im_j,\quad
{s'_{ij}}^{+}=\frac{1-x_A}{x_A} Q^2 +J^2_j -m^2_i -m^2_j.&
\end{eqnarray}
with $x_A$ defined in \Eq{eq:ficonv}.
The generation of the real phase space proceeds as in section \ref{sec:fim} with
\begin{equation}
 x=\frac{Q^2}{Q^2-J^2_j+s_{ij}},\quad
 z=\frac{s'_{ia}}{Q^2-J^2_j+s_{ij}}.
\end{equation}
The respective modifications for the massless case follow from setting $m_i=m_j=0$.

\subsection{Initial-state clustering with final-state spectator }
\subsubsection{Massless particles}\label{sec:if}
The phase space of $n+1$ massless partons can be expressed as a convolution of the phase space of $n$ massless jets with the dipole phase-space measure $dR_{ia,k}$ for the emission of an additional massless parton $i$ from the initial state $a$ with a massless final-state spectator $k$. As stated in Ref. \cite{Catani:1996vz}, this can be achieved by the replacements $a\rightarrow k$ and $j\rightarrow a$ in the formulae from section \ref{sec:fi}. Collinear initial-state singularities which now appear due to the clustering in the initial state have to be factorised into the parton distribution functions (see section \ref{sec:factorisation}). Now $x$ behaves in the initial-state collinear limit as (cf. from ref.~\cite[(5.63)]{Catani:1996vz}) 
\begin{subequations}\label{eq:colfacif}
\begin{align}
x=\frac{p_{a}\cdot\left(p_{i}+p_{k}\right)-p_{i}\cdot p_{k}}{p_{a}\cdot\left(p_{i}+p_{k}\right)}\xrightarrow[p_i\rightarrow (1-z)p_a]{}z.
\end{align}
A jet function defined with this initial-state clustering in combination with a final-state spectator automatically fulfils the condition for factorisability of initial-state collinear singularities (cf. \Eq{eq:collfactable})
\begin{align}
F^{n+1}_{J_1,\ldots,J_n}\left(p_a,p_b;p_1,\ldots,p_{n-1},p_k,p_i\right)\xrightarrow[p_i\rightarrow (1-z)p_a]{}F^{n}_{J_1,\ldots,J_n}\left(zp_a,p_bp_1,\ldots,p_{n-1},p_k\right).
\end{align}
\end{subequations}

\subsubsection{Massive particles}
As in the massless case, the phase space of $n$ massive partons and one massless parton can be expressed as a convolution of the phase space of $n$ massive jets with the dipole phase-space measure $dR_{ia,k}$ for the emission of an additional massless
parton $i$ from the initial state $a$ with a massive final-state spectator $k$. Again, this can be achieved by the replacements $a\rightarrow k$ and $j\rightarrow a$, $m_i\rightarrow 0$ together with $m_{ij}\rightarrow m_k$ in the formulae from section \ref{sec:fim} (see \Ref{Catani:2002hc}). The argument guaranteeing factorisability of initial-state collinear singularities from \Eq{eq:colfacif} also holds after these replacements. Upon setting $m_k=0$ the massless case (cf. section \ref{sec:if}) is recovered.

\subsection{Initial-state clustering with initial-state spectator}\label{sec:ii}
When choosing an initial-state spectator $b$ for the emission of an additional parton $i$ from the initial state $a$ the phase space for $n$ massless or massive partons and one additional massless parton can again be written as a convolution of the phase space for $n$ massless or massive jets with the dipole phase-space measure $dR_{ia,b}$ (see \Ref{Catani:1996vz})
\begin{equation}\label{eq:psii}
  dR_{n+1}\left(p_a,p_b;p_1,\ldots,p_n,p_i\right)
  =dR_{n}\left(xp_a,p_b;J_1,\ldots,J_{n}\right)dR_{ia,b}.
\end{equation}
In four space-time dimensions the dipole phase-space measure $dR_{ia,b}$ can be parameterised by the momenta of the incoming partons $p_a,p_b$ and three integration variables ${\bf\Phi} = \{\phi,v,x\}$ (cf. ref.~\cite[(5.151)]{Catani:1996vz})
\begin{align}\label{eq:dippsii}
  dR_{ia,b}=\frac{p_a\cdot p_b}{2\left(2\pi\right)^3}
  d\phi\;dv\;dx\;\Theta\left(\phi\left(2\pi-\phi\right)\right)
  \Theta\left(v\right)\Theta\left(1-\frac{v}{1-x}\right)
  \Theta\left(x\left(1-x\right)\right)
\end{align}
 including an integration over $x$ leading to a convolution of the phase spaces with the PDFs  and the flux factor
\begin{eqnarray}\label{eq:iiconv}
\nn& &dx_adx_b\frac{f_{a}\left(x_a\right)f_{b}\left(x_b\right)}{2x_ax_bs}dR_{n+1}\left(p_a,p_b; p_1,\ldots,p_{n},p_i\right)\\
&\overset{x_A=xx_a}{=}&dx_Adx_b\frac{f_{a}\left(\frac{x_A}{x}\right)f_{b}\left(x_b\right)}{2x_Ax_bs}dR_{n}\left(xp_a,p_b;J_1,\ldots,J_{n}\right)dR_{ia,b}(\phi,v,x).
\end{eqnarray}
The $n$ jet momenta are defined by the corresponding clustering of $n+1$ (massless/massive) partonic momenta to $n$ (massless/massive) jet momenta  (see \Eq{eq:CSpsmap})
\begin{equation}\label{eq:psmapii}
(p_1,\ldots,p_i,\ldots,p_{n+1})\mapsto(J_1,\ldots,J_n)
\end{equation}
with
\begin{equation}\label{eq:iimap}
J_m=\;{\Lambda_{ia,b}}\;p_m, \quad m\neq i.
\end{equation}
The Lorentz transformation of all outgoing momenta $p_m$ ($m\neq i$) to balance the transverse momentum is given by (see ref.~\cite[(5.144)]{Catani:1996vz})
\begin{eqnarray}\label{eq:ltrsfii}
  {\left[\Lambda_{ia,b}\right]^{\mu}}_{\nu}&=& {g^{\mu}}_{\nu}
  -\frac{2\left(K+\widetilde{K}\right)^{\mu}\left(K+\widetilde{K}\right)_{\nu}}
  {\left(K+\widetilde{K}\right)^2}
  +\frac{2\widetilde{K}^{\mu}K_{\nu}}{K^2}
\end{eqnarray}
and transforms $K$ into $\widetilde K$ 
\begin{equation}
K=\;p_a+p_b-p_i,\quad \widetilde{K}=\;xp_{a}+p_b.
\end{equation}
To obtain the inverse of the boost $K$ and $\widetilde K$ have to be exchanged.  Defining $x$ as (see ref.~\cite[(5.138)]{Catani:1996vz}))
\begin{equation}\label{eq:xii}
x={p_a\cdot p_b-p_i\cdot p_a-p_i\cdot p_b\over p_a\cdot p_b}
\end{equation}
guarantees that the transformation of the whole final state in \Eq{eq:iimap} does not affect momentum conservation $\sum\limits_{i=1}^n J_i=x p_a+p_b$ and the on-shell conditions $J^2_l=m^2_l$ ($l=1,\ldots,n$).

Again, a parameterisation of the unresolved partonic momenta is needed in this work.
The preimage of the mapping in \Eq{eq:psmapii} is parameterised in terms of the $n$ jet momenta and the integration variables ${\bf\Phi} = \{\phi,v,x\}$
 \begin{equation}\label{eq:psinvmapii}
(J_1,\ldots,J_n;{\bf \Phi})\mapsto(p_1,\ldots,p_i,\ldots,p_{n+1}).
\end{equation} 
In \Ref{Catani:1996vz} $v$ is defined through
\begin{equation}
   s_{ia}=\;v s_{ab}. 
\end{equation}
From the definition of $x$ in \Eq{eq:xii} follows
\begin{equation}
  s_{ib}=\left(1-x-v\right)s_{ab}
\end{equation}
which leads to
\begin{equation}\label{eq:pipapb}
  s_{ia}+s_{ib} = (1-x)s_{ab}.
\end{equation}
It is straightforward to express the momentum $p_i$ in the rest frame of $p_a+p_b$ rotated such that $p_a$ is aligned with the positive $z$-axis
\begin{equation}
  p'_a = \frac{\sqrt{s_{ab}}}{2}\left(1,0,0,1\right).
\end{equation}
In this specific frame the momentum $p_i$ can be parameterised in terms of the three integration variables ${\bf\Phi} = \{\phi,v,x\}$ in \Eq{eq:dippsii} as
\begin{equation}
  p'_i=\left(1-x\right)\frac{\sqrt{s_{ab}}}{2}(1,\sin\theta'_i\cos\phi,
  \;\sin\theta'_i\sin\phi,\;\cos\theta'_i)
\end{equation}
with the angle $\theta'$ given by
\begin{equation}
\cos\theta'_i=1-\frac{2v}{1-x}.
\end{equation}
The momentum $p_i$  in the rest frame of $p_a+p_b$ is obtained by inverting the rotations of $p'_i$
\begin{equation}
  p_i=\Lroty{-\theta_y}\Lrotx{-\phi_x}p'_i
\end{equation}
with the rotation angles defined in \Eq{eq:thetaangle} and \Eq{eq:phiangle} but with $J_j \to p_a$. Note that no boost is needed here since $p'_i$ is already defined in the rest frame of $p_a+p_b$. For each unresolved momentum $p_i$ the inversion of the clustering in \Eq{eq:ltrsfii} yields the final-state momenta $p_m$ $(m=1,\ldots,n$) by boosting the jet momenta back to the partonic frame
\begin{equation}
  p_m=\Lambda_{ia,b}^{-1}\;J_m.
\end{equation}
Since the two incoming partons and the parton which is clustered with the initial state are always assumed to be massless, the massive case does not need to be studied.

Note that $x$ from \Eq{eq:xii} behaves in the initial-state collinear limit as
\begin{subequations}\label{eq:colfacii}
\begin{align}
x={p_a\cdot p_b-p_i\cdot p_a-p_i\cdot p_b\over p_a\cdot p_b}\xrightarrow[p_i\rightarrow (1-z)p_a]{}z.
\end{align}
A jet function defined with this initial-state clustering in combination with an initial-state spectator automatically fulfils the condition for factorisability of initial-state collinear singularities (cf. \Eq{eq:collfactable})
\begin{align}
F^{n+1}_{J_1,\ldots,J_n}\left(p_a,p_b;p_1,\ldots,p_{n},p_i\right)\xrightarrow[p_i\rightarrow (1-z)p_a]{}F^{n}_{J_1,\ldots,J_n}\left(zp_a,p_bp_1,\ldots,p_{n}\right).
\end{align}
\end{subequations}

\subsubsection{Modifications beneficial for \PSS}
For the application of \PSS it proves again beneficial to express the phase space to be convoluted with the PDFs and the flux factor in terms of the invariants
\begin{equation}
Q^2=2xp_a\cdot p_b,\quad 
s_{ia}=2p_i\cdot p_a, \quad
s_{ib}=2p_i\cdot p_b.  
\end{equation}
The phase-space measure parameterised by the invariants reads
\begin{equation}
dR_{ia,b}=\frac{Q^2}{32\pi^3(Q^2+s_{ia}+s_{ib})^2}\; d\phi\;ds_{ia}\;ds_{ib}\Theta\left(\phi\left(2\pi-\phi\right)\right)\Theta\left(s_{ia}\left(s_{ia}^+-s_{ia}\right)\right)\Theta\left(s_{ik}\left(s_{ik}^+-s_{ik}\right)\right).
\end{equation}
The integration boundaries are given by
\begin{eqnarray}
&s_{ia}^+=\frac{1-x_A}{x_A} Q^2, \quad
s_{ib}^+=s_{ia}^+-s_{ia}&
\end{eqnarray}
 with $x_A$ defined in \Eq{eq:iiconv}.
The parameterisation of the real phase space in terms of the integration variables proceeds as depicted above but with
 \begin{equation}
   x=\frac{Q^2}{Q^2+s_{ia}+s_{ib}}, \quad
 v=\frac{s_{ia}}{Q^2+s_{ia}+s_{ib}}. 
\end{equation}

\subsection{Impact of the new $3\to2$ clusterings}
As stated in the introduction to this chapter, once the resolution criterion of the jet algorithm has singled out an unresolved pair of final-state partons or unresolved radiation associated with the beam, any of the remaining partons in the final state can be chosen as the spectator. Moreover, an initial-state parton might also be chosen to be the spectator to define the clustering. As already indicated, this freedom can be exploited to minimise the difference of the $3\to2$ clustering prescription with respect to the recombination procedures used in conventional $2\to1$  jet algorithms (see section~\ref{sec:jets}).

\subsubsection{Spectator influence in final-state clusterings}
In conventional $2\to1$ jet algorithms the momenta of recombined jets of two unresolved final-state partons $ij$ can be defined by adding the partonic $4$-momenta to form the jet momentum $p_i+p_j$ (see `E-scheme' in section \ref{sec:jets}). The $3\to2$ final-state clustering prescriptions presented in section~\ref{sec:ff} and section~\ref{sec:fim} on the other hand recombine the unresolved partons $ij$ in combination with a spectator from the final or the initial state. In order to quantify the influence of the different clusterings and the choice of the spectator on the final-state jet momenta with respect to the $2\to1$ prescriptions, the following norm measuring the difference of two $4$-momenta $p_m=(p_m^0,\vec{p}_m)$ and $p_n=(p_n^0,\vec{p}_n)$ can be used:
\begin{equation}
 ||p_m-p_n||=\max\left(\left|p_m^0-p_n^0\right|,\left|\vec{p}_m-\vec{p}_n\right|\right).
\end{equation}
From \Eq{eq:ffmapmassive} follows that the difference of the clustered jet momentum $J_j$ and the $4$-momentum sum $p_i+p_j$ for the final-state clustering with a final-state spectator depends on the spectator parton $k$:
\begin{eqnarray}\label{eq:ffnorm}
  \nn &&||J_j-(p_i+p_j)||=
  \max\left(\left|J^0_j-(p^0_i+p^0_j)\right|,\left|\vec{J}_j-(\vec{p}_i+\vec{p}_j)\right|\right)\\
  &=&\max\left(\left|(A_k-1)\; p^0_k+\frac{A_{ij}}{2}\;(p^0_i+p^0_j)\right|,\left|(A_k-1)\; \vec{p}_k+\frac{A_{ij}}{2}\;(\vec{p}_i+\vec{p}_j)\right|\right).
\end{eqnarray}
Different final-state partons can thus be tried as spectator $k$ to choose one which minimises the quantity $||J_j-(p_i+p_j)||$. 

In addition, also an initial-state spectator can be chosen. Again, the difference of the clustered jet momentum and the $4$-momentum sum depends on the choice for the initial-state spectator $a$ (see \Eq{eq:fimapmassless}):
\begin{equation}\label{eq:finorm}
||{J}_j - (p_i+p_j)||=(1-x)p^0_a.
\end{equation}
For a given unresolved final-state pair $ij$ to be combined, the spectator from the initial or final state which minimises the quantity defined in \Eq{eq:ffnorm} and \Eq{eq:finorm} is chosen.

\subsubsection{Spectator influence in initial-state clusterings}
In conventional $2\to1$ jet algorithms unresolved additional radiation which is associated with the beam (`lost in the beam pipe' due to the experimental setup) is removed from the list of momenta. The other final-state momenta remain unchanged. The situation is different for the $3\to2$ initial-state clusterings presented in section~\ref{sec:if} and section~\ref{sec:ii}. The unresolved radiation $i$ is clustered with an initial-state parton $a$ in combination with a final- or initial-state spectator resulting in non-trivial clustering prescriptions for the final-state momenta. From section \ref{sec:if} follows that the difference in the final state between the $3\to 2$ initial-state clustering with a final-state spectator $k$ and the conventional $2\to1$ prescription depends on the initial-state parton $a$ to be clustered with and the final-state spectator $k$ (through $x$, cf. \Eq{eq:colfacif}):
\begin{eqnarray}\label{eq:ifnorm}
||{J}_k - p_k||=||(1-x)p_a-p_i||=\max\left(\left|(1-x)p^0_a-p^0_i\right|,\left|(1-x)\vec{p}_a-\vec{p}_i\right|\right).
\end{eqnarray}
When choosing an initial-state spectator $b$ in the $3\to2$ clustering, the whole final state has to be boosted in order to balance the transverse momentum of the unresolved parton $i$ (see \Eq{eq:iimap}). The difference of the final state obtained with this $3\to2$ clustering and the conventional $2\to1$ prescription therefore reads
\begin{equation}\label{eq:iinorm}
  ||\sum\limits_m {J}_m - \sum\limits_{m\neq i} p_m||=|| \widetilde{K} -K||=\max\left(\left|(1-x)p^0_a-p^0_i\right|,\left|(1-x)\vec{p}_a-\vec{p}_i\right|\right).
\end{equation}
Note that \Eq{eq:iinorm} also depends on the initial-state parton $a$ to be clustered with and the initial-state spectator $b$ (through $x$, see \Eq{eq:xii}).

For a given unresolved final-state parton $i$ associated with the beam, the spectator has to be chosen either from the final or the initial state. Additionally, also the initial-state parton $a$ to be clustered with might be selected. Both are chosen to find the minimum of the quantities defined in \Eq{eq:ifnorm} and \Eq{eq:iinorm}.

\section{Validation of the phase space parameterisations}\label{sec:validphsp}
To simplify the validation, initial-state radiation and final-state radiation are studied separately. This allows to validate complementary aspects of the phase space parameterisations presented in chapter \ref{sec:PhaseSpaceparameterisation}. In particular, Drell-Yan production at a proton collider 
\begin{equation}\label{eq:drellyan}
pp\rightarrow e^+e^-
\end{equation}
and the production of top-quark pairs at a lepton collider
\begin{equation}\label{eq:toppair}
 e^+e^-\rightarrow t\bar{t}
\end{equation}
are studied as exemplary processes. Compact analytic results for the NLO corrections to these processes which are calculated using \PSS are available in the literature (see \Refs{Harris:2001sx,Brandenburg:1998xw}). Thus, the implementation of the method presented here is straightforward and not very error-prone. Since these two processes are fairly simple, the Monte Carlo integrations performed in the calculations are comparatively stable allowing thorough numerical validations. Additionally, scrutiny of the approach for the processes in \Eq{eq:drellyan} and \Eq{eq:toppair} lays the foundation to finally apply the machinery to the production of single top-quarks at the LHC (see chapter~\ref{sec:sngltp})
\begin{equation}
pp\rightarrow tj
\end{equation}
combining the complementary aspects featured by the two example processes studied here.

For the exclusive production of $n$ jets in the final state a differential $n$-jet event weight at NLO accuracy for jets obtained with a $3\to2$ jet algorithm is defined as (cf. \Eq{eq:diffnjetxs})
\begin{equation}\label{eq:diffwgt}
  {d\sigma^{\NLO}\over dR_n(J_1\ldots,J_{n})}
  ={d\sigma^{\text{BV}}\over dR_n(J_1\ldots,J_{n})}
  + {d\sigma^{\text{R}}\over dR_n(J_1\ldots,J_{n})}.
\end{equation}
As before the superscripts $\text{BV}$ and $\text{R}$ indicate the contributions from the Born matrix elements together with the virtual corrections and the real contribution. The soft and collinear singularities in the real contribution are regularised using the \PSS method. The regularised soft/collinear contributions are combined with the virtual corrections to cancel the respective soft and collinear singularities (cf. section~\ref{sec:PSS}). Since the real corrections are calculated using the factorised jet phase space as in
\Eq{eq:unres-real-fact} an additional three dimensional integration is required per jet event $J_1,\ldots,J_n$ when calculating $d\sigma^{\textrm{R}}$. The total cross section at NLO accuracy can be calculated by integrating \Eq{eq:diffwgt} over the full phase space $dR_n(J_1\ldots,J_{n})$. 

To check the implementation, the results are compared to the outcome of a standard parton-level Monte Carlo program. The results for the total cross sections obtained from integrating over \Eq{eq:diffnjetxs} with the factorised jet phase space perfectly agree with results obtained with a conventional partonic phase space. However, the numerical comparison of total cross sections might be rather insensitive to calculational details and depending on the phase space region possible inconsistencies could easily hide in the statistical uncertainty. To be more thorough, differential distributions, allowing to study different regions of the phase space independently, have to be compared. Arbitrary distributions of $n$-jet observables $O\left(J_{1},\dots,J_{n}\right)$ can be calculated at NLO accuracy by integrating \Eq{eq:diffwgt} over the $n$ jet momenta
\begin{equation}\label{eq:diffobs}
\frac{d\sigma^{\NLO}}{dO\left(\widetilde{J}_{1},
    \dots,\widetilde{J}_{n}\right)}
  =\int dR_n(J_1,\ldots,J_n)
  \frac{d\sigma^{\NLO}}{dR_n(J_1,\ldots,J_n)}
  \delta\left(O\left(J_{1},\dots,J_{n}\right)
    -O\left(\widetilde{J}_{1},\dots,\widetilde{J}_{n}\right)\right).
\end{equation}
These distributions can be compared to respective distributions calculated with a conventional parton level Monte Carlo employing the same $3\to 2$ jet algorithm (cf. \Eq{eq:diffjetxs})
\begin{equation}\label{eq:3to2jetalg}
F^{n+1}_{J_n,\ldots,J_n}:\;R_{n+1}\rightarrow R_n,\quad (p_1,\ldots,p_{n+1})\xmapsto{3\to2} (J_1,\ldots,J_n).
\end{equation}
The comparison of different variables allows to perform a detailed check of the phase space factorisations and the equivalence of the jet momenta used in \Eq{eq:diffobs} and the ones defined by \Eq{eq:3to2jetalg} over the entire phase space. 

\subsection{Coloured Initial-state partons: Drell-Yan production}\label{sec:DY}
\begin{figure}[htbp]
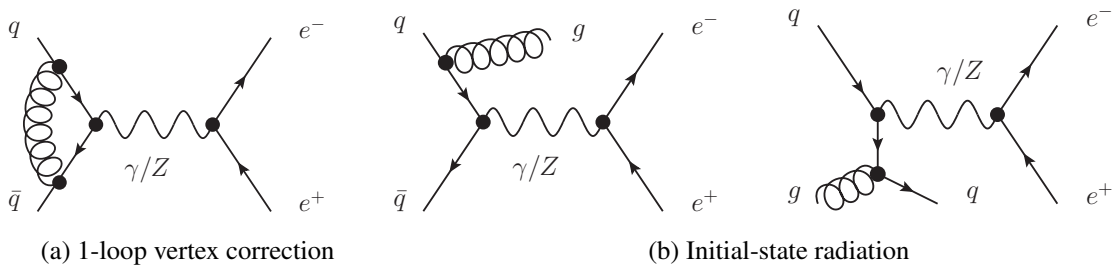

  \begin{center}
    \leavevmode
    \hfill\begin{subfigure}[b]{0.32\textwidth}
     \includegraphics[width=1\textwidth]{{{%
          dytt/qqeev}}}
     \caption{$1$-loop  vertex correction}
     \label{fig:qqeev}
    \end{subfigure}
    \begin{subfigure}[b]{0.67\textwidth}
\leavevmode
     \includegraphics[width=0.49\textwidth]{{{%
          dytt/qqeer1}}}
     \includegraphics[width=0.49\textwidth]{{{%
          dytt/qqeer2}}}
     \caption{Initial-state radiation}
     \label{fig:qqeer}
    \end{subfigure}
     \caption{Drell-Yan at NLO: Examples for (a) virtual and (b) for real corrections.}
    \label{fig:qqeeNLO}
  \end{center}
\end{figure}
\begin{figure}[htbp]
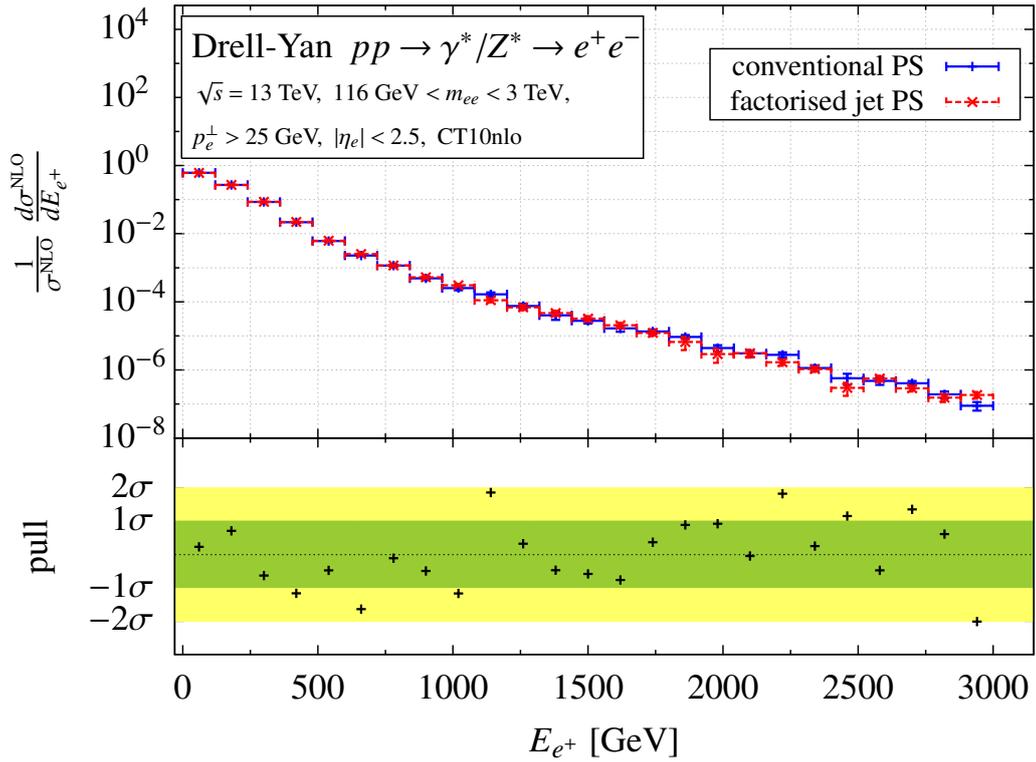

  \begin{center}
    \leavevmode
    \makebox[\textwidth]{\includegraphics[width=\ftw\textwidth]{{{%
          dytt/qqee3-2ptmn25etamx2.5compE1-50x1e7-crop}}}}
    \caption{Energy distribution of the positron from Drell-Yan production
      calculated at NLO accuracy using a conventional parton level MC (solid blue) compared to 
      results from the factorised jet phase space as described in
      section~\ref{sec:ii} (red dashed).}
    \label{fig:partonMCvsjetMC_DrellYan_e1}
  \end{center}
\end{figure}
Drell-Yan production in proton collisions at the LHC with a centre-of-mass energy of $\sqrt{s}=13$~TeV is considered. Since the case of unresolved radiation in the real corrections has to be checked, any jet in the final state is vetoed by exclusively allowing an electron-positron pair $e^+e^-$ in the final state. By being more inclusive, e.g. by dropping the jet veto, potential differences could be washed out and the check would be less sensitive. Cuts to the final state to mimic Drell-Yan measurements at the LHC (cf. e.g. \Ref{Aad:2016zzw}) are applied: The invariant mass of the electron-positron pair $m_{ee}$ is required to be in the mass range of $116$ GeV$<m_{ee}<3$ TeV. Any resolved final state must have a transverse momentum of at least $p^{\perp}_{\tmin}=25$ GeV and a pseudo rapidity not exceeding $|\eta_{\text{max}}|=2.5$. The detector is assumed to be blind outside these cuts. 
Because of the colourless final state unresolved radiation from the initial state (cf. \Fig{fig:qqeer}) is associated with the beam and has to be clustered with an initial-state parton in combination with an initial-state spectator according to section \ref{sec:ii}. In the NLO calculation the PDF set `CT10nlo' (see \Ref{Lai:2010vv}) is used to relate the partonic and hadronic initial states (cf. section~\ref{sec:factorisation}). The remaining parameters are set to the values listed in section~\ref{sec:sngltpsetup}. Respective amplitudes for the Born, (renormalised) virtual and real contributions are given in \Ref{Harris:2001sx}. Corresponding tree-level matrix elements generated with MadGraph5 (see \Ref{Alwall:2011uj}) are used in the amplitude calculation. The `two cut-off \PSS technique' (cf. section~\ref{sec:PSS}) to mediate the cancellation of the infrared divergences in the virtual and real corrections as presented in \Ref{Harris:2001sx} is employed for the process considered here.

In \Fig{fig:partonMCvsjetMC_DrellYan_e1} the distribution of energy of the positron from Drell-Yan production calculated at NLO accuracy is shown as an example. The results from the conventional parton level Monte Carlo are shown as blue solid lines and the red dashed lines show the results using the factorised jet phase space as presented in section \ref{sec:ii}. The error bars in the histograms correspond to the limited statistics of the Monte Carlo integrations. The discrepancy between the two approaches in terms of
standard deviations (`pull') is shown in the lower part of the plots. By looking at the pull distributions it is obvious that the two approaches perfectly agree within the statistical uncertainties of the numerical integration with the discrepancy being less than one standard deviation in most cases. 
Analogous figures for the polar angle distribution of the electron together with the invariant mass and rapidity distributions of the $e^+e^-$-system can be found in \Fig{fig:partonMCvsjetMC_DrellYan_cth2} to \Fig{fig:partonMCvsjetMC_DrellYan_yee} in appendix~\ref{app:phspval_DY} yielding similar conclusions. This study validates the initial-state clustering in combination with an initial-state spectator as presented in section~\ref{sec:ii}.

\subsection{Coloured final-state partons: Leptonic production of top-quark pairs}\label{sec:ttprod}
\begin{figure}[htbp]
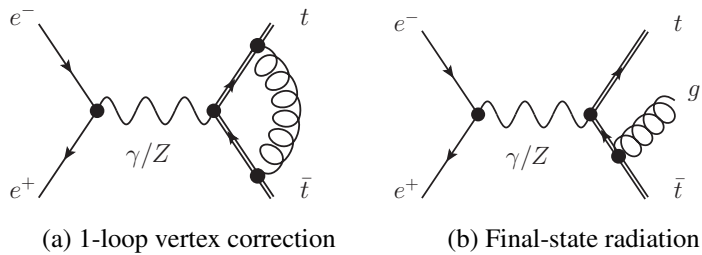

  \begin{center}
    \leavevmode
    \begin{subfigure}[b]{0.32\textwidth}
     \includegraphics[width=1\textwidth]{{{%
          dytt/eettVi}}}
     \caption{$1$-loop  vertex correction}
     \label{fig:eettv}
    \end{subfigure}
    \begin{subfigure}[b]{0.32\textwidth}
\leavevmode
     \includegraphics[width=1\textwidth]{{{%
          dytt/eettRe}}}
     \caption{Final-state radiation}
     \label{fig:eettr}
    \end{subfigure}
     \caption{Top-pair production at NLO: Examples for (a) virtual and (b) real corrections.}
    \label{fig:eettNLO}
  \end{center}
\end{figure}
\begin{figure}[htbp]
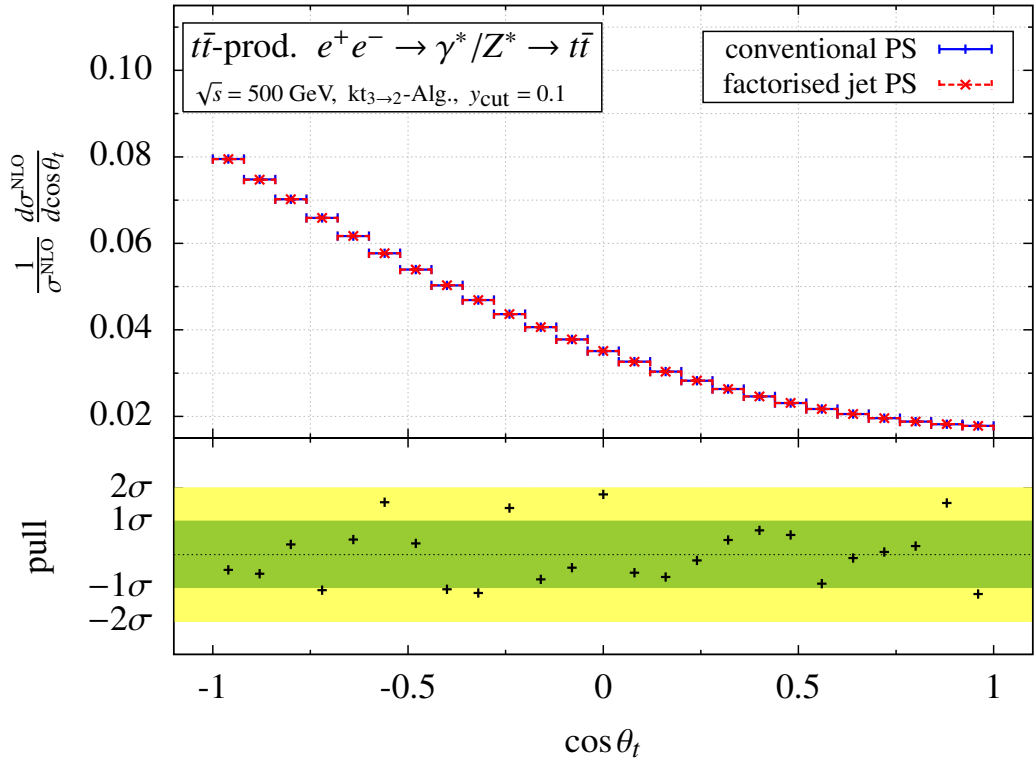

  \begin{center}
    \leavevmode
    \makebox[\textwidth]{\includegraphics[width=\ftw\textwidth]{{{%
          dytt/eeQQKT3-2ycut0.1compCTH1-50x1e7-crop}}}}
    \caption{Polar angle distribution of the top-tagged jet from top-quark pair production 
      in $e^+e^-$ annihilation calculated at NLO accuracy using a conventional parton
      level MC (solid blue) compared to results from the factorised jet
      phase space as described in section~\ref{sec:ffm} (red dashed).}
    \label{fig:partonMCvsjetMC_ttprod_cth1}
  \end{center}
\end{figure}
In top-quark pair production from $e^+e^-$ annihilation the initial state is colourless. When vetoing additional resolved jets in the final state, unresolved final-state radiation (cf. \Fig{fig:eettr}) has to be clustered with one of the top quarks with the other top quark playing the role of the spectator (cf. section  \ref{sec:ffm}). For the NLO calculation a top-quark mass of $m_t=174$~GeV is used with the remaining parameters set to the values listed in section~\ref{sec:sngltpsetup}. Having a future linear collider in mind, the centre-of-mass energy is set to $\sqrt{s}=500$~GeV corresponding to top-quark studies in the continuum. Respective amplitudes for the Born, (renormalised) virtual and real corrections can be taken from \Ref{Harris:2001sx} and \Ref{Brandenburg:1998xw}. The amplitudes are calculated using corresponding tree-level matrix elements generated with Madgraph5 (see \Ref{Alwall:2011uj}). Following the example given in \Ref{Harris:2001sx} the cancellation of the infrared divergences is carried out in terms of the `two cut-off \PSS technique' (cf. section~\ref{sec:PSS}).
The decay of the top quarks is not included but they are treated as tagged top-jets. These jets are defined by an augmented $3\to2$ $kt$-jet algorithm (cf. section~\ref{sec:3to2jetalg}) with a resolution criterion for lepton colliders given by (cf. \Ref{Salam:2009jx})
\begin{equation}\label{eq:reslepcoll}
  \yij=2\;\frac{\min\left(E^2_i,E^2_j\right)
    \left(1 - \cos(\theta_{ij})\right)}{s}
\end{equation}
where $E_i$ denotes the energy of parton $i$ and $\theta_{ij}$ is the angle between the $3$-momenta of partons $i$ and $j$.
The resolution \ycut, defining two particles as being unresolved when $\yij<\ycut$, is set to $\ycut=0.1$. 

The polar angle distribution of the top quark from leptonic top-quark pair production is shown in \Fig{fig:partonMCvsjetMC_ttprod_cth1} as an example. As before, the blue solid curves correspond to the results of a conventional parton level Monte Carlo while the red dashed curves are obtained using the factorised jet phase space as presented in section \ref{sec:ffm}. The pull distributions in the lower parts of the plots again confirm perfect agreement of the two approaches within the statistical uncertainties with discrepancies of less than one standard deviation in most cases.
Analogous distributions with respect to the the azimuthal angle, the transverse momentum and the rapidity of the outgoing top quark can be found in \Fig{fig:partonMCvsjetMC_ttprod_ph1} to \Fig{fig:partonMCvsjetMC_ttprod_y1} in appendix~\ref{app:phspval_tt} underpinning the conclusion drawn above. Hence, the clustering of the additional radiation with massive final-state partons in combination with a massive final-state spectator is successfully validated (cf. section~\ref{sec:ffm}).

\subsection{Coloured initial- and final-state partons: Hadronic single top-quark production}\label{sec:sngltpphsp}
The developed algorithm has been successfully validated separately for the initial-state clusterings in section~\ref{sec:DY} and the final-state clusterings in section~\ref{sec:ttprod} with rather simple example processes. 
As a first application to the hadronic production of coloured particles forming jets, the production of single top quarks in proton-proton collisions at the LHC running at a centre-of-mass energy of $\sqrt{s}=13$~TeV is considered in this section. In the following, a brief introduction to single top-quark production at the LHC with an emphasis on why it is paradigmatic for the application of the \MEM is given.

\subsubsection{Single top-quark production at the LHC}\label{sec:sngltp}
Within the six quark flavours of the \SM the top quark takes on a special role because of its high mass ($m_t\approx 173\text{ GeV}$) which is comparable to the mass of a gold atom and of the order of the electroweak scale. It is the heaviest elementary
particle currently known. Its partner the bottom quark being the second heaviest quark is $\approx 35$ times lighter than the top quark. In the \SM the fermions' masses are encoded in the dimensionless `Yukawa couplings' which quantify the coupling of the fermions to the Higgs field. They are given by the ratios of the particle masses and the vacuum expectation value of the Higgs field. Because of its high mass the top quark is the only elementary particle in the \SM with a Yukawa coupling close to unity, in contrast to all other elementary particles whose Yukawa couplings seem `unnaturally small'. Because of its high mass the \SM predicts a very small mean life-time for the top quark of around $\approx 10^{-25}\text{ s}$ which is about one order of magnitude smaller than the time scale of hadronisation. Thus, top quarks are the only quarks that decay before hadronising. They have to be produced through high energetic particle collisions and allow to be studied as quasi-free quarks. Because of the direct decay of the top quark itself without an intermediate bound state, the entire information carried by the top quark is transferred to its decay products, making it a promising laboratory for precision tests of the \SM or finding hints for New Physics beyond the \SM.  Precise measurements of the parameters associated with the top quark like the electroweak gauge couplings or its mass are essential to test the consistency of the \SM and constrain New Physics models (see e.g. \Ref{Kroninger:2015oma} for a review of top-quark physics at the LHC).

Top-quark pairs ($t\bar{t}$) can be produced via the strong interaction which is the dominant source of top quarks at the LHC. Electroweak charged-current interactions also allow the production of single top quarks in hadron collisions. Although suppressed compared to top-quark pair production due to the smaller coupling strength of the weak interaction, single top-quark production offers a complementary laboratory to study the electroweak top-quark properties: In the \SM single top quarks are produced via the electroweak $Wtb$ vertex depending on the top-quark mass $m_t$, the CKM matrix element $V_{tb}$ and the weak coupling 
\begin{equation}\label{eq:alweak}
\alpha_W={\ae\over\sin^2{\theta_W}}.
\end{equation}
The `Weinberg angle' in the on-shell scheme is given by the ratio of the weak boson masses
\begin{equation}
\cos{\theta_W}={m_W\over m_Z}
\end{equation}
and parameterises the connection between the weak coupling and the electro-magnetic coupling $\ae$.
The top quark exhibits a $V-A$ coupling structure to the $W$ boson resulting in a high degree of top-quark polarisation which is transferred to its decay products making it accessible in the experiments.

Based on the virtuality of the $W$ boson interacting with the top quark three partonic production channels can be distinguished at lowest order in hadron collisions:
\begin{subequations}
\begin{eqnarray}\label{eq:topchs}
qq'\rightarrow t\bar{b},& q^2_W>m^2_t &s\textrm{-channel production},\\
\label{eq:topcht}
qb\rightarrow tq',& q^2_W<0  &t\textrm{-channel production},\\
\label{eq:topchtW}
gb\rightarrow tW^-,&q^2_W=m^2_W &\textrm{$tW$-channel}.
\end{eqnarray}
\end{subequations}

\begin{figure}[htbp]
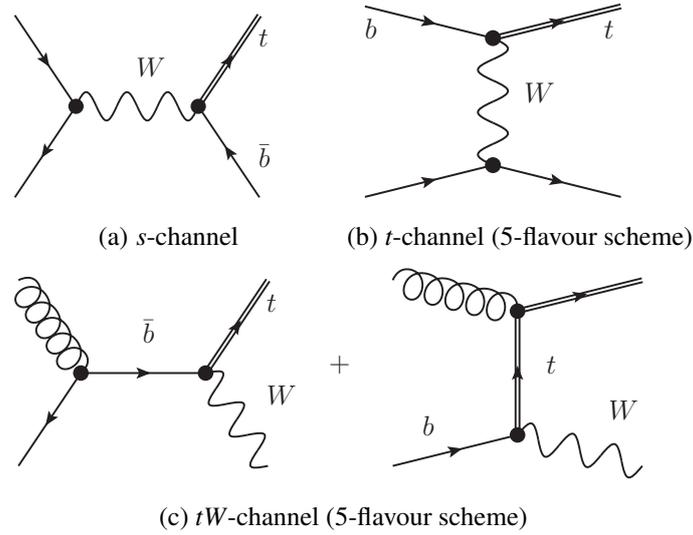

  \begin{center}
    \leavevmode
    \begin{subfigure}[b]{0.29\textwidth}
     \includegraphics[width=1\textwidth]{{{%
          sgt/sgts}}}
     \caption{$s$-channel}
     \label{fig:sgts}
    \end{subfigure}
    \begin{subfigure}[b]{0.29\textwidth}
     \includegraphics[width=1\textwidth]{{{%
           sgt/sgtt}}}
     \caption{\mbox{$t$-channel ($5$-flavour scheme)}}
     \label{fig:sgtt}
    \end{subfigure}
    \begin{subfigure}[b]{0.59\textwidth}
     \includegraphics[width=1\textwidth]{{{%
           sgt/sgtW}}}
     \caption{$tW$-channel ($5$-flavour scheme)}
     \label{fig:sgtW}
    \end{subfigure}
     \caption{Single top-quark production channels at lowest order: (a) $s$-channel, (b) $t$-channel and (c) $tW$-channel.}
    \label{fig:singletopprod}
  \end{center}
\end{figure}
At the LHC the $t$-channel, with a space-like $W$ boson (see \Eq{eq:topcht} and \Fig{fig:sgtt}), contributes the biggest share of $\approx80\%$ to the production of single top quarks (cf. table~\ref{tab:inclxs}). The notation in \Eq{eq:topcht} suggests that the $b$-quark is assumed to be an active flavour inside the proton and has to be accounted for by PDFs in the `$5$-flavour scheme' making  \Eq{eq:topcht} a purely electroweak process at leading order $\ord(\alpha^2_W)$. When assuming the `$4$-flavour scheme' (without a $b$-quark PDF) the $t$-channel production is formally a higher-order process (of order $\ord(\as\alpha^2_W)$) because the $b$-quark has to come from an initial-state gluon splitting 
\begin{equation}
qg\rightarrow tq'\bar{b}.
\end{equation}
Consistently, calculations employing both schemes should give similar results because the initial-state gluon splittings needed in the $4$-flavour scheme are accounted for by the evolution of the $b$-quark PDF in the $5$-flavour scheme (see e.g. \Refs{Campbell:2009ss,Campbell:2009gj}). 

The $s$-channel, with a time-like $W$ boson (see \Eq{eq:topchs} and \Fig{fig:sgts}), is the second purely electroweak production mode at leading order and only amounts to a few percent ($\approx4\%$) of the total single top-quark production cross section at the LHC  (cf. table~\ref{tab:inclxs}). The second most important source for single top quarks at the LHC is the production in association with an on-shell $W$ boson also referred to as $tW$-channel (see \Eq{eq:topchtW} and \Fig{fig:sgtW}). Because the $b$-quark is needed in the initial state, this process can also be described in the $5$- or $4$-flavour scheme but is already also a QCD process at leading order $\ord(\as\alpha_W)$ in the $5$-flavour scheme. 

This work focuses on the $s$- and $t$-channel production modes described in the $5$-flavour scheme.

\begin{table}[htbp]
\centering
\def\arraystretch{1.8}
\begin{tabular}{|l|l|l|l|l|}
 \hline
$\sqrt{s}=13$ TeV& {$t$-channel} & {$s$-channel} & {$tW$-channel} & pair production  \\ \hline
{${\sigma^{\Born}}^{\Delta_{2\mu}}_{\Delta_{\mu/2}}$ [pb]} & $141.50(9)_{-14.41}^{+11.36}$ & $4.707(9)_{-0.156}^{+0.119}$ & $26.613(5)_{+0.482}^{-0.988}$ &  $474.63(5)_{+138.58}^{-100.48}$ \\ \hline
{${\sigma^{\NLO}}^{\Delta_{2\mu}}_{\Delta_{\mu/2}}$ [pb]} & $135.79(8)_{-2.1}^{+3.8}$   & $6.340(8)_{+0.092}^{-0.064}$ & $28.339(6)_{-1.20}^{+0.91}$ & $708.90(8)_{+82.06}^{-82.89}$  \\ \hline
\end{tabular}
\caption{Total inclusive top-quark production cross section predictions at the LHC}
\label{tab:inclxs}
\end{table}

Differential cross section predictions for the $t$-channel process neglecting colour exchange between the two quark lines are known to next-to-next-to-leading order (NNLO) accuracy (see  \Refs{Brucherseifer:2014ama,Berger:2016oht,Berger:2017zof}). The approximation of full NNLO results for the $s$-channel process by next-to-next-to-leading-logarithm (NNLL) resummation of soft and collinear gluon corrections is studied in \Ref{Kidonakis:2010tc}. Full next-to-leading order (NLO) results for both $s$- and $t$-channel are known (see e.g. \Refs{Harris:2002md,Cao:2004ky} and are implemented in public numerical tools like MCFM (see \Ref{Campbell:2010ff}) or HATHOR (see \Ref{Kant:2014oha}). For the $t$-channel with top-quark decay including full off-shell effects NLO results matched to parton showers are presented in \Ref{Frederix:2016rdc}.
Table \ref{tab:inclxs} shows the total inclusive cross section predictions at Born and NLO accuracy of the three single top-quark production channels at the LHC running at $13$ TeV. For comparison, the top-quark pair production cross section is also shown. The cross section values are obtained with HATHOR assuming an on-shell top-quark mass of $m_t=173.2$ GeV and using the PDF set `MSTW2008nnlo68cl' (see \Ref{Martin:2009iq}, all other parameters correspond to the default values given in  \Ref{Kant:2014oha}). The top-quark mass is chosen as the central scale $\mu=\mur=\muf=m_t$ and the impact of its up- and downwards variation by a factor of $2$ is represented by the super- and subscripts. 

As can be seen from table \ref{tab:inclxs}, the rates for single top quarks, especially produced in the $s$-channel, expected at the LHC are rather small compared to the production of top-quark pairs. Since top-quark pair production poses a large fraction of the background, its abundance makes it even harder to measure single top-quark production in the experiments. However, its aforementioned dependence on the electroweak top-quark properties makes experimental single top-quark studies a worthwhile endeavour (see e.g. \Ref{Giammanco:2017xyn}). In particular regarding top-quark mass determinations, the authors of \Ref{Alekhin:2016jjz} find some tension between mass values extracted from top-quark pair production and single top-quark production. Because of the combination of small event rates, large background and the variety of model parameters (top-quark mass, CKM matrix elements, coupling structure, ...) to be studied, electroweak single top-quark production at the LHC represents a prime example for the application of the \MEM.

Despite all the difficulties there has been quite some experimental progress regarding $s$- and $t$-channel single top-quark production at the LHC:
Both CMS and ATLAS have reported precise measurements of the $t$-channel production of single top quarks at the LHC running at $7$, $8$ and $13$ TeV (see e.g \Refs{Chatrchyan:2012ep,Aad:2014fwa,Khachatryan:2014iya,Aaboud:2017pdi,Sirunyan:2016cdg,Aaboud:2016ymp,CMS:2015jca,CMS:2014ika,CMS:2016xnv}).
However for the $s$-channel production at the LHC, only ATLAS has found first evidence in the $8$ TeV data set.  This has been achieved by using a \MEM technique to separate signal from background (see \Ref{Aad:2015upn}).

\subsubsection{Calculational setup}\label{sec:sngltpsetup}
\begin{figure}[htbp]
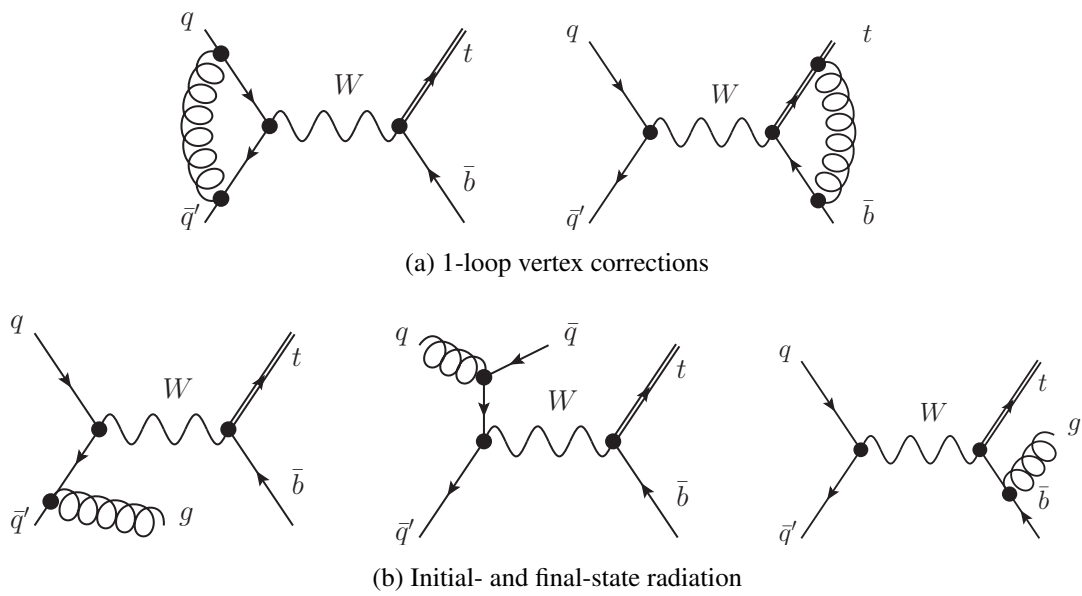

  \begin{center}
    \leavevmode
    \begin{subfigure}[b]{0.67\textwidth}
     \includegraphics[width=0.49\textwidth]{{{%
           sgt/sgtsv}}}
     \includegraphics[width=0.49\textwidth]{{{%
           sgt/sgtsvf}}}
     \caption{$1$-loop vertex corrections}
     \label{fig:sgtsv}
    \end{subfigure}
    \begin{subfigure}[b]{1\textwidth}
     \hfill\includegraphics[width=0.32\textwidth]{{{%
           sgt/sgtsri1}}}
     \includegraphics[width=0.32\textwidth]{{{%
           sgt/sgtsri2}}}
     \includegraphics[width=0.32\textwidth]{{{%
           sgt/sgtsrf}}}
     \caption{Initial- and final-state radiation}
     \label{fig:sgtsr}
    \end{subfigure}
     \caption{$s$-channel single top-quark production at NLO: (a) virtual and (b) examples of real corrections.}
    \label{fig:sgtsNLO}
  \end{center}
\end{figure}

\begin{figure}[htbp]
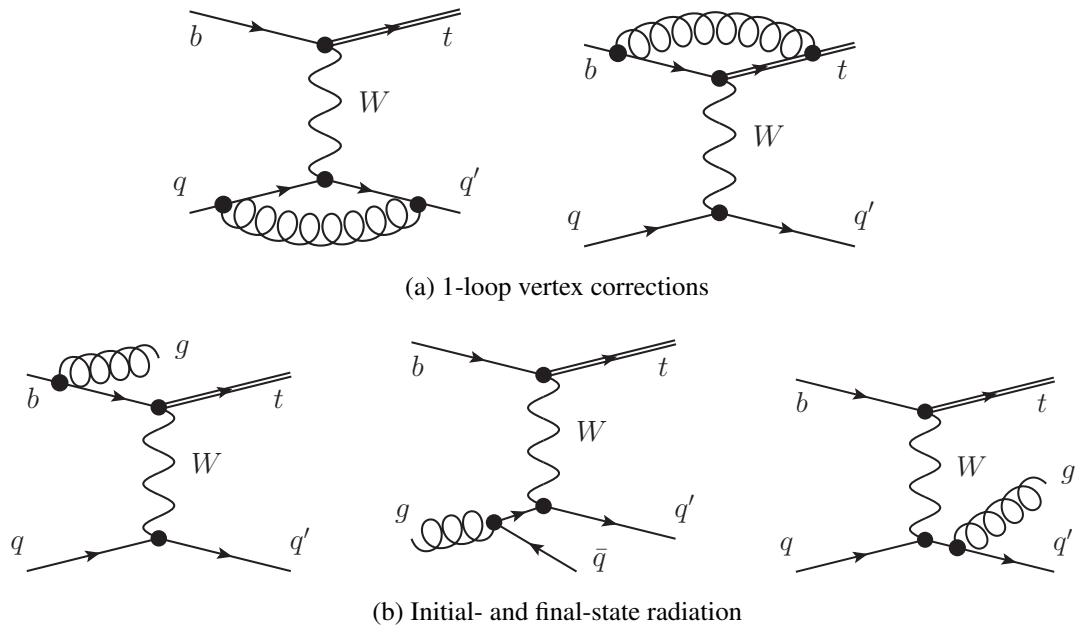

  \begin{center}
    \leavevmode
    \begin{subfigure}[b]{0.67\textwidth}
     \includegraphics[width=0.49\textwidth]{{{%
           sgt/sgttvl}}}
     \includegraphics[width=0.49\textwidth]{{{%
           sgt/sgttvh}}}
     \caption{$1$-loop vertex corrections}
     \label{fig:sgttv}
    \end{subfigure}
    \begin{subfigure}[b]{1\textwidth}
     \hfill\includegraphics[width=0.32\textwidth]{{{%
           sgt/sgttrh}}}
     \includegraphics[width=0.32\textwidth]{{{%
           sgt/sgttrl1}}}
     \includegraphics[width=0.32\textwidth]{{{%
           sgt/sgttrl2}}}
     \caption{Initial- and final-state radiation}
     \label{fig:sgttr}
    \end{subfigure}
     \caption{$t$-channel single top-quark production at NLO: (a) virtual and (b) examples of real corrections.}
    \label{fig:sgttNLO}
  \end{center}
\end{figure}

Representative Feynman diagrams for the NLO corrections are shown in \Fig{fig:sgtsNLO} for the $s$-channel and in \Fig{fig:sgttNLO} for the $t$-channel. Note that no $1$-loop diagrams with both quark lines connected by a gluon (`box diagrams') contribute to the virtual corrections  because the interference with the Born diagrams (see \Fig{fig:sgts} and \Fig{fig:sgtt}) vanishes due to the colour algebra (cf. \Eq{eq:colalg}). The remaining real diagrams are obtained by attaching the additional gluon to the other fermion lines in \Fig{fig:sgtsr} and \Fig{fig:sgttr} in all possible ways. The ultraviolet divergences in the virtual corrections are regulated by performing the loop integrals in $d=4-2\varepsilon$ dimensions. The resulting $\varepsilon$ poles are absorbed into the bare parameters by introducing renormalised quantities (cf. section~\ref{sec:regren}). To achieve this, the quark fields are renormalised by imposing the on-shell renormalisation conditions (cf. \Eq{eq:osren}). See e.g. \Ref{Harris:2002md} for details on the renormalisation procedure for the calculation of $s$- and $t$-channel single top-quark production at NLO. The remaining singularities in the renormalised virtual corrections are infrared divergences which have to be cancelled with the corresponding singularities occurring in the phase space integration of the real corrections (cf. section~\ref{sec:IRdivcanc}). This mutual cancellation of the infrared divergences between the virtual and the real corrections is carried out by applying a \PSS technique (cf. section~\ref{sec:PSS}). Detailed NLO calculations of single top-quark production in the $s$- and in the $t$-channel employing the `one cut-off \PSS technique' are published in \Ref{Cao:2004ky}. 

If not stated otherwise, the following setting for the cross section calculations for single top-quark production processes is used in this work: 
All predictions are done for proton-proton collisions at the LHC running at a centre-of-mass energy of $\sqrt{s}=13$ TeV. To relate the partonic calculations to the proton-proton initial state, the PDF set `MSTW2008nnlo68cl' (see \Ref{Martin:2009iq}) is employed. The running QCD coupling $\alpha_s(\mur)$ (see section~\ref{sec:QCD}) is taken as provided by the PDF set to be consistent with the evolution of the PDFs (cf. section~\ref{sec:factorisation}). The electro-magnetic coupling evaluated at the $Z$-mass is set to
\begin{equation}
  \ae(m_Z)=1/132.2332298
\end{equation}
but it only appears as an overall factor which cancels in ratios of cross sections. The masses of the electroweak gauge bosons are given in \Ref{Olive:2016xmw} as
\begin{equation}
  m_Z=91.1876\ \mbox{GeV}\quad \mbox{and}\quad  m_W=80.385 \mbox{ GeV}.
\end{equation}
The top-quark mass renormalised in the pole-mass scheme is set to
\begin{equation}
  m_t=173.2\  \mbox{GeV}.
\end{equation}
The jets are defined according to a $kt$-jet algorithm with a resolution criterion defined in \Eq{eq:jetsrescrit} (with $\alpha=2$ and $R=1$). Even though the anti-$kt$-jet algorithm (see \Ref{Cacciari:2008gp}) is more widely used by the experiments, the $kt$-jet algorithm offers the possibility to be formulated in an `exclusive variant'. Whereas exactly requiring a desired number of jets to be resolved is strongly discouraged for the anti-$k_t$-jet algorithm\footnote{Ref~\cite[p. 21]{Cacciari:2011ma}: ``We advise against the use of exclusive jets
in the context of the anti-$kt$ algorithm, because of the lack of physically meaningful hierarchy in the
clustering sequence.''} . For final-state objects to be resolved they are required to pass the following cuts
\begin{equation}\label{eq:jetcuts}
p^{\perp}>\sqrt{d_{\text{cut}}}=30\  \mbox{GeV},\quad |\eta|<3.5
\end{equation}
and  the detector is assumed to be blind outside these cuts. An `exclusive' formulation of the jet algorithm is defined by
demanding one resolved top-tagged jet ($t$) accompanied by exactly one resolved light jet ($j$)
\begin{equation}
p+p\rightarrow t+j.
\end{equation}
Not vetoing additional resolved jets ($X$) is referred to as `inclusive'
\begin{equation}
p+p\rightarrow t+j\;\left(+X\right).
\end{equation}
In the inclusive case the signal signature is still defined to consist of one resolved top-tagged jet ($t$) and one resolved light jet ($j$) which is taken to be the hardest of possibly two resolved jets ($p^\perp_j>p^\perp_X$). This way all signal predictions are accurate to NLO.\\
The same dynamical scale for the renormalisation and factorisation scale $\mur$ and $\muf$ is chosen. The central scale choice is defined by the total transverse energy $E^{\perp}=E \sin(\theta)$ of the resolved final state
\begin{equation}
\label{eq:scale-definition}
\mur=\muf=\mu_0=E^{\perp}_t+E^{\perp}_j\;\left(+E^{\perp}_X\right).
\end{equation} 

\subsubsection{Validation of the phase space parameterisations}
Apart from testing all the clusterings from section~\ref{sec:PhaseSpaceparameterisation} and their interplay at the same time, the validations presented here also serve as a consistency check of the implementation of the non-trivial NLO calculation, especially regarding the \PSS. 

\begin{figure}[htbp]
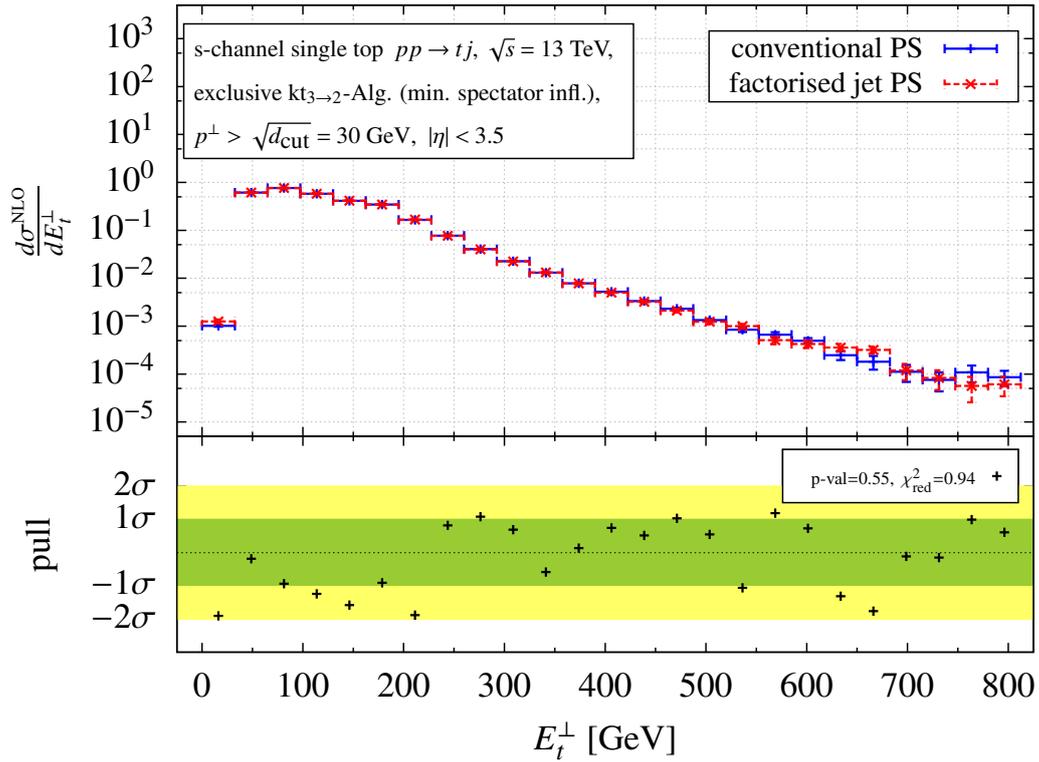

  \begin{center}
    \leavevmode
    \makebox[\textwidth]{\includegraphics[width=\ftw\textwidth]{{{%
          sgt/phsp/sgtsKT3-2ycut30compmnspecET1-50x1e7-crop}}}}    
   \caption{Transverse energy distribution of the top-tagged jet from exclusive $s$- single top-quark production
      calculated at NLO accuracy using a conventional parton level MC (solid blue) and the factorised jet phase space (dashed red) as described in section~\ref{sec:PhaseSpaceparameterisation}.}
    \label{fig:partonMCvsjetMC_sgts_et1}
  \end{center}
\end{figure}
\begin{figure}[htbp]
  \begin{center}
    \leavevmode
    \makebox[\textwidth]{\includegraphics[width=\ftw\textwidth]{{{%
          sgt/phsp/sgttKT3-2ycut30compmnspecETA2-50x1e7-crop}}}}
     \caption{Same as \Fig{fig:partonMCvsjetMC_sgts_et1} but for the pseudo rapidity of the light jet from the $t$-channel.}
    \label{fig:partonMCvsjetMC_sgtt_eta2}
  \end{center}
\end{figure}
The factorisation of the real phase space in terms of the phase space of the resolved jets and the dipole phase-space measures corresponding to the unresolved partonic configurations is validated in the following (cf. \Eq{eq:unres-real-fact} and sections~\ref{sec:ff} to \ref{sec:ii}). For this purpose it is sufficient to study the exclusive formulation of the jet algorithm with non-trivial recombinations of unresolved partonic configurations in the real corrections. In principle, contributions from kinematic configurations with an additional resolved jet can be calculated according to the last line of \Eq{eq:diffjetxs} but do not pose any conceptual problems. Since in single top-quark production there are coloured partons in the initial and in the final state, all phase space parameterisations given in section~\ref{sec:PhaseSpaceparameterisation} enter and can be validated at the same time.

In \Fig{fig:partonMCvsjetMC_sgts_et1} and \Fig{fig:partonMCvsjetMC_sgtt_eta2} distributions of the transverse energy of the top-tagged jet and pseudo rapidity of the light jet from $s$- and $t$-channel single top-quark production calculated at NLO accuracy are shown as examples. 
The distributions calculated with the conventional parton level Monte Carlo employing the $3\to2$ jet clustering (see \Eq{eq:3to2jetalg}) are shown as solid blue lines. The distributions obtained according to \Eq{eq:diffobs} are shown as dashed red lines. The difference of both distributions normalised to its statistical error (`pull') is shown at the bottom of the plots. Additionally, the p-value and reduced $\chi^2$ of the comparison of the two histograms as described in \Ref{Gagunashvili:2007zz}  and implemented in \Refs{Brun:1997pa,Antcheva:2009zz,Moneta:2008zza} are given in each plot. 
By examining the pull distributions in \Fig{fig:partonMCvsjetMC_sgts_et1} and \Fig{fig:partonMCvsjetMC_sgtt_eta2}, perfect agreement between the two approaches within the statistical uncertainties for both the $s$- and the $t$-channel is obvious. The results for the corresponding p-values and reduced $\chi^2$ confirm this observation. Note that the compared distributions are not normalised. Thus, this study also serves as a validation in terms of the fiducial cross sections.
Additional distributions of the (transverse) energy and pseudo rapidity of the top-tagged and the light jet and distributions of the invariant mass and rapidity of the top-light jet system from the $s$- and the $t$-channel are collected in \Fig{fig:partonMCvsjetMC_sgts_eta1} to \Fig{fig:partonMCvsjetMC_sgtt_yee} in appendix~\ref{app:phspval_sgt}. This comparison of various distributions of jet variables verifies the identification of the factorised jet phase space in \Eq{eq:diffobs} with the clustered jets defined by \Eq{eq:3to2jetalg} for processes involving all of the clusterings worked out in section~\ref{sec:PhaseSpaceparameterisation}. 

\subsubsection{Dependence on the \PSS parameter}
As mentioned in section~\ref{sec:sngltpsetup}, the one cut-off \PSS method as presented in \Ref{Cao:2004ky} is utilised to mutually cancel the infrared divergences appearing in the virtual and real contributions of the NLO corrections. The slicing parameter $s_\tmin$ is introduced to control the separation of the real phase space into soft/collinear and hard regions (cf. section~\ref{sec:PSS}). When considering jet production processes it is crucial that this phase space separation does not interfere with the jet clustering (i.e. the soft/collinear regions of the real phase space have to always lie safely within the jet cones defined by the jet algorithm with $s_\tmin$ much smaller than any physical scale of the process). 

The numerical Monte Carlo integration of the hard contribution produces logarithmic $s_\tmin$ dependences (cf. \Eq{eq:PSSlogbuild}). These numerical logarithmic $s_\tmin$ dependences cancel with the corresponding analytic logarithmic $s_\tmin$ dependences of the soft/collinear contribution when both contributions are combined. Since both intermediate results scale logarithmically with $s_\tmin$, their combination results in a loss of significant digits and a growing relative statistical integration error for small $s_\tmin$.

As described in section~\ref{sec:PSS},  a systematic error is introduced when soft and collinear approximations of the cross section are applied in the soft/collinear regions. This systematic error scales with the phase-space volume of these regions which itself scales with the slicing parameter $s_\tmin$ making the result no longer independent of $s_\tmin$. For sufficiently small values of the slicing parameter the systematic error from the soft/collinear contribution can be neglected with respect to the statistical error from the hard contribution. By finding a compromise between statistical and systematic uncertainties an approximate $s_\tmin$ independence of the final result can be established. This approximate independence of the final result of the slicing parameter is an important test of the successful application of the \PSS method.

\begin{figure}[htbp]
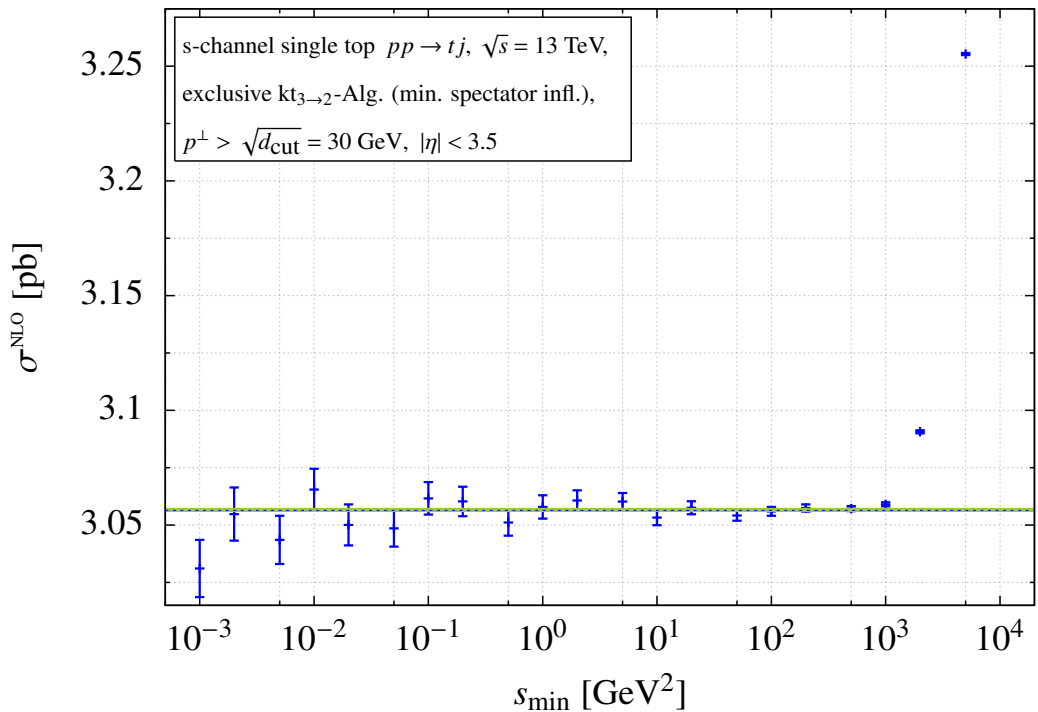

  \begin{center}
   \includegraphics[width=\ftw\textwidth]{{{%
          sgt/smin/sgtsKT3-2ycut30sminc-50x1e7-crop}}}
     \caption{Fiducial NLO cross section for exclusive $s$-channel single top-quark production as a function of the slicing parameter $s_\tmin$.}
    \label{fig:sminc_sgts}
  \end{center}
\end{figure}
\begin{figure}[htbp]
  \begin{center}
   \includegraphics[width=\ftw\textwidth]{{{%
          sgt/smin/sgttKT3-2ycut30sminc-50x1e7-crop}}}
     \caption{Same as \Fig{fig:sminc_sgts} but for the $t$-channel.}
    \label{fig:sminc_sgtt}
  \end{center}
\end{figure}
The calculation of the NLO corrections to $s$- and $t$-channel single top-quark production involves several different partonic sub-processes with initial- as well as final-state radiation (cf. \Fig{fig:sgtsNLO}) requiring a non-trivial cancellation of the infrared divergences between the virtual and the real contributions. Furthermore, initial-state collinear divergences have to be absorbed into bare PDFs by collinear factorisation (cf. section~\ref{sec:factorisation}). The application of the \PSS method to single top-quark production as described in \Ref{Cao:2004ky} is therefore non-trivial and error-prone. To rule out possible inconsistencies the $s_\tmin$ dependence of the implementation is thoroughly studied. Because the cancellation of the \IR divergences in the soft/collinear regions of the real phase space has to be tested, it is sufficient to study the exclusive formulation of the jet algorithm. NLO contributions with an additional resolved jet are already rendered \IR safe by the jet algorithm. In \Fig{fig:sminc_sgts} and \Fig{fig:sminc_sgtt} results for the fiducial cross section for the $s$- and $t$-channel production of single top quarks calculated at NLO accuracy for  $21$ different values of the slicing parameter in the range between $s_\tmin=10^{-3}\GeV^2$ and $s_\tmin=5\times10^{3}\GeV^2$ are shown. The results for the $s$-channel cross section are compatible with a constant within their statistical uncertainties for $s_\tmin<1000\GeV^2$. For $s_\tmin>1000\GeV^2$ the $s$-channel cross section shows a strong dependence on the slicing parameter. This is not a surprise since at this scale the slicing parameter $s_\tmin$ clearly interferes with the transverse-momentum cut of the $kt$-jet algorithm of ${p^{\perp2}_\tmin}=900\GeV^2$ (cf. \Eq{eq:jetcuts}). The result of a fit to the first $17$ cross section values assuming a constant cross section yields $\sigma^\NLO_\text{$s$-chan}=(3.0564\pm0.0008)~\text{pb}$ and is shown by the straight dashed line in the upper plot. The analogous study for the $t$-channel yields similar results but with an approximate $s_\tmin$ independence for $s_\tmin< 100\GeV^2$. The first $13$ cross section values are compatible with a constant cross section obtained by a fit as $\sigma^\NLO_\text{$t$-chan}=(60.04\pm0.02)~\text{pb}$ (see straight dashed line \Fig{fig:sminc_sgtt}).

Since differential jet cross sections at NLO accuracy are needed for the application within the framework of the \MEM, merely investigating the dependence of the fiducial cross section on the slicing parameter is not sufficient. Differential distributions may introduce additional scales while probing different regions of the phase space individually. They may be more sensitive to the choice of $s_\tmin$. Moreover, for the aforementioned NLO calculations within the approach presented in this work, the factorised phase space parameterisations as described in section~\ref{sec:PhaseSpaceparameterisation} are used.  The calculation of $s$- and $t$-channel production of single top quarks at NLO accuracy introduces different parameterisations in different phase space regions according to the $3\to2$ clusterings. As mentioned before, this makes the implementation especially error-prone. Validation of the approximate $s_\tmin$ independence of differential distributions serves as an additional strong check of the implementation. 
\begin{figure}[htbp]
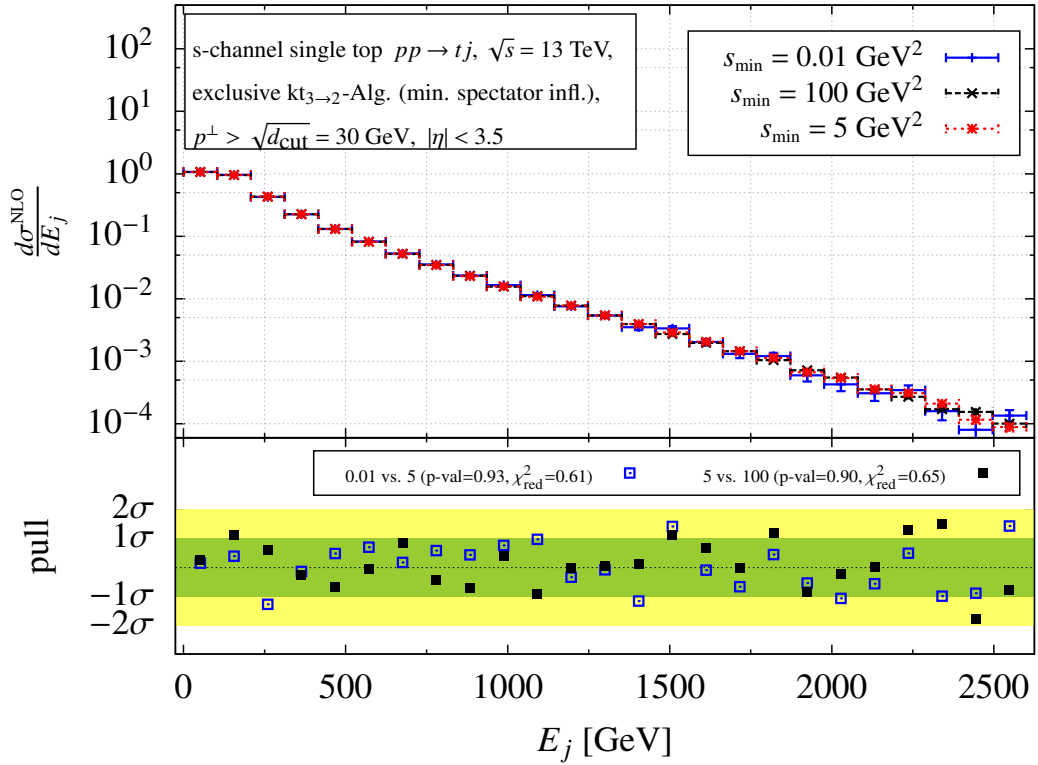

  \begin{center}
    \leavevmode
     \makebox[\textwidth]{\includegraphics[width=\ftw\textwidth]{{{%
          sgt/smin/sgtsKT3-2ycut30compmnspecE2-50x1e7sminc-crop}}}}
    \caption{Energy distribution of the light jet from exclusive $s$-channel single top-quark production
      calculated at NLO accuracy with three different values of the slicing parameter $s_\tmin$.}
    \label{fig:sminc_sgts_e2}
  \end{center}
\end{figure}
\begin{figure}[htbp]
  \begin{center}
    \leavevmode
     \makebox[\textwidth]{\includegraphics[width=\ftw\textwidth]{{{%
          sgt/smin/sgttKT3-2ycut30compmnspecETA1-50x1e7sminc-crop}}}}
    \caption{Same as \Fig{fig:sminc_sgts_e2} but for the pseudo rapidity distribution of the top-tagged jet from the $t$-channel.}
    \label{fig:sminc_sgtt_eta1}
  \end{center}
\end{figure}
From \Fig{fig:sminc_sgts} and \Fig{fig:sminc_sgtt} a value of $s_\tmin=5\GeV^2$ for the $s$-channel and a value of $s_\tmin=1\GeV^2$ for the $t$-channel is found to represent a compromise between systematic and statistical uncertainties. These values lie within the respective plateaus of the $s_\tmin$ dependence of the fiducial cross sections. They are chosen well below all obvious physical scales but are still large enough to enable satisfactory statistical precision in the numerical integration in a reasonable amount of time. 

The validation of the approximate $s_\tmin$ independence of differential distributions in the proximity of the chosen $s_\tmin$ values is exemplified in \Fig{fig:sminc_sgts_e2} and \Fig{fig:sminc_sgtt_eta1} by comparing them to distributions calculated with higher and lower values of the slicing parameter. The distribution of the energy of the light jet produced in the $s$-channel process calculated at NLO accuracy for $s_\tmin=0.01\GeV^2$ (solid blue), $s_\tmin=5\GeV^2$ (dotted red) and $s_\tmin=100\GeV^2$ (dashed black) is shown in \Fig{fig:sminc_sgts_e2}. At the bottom of the plot the pull distributions together with the p-values and reduced $\chi^2$  for the comparisons of the histograms for $s_\tmin=5\GeV^2$ and the lower (open blue squares) and higher values (solid black squares) are displayed. The distributions obtained with different values of the slicing parameter agree with each other within the statistical uncertainties. Similar comparisons for the distributions of transverse energy and pseudo rapidity of the top-tagged jet, the pseudo rapidity of the light jet as well as the invariant mass and rapidity of the top-light jet system can be found in \Fig{fig:sminc_sgts_et1} to \Fig{fig:sminc_sgts_yee} in appendix~\ref{app:smindep_sgt}. Altogether, the value $s_\tmin=5\GeV^2$ is chosen sufficiently small for the NLO calculation of differential cross sections in the $s$-channel. In \Fig{fig:sminc_sgtt_eta1} an analogous analysis for the pseudo rapidity of the top-tagged jet from the $t$-channel but with respective values for the slicing parameter of $s_\tmin=0.001\GeV^2$ (solid blue), $s_\tmin=1\GeV^2$ (dotted red) and $s_\tmin=50\GeV^2$ (dashed black) is presented. Respective distributions of transverse energy of the top-tagged jet, the energy and pseudo rapidity of the light jet as well as the invariant mass and rapidity of the top-light jet system are collected in \Fig{fig:sminc_sgtt_e2} to \Fig{fig:sminc_sgtt_mee} in appendix~\ref{app:smindep_sgt}.  A similar conclusion regarding the chosen value $s_\tmin=1\GeV^2$ in the $t$-channel calculations can be drawn. This validation of the approximate $s_\tmin$ independence of the distributions also nicely demonstrates the practicality of the \PSS method for the cancellation of the \IR divergences in the differential jet cross sections defined in terms of the $3\to2$ jet algorithm. 
\clearpage

\chapter{Application and results}\label{sec:applres}
The applicability of differential jet cross sections presented here is twofold: Interpreting the differential cross section as an event weight for events defined in terms of the final-state variables allows to generate corresponding unweighted events. For event weights calculated at NLO accuracy the distribution of the events in the generated sample follows the NLO cross section. Thus, the generated events can be treated as the outcome of a toy experiment and can be used to make NLO predictions for the distribution of variables by filling respective histograms.  Beyond that, the ability to calculate the event weight at NLO accuracy for a given event enables the calculation of the cumulative likelihood for the whole event sample including full NLO corrections. Thereby, the powerful \MEM is elevated to a sound theoretical foundation at NLO accuracy. 

In the following sections the generation of unweighted jet events by means of appropriate NLO event weights is performed. The same weights are, in turn, used in the \MEM to analyse the generated jet event samples. In particular, specific event definitions, based on studies regarding the dependence of distributions of jet variables on the clustering prescriptions, are given in section~\ref{sec:evgen}. The impact of the NLO corrections on the distribution of these events is investigated anticipating the application of the \MEM to the events which follow the NLO cross section. In section~\ref{sec:MEMNLO} the impact of the NLO corrections on the analyses is demonstrated by showing that the application of the \MEM with likelihoods based on the Born approximation only to these events yields biased estimators with respect to the analogous analyses based on the NLO predictions.

\section{Generating unweighted jet events distributed according to NLO cross sections}\label{sec:evgen}
With an NLO event weight like in \Eq{eq:diffnjetxs} it is straightforward to generate unweighted jet events $\{\vec{x}_i\}$ distributed according to the NLO cross section using a `acceptance-rejection' algorithm (see \Ref{von195113}). For a specific jet algorithm (defined by resolution criterion, clustering prescription and experimental cuts) the respective differential jet cross section (see \Eq{eq:diffnjetxs}) can be interpreted as an NLO event weight for the event $\vec{x}_i$ 
\begin{equation}\label{eq:evwgtn}
  \rho\left(\vec{x_i}\right)=
  \frac{d\sigma^{\NLO}}{d{x_i}_1\ldots d{x_i}_r}.
\end{equation}
The application of the acceptance-rejection method works as follows: An upper boundary for the weights has to be determined
\begin{equation}
\rho_{\text{max}}=\sup\limits_{i}\rho\left(\vec{x}_i\right)
\end{equation}
for example within the phase space integration of the total cross section or on-the-fly while evaluating the weights for possible event candidates. A candidate for a jet event $\vec{x}$ is constructed using $r$ random numbers. The weight introduced in \Eq{eq:evwgtn} is
calculated for the candidate event and interpreted as a measure for the probability of occurrence of the event. Note that according to \Eq{eq:diffnjetxs}, additional integrations must be performed to evaluate $\rho\left(\vec{x_i}\right)$. This weighted event can be `unweighted' with an additional uniformly distributed random number 
\begin{equation}
r_{\textrm{u}}\in[0,\rho_{\textrm{max}}].
\end{equation} 
The candidate event is accepted if
\begin{equation}
r_{\textrm{u}}<\rho\left(\vec{x_i}\right)
\end{equation}
and discarded otherwise. Unweighted events obtained by this method are distributed according to the probability distribution that is given by the NLO cross section. Filling histograms with these unweighted events and comparing them to the respective differential distributions calculated at NLO accuracy with a conventional parton level Monte Carlo serves as a consistency check of the event generation.

\subsection{Application I: Top-quark pair events from $e^+e^-$ annihilation following the NLO cross section}
\label{sec:evgentt}
The event generation for $t\bar{t}$ events produced at an $e^+e^-$ collider running at $\sqrt{s}=500$~GeV is validated as a first example. The setup given in section~\ref{sec:ttprod} is used.
For simplicity, the top-quark decay is not included but it is assumed that top-tagged jets defined by the $3\to2$ jet algorithm based on \Eq{eq:reslepcoll} are observed. Vetoing additional resolved jets by exclusively requiring a pair of top-tagged jets in the final state allows to parameterise their jet momenta $J_t,J_{\bar{t}}$ dependent on the top-quark mass $m_t$ as
\begin{eqnarray}\label{eq:jmomtt}
\nn 
J_{t}&=&\left({\sqrt{s}\over2},\;\sqrt{{s\over4}-m_t^2}\;\cos{\phi_{t}}\sin{\theta_{t}},
  \;\sqrt{{s\over4}-m_t^2}\;\sin{\phi_{t}}\sin{\theta_{t}},
  \;\sqrt{{s\over4}-m_t^2}\;\cos{\theta_{t}}\right),\\
J_{\bar{t}} &=& \left({\sqrt{s}\over2},-\vec{J}_t\right).
\end{eqnarray}
\begin{figure}[htbp]
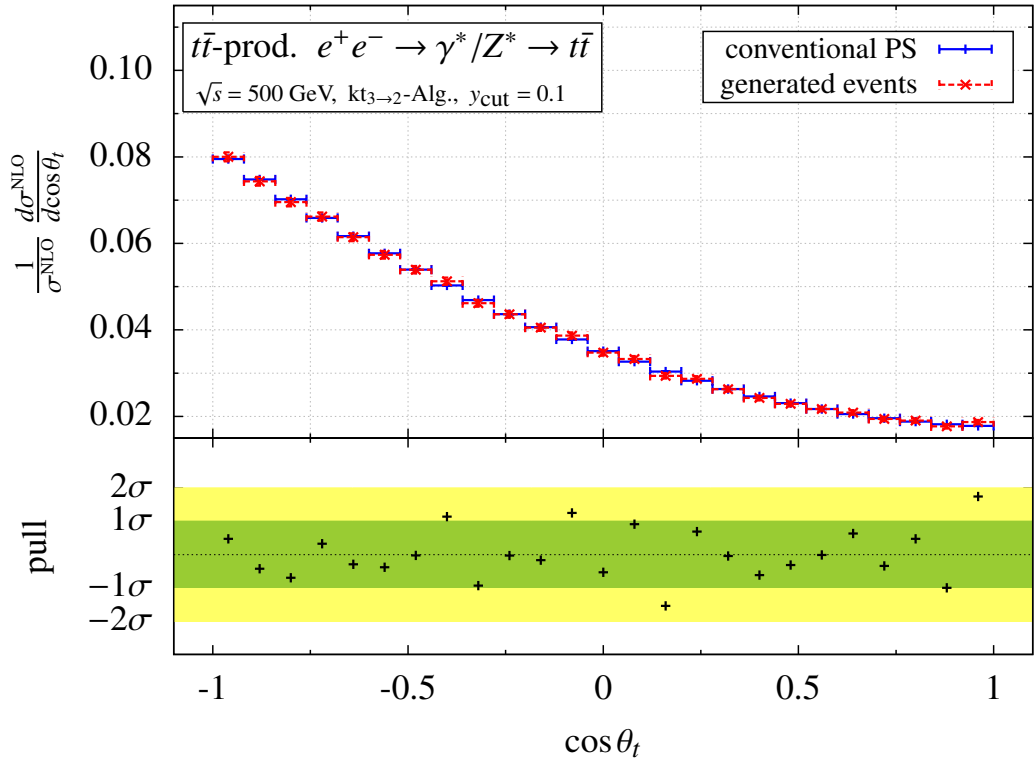

  \begin{center}
    \leavevmode
    \makebox[\textwidth]{\includegraphics[width=\ftw\textwidth]{{{%
          dytt/eeQQKT3-2ycut0.1compCTH1-150evts-crop}}}}
    \caption{Validation of the generation of unweighted NLO 
      top-quark pair events by comparing the polar angle distribution of the top-tagged jet calculated with a conventional parton level MC (solid blue) to histograms filled with the unweighted events (dashed red).}
    \label{fig:partonMCvsunwgtEv_ttprod_cth1}
  \end{center}
\end{figure}
A $t\bar{t}$ event is defined by the angles of the observed top-tagged jet $\vec{x}=(\cos{\theta_{t}},\;\phi_{t})$.
Note that the $3\to2$ clustering prescriptions keep the top-quark jets on shell $J^2_t=J^2_{\bar{t}}=m^2_t$ and the centre-of-mass energy of the colliding leptons $s$ is fixed. From \Eq{eq:jmomtt} it is obvious that momentum conservation, on-shell conditions and symmetry around the $z$-axis essentially constrain the kinematics of the interaction to a dependence on the polar angle $\theta_t$ only (cf. the flat $\phi_t$-distribution in \Fig{fig:partonMCvsjetMC_ttprod_ph1}). An NLO jet event weight for $\vec{x}=(\cos{\theta_{t}},\;\phi_{t})$ can be obtained from the differential jet cross section according to
\begin{equation}\label{eq:jetwgttt}
\frac{d\sigma^\NLO}{d\vec{x}}
  =\frac{d\sigma^\NLO}{d\cos{\theta_{t}}
    \;d\phi_{t}}\\
  =
  \frac{\beta_t}{32\pi^2}\;\frac{d\sigma^\NLO}
  {dR_{2}(J_1,J_2)}\bigg|_{\cos{\theta_{1}}=\cos{\theta_{t}},
    \;\phi_{1}=\phi_{t}}
\end{equation}
with $\beta_t = \sqrt{1-\frac{4m^2_t}{s}}$.

With \Eq{eq:jetwgttt} $73128$ unweighted $t\bar{t}$ events distributed according to the NLO cross section are generated for a top-quark mass of $m_t=174$~GeV. Because the jet momenta in \Eq{eq:jmomtt} are completely determined by the generated variables $\vec{x}=(\cos{\theta_{t}},\;\phi_{t})$ also histograms of arbitrary jet variables constructed from these jet momenta can be filled with these unweighted events.

The comparison of the filled histograms (red dashed lines) with distributions calculated at NLO accuracy using a conventional parton level Monte Carlo (blue solid lines) is exemplified in \Fig{fig:partonMCvsunwgtEv_ttprod_cth1} for the polar angle of the top-tagged jet.
Distributions of the azimuthal angle, the transverse momentum and the rapidity of the top-quark jet are collected in \Fig{fig:partonMCvsunwgtEv_ttprod_ph1} to \Fig{fig:partonMCvsunwgtEv_ttprod_y1} in appendix~\ref{app:unwgtev_tt}. Again, in the lower parts of the plots the pull distributions are shown. The pull distributions show perfect agreement between the respective histograms within
the statistical uncertainties. In conclusion, the generated events are distributed according to the NLO predictions.

\subsection{Application II: Single top-quark events from proton-proton collisions following the NLO cross section}\label{sec:sgtevgen}
In this section jet events comprised of a resolved top-tagged and a light jet that are unweighted using the differential cross section defined in \Eq{eq:diffjetxs} are generated. In what follows, a jet event is defined as a set of measured jet variables describing the signal signature and constraining the $4$-momenta in the final state. To consistently maintain universality in the event definition in terms of generic jet variables, the differential jet cross section from \Eq{eq:diffjetxs} is defined for a jet algorithm employing the $3\to2$ clustering prescriptions given in section~\ref{sec:PhaseSpaceparameterisation} (see \Eq{eq:diffnjetxs}). However, current experiments typically rely on jet algorithms using conventional $2\to1$ recombination schemes. It is thus important to examine the impact of the modified recombination procedure on various jet variables entering a potential event definition. Performing this study might help to choose a sensible event definition with minimal dependence on the details of the clustering prescriptions used in the jet algorithms.

\subsubsection{Impact of the new jet clustering}\label{sec:sgtimpac}
Since the clustering prescriptions only enter the calculation if unresolved radiation is recombined to jets, it is sufficient to study the exclusive formulation of the jet algorithm. Contributions stemming from additional resolved jets in the final state are not sensitive to the recombination procedure at NLO accuracy. Note that only normalised distributions are studied because the \MEM is only sensitive to the shapes and not the total event number. In case the extended \MEM is used any difference in the normalisation can  be achieved by an appropriate rescaling.
\begin{figure}[htbp]
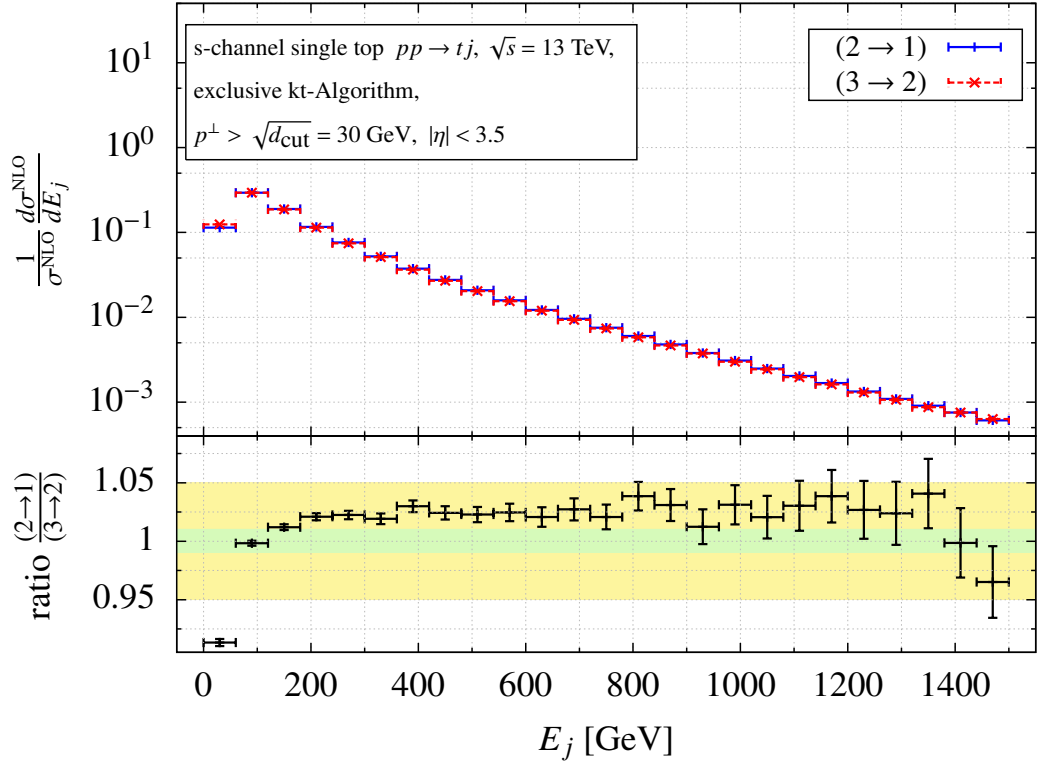

  \begin{center}
    \leavevmode
    \makebox[\textwidth]{\includegraphics[width=\ftw\textwidth]{{{%
          sgt/jetcomp/sgtsKT3-2ycut30compjetE2-200x1e7-crop}}}}
    \caption{Energy distributions of the light jet from exclusive $s$-channel single top-quark production calculated at NLO accuracy with $3\rightarrow 2$ and $2\rightarrow 1$ jet clusterings.}
    \label{fig:jetcom_sgts_e2}
  \end{center}
\end{figure}
\begin{figure}[htbp]
  \begin{center}
    \leavevmode
    \makebox[\textwidth]{\includegraphics[width=\ftw\textwidth]{{{%
          sgt/jetcomp/sgtsKT3-2ycut30compjetETA2-200x1e7-crop}}}}
    \caption{Same as \Fig{fig:jetcom_sgts_e2} but for the pseudo rapidity of the light jet jet from the $s$-channel.}
    \label{fig:jetcom_sgts_eta2}
  \end{center}
\end{figure}
\begin{figure}[htbp]
  \begin{center}
    \leavevmode
    \makebox[\textwidth]{\includegraphics[width=\ftw\textwidth]{{{%
          sgt/jetcomp/sgttKT3-2ycut30compjetET1-200x1e7-crop}}}}
    \caption{Same as \Fig{fig:jetcom_sgts_e2} but for the transverse energy of the top-tagged jet jet from the $t$-channel.}
    \label{fig:jetcom_sgtt_et1}
  \end{center}
\end{figure}
\begin{figure}[htbp]
  \begin{center}
    \leavevmode
    \makebox[\textwidth]{\includegraphics[width=\ftw\textwidth]{{{%
          sgt/jetcomp/sgttKT3-2ycut30compjetETA1-200x1e7-crop}}}}
    \caption{Same as \Fig{fig:jetcom_sgts_e2} but for the pseudo rapidity of the top-tagged jet jet from the $t$-channel.}
    \label{fig:jetcom_sgtt_eta1}
  \end{center}
\end{figure}
\begin{figure}[htbp]
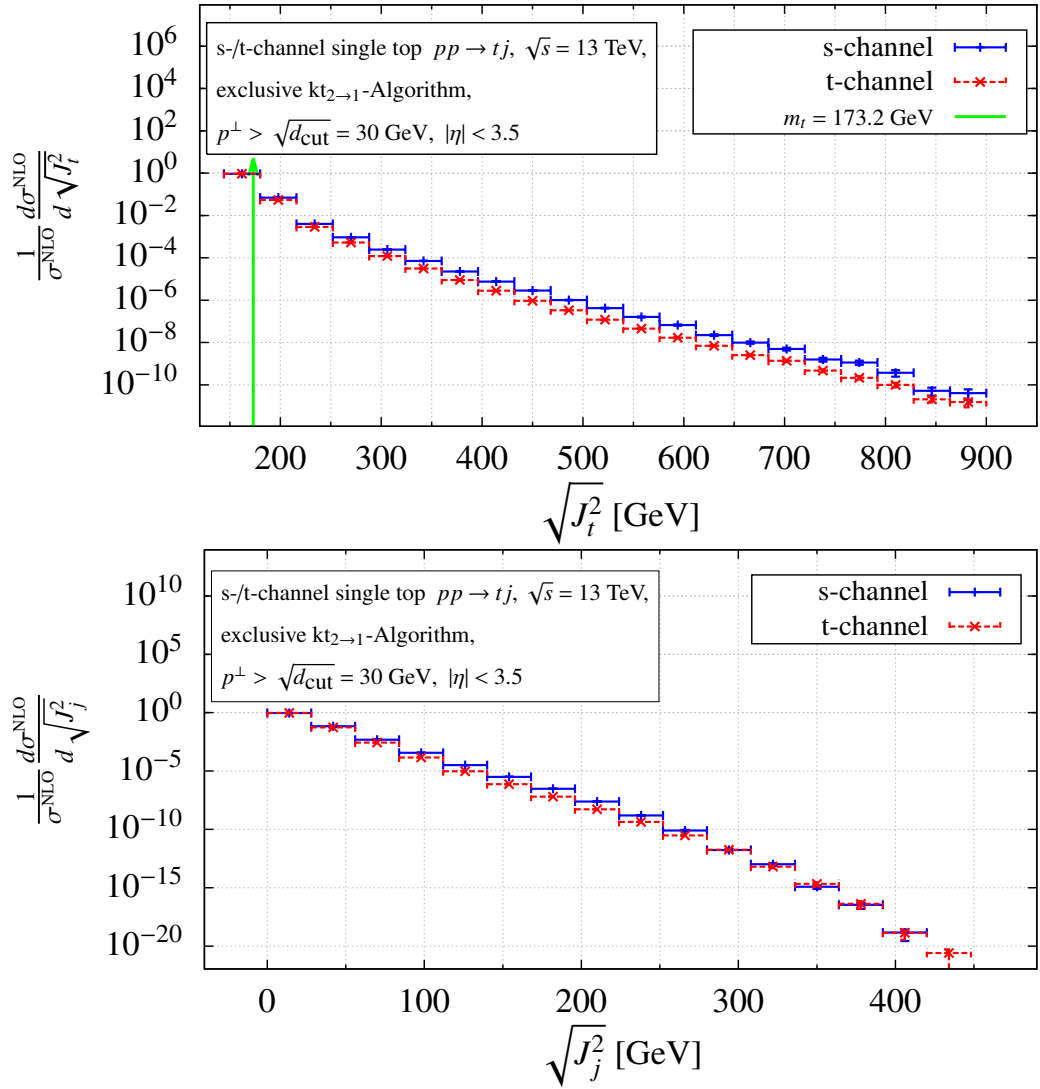

  \begin{center}
    \leavevmode
    \includegraphics[width=0.87\textwidth]{{{%
          sgt/jetcomp/sgtKT3-2ycut30compjetM1-200x1e7-crop}}}\\
    \includegraphics[width=0.87\textwidth]{{{%
          sgt/jetcomp/sgtKT3-2ycut30compjetM2-200x1e7-crop}}}
    \caption{Distribution of the jet masses from exclusive $s$- and $t$-channel single top-quark production calculated at NLO accuracy with $2\rightarrow 1$ jet clusterings.}
    \label{fig:jetcom_sgtm}
  \end{center}
\end{figure}
In \Fig{fig:jetcom_sgts_e2} to \Fig{fig:jetcom_sgtt_eta1} distributions calculated at NLO accuracy with a $3\to2$ (red dashed) and $2\to1$ jet algorithm (solid blue) are compared. In particular, distributions of the energy and pseudo rapidity of the light jet from the $s$-channel and the transverse energy and the pseudo rapidity of the top-tagged jet from the $t$-channel are studied as examples. Similar studies for various energy and angular variables are collected in \Fig{fig:jetcom_sgts_et1} to \Fig{fig:jetcom_sgtt_eta2} in appendix~\ref{app:impact3to2_sgt}. The bottom of the plots shows the ratio of the respective distributions. 

For the majority of the studied variables the difference between the two recombination schemes is at the level of a few per cent in most regions of phase space. However, also pronounced differences can be observed especially in the energy distributions. In bins near the kinematical threshold their shapes can differ by more than $50\%$ (see e.g. \Fig{fig:jetcom_sgtt_et1}). To explain these differences it should be noted that jets defined by the $3\to2$ clustering are strictly kept on their mass shell while the masses of jets defined by a $2\to1$ clustering might differ significantly from the masses of their parent partons. This is illustrated in \Fig{fig:jetcom_sgtm} where the mass distributions of the jet containing the top quark and the jet comprised out of light quarks only are shown. The invariant masses of the jets defined by the $2\to1$ jet algorithm largely overshoot the on-shell values of the $3\to2$ clusterings given by $J_t^2=m_t^2$  and $J_j^2=0$. Phase space boundaries and especially thresholds in energy distributions are affected by these mass differences.

In addition, exact $4$-momentum conservation is guaranteed in all $3\to2$ clusterings given in section~\ref{sec:PhaseSpaceparameterisation}. For example, in the E-scheme the recombined momentum is defined as the sum of the $4$-momenta of the recombined particles. In this case the overall $4$-momentum is also conserved. However, unresolved particles which are associated with the beam are usually simply dropped from the list of particles without any recombination in conventional $2\to1$ jet algorithms. This clearly causes an overall imbalance of the momentum. Regarding the angular distributions, with only minor differences of a few per cent in the bins where most of the events are expected, not such a pronounced dependence on the clustering prescription is observed (see e.g. \Fig{fig:jetcom_sgts_eta2} and \Fig{fig:jetcom_sgtt_eta1}). 

Finally it should be noted that the observed differences are solely stemming from the real contribution to the NLO corrections. Observing large differences in specific regions of phase space hints at large NLO corrections signalling potential unreliability of fixed order perturbation theory in these regions. Defining the events in terms of variables which only show a weak dependence on the clustering prescription might help to improve the reliability of perturbation theory.

\subsubsection{Jet events for the exclusive production of one resolved top-tagged jet and one resolved light jet}
\label{sec:event-definition-exclusive-case}
In the exclusive case the observed final state is required to contain precisely one resolved jet containing the top-quark and one resolved jet formed by light quarks only: $pp\rightarrow tj$. Events with more than one resolved light jet are vetoed. Since the decay of the top quark is not included, the single top-quark events are defined as far as possible in terms of variables related to the light jet. Additionally, the findings from the study of the impact of the $3\to2$ clustering prescription given above suggest to favour angular variables in the event definition. An exclusive single top-quark event $\vec{x}$ is defined by the pseudo rapidity of the resolved top-tagged jet ($t$) and the energy, pseudo rapidity and azimuthal angle of the light jet ($j$)
\begin{equation}\label{eq:eventdef}
\vec{x}=(\eta_{t},E_{j}, \eta_{j}, \phi_{j}).
\end{equation}
The jet momenta of the two resolved jets can be parameterised with these variables as
\begin{eqnarray}\label{eq:eventparexcl}
\nn J_t&=&\left(E_{t},\;-J^{\perp}\cos{\phi_{j}},\;-J^{\perp}\sin{\phi_{j}},
  \;J^{\perp}\sinh{\eta_{t}}\right),\\
J_{j}&=&\left(E_{j},\;J^{\perp}\cos{\phi_{j}},\;J^{\perp}\sin{\phi_{j}},
  \;J^{\perp}\sinh{\eta_{j}}\right)
\end{eqnarray}
with
\begin{displaymath}
  J^{\perp}=\frac{E_{j}}{\cosh{\eta_j}}
  \quad\mbox{and}\quad E_{t}=\sqrt{{J^{\perp}}^2\cosh^2{\eta_t}+m^2_t}
\end{displaymath}
and the top-quark mass $m_t$ as a free parameter. Transforming the variables in the phase-space measure
\begin{equation}
  \label{eq:OneParticlePhaseSpaceMeasure}
  d^4p\;\delta(p^2-m^2)=\frac{d^3p}{2E}
  =\frac{1}{2}\sqrt{E^2-m^2}\;dE\;d\cos{\theta}\;d\phi
  = \frac{p^{\perp} }{2\cosh{\eta}}\;dE\;d\eta\;d\phi
\end{equation}
and rewriting the delta functions assuring momentum conservation
\begin{eqnarray}
 dx_a\;dx_b\;dR_2(J_t,J_j)
 &=&dx_a\;dx_b\;\frac{d^3J_j}{2E_j}
 \frac{d\vec{J}^{\perp}_tdJ^z_t}{2E_t}\;
 \delta\left(\vec{J}^{\perp}_j+\vec{J}^{\perp}_t\right) \nonumber\\
\nonumber&\times&\delta\left((x_a+x_b)\frac{\sqrt{s}}{2}-E_j-E_t\right)
 \delta\left((x_a-x_b)\frac{\sqrt{s}}{2}-J^z_j-J^z_t\right)\\
 \nonumber
 &=&\frac{1}{2s}\;\frac{{J^{\perp}}^2\cosh{\eta_t}}{E_t\;\cosh{\eta_j}}\;
 dE_j\;d\eta_j\;d\phi_jd\eta_t\;d\vec{J}^{\perp}_t\;dx_a\; dx_b\;
 \delta\left(\vec{J}^{\perp}_j+\vec{J}^{\perp}_t\right)\\
 &\times& 
 \delta\left(x_a-\frac{1}{\sqrt{s}}(E_j+E_t+J^z_j+J^z_t)\right)\;
 \delta\left(x_b-\frac{1}{\sqrt{s}}(E_j+E_t-J^z_j-J^z_t)\right)
\end{eqnarray}
allows to write the exclusive jet event weight including NLO corrections in terms of the differential $2$-jet cross section at NLO (cf. \Eq{eq:diffnjetxs})
\begin{equation}\label{eq:diffjxsex}
  \frac{d\sigma_\excl^{\text{NLO}}}{d\vec{x}}= 
 \frac{d\sigma_\excl^{\text{NLO}}}{d\eta_{t}\; dE_{j}\; d\eta_{j}\;
   d\phi_{j}}
 =\frac{{J^{\perp}}^2\cosh{\eta_t}}{2\;s\;E_t\;\cosh{\eta_j}} \;
 \frac{d\sigma_\excl^{\text{NLO}}}{dR_2(J_{t},J_{j})}.
\end{equation}
\begin{figure}[htbp]
  \begin{center}
    \leavevmode
    \makebox[\textwidth]{\includegraphics[width=\ftw\textwidth]{{{%
          sgt/evgen/sgtsKT3-2ycut30compmnspecETA1-94evts-crop}}}}
    \caption{Pseudo rapidity distribution of the top-tagged jet from exclusive $s$-channel single top-quark production
      calculated at NLO accuracy using a conventional parton level MC (solid blue) compared to histograms filled with generated NLO events (dashed red).}
       \label{fig:partonMCvsgenev_sgts_eta1}
  \end{center}
\end{figure}
\begin{figure}[htbp]
  \begin{center}
    \leavevmode
    \makebox[\textwidth]{\includegraphics[width=\ftw\textwidth]{{{%
          sgt/evgen/sgttKT3-2ycut30compmnspecE2-90evts-crop}}}}
    \caption{Same as \Fig{fig:partonMCvsgenev_sgts_eta1} but for the energy of the light jet from the $t$-channel.}
    \label{fig:partonMCvsgenev_sgtt_e2}
  \end{center}
\end{figure}
This exclusive NLO jet event weight can be used to generate unweighted jet events $\{\vec{x}_i\}$ for single top-quark production which are distributed according to the NLO cross section.

A total of $N_\excl=12755$ unweighted jet events produced for the $s$-channel and $N_\excl=24088$ produced for the $t$-channel are generated with a top-quark mass of $m_t=173.2\GeV$. Because the set of variables in the exclusive event definition (see \Eq{eq:eventdef}) unambiguously determines the jet momenta in \Eq{eq:eventparexcl}, the generated events can be used to fill histograms of arbitrary jet variables corresponding to these jet momenta. 

To check the distribution of the unweighted events according to the NLO cross section they are filled in histograms and compared to distributions calculated at NLO accuracy using a conventional parton level Monte Carlo (cf. \Eq{eq:diffobs}) employing a $3\to2$ jet algorithm (cf. \Eq{eq:3to2jetalg}).
In \Fig{fig:partonMCvsgenev_sgts_eta1} and \Fig{fig:partonMCvsgenev_sgtt_e2} exemplary distributions of the pseudo rapidity of the top-tagged jet from $s$-channel single top-quark production and the energy of the light jet from $t$-channel single top-quark production are shown. Additional histograms of the (transverse) energies and rapidities of the top-tagged and light jet as well as of the invariant mass and rapidity of the top-light jet system from the $s$- and $t$-channel can be found in \Fig{fig:partonMCvsgenev_sgts_et1} to \Fig{fig:partonMCvsgenev_sgtt_mee} in appendix~\ref{app:unwgtev_sgt}. 
The distributions calculated with the conventional parton level Monte Carlo with subsequent $3\to2$ jet clustering are shown as blue lines while the histograms filled with the unweighted NLO events are shown as dashed red lines. Again, at the bottom of each plot the pull distributions, p-values and reduced $\chi^2$ of the histogram comparisons are shown.
The pull distributions illustrate that the generated events are in fact distributed according to the respective NLO cross sections of $s$-channel and $t$-channel single top-quark production. The p-values and reduced $\chi^2$ confirm this observation. 

\subsubsection{Impact of NLO corrections on the distribution of the exclusive events}\label{sec:kfacexcl}
In this section the impact of the NLO corrections on the theoretical predictions for single top-quark production are investigated. This study is especially instructive in view of the application of the \MEM to events distributed according to the NLO cross section.
In table~\ref{tab:fidxsex} the Born and NLO cross sections for the $s$- and $t$-channel production of a single resolved top quark in association with exactly one resolved light jet in the fiducial phase space are summarised.
\begin{table}[htbp]
  \centering
  \caption{Fiducial exclusive cross sections.}
  \label{tab:fidxsex}
  \def\arraystretch{1.8}
  \begin{tabular}{|c|c|c|}
   \hline
    $\substack{p^\perp>30\GeV,\; \eta<3.5 \\ 3\to2 \text{ clustering}}$
    & $s$-channel                         
    & $t$-channel                        \\ \hline
    ${\sigma^{\text{Born}}}^{\Delta_{2\mu_0}}_{\Delta_{\mu_0/2}}$ [pb]
    & $3.093(2)_{-0.099\;(-3.2\%)}^{+0.075\;(+2.4\%)}$ 
    & $80.07(3)_{-8.37\;(-10.4\%)}^{+6.71\;(+8.4\%)}$ \\ \hline
    ${\sigma^{\text{NLO}}_\excl}^{\Delta_{2\mu_0}}_{\Delta_{\mu_0/2}}$ [pb]
    & $3.05(7)_{-0.02\;(-0.7\%)}^{+0.03\;(+1.0\%)}$  
    & $60.0(3)_{-2.3\;(-3.9\%)}^{+3.3\;(+5.5\%)}$ \\ \hline
\end{tabular}
\end{table}
The fiducial exclusive $s$-channel cross section receives only a small negative NLO correction of about $-1.1\%$ but the NLO corrections reduce the fiducial exclusive $t$-channel cross section by about $-25.1\%$. The impact of the simultaneous variation of the renormalisation and factorisation scale between $\mur=\muf=\mu_0/2$ and $\mur=\muf=2\mu_0$ is reduced from $-3.2\%$ and $+2.4\%$ to $-0.7\%$ and $+1.0\%$ in the $s$-channel and from $-10.4\%$ and $+8.4\%$ to $-3.9\%$ and $+5.5\%$ in the $t$-channel (see sub- and superscript in table~\ref{tab:fidxsex}). At least the impact of the NLO corrections to the fiducial cross section for the exclusive $s$-channel single top-quark production is covered by the scale variation in the Born result while for the $t$-channel it is not covered at all.

\begin{figure}[htbp]
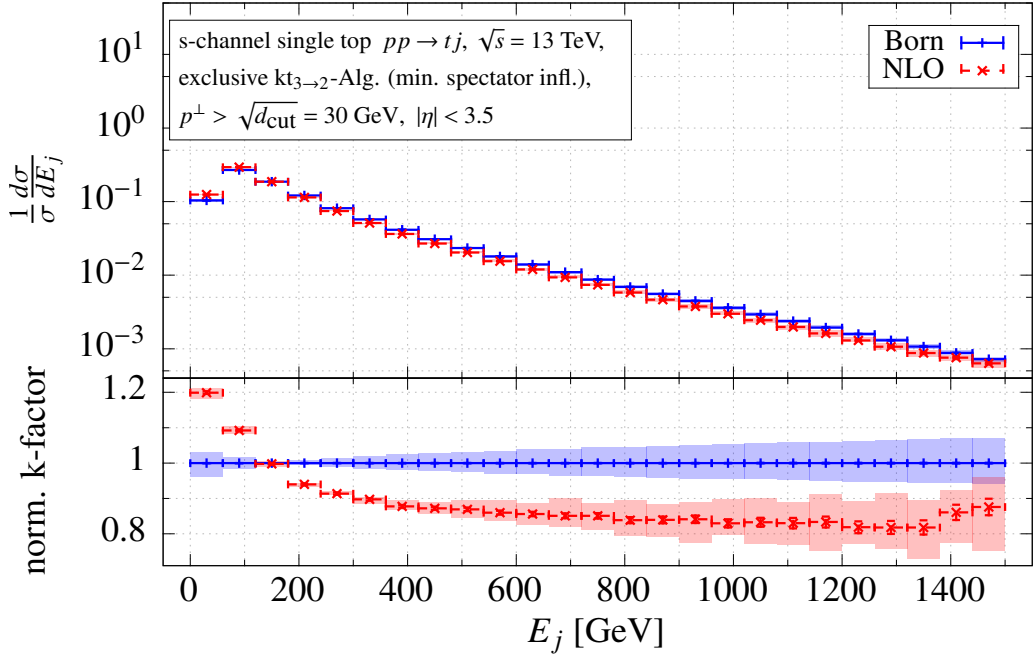

  \begin{center}
    \leavevmode
    \makebox[\textwidth]{\includegraphics[width=\ftw\textwidth]{{{%
          sgt/kfmu/sgtsKT3-2ycut30compmumnspecE2-200x1e7-crop}}}}
    \caption{Normalised energy distribution and k-factors of the energy of the light jet from $s$-channel
      single top-quark production using the exclusive event
      definition.}  
    \label{fig:kfac_sgts_e2}
  \end{center}
\end{figure}
\begin{figure}[htbp]
  \begin{center}
    \leavevmode
    \makebox[\textwidth]{\includegraphics[width=\ftw\textwidth]{{{%
          sgt/kfmu/sgtsKT3-2ycut30compmumnspecETA2-200x1e7-crop}}}}
    \caption{Same as \Fig{fig:kfac_sgts_e2} but for the pseudo rapidity of the light jet from the $s$-channel.}  
    \label{fig:kfac_sgts_eta2}
  \end{center}
\end{figure}
\begin{figure}[htbp]
  \begin{center}
    \leavevmode
    \makebox[\textwidth]{\includegraphics[width=\ftw\textwidth]{{{%
          sgt/kfmu/sgtsKT3-2ycut30compmumnspecETA1-200x1e7-crop}}}}
    \caption{Same as \Fig{fig:kfac_sgts_e2} but for the pseudo rapidity of the top-tagged jet from the $s$-channel.}  
    \label{fig:kfac_sgts_eta1}
  \end{center}
\end{figure}
\begin{figure}[htbp]
  \begin{center}
    \leavevmode
    \makebox[\textwidth]{\includegraphics[width=\ftw\textwidth]{{{%
          sgt/kfmu/sgttKT3-2ycut30compmumnspecE2-200x1e7-crop}}}}
    \caption{Same as \Fig{fig:kfac_sgts_e2} but for the energy of the light jet from the $t$-channel.}  
    \label{fig:kfac_sgtt_e2}
  \end{center}
\end{figure}
\begin{figure}[htbp]
  \begin{center}
    \leavevmode
    \makebox[\textwidth]{\includegraphics[width=\ftw\textwidth]{{{%
          sgt/kfmu/sgttKT3-2ycut30compmumnspecETA2-200x1e7-crop}}}}
    \caption{Same as \Fig{fig:kfac_sgts_e2} but for the pseudo rapidity of the light jet from the $t$-channel.}  
    \label{fig:kfac_sgtt_eta2}
  \end{center}
\end{figure}
\begin{figure}[htbp]
  \begin{center}
    \leavevmode
    \makebox[\textwidth]{\includegraphics[width=\ftw\textwidth]{{{%
          sgt/kfmu/sgttKT3-2ycut30compmumnspecETA1-200x1e7-crop}}}}
    \caption{Same as \Fig{fig:kfac_sgts_e2} but for the pseudo rapidity of the top-tagged jet from the $t$-channel.}  
    \label{fig:kfac_sgtt_eta1}
  \end{center}
\end{figure}
\begin{figure}[htbp]
  \begin{center}
    \leavevmode
    \makebox[\textwidth]{\includegraphics[width=\ftw\textwidth]{{{%
          sgt/kfmu/sgttKT3-2ycut30compmumnspecETA2-200x1e7nonorm-crop}}}}
    \caption{Same as \Fig{fig:kfac_sgtt_eta2} but unnormalised.}  
    \label{fig:kfac_sgtt_eta2nonorm}
  \end{center}
\end{figure}
The impact of the NLO corrections is illustrated in more detail by comparisons of distributions calculated in the Born approximation (solid blue) and including NLO corrections (dashed red) in \Fig{fig:kfac_sgts_e2} to \Fig{fig:kfac_sgts_eta1}. Respective plots for the $t$-channel are given in \Fig{fig:kfac_sgtt_e2} to \Fig{fig:kfac_sgtt_eta1}. At the bottom of the plots the ratio of the NLO results with respect to the Born results (`k-factor' distributions) are shown. The simultaneous down- and upwards variation of the renormalisation and factorisation scale is presented by the shaded areas (including statistical errors). From the normalised k-factor distributions it can be seen that the impact of the NLO corrections to the shapes of the distributions ranges from $-20\%$ to $+20\%$ in the $s$-channel and from $-10\%$ to $+25\%$ in the $t$-channel. 
The general trend in a wide range of phase space inferred from \Fig{fig:kfac_sgts_e2} to \Fig{fig:kfac_sgtt_eta1} is that varying the renormalisation and factorisation scale results in more pronounced changes in the normalised NLO than in the normalised Born distributions. The largest effect can be seen for large values of the pseudo rapidity of the light jet in the $s$-channel where very few events are expected though (see \Fig{fig:kfac_sgts_eta2}). This rather unexpected observation may be traced back to an interplay of the electroweak nature of the considered process in the Born approximation and the normalisation of the distributions: In the one-loop QCD corrections QCD interactions appear for the first time introducing a leading-order scale dependence on the renormalisation scale $\mur$. The only scale dependence of the Born cross section is the dependence of the PDFs on the factorisation scale $\muf$ which cancels to some degree in the cross section ratio. 
Because of that and since the MEM utilising extended likelihoods (cf. section~\ref{sec:MLext}) is also considered, not only normalised distributions should be studied but also unnormalised ones.
By comparison, the impact of the NLO corrections on the respective unnormalised distributions scales with the corresponding ratio of the fiducial cross sections $\sigma^{\text{NLO}}_\excl/\sigma^{\text{Born}}$ given in table~\ref{tab:fidxsex} (see the example given in \Fig{fig:kfac_sgtt_eta2nonorm} and \Fig{fig:kfac_sgts_e2nonorm} to \Fig{fig:kfac_sgtt_eta1nonorm} in appendix~\ref{app:impNLO_sgtexcl}).
Indeed, when examining the unnormalised NLO distributions it can be seen that the impact of the scale variation is mostly reduced with respect to the unnormalised Born distributions particularly in areas of the phase space where most events are expected. This holds especially true for the $t$-channel. 

Overall, this study of the impact of the NLO corrections to exclusive single top-quark production suggests that theoretical calculations based on the Born approximation do not only give inaccurate predictions for differential distributions but also do not allow to reliably estimate the theoretical uncertainty due to missing higher orders by a customary variation of the scales. This observation emphasises the importance of higher-order corrections when calculating predictions for (normalised) differential distributions of the production of a single top-quark in association with exactly one light jet.

\subsubsection{Jet events for the inclusive production of one resolved top-tagged jet and at least one resolved light jet}
\label{sec:event-definition-inclusive-case}
Taking NLO corrections into account allows to theoretically describe the emission of an additional resolved jet ($X$) in the final state
\begin{equation}
pp\rightarrow tjX.
\end{equation}
In the exclusive case, phase space cuts on the real radiation are required when vetoing these additional resolved jets. With phase space cuts additional scales are introduced into the NLO calculation resulting in potentially uncancelled logarithms which might spoil the convergence of the perturbative series  (cf. the discussion of the slicing parameter dependence in the \PSS method in section~\ref{sec:PSS}).

Without a veto on this additional jet activity more inclusive events can be defined by the same set of variables as in \Eq{eq:eventdef} consisting of the pseudo rapidity of the top-tagged jet ($t$) and now the energy, pseudo rapidity and azimuthal angle of the hardest resolved light jet ($j$). If only one top-tagged jet and one light jet are resolved their jet momenta can again be parameterised as given by \Eq{eq:eventparexcl}. On the other hand, when an additional jet $X$ is resolved (which has to be softer than the light jet $j$ by construction: $J^{\perp}_X<J^{\perp}_j$) the jet momenta of the three resolved jets can be parameterised as 
\begin{eqnarray}\label{eq:eventparincl}
\nn  J_t&=&\left(E_{t},\;-J^{\perp}_t\cos{\phi_{t}},\;
    -J^{\perp}_t\sin{\phi_{t}},\;J^{\perp}_t\sinh{\eta_{t}}\right),\\
\nn  J_{j}&=&\left(E_{j},\;J^{\perp}_j\cos{\phi_{j}},\;J^{\perp}_j
    \sin{\phi_{j}},\;J^{\perp}_j\sinh{\eta_{j}}\right),\\
  J_{X}&=&\left(E_{X},\;J^{\perp}_{X}\cos{\phi_{X}},\;J^{\perp}_{X}
    \sin{\phi_{X}},\;J^{\perp}_{X}\sinh{\eta_{X}}\right)
\end{eqnarray}
with
\begin{eqnarray*}
  &J^{\perp}_j=\frac{E_{j}}{\cosh{\eta_j}},\quad J^{\perp}_t
  =\sqrt{(J^{\perp}_j\cos{\phi_{j}}+J^{\perp}_{X}\cos{\phi_{X}})^2
    +(J^{\perp}_j\sin{\phi_{j}}+J^{\perp}_{X}\sin{\phi_{X}})^2},&\\
  &\tan\phi_t=\frac{J^{\perp}_j\sin{\phi_{j}}+J^{\perp}_{X}
    \sin{\phi_{X}}}{J^{\perp}_j\cos{\phi_{j}}+J^{\perp}_{X}\cos{\phi_{X}}},
  \quad
  E_{t}=\sqrt{{J^{\perp}_t}^2\cosh^2{\eta_t}+m^2_t},\quad
  E_{X}=J^{\perp}_{X}\cosh{\eta_X}.&
\end{eqnarray*}
Since the additional jet $X$ does not enter the event definition, the related variables have to be integrated over the allowed range (given by the inclusive event definition)
\begin{displaymath}
  p^{\perp}_{\text{min}}<J^{\perp}_{X}<J^{\perp}_{j},\quad
  0<\phi_{X}<2\pi,\quad 
  -\eta_{\text{max}}<\eta_{X}<\eta_{\text{max}}.
\end{displaymath}
Using again the transformation from \Eq{eq:OneParticlePhaseSpaceMeasure} and rewriting the delta functions as
\begin{eqnarray}\label{eq:realpsincl}
\nonumber dx_a\;dx_b\;dR_3(J_t,J_j,J_X)&=&dx_a\;dx_b\;\frac{d^3J_j}{2E_j}\frac{d^3J_X}{2E_X}
  \frac{d\vec{J}^{\perp}_tdJ^z_t}{2E_t}
  \delta\left(\vec{J}^{\perp}_j+\vec{J}^{\perp}_t+\vec{J}^{\perp}_X\right)\\
  \nonumber&&\times\;
  \delta\left((x_a+x_b)\frac{\sqrt{s}}{2}-E_j-E_t-E_X\right)
  \delta\left((x_a-x_b)\frac{\sqrt{s}}{2}-J^z_j-J^z_t-J^z_X\right)\\
  \nonumber
  &=& \frac{1}{4}\; \frac{1}{s}\;
  \frac{{J^{\perp}_j}{J^{\perp}_X}{J^{\perp}_t}\cosh{\eta_t}}
  {E_t\;\cosh{\eta_j}}\;
  dE_j\;d\eta_j\;d\phi_j\;dJ^{\perp}_X\;d\eta_X\;d\phi_X\;d\eta_t\;
  d\vec{J}^{\perp}_t\;dx_a\;dx_b\;
  \\
  \nonumber&\times&\;
  \delta\left(\vec{J}^{\perp}_j+\vec{J}^{\perp}_t+\vec{J}^{\perp}_X\right)\;  
  \delta\left(x_a-\frac{1}{\sqrt{s}}(E_j+E_t+E_X+J^z_j+J^z_t+J^z_X)\right)\\
  &\times&
  \delta\left(x_b-\frac{1}{\sqrt{s}}(E_j+E_t+E_X-J^z_j-J^z_t-J^z_X)\right)
\end{eqnarray}
allows to obtain the the $3$-jet contribution to the NLO jet event weight for $\vec{x}$ from the differential $3$-jet cross section as
\begin{eqnarray}\label{eq:diffjxs3j}
  \nonumber \frac{d\sigma_{3\text{-jet}}}{d\vec{x}}&=& 
  \frac{d\sigma_{3\text{-jet}}}{d\eta_{t}\; dE_{j}\; d\eta_{j}\; d\phi_{j}}\\
  &=&
  \int\limits_{p^{\perp}_{\text{min}}}^{J^{\perp}_{j}}dJ^{\perp}_{X}
  \int\limits_{-\eta_{\text{max}}}^{\eta_{\text{max}}}d\eta_{X}
  \int\limits_{0}^{2\pi}d\phi_{X}\;
  \frac{{J^{\perp}_j}{J^{\perp}_X}{J^{\perp}_t}
    \cosh{\eta_t}}{4s\;E_t\;\cosh{\eta_j}} \;
  \frac{d\sigma_{3\text{-jet}}}{ dR_3(J_t,J_j,J_X)}.
\end{eqnarray}
Here, the $3$-jet cross section $d\sigma_{3\text{-jet}}$ is given in terms of the real emission matrix element evaluated for partonic configurations leading to three resolved jets (cf. the last line in \Eq{eq:diffjetxs}).
The NLO weight for the inclusive jet events is given by the sum of the $2$-jet (see \Eq{eq:diffjxsex}) and $3$-jet contribution (see \Eq{eq:diffjxs3j})
\begin{equation}\label{eq:diffjxsin}
 \frac{d\sigma_\incl^{\text{NLO}}}{d\vec{x}}= 
 \frac{d\sigma^{\text{NLO}}_\excl}{d\vec{x}}
 + \frac{d\sigma_{3\text{-jet}}}{d\vec{x}}.
\end{equation}
\begin{figure}[htbp]
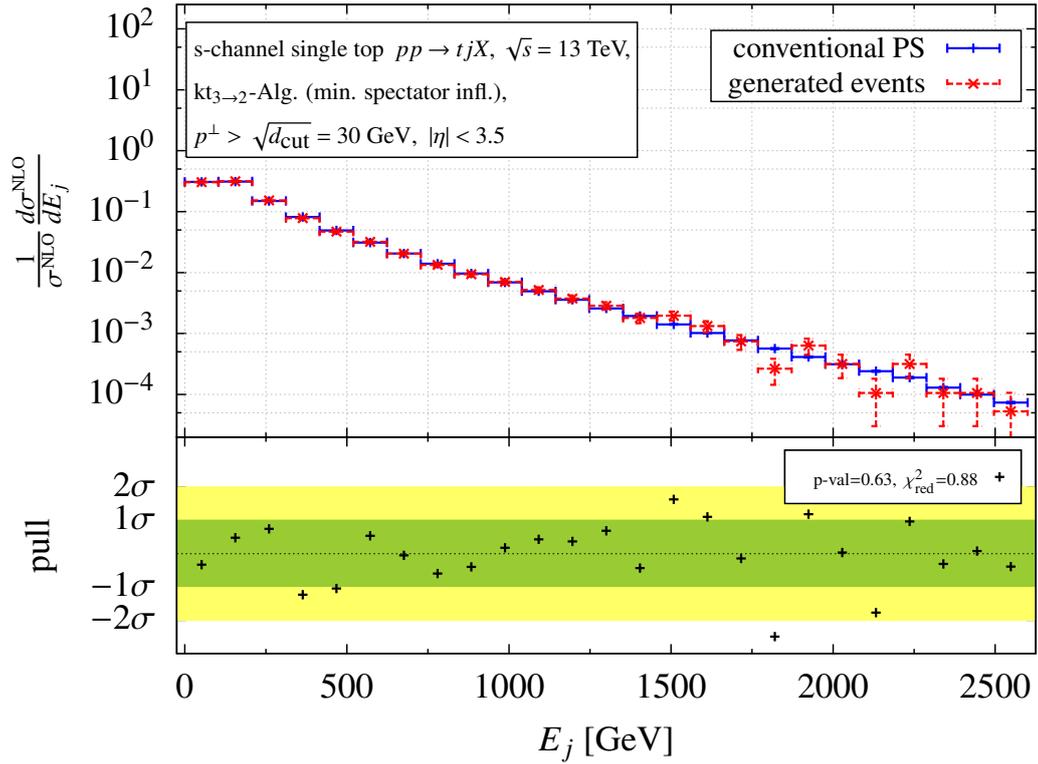

  \begin{center}
    \leavevmode
    \makebox[\textwidth]{\includegraphics[width=\ftw\textwidth]{{{%
          sgt/evgen/sgtsKT3-2ycut30compmnspecincE2-100evts-crop}}}}
    \caption{Energy distribution of the light jet from inclusive $s$-channel single top-quark production
      calculated at NLO accuracy using a conventional parton level MC (solid blue) compared to histograms filled with generated NLO events (dashed red).}
       \label{fig:partonMCvsgenev_sgtsinc_e2}
  \end{center}
\end{figure}
\begin{figure}[htbp]
  \begin{center}
    \leavevmode
    \makebox[\textwidth]{\includegraphics[width=\ftw\textwidth]{{{%
          sgt/evgen/sgttKT3-2ycut30compmnspecincETA1-114evts-crop}}}}
    \caption{Same as \Fig{fig:partonMCvsgenev_sgtsinc_e2} but for the pseudo rapidity of the top-tagged jet from the $t$-channel.}
    \label{fig:partonMCvsgenev_sgttinc_eta1}
  \end{center}
\end{figure}
Again, the distribution of the generated inclusive events according to the respective NLO cross section is validated by reproducing differential
distributions determined via a conventional Monte Carlo integration by filling histograms with the unweighted events. When not vetoing events with an additional resolved jet in the final state, more events pass the event selection criteria increasing the available statistics of the analysis. To  proportionally account for these extra $3$-jet events in the inclusive samples the number of unweighted exclusive events is scaled by the ratio of the inclusive and exclusive fiducial NLO cross sections
\begin{equation}\label{eq:Ninclscale}
N_\incl={\sigma^{\text{NLO}}_\incl\over\sigma^{\text{NLO}}_\excl}\;N_\excl.
\end{equation}
Accordingly, $N_\incl=16964$ unweighted $s$-channel events and  $N_\incl=32278$ unweighted $t$-channel events are generated. The top-quark mass is again set to $m_t=173.2\GeV$. Because the set of variables used in the inclusive event definition $\vec{x}$ might not determine the observed final state completely (cf. \Eq{eq:eventparincl}), only distributions of the variables in the event definition can be compared. This is in contrast to the case of the exclusive event definition, where the final state is completely determined by the jet variables in $\vec{x}$, allowing to also compare arbitrary jet variables. 

In \Fig{fig:partonMCvsgenev_sgtsinc_e2} and \Fig{fig:partonMCvsgenev_sgttinc_eta1} comparisons of the NLO distributions of the energy of the light jet from inclusive $s$-channel  production of single top quarks and the pseudo rapidity of the top-tagged jet from the $t$-channel are shown as examples. Comparisons of other variables in $\vec{x}$ for the $s$- and $t$-channel can be found in \Fig{fig:partonMCvsgenev_sgtsinc_eta2} to \Fig{fig:partonMCvsgenev_sgttinc_eta2} in appendix~\ref{app:unwgtev_sgtinc}. As one can see from the pull distributions, the p-values and reduced $\chi^2$ given at the bottom of each plot, both results are in perfect agreement with each other. This leads to the conclusion that the generated events which are unweighted using \Eq{eq:diffjxs3j} are distributed according to the inclusive NLO cross section.

\subsubsection{Impact of NLO corrections on the distribution of the inclusive events}\label{sec:kfacincl}
In analogy to the exclusive case, a study of the impact of NLO corrections on the distributions of the inclusive single top-quark events is performed because the generated events following the NLO cross section serve as input to the \MEM in section~\ref{sec:MEMNLO}. In table~\ref{tab:fidxsin} the $s$- and $t$-channel cross sections for the production of a single top quark in association with at least one light jet in the fiducial phase space region calculated in the Born approximation and including NLO corrections are shown.
\begin{table}[htbp]
  \centering
  \caption{Fiducial inclusive cross sections}
  \label{tab:fidxsin}
  \def\arraystretch{1.8}
  \begin{tabular}{|c|c|c|c|c|}
     \hline
    $\substack{p^\perp>30\GeV,\; \eta<3.5 \\ 3\to2 \text{ clustering}}$
    &{$s$-channel}                         
    &{$t$-channel}                          \\ \hline
    {${\sigma^{\text{Born}}}^{\Delta_{2\mu_0}}_{\Delta_{\mu_0/2}}$ [pb]} 
    &{ $3.093(2)_{-0.099\;(-3.2\%)}^{+0.075\;(+2.4\%)}$ } 
    &{ $80.07(3)_{-8.37\;(-10.4\%)}^{+6.71\;(+8.3\%)}$} \\ \hline
    {${\sigma^{\text{NLO}}_\incl}^{\Delta_{2\mu_0}}_{\Delta_{\mu_0/2}}$ [pb]}
    & $4.07(1)_{+0.08\;(+2.2\%)}^{-0.06\;(-1.5\%)}$ 
    & $80.4(5)_{-0.3\;(-0.4\%)}^{+1.5\;(+1.9\%)}$ \\ \hline
\end{tabular}
\end{table}
The results in table~\ref{tab:fidxsin} show that in the $s$-channel the contribution of the additional resolved jet adds about $+33.0\%$ to the fiducial exclusive NLO cross section from table~\ref{tab:fidxsex} increasing the NLO corrections to $+31.6\%$.  In the $t$-channel taking also the $3$-jet contribution into account amounts to an increase of the fiducial exclusive NLO cross section from table~\ref{tab:fidxsex} by about  $+34.2\%$ and reduces the size of the NLO corrections to about $+1.0\%$. Simultaneous down- and upwards variation of the renormalisation and factorisation scale $\mur=\muf=\mu_0$ by a factor of $2$ results in reduced shifts of the fiducial inclusive NLO cross section of about $+2.2\%$ and $-1.5\%$ in the $s$-channel and about $-0.4\%$ and $+1.9\%$ in the $t$-channel (see sub- and superscripts in table~\ref{tab:fidxsin}). In contrast to the observation in the exclusive case, not vetoing an additional resolved jet results in NLO corrections to the fiducial cross section whose impact is now covered by the scale variation in the $t$-channel Born result but not anymore in the $s$-channel.

\begin{figure}[htbp]
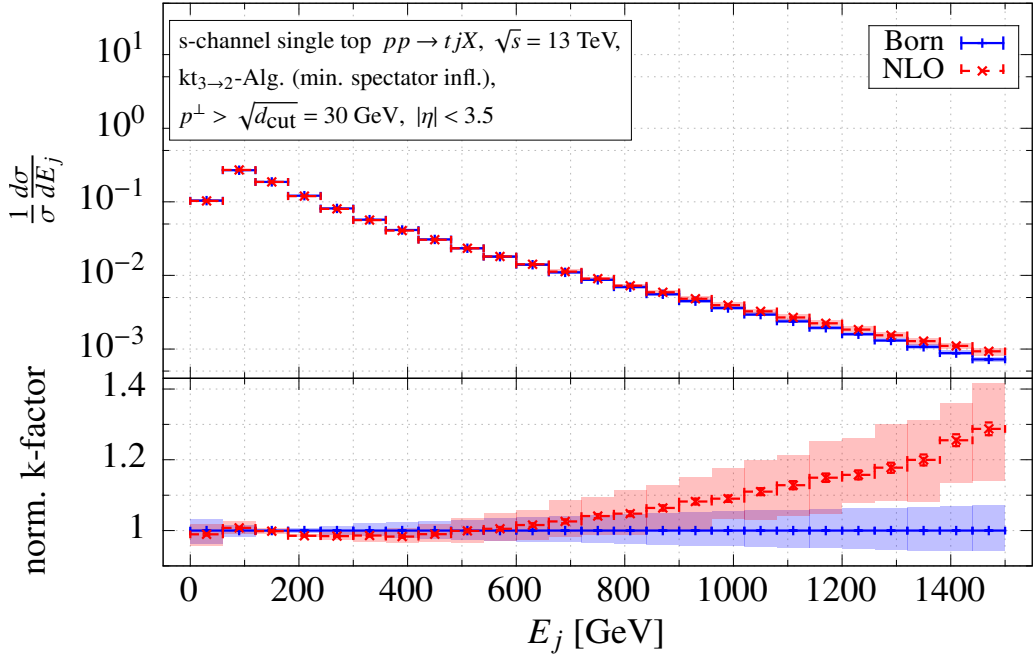

  \begin{center}
    \leavevmode
    \makebox[\textwidth]{\includegraphics[width=\ftw\textwidth]{{{%
          sgt/kfmu/sgtsKT3-2ycut30compmumnspecincE2-200x1e7-crop}}}}
    \caption{Normalised energy distribution and k-factors of the energy of the light jet from $s$-channel
      single top-quark production using the exclusive event
      definition.}  
    \label{fig:kfac_sgtsinc_e2}
  \end{center}
\end{figure}
\begin{figure}[htbp]
  \begin{center}
    \leavevmode
    \makebox[\textwidth]{\includegraphics[width=\ftw\textwidth]{{{%
          sgt/kfmu/sgtsKT3-2ycut30compmumnspecincETA2-200x1e7-crop}}}}
    \caption{Same as \Fig{fig:kfac_sgtsinc_e2} but for the pseudo rapidity of the light jet from the $s$-channel.}  
    \label{fig:kfac_sgtsinc_eta2}
  \end{center}
\end{figure}
\begin{figure}[htbp]
  \begin{center}
    \leavevmode
    \makebox[\textwidth]{\includegraphics[width=\ftw\textwidth]{{{%
          sgt/kfmu/sgtsKT3-2ycut30compmumnspecincETA1-200x1e7-crop}}}}
    \caption{Same as \Fig{fig:kfac_sgtsinc_e2} but for the pseudo rapidity of the top-tagged jet from the $s$-channel.}  
    \label{fig:kfac_sgtsinc_eta1}
  \end{center}
\end{figure}
\begin{figure}[htbp]
  \begin{center}
    \leavevmode
    \makebox[\textwidth]{\includegraphics[width=\ftw\textwidth]{{{%
          sgt/kfmu/sgttKT3-2ycut30compmumnspecincE2-200x1e7-crop}}}}
    \caption{Same as \Fig{fig:kfac_sgtsinc_e2} but for the energy of the light jet from the $t$-channel.}  
    \label{fig:kfac_sgttinc_e2}
  \end{center}
\end{figure}
\begin{figure}[htbp]
  \begin{center}
    \leavevmode
    \makebox[\textwidth]{\includegraphics[width=\ftw\textwidth]{{{%
          sgt/kfmu/sgttKT3-2ycut30compmumnspecincETA2-200x1e7-crop}}}}
    \caption{Same as \Fig{fig:kfac_sgtsinc_e2} but for the pseudo rapidity of the light jet from the $t$-channel.}  
    \label{fig:kfac_sgttinc_eta2}
  \end{center}
\end{figure}
\begin{figure}[htbp]
  \begin{center}
    \leavevmode
    \makebox[\textwidth]{\includegraphics[width=\ftw\textwidth]{{{%
          sgt/kfmu/sgttKT3-2ycut30compmumnspecincETA1-200x1e7-crop}}}}
    \caption{Same as \Fig{fig:kfac_sgtsinc_e2} but for the pseudo rapidity of the top-tagged jet from the $t$-channel.}  
    \label{fig:kfac_sgttinc_eta1}
  \end{center}
\end{figure}
\begin{figure}[htbp]
  \begin{center}
    \leavevmode
    \makebox[\textwidth]{\includegraphics[width=\ftw\textwidth]{{{%
          sgt/kfmu/sgtsKT3-2ycut30compmumnspecincETA1-200x1e7nonorm-crop}}}}
    \caption{Same as \Fig{fig:kfac_sgtsinc_eta1} but unnormalised.}  
    \label{fig:kfac_sgtsnc_eta1nonorm}
  \end{center}
\end{figure}
As in the exclusive case, the impact of the NLO corrections is illustrated in more detail by studying normalised and unnormalised distributions. Distributions calculated in the Born approximation and including NLO corrections are shown in \Fig{fig:kfac_sgtsinc_e2} to \Fig{fig:kfac_sgtsinc_eta1} for the $s$-channel and in \Fig{fig:kfac_sgttinc_e2} to \Fig{fig:kfac_sgttinc_eta1} for the $t$-channel. By comparing the NLO and Born results similar features as in the exclusive case become apparent: The impact of the NLO corrections to the normalised distributions ranges from $-5\%$ to $+30\%$ in the $s$-channel and from $-10\%$ to $+30\%$ in the $t$-channel.
Overall, allowing to resolve an additional jet results in flatter k-factor distributions especially in regions where a significant fraction of the events is expected. The $t$-channel pseudo rapidity distribution of the light jet poses an exception to that by receiving NLO corrections of up to $+30\%$ in the
central region (see \Fig{fig:kfac_sgttinc_eta2}). This can be explained by recalling the definition of the light jet $j$ in the inclusive setting as the hardest of the resolved light jets (i.e. $J^\perp_j>J^\perp_X$). If two light jets are resolved, only the sum of their transverse momenta has to balance the transverse momentum of the top-tagged jet. Jets with a high transverse momentum component are more likely to also have a small pseudo rapidity value. By always selecting the hardest jet the in the $t$-channel otherwise depleted central pseudo rapidity region is populated. 

Similar to the exclusive case, significant cancellations of the (factorisation) scale dependence in the normalised Born distributions, especially in regions of phase space where many events are expected, are apparent. 
In \Fig{fig:kfac_sgtsnc_eta1nonorm} an example for the unnormalised distributions is given for the pseudo rapidity of the top-tagged jet from the $s$-channel. The unnormalised distributions for the remaining jet variables are collected in \Fig{fig:kfac_sgtsinc_e2nonorm} to \Fig{fig:kfac_sgtsinc_eta1nonorm} in appendix~\ref{app:impNLO_sgtincl}.
Again, the unnormalised k-factor distributions scale with the ratio of the fiducial cross sections $\sigma^\NLO_\incl/\sigma^\Born$ with respect to the normalised distributions.
Note that if in the real corrections a new partonic channel opens this may amount to large contributions to the NLO corrections, especially if additional resolved radiation is not vetoed. Since these newly appearing contributions are essentially of leading order they exhibit a leading-order scale dependence without proper compensation. Hence, the scale dependence of the NLO predictions can be stronger than in the Born approximation also for unnormalised distributions (cf. \Fig{fig:kfac_sgtsnc_eta1nonorm}).

On the one hand, not vetoing the additional resolved jet results in somewhat flatter k-factor distributions and more overlap of the Born and NLO uncertainty bands particularly in highly populated phase space regions. However, in several regions of phase space the distributions receive sizeable NLO corrections together with still distinct theoretical uncertainty bands. This is a sign of the unreliability of the predictions based on the Born approximation and especially their estimated theoretical uncertainty. This is best demonstrated in the normalised distributions where the customary practice of scale variation in the Born results often underestimates the impact of the NLO corrections in many regions of the phase space. This study shows that also without vetoing the emission of additional resolved jets, higher-order corrections have an important impact on the theoretical predictions for differential distributions for both the $s$- and $t$-channel production of single top quarks together with at least one light jet.

\subsection{Application III: Retaining the $2\to1$ clustering prescriptions for the specific exclusive single top-quark event definition}
\label{sec:21event-definition}
By defining events in terms of variables which do not completely fix the $4$-momenta of the observed final state objects---like in the case of the inclusive event definition (cf. section~\ref{sec:event-definition-inclusive-case})---it is possible to consistently construct weights including NLO corrections for jet events defined by $2\to1$ jet clustering algorithms. Inspired by the mass determinations within the \MEM, it is demonstrated how this can be achieved for the special case that variables, which do not fix the masses of the final-state objects, are used in the event definition. Since the choice of the jet clustering  prescription only influences the part of the real corrections where the additional radiation is unresolved, the focus is on the exclusive event definition.

For $2\to1$ jet algorithms a widely used recombination scheme is the so-called E-scheme where the jet momentum of jet $J_{(kl)}$ formed by the clustering of the unresolved parton pair $k,l$ is given by the $4$-momentum sum of the partonic momenta (cf. section~\ref{sec:jets})
\begin{equation}\label{eq:2to1rec}
J_{(kl)}=p_k+p_l.
\end{equation}
When recombining jets according to the E-scheme, the overall $4$-momentum is conserved but the on-shell condition is violated since the clustering introduces an invariant jet mass depending on the partonic momenta $p_k$ and $p_l$
\begin{equation}
J^2_{(kl)}=m^2_k+m^2_l+2p_k\cdot p_l.
\end{equation}
In the strict soft and collinear limits ($p_k\cdot p_l\rightarrow0$) also the jets defined by the $2\to1$ jet algorithm go on shell.

In the differential jet cross section in \Eq{eq:diffjetxs} the function $\vec{x}(J_1,\ldots,J_n)$ maps the jet momenta to the variables $\vec{x}$ used to describe the jet event. If the jet masses are fixed by these variables, there is a mismatch with the jet event variables used in the corresponding Born and virtual contributions in \Eq{eq:diffjetxs} which are obtained from on-shell jet momenta.
However, it is possible to parameterise the $n$-body phase space measure $dR_n(\vec{y})$ in terms of a subset of variables $\vec{\tilde{y}}$ which do not fix the masses of the final-state objects and their explicit masses as free parameters
\begin{equation}
\vec{y}=\left(m_1,\ldots,m_n,{\vec{\tilde{y}}}\right).
\end{equation}
The remaining variables are determined by the relations
\begin{equation}
p^2_i=m^2_i,\; i=1,...,n.
\end{equation}
For example, the $4$-momentum of a particle is completely specified by values for its energy, pseudo rapidity and azimuthal angle and arbitrary mass values $m\leq E$
\begin{equation}
\left(m,E,\eta,\phi\right)\rightarrow\left(E,\;p_x=p^\perp\cos\phi,\;p_y=p^\perp\sin\phi,\;p_z=p^\perp\sinh\eta\right)\quad\text{with } p^\perp={\sqrt{E^2-m^2}\over\cosh\eta}.
\end{equation}
\begin{figure}[htbp]
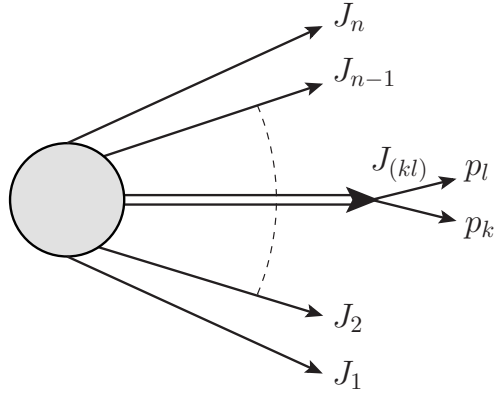

  \begin{center}
    \leavevmode
   \includegraphics[width=0.49\textwidth]{{{%
         sgt/2to1/21clusn}}}
    \caption{Factorisation of the real phase space for the $2\to1$ clustering of partons $k,l$.}
    \label{fig:21psfac}
  \end{center}
\end{figure}
The real phase space of $n+1$ partons can be factorised into a phase space of $n$ jets (parameterised by $\vec{y}$) of which the jet $(kl)$ follows from the $2\to1$ clustering of partons $k,l$ with momentum $J_{(kl)}=p_k+p_l$ (see \Fig{fig:21psfac})
\begin{equation}\label{eq:2to1psfac}
dR_{n+1}(P\rightarrow p_1,\ldots,p_i,\ldots,p_r,\ldots,p_{n+1})=(2\pi)^3dJ^2_{(kl)}\; dR_n\left(m_1,\ldots,\sqrt{J^2_{(kl)}},\ldots,m_n,\vec{\tilde{y}}\right)\;dR_2(J_{(kl)}\rightarrow p_k,p_l).
\end{equation}
By limiting the variables $\vec{x}$ defining the differential cross section in \Eq{eq:diffjetxs} to the subset not fixing the masses of the observed objects $\vec{x}\rightarrow\vec{\tilde{x}}$ and by identifying the variables $\vec{\tilde{y}}$ parameterising the $n$ jets in \Eq{eq:2to1psfac} with the ones used in \Eq{eq:diffjetxs} for the Born and virtual contribution, the differential exclusive $n$-jet cross section at NLO accuracy can be written as
\begin{eqnarray}
  \label{eq:NLO-diff-xs-massind}
  {d\sigma_\excl^\NLO\over
    d\tilde{x}_1\ldots d\tilde{x}_r} &=& \int dR_n(\vec{y})
  {d\sigma^{BV}\over dy_1\ldots dy_s} {\cal F}^{n}_{J_1,\ldots,J_n}(y_1,\ldots, y_s)
  \;\delta(\vec{\tilde{x}} - \vec{\tilde{x}}(J_1,\ldots,J_n))\nn\\
\nn  &+& \sum\limits_{(kl)}\sum\limits_{\substack{l,k\\ k\neq l}}\Bigg\{\int(2\pi)^3dJ^2_{(kl)}\; dR_n\left(m_1,\ldots,\sqrt{J^2_{(kl)}},\ldots,m_n,\vec{\tilde{y}}\right)\;dR_2(J_{(kl)}\rightarrow p_k,p_l)\;\Theta^{kl}_{(kl)}  \\
\nn&&\qquad\qquad\times\;{d\sigma^R\over dz_1\ldots dz_{t}}\;
{\cal F}^{n}_{J_1,\ldots,J_{(kl)},\ldots,J_n}(y_1,\ldots, y_t)
\; \delta(\vec{\tilde{x}} - \vec{\tilde{x}}(J_1,\ldots,J_{(kl)},\ldots,J_n))\\
  &&\qquad+ \int dR_{n+1}(\vec{z'})
  \;\left({1\over n^2}-\Theta^{kl}_{(kl)}\right)\left(1-\Theta(p^\perp_{l}-p^\perp_{\text{min}})\Theta(\eta_{\text{max}}-|\eta_{l})|\right)\nn\\
&&\qquad\qquad\times\; {d\sigma^R\over dz'_1\ldots dz'_{t}}\;
  {\cal F}^{n+1}_{J_1,\ldots,J_n}(z'_1,\ldots, z'_t)\;\delta(\vec{\tilde{x}} - \vec{\tilde{x}}(J_1,\ldots,J_n))\Bigg\}.\nn\\
\end{eqnarray}
Here, the function $\vec{\tilde{x}}(J_1,\ldots,J_n)$ maps the jet momenta to the subset of variables in the jet event definition.
The variables $z_i$ depend on $m_1,\ldots,\sqrt{J^2_{(kl)}},\ldots,m_n,\vec{\tilde{y}}$ and $p_k,p_l$. The sums over $l$ and $k$ run over all $n+1$ final-state partons and the sum over $(kl)$ runs over all $n$ jets. In \Eq{eq:NLO-diff-xs-massind}, $\Theta^{kl}_{(kl)}=1$ only if the parton pair $k,l$ would be clustered by the jet algorithm to form the jet $(kl)$ in the final state ($\Theta^{kl}_{(kl)}=0$ otherwise). Note that the invariant mass squared of the clustered jet is not observed according to the event definition $\tilde{x}$ and is therefore integrated over in the second line of \Eq{eq:NLO-diff-xs-massind}. However, this integration does not affect the variables in $\vec{\tilde{x}}(J_1,\ldots,J_n)$.
The last two lines of \Eq{eq:NLO-diff-xs-massind} account for additional real radiation $l$ which is not clustered with any of the other partons to form a final-state jet but exceeds the experimental cuts on $p^\perp$ and $\eta$ and is therefore associated with the beam. Since in $2\to1$ jet algorithms this kind of radiation is omitted in the observed final state and does not contribute to the event as a resolved jet, it has to be integrated out.

The application of \Eq{eq:NLO-diff-xs-massind} to the exclusive production of a single resolved top-tagged jet in association with exactly one resolved light jet at a hadron collider is illustrated in the following.
The particular exclusive event definition $\vec{x}=\vec{\tilde{x}}=(\eta_{t},E_{j}, \eta_{j}, \phi_{j})$ (cf. \Eq{eq:eventdef}) does not fix the invariant masses of the jets. Instead, the $2$-jet parameterisation in terms of $\vec{\tilde{x}}$ depends on the squared jet masses  ${J^2_t},{J^2_j}$ as free parameters (cf. \Eq{eq:eventparexcl})
\begin{eqnarray}\label{eq:eventpar21excl}
\nn J_t&=&\left(E_{t},\;-J^{\perp}\cos{\phi_{j}},\;-J^{\perp}\sin{\phi_{j}},
  \;J^{\perp}\sinh{\eta_{t}}\right),\\
J_{j}&=&\left(E_{j},\;J^{\perp}\cos{\phi_{j}},\;J^{\perp}\sin{\phi_{j}},
  \;J^{\perp}\sinh{\eta_{j}}\right)
\end{eqnarray}
with
\begin{displaymath}
  J^{\perp}={\sqrt{E^2_{j}-J^2_j}\over\cosh{\eta_j}}
  \quad\mbox{and}\quad E_{t}=\sqrt{{J^{\perp}}^2\cosh^2{\eta_t}+J^2_t}.
\end{displaymath}
This allows to drop the mass-shell requirement for the clustered jets by integrating over all allowed jet masses in the clustered real contribution. An unambiguous interrelation with the Born and virtual contributions of the NLO event weight which are defined for the jet momenta from \Eq{eq:eventpar21excl} set on shell ($J^2_t=m^2_t$, $J^2_j=0$) as
\begin{equation}
\frac{d\sigma^{\text{BV}}}{d\vec{\tilde{x}}}=
 \frac{d\sigma^{\text{BV}}}{d\eta_{t}\; dE_{j}\; d\eta_{j}\;
   d\phi_{j}}
 =\left.\frac{{J^{\perp}}^2\cosh{\eta_t}}{2\;s\;E_t\;\cosh{\eta_j}} \;
 \frac{d\sigma^{\text{BV}}}{dR_2(J_{t},J_{j})}\right|_{{{J^2_t=m^2_t,\atop J^2_j=0}}}
\end{equation} 
is maintained at the same time. 

The partonic reaction for the real corrections to the production of a single top quark ($t$) in association with a light quark ($j$) is denoted by
\begin{equation}
p_a+p_b\rightarrow p_t+p_j+p_r,
\end{equation}
where $r$ denotes additional radiation.
In order to contribute to the exclusive signal signature, the additional radiation $r$ has to be unresolved.
Thus, it has to be either clustered in the final state by the jet algorithm or associated with the beam due to exceeding the experimental cuts.
\begin{table}[htbp]
  \centering
  \caption{Unresolved radiation in single top-quark production according to the $2\to1$ jet algorithm}
  \label{tab:2to1clus_sgt}
  \def\arraystretch{1.8}
  \begin{tabular}{|l|l|l|l|}
   \hline
    case
    &radiation $r$ unresolved because
    & jet momenta 
    &squared jet masses                        \\ \hline
    `light'
    &    clustered with light jet
    & $J_t=p_t$,
    & $J^2_t=m^2_t$, \\
    &$\Theta^{jr}_{(jr)}=1$
    & $J_{(jr)}=p_j+p_r$  
    & $J^2_{(jr)}=2p_j\cdot p_r$ \\ \hline
    `top'
    &clustered with top-tagged jet
    & $J_{(tr)}=p_t+p_r$,
    & $J^2_{(tr)}=m^2_t+2p_t\cdot p_r$, \\
    &$\Theta^{tr}_{(tr)}=1$
    & $J_j=p_j$
    & $J^2_j=0$ \\ \hline
    `beam'
    &associated with beam
    & $J_t=p_t$,
    & $J^2_t=m^2_t$\\
    &$p^{\perp}_r<p^{\perp}_{\text{min}}\;\text{ or }\; |\eta_r|>\eta_{\text{max}}$
    & $J_j=p_j$  
    & $J^2_j=0$ \\ \hline
\end{tabular}
\end{table}
\begin{figure}[htbp]
  \begin{center}
    \leavevmode
    \begin{subfigure}[b]{0.49\textwidth}
 \includegraphics[width=0.87\textwidth]{{{%
          sgt/2to1/21cluslight}}}
    \caption{Clustering in light jet.}
    \label{fig:21psfacflight}
   \end{subfigure}
    \begin{subfigure}[b]{0.49\textwidth}
 \includegraphics[width=0.87\textwidth]{{{%
          sgt/2to1/21clustop}}}
    \caption{Clustering in top-tagged jet.}
    \label{fig:21psfacftop}
   \end{subfigure}
    \caption{Possible $2\to1$ final-state clusterings.}
    \label{fig:21psfacf}
  \end{center}
\end{figure}
In table~\ref{tab:2to1clus_sgt} the three different cases of unresolved radiation for this process are listed together with the resulting jet momenta and their squared masses.
For all three cases the real phase space can be factorised in a phase space of resolved jets corresponding to the given jet variables $\eta_t,E_j,\eta_j,\phi_j$ and a phase space describing the unresolved configurations which can be integrated out numerically.

Considering the case where a pair of light partons $j,r$ are clustered to form the light jet with momentum $J_{(jr)}=p_j+p_r$ (see \Fig{fig:21psfacflight} and case `light' in table~\ref{tab:2to1clus_sgt}), the respective real phase space $dR_3(p_a+p_b\rightarrow J_t,p_j,p_r)$ can be expressed as a convolution of two phase spaces by treating the clustered jet as an intermediate particle decaying into the unresolved parton pair
\begin{eqnarray}\label{eq:phspacfac2to1lr}
\nn dx_a\;dx_b\;dR_3(p_a,p_b\rightarrow J_t,p_j,p_r)&=&dx_a\;dx_b\;(2\pi)^3d{J}^2_{(jr)}\;dR_2(p_a,p_b\rightarrow J_t,{J}_{(jr)})\;dR_2({J}_{(jr)}\rightarrow p_j,p_r),\\
\nonumber&=&(2\pi)^3d{J}^2_{(jr)}\;{{{{J}}^{\perp}}^2\cosh{\eta_t}\over2s\;E_t\;\cosh{{\eta}_j}}\;d{E}_j\;d{\eta}_j\;d{\phi}_j\;d\eta_t\; dx_a\;dx_b\; d\vec{J}^{\perp}_t\\
\nn&\times&\delta\left(x_a-\frac{1}{\sqrt{s}}(E_j+E_t+J^z_{(jr)}+J^z_t)\right)\;
 \delta\left(x_b-\frac{1}{\sqrt{s}}(E_j+E_t-J^z_{(jr)}-J^z_t)\right)\\
&\times&\;\delta\left(\vec{{J}}^{\perp}_{(jr)}+\vec{J}^{\perp}_t\right)\;dR_2({J}_{(jr)}\rightarrow p_j,p_r).
\end{eqnarray}
The overall $4$-momentum is conserved by the two-body decay while the squared jet mass ${J}^2_{(jr)}=(p_j+p_r)^2$ has to be integrated over. For this clustering the jet momenta are given by the parameterisation in \Eq{eq:eventpar21excl} in terms of the event variables $\eta_t,E_j,\eta_j,\phi_j$ and the squared jet masses $J^2_t=m^2_t$ and $J^2_{(jr)}$.
Using \Eq{eq:phspacfac2to1lr} the respective real contribution corresponding to the clustering in the light jet to the exclusive event weight reads
\begin{eqnarray}\label{eq:evwgt21light}
\nn \left.\frac{d\sigma^{\text{R}}}{d\vec{\tilde{x}}}\right|_{\text{light}}&=& 
 \left.\frac{d\sigma^{\text{R}}}{d\eta_{t}\; d{E}_{j}\; d{\eta}_{j}\;
   d{\phi}_{j}}\right|_{\text{light}}\\
 &=&(2\pi)^3\int d{J}^2_{(jr)}\;{{{{J}}^{\perp}}^2\cosh{\eta_t}\over2s\;E_t\;\cosh{{\eta}_j}}
\int dR_2({J}_{(jr)}\rightarrow p_j,p_r)\;\Theta^{jr}_{(jr)} \;
 \left.\frac{d\sigma^{\text{R}}}{dR_3(J_t,p_j,p_r)}\right|_{J^2_t=m^2_t,\atop J^2_{(jr)}=2p_j\cdot p_r}.\nn\\
\end{eqnarray}
The additional radiation $r$ can also be clustered with the top quark to form the top-tagged jet with momentum $J_{(tr)}=p_t+p_r$ (see \Fig{fig:21psfacftop} and case `top' in table~\ref{tab:2to1clus_sgt}). The jet momenta can be parameterised with the variables in the event definition $\vec{\tilde{x}}$ according to \Eq{eq:eventpar21excl} by demanding an on-shell light jet $J^2_j=0$. Again, the respective real phase space $dR_3(p_a+p_b\rightarrow p_t,J_j,p_r)$ corresponds to a phase-space convolution similar to \Eq{eq:phspacfac2to1lr} but with a decay of the off-shell top-tagged jet to the unresolved parton pair.
As before, the squared mass of the clustered top-tagged jet ${J}^2_{(tr)}=(p_t+p_r)^2$ has to be integrated over.
The real contribution to the exclusive event weight  corresponding to the clustering in the top-tagged jet follows as
\begin{eqnarray}\label{eq:evwgt21top}
\nn \left.\frac{d\sigma^{\text{R}}}{d\vec{\tilde{x}}}\right|_{\text{top}}&= &
 \left.\frac{d\sigma^{\text{R}}}{d{\eta}_{t}\; dE_{j}\; d\eta_{j}\;
   d\phi_{j}}\right|_{\text{top}}\\
 &=&(2\pi)^3\int d{J}^2_{(tr)}\;{{{J}^{\perp}}^2\cosh{{\eta}_t}\over2s\;{E}_t\;\cosh{\eta_j}}
\int dR_2({J}_{(tr)}\rightarrow p_t,p_r)\;\Theta^{tr}_{(tr)} \;
 \left.\frac{d\sigma^{\text{R}}}{dR_3(p_t,J_j,p_r)}\right|_{J^2_{(tr)}=m^2_t+2p_t\cdot p_r,\atop J^2_j=0}\nn\\
&+&(2\pi)^3\int d{J}^2_{(tj)}\;{{{J}^{\perp}}^2\cosh{{\eta}_t}\over2s\;{E}_t\;\cosh{\eta_j}}
\int dR_2({J}_{(tj)}\rightarrow p_t,p_j)\;\Theta^{tj}_{(tr)} \;
 \left.\frac{d\sigma^{\text{R}}}{dR_3(p_t,p_j,J_j)}\right|_{J^2_{(tj)}=m^2_t+2p_t\cdot p_j,\atop J^2_j=0}.\nn\\
\end{eqnarray}
Also the light quark $j$ can be clustered with the top quark resulting in the clustered jet's momentum $J_{(tj)}=p_t+p_j$ while the additionally radiated quark $r$ constitutes the resolved light jet with momentum $J_j=p_r$ (see the last line of \Eq{eq:evwgt21top}).

The real phase space according to a top-tagged jet, a light jet and additional real radiation can be parameterised in terms of the variables in the exclusive event definition $\vec{\tilde{x}}$ describing the resolved on-shell jets and additional integrations over the momentum of the real radiation $p_r$ as (cf. \Eq{eq:realpsincl} with $J_X\rightarrow p_r$)
\begin{eqnarray}\label{eq:phspacfac2to1i}
\nonumber dx_a\;dx_b\;  dR_3(p_a,p_b\rightarrow J_t,J_j,p_r)&=&\frac{{J^{\perp}_j}{p^{\perp}_r}{J^{\perp}_t}\cosh{\eta_t}}{4s\;E_t\;\cosh{\eta_j}}\;dE_j\;d\eta_j\;d\phi_j\;d\eta_t\;dp^{\perp}_r\;d\eta_r\;d\phi_r\; dx_a\;dx_b\;  d\vec{J}^{\perp}_t\\&\times&\nn \delta\left(\vec{J}^{\perp}_j+\vec{J}^{\perp}_t+\vec{p}^{\perp}_r\right)
  \;\delta\left(x_a-\frac{1}{\sqrt{s}}(E_j+E_t+E_r+J^z_j+J^z_t+p^z_r)\right)\\
&\times&
  \delta\left(x_b-\frac{1}{\sqrt{s}}(E_j+E_t+E_r-J^z_j-J^z_t-p^z_r)\right).
\end{eqnarray}
In order to be associated with the beam, the momentum of the real radiation which is parameterised by $p^\perp_r$, $\phi_r$ and $\eta_r$ has to be integrated out such that the additional radiation ends up in the beam pipe ($p^{\perp}_{r}<p^{\perp}_{\text{min}}$ or $|\eta_{r}|>\eta_{\text{max}}$, cf. case `beam' in table~\ref{tab:2to1clus_sgt}).
The jet momenta of the resolved top-tagged $J_t$ and resolved light jet $J_j$ together with the momentum of the additional radiation $p_r$ can be parameterised with the variables in the event definition $\eta_t,E_j,\eta_j,\phi_j$ and the integration variables of the unresolved radiation in \Eq{eq:phspacfac2to1i} by the parameterisation given in \Eq{eq:eventparincl} with $J_X\rightarrow p_r$. The respective real contribution for unresolved radiation  associated with the beam to the exclusive event weight reads
\begin{eqnarray}\label{eq:evwgt21ini}
  \nn \left.\frac{d\sigma^{\text{R}}}{d\vec{\tilde{x}}}\right|_{\text{beam}}&= &
  \left.\frac{d\sigma^{\text{R}}}{d\eta_{t}\; dE_{j}\; d\eta_{j}\; d\phi_{j}}\right|_{\text{beam}}\\
  &=&\int dp^{\perp}_r
  \int d\eta_r
  \int d\phi_{r}\;
  \frac{{J^{\perp}_j}{p^{\perp}_r}{J^{\perp}_t}\cosh{\eta_t}}{4s\;E_t\;\cosh{\eta_j}} \;\left(1-\Theta^{jr}_{(jr)}-\Theta^{tr}_{(tr)}\right)\;
  \frac{d\sigma^{\text{R}}}{dR_3(J_t,J_j,p_r)}\nn\\
&&\times\;\left(\Theta(p^\perp_{\text{min}}-p^\perp_{r})+\Theta(|\eta_{r}|-\eta_{\text{max}})\right)\nn\\
&+&\int dp^{\perp}_j
  \int d\eta_j
  \int d\phi_{j}\;
  \frac{{J^{\perp}_j}{p^{\perp}_j}{J^{\perp}_t}\cosh{\eta_t}}{4s\;E_t\;\cosh{\eta_j}} \;\left(1-\Theta^{jr}_{(jr)}-\Theta^{tj}_{(tj)}\right)\;
  \frac{d\sigma^{\text{R}}}{dR_3(J_t,p_j,J_j)}\nn\\
&&\times\;\left(\Theta(p^\perp_{\text{min}}-p^\perp_{j})+\Theta(|\eta_{j}|-\eta_{\text{max}})\right).\nn\\
\end{eqnarray}
Again, also the light quark $j$ might be unresolved because of $p^{\perp}_{j}<p^{\perp}_{\text{min}}$ or $|\eta_{j}|>\eta_{\text{max}}$ while the real radiation $r$ is resolvable and constitutes the light jet with momentum $J_j=p_r$ (see last two lines of \Eq{eq:evwgt21ini}).

The exclusive NLO weight for the jet events $\vec{\tilde{x}}=(\eta_{t},E_{j}, \eta_{j}, \phi_{j})$ obtained with a $2\to1$ jet algorithm is given by the sum of the exclusive real contributions from \Eq{eq:evwgt21light}, \Eq{eq:evwgt21top} and \Eq{eq:evwgt21ini} and the Born and virtual parts
\begin{equation}\label{eq:2to1evwgtexcl}
  \frac{d\sigma_\excl^{\NLO}}{d\vec{\tilde{x}}}
  =
\frac{d\sigma^{\text{BV}}}{d\vec{\tilde{x}}}
  + \left.\frac{d\sigma^{\text{R}}}{d\vec{\tilde{x}}}\right|_{\text{light}}+\left.\frac{d\sigma^{\text{R}}}{d\vec{\tilde{x}}}\right|_{\text{top}}+\left.\frac{d\sigma^{\text{R}}}{d\vec{\tilde{x}}}\right|_{\text{beam}}.
\end{equation}
\begin{figure}[htbp]
  \begin{center}
    \leavevmode
     \makebox[\textwidth]{\includegraphics[width=\ftw\textwidth]{{{%
          sgt/2to1/evgen/sgtsKT3-2ycut30comptwoETA1-90evts-crop}}}}
    \caption{Pseudo rapidity distribution of the top-tagged jet from exclusive $s$-channel single top-quark production calculated at NLO accuracy using a conventional parton level MC (solid blue) with $2\to1$ jet clustering compared to histograms filled with generated NLO events (dashed red).}
    \label{fig:partonMCvsgenev_sgts21_eta1}
  \end{center}
\end{figure}
\begin{figure}[htbp]
  \begin{center}
    \leavevmode
     \makebox[\textwidth]{\includegraphics[width=\ftw\textwidth]{{{%
          sgt/2to1/evgen/sgttKT3-2ycut30comptwoE2-100evts-crop}}}}
    \caption{Same as \Fig{fig:partonMCvsgenev_sgts21_eta1} but for the energy of the light jet from the $t$-channel.}
    \label{fig:partonMCvsgenev_sgtt21_e2}
  \end{center}
\end{figure}
This event weight is used to generate samples of $N_\excl=12775$ ($N_\excl=24088$) exclusive $s$-channel ($t$-channel) single top-quark events which are obtained with the $2\to1$ jet algorithm. 

To validate that the distribution of these events follows the respective NLO cross section, they are used to fill histograms of jet variables. These histograms are compared to distributions according to the $2\to1$ jet algorithm calculated at NLO accuracy. The pseodo rapidity distribution of the top-tagged jet from the $s$-channel and the energy distribution of the light jet from the $t$-channel are shown in \Fig{fig:partonMCvsgenev_sgts21_eta1} and \Fig{fig:partonMCvsgenev_sgtt21_e2} as examples. Respective comparisons of other variables in $\vec{\tilde{x}}$ can be found in \Fig{fig:partonMCvsgenev_sgts21_e2} to \Fig{fig:partonMCvsgenev_sgtt21_eta2} in appendix~\ref{app:unwgtev_sgt21}. Note that the calculated distributions (shown as solid blue lines) correspond to the ones calculated in the context of the impact study of the clustering prescription in section~\ref{sec:sgtimpac}. 
The plots illustrate that the event weight from \Eq{eq:2to1evwgtexcl} can be used to generate unweighted events $\{\vec{\tilde{x}}_i\}$ that are distributed according to the exclusive cross section calculated at NLO accuracy relying on the $2\to1$ clusterings. The respective pull distributions, p-values and reduced $\chi^2$ given at the bottom of the plots confirm this observation.

\subsubsection{Impact of NLO corrections}
\begin{figure}[htbp]
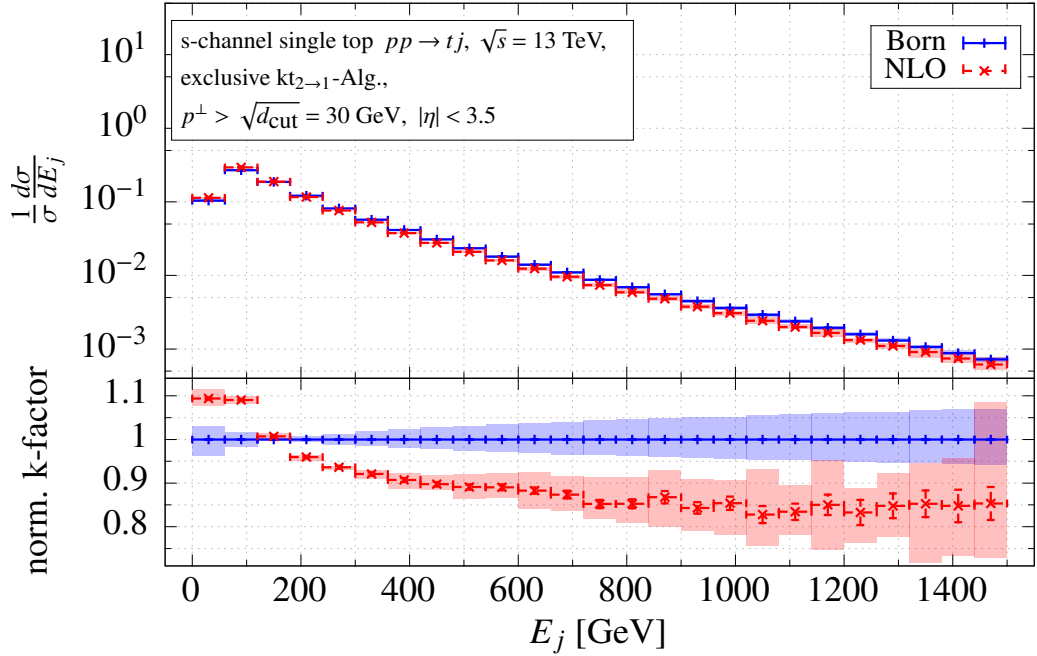

  \begin{center}
    \leavevmode
    \makebox[\textwidth]{\includegraphics[width=\ftw\textwidth]{{{%
          sgt/2to1/kfmu/sgtsKT3-2ycut30compmutwoE2-200x1e7-crop}}}}
    \caption{Energy distribution and k-factors of the energy of the light jet from $s$-channel single top-quark production employing the $2\to1$ jet algorithm with an exclusive event definition.}
    \label{fig:kfac_sgts21_e2}
  \end{center}
\end{figure}
\begin{figure}[htbp]
  \begin{center}
    \leavevmode
    \makebox[\textwidth]{\includegraphics[width=\ftw\textwidth]{{{%
          sgt/2to1/kfmu/sgtsKT3-2ycut30compmutwoETA2-200x1e7-crop}}}}
    \caption{Same as \Fig{fig:kfac_sgts21_e2} but for the pseudo rapidity of the light jet from the $s$-channel.}
    \label{fig:kfac_sgts21_eta2}
  \end{center}
\end{figure}
\begin{figure}[htbp]
  \begin{center}
    \leavevmode
    \makebox[\textwidth]{\includegraphics[width=\ftw\textwidth]{{{%
          sgt/2to1/kfmu/sgtsKT3-2ycut30compmutwoETA1-200x1e7-crop}}}}
    \caption{Same as \Fig{fig:kfac_sgts21_e2} but for the pseudo rapidity of the top-tagged jet from the $s$-channel.}
    \label{fig:kfac_sgts21_eta1}
  \end{center}
\end{figure}
\begin{figure}[htbp]
  \begin{center}
    \leavevmode
    \makebox[\textwidth]{\includegraphics[width=\ftw\textwidth]{{{%
          sgt/2to1/kfmu/sgttKT3-2ycut30compmutwoE2-200x1e7-crop}}}}
    \caption{Same as \Fig{fig:kfac_sgts21_e2} but for the energy of the light jet from the $t$-channel.}
    \label{fig:kfac_sgtt21_e2}
  \end{center}
\end{figure}
\begin{figure}[htbp]
  \begin{center}
    \leavevmode
    \makebox[\textwidth]{\includegraphics[width=\ftw\textwidth]{{{%
          sgt/2to1/kfmu/sgttKT3-2ycut30compmutwoETA2-200x1e7-crop}}}}
    \caption{Same as \Fig{fig:kfac_sgts21_e2} but for the pseudo rapidity of the light jet from the $t$-channel.}
    \label{fig:kfac_sgtt21_eta2}
  \end{center}
\end{figure}
\begin{figure}[htbp]
  \begin{center}
    \leavevmode
    \makebox[\textwidth]{\includegraphics[width=\ftw\textwidth]{{{%
          sgt/2to1/kfmu/sgttKT3-2ycut30compmutwoETA1-200x1e7-crop}}}}
    \caption{Same as \Fig{fig:kfac_sgts21_e2} but for the pseudo rapidity of the top-tagged jet from the $t$-channel.}
    \label{fig:kfac_sgtt21_eta1}
  \end{center}
\end{figure}
The impact of the NLO corrections on the distributions corresponding to the $2\to1$ jet algorithm is illustrated in \Fig{fig:kfac_sgts21_e2} to \Fig{fig:kfac_sgts21_eta1} for the $s$-channel and in \Fig{fig:kfac_sgtt21_e2} to \Fig{fig:kfac_sgtt21_eta1} for the $t$-channel. The NLO predictions are shown in red and the Born predictions in blue. At the bottom of the plots the k-factor distributions are shown. The theoretical uncertainties of the predictions are estimated by variations of the renormalisation and the factorisation scale corresponding to the red and blue bands. Since the choice of the clustering prescription in the jet algorithm only influences the distributions at NLO accuracy, it is instructive to compare the impact of the NLO corrections corresponding to the $2\to1$ and the $3\to2$ clusterings. Differences between the two should be also be apparent in the impact studies of the different clustering prescriptions presented in section~\ref{sec:sgtimpac}.

Comparing \Fig{fig:kfac_sgts21_e2} to \Fig{fig:kfac_sgtt21_eta1} to the respective studies employing the $3\to2$ jet algorithm presented in \Fig{fig:kfac_sgts_e2} to \Fig{fig:kfac_sgtt_eta1} in section~\ref{sec:kfacexcl}, a similar qualitative picture of the k-factor distributions is visible at large. However, the first bins in the energy distributions are affected differently by the NLO corrections conforming to the two clustering prescriptions (cf. \Fig{fig:kfac_sgts21_e2} and \Fig{fig:kfac_sgts_e2} or \Fig{fig:kfac_sgtt21_e2} and \Fig{fig:kfac_sgtt_e2}). This observation is in accordance with the impact studies of the $3\to2$ clustering. As pointed out in section~\ref{sec:sgtimpac}, production thresholds in energy distributions are particularly sensitive to the mass of the studied object which can differ significantly within the two clustering prescriptions (cf. \Fig{fig:jetcom_sgtm}).  Apart from that, the $s$-channel does not seem to be very sensitive to the choice of the clustering prescription. Thus, similar findings regarding the impact of the NLO corrections as deduced from \Fig{fig:kfac_sgts_e2} to \Fig{fig:kfac_sgtt_eta1} in section~\ref{sec:kfacexcl} apply. In the $t$-channel though, the k-factor distribution of the pseudo-rapidity distribution of the top-tagged jet obtained with the $2\to1$ jet algorithm shows qualitatively different behaviour to the respective one corresponding to the $3\to2$ clusterings. The impact is biggest at the phase space boundaries. Again, this observation matches the findings in section~\ref{sec:sgtimpac} regarding the impact of the new clustering on the pseudo-rapidity distribution of the top-tagged jet (cf. \Fig{fig:jetcom_sgtt_eta1}).
Overall, the sizes of the theoretical uncertainty bands are slightly larger in NLO distributions obtained with the $2\to1$ jet algorithm compared to the ones obtained with the $3\to2$ jet algorithm presented in section~\ref{sec:sgtevgen}.

\section{The \MEM at next-to-leading order QCD}\label{sec:MEMNLO}
\subsection{Likelihood at NLO accuracy}
The definition of a differential $n$-jet cross section at NLO (see \Eq{eq:diffnjetxs}) enables the assignment of NLO weights to events with $n$ observed jets. With these NLO weights an NLO likelihood for a set of $N$ events $\{\vec{x}_i\}$ each comprised of $r$ $n$-jet variables can be constructed as (cf. \Eq{eq:MLlikeli} and \Eq{eq:likeli})
\begin{eqnarray}\label{eq:likeliNLO}
  \Lik^\NLO\left(\omega\right)
  &=& 
  \nn\prod\limits_{i=1}^{N} \Lik^\NLO
  \left(\omega|\vec{x}_i\right)=\left(\frac{1}{\sigma^\NLO(\omega)}\right)^N
  \prod\limits_{i=1}^{N}\frac{d\sigma^\NLO(\omega)}{d\vec{x}_i}\\
  &=& \left(\frac{ 1}{\sigma^\NLO(\omega)}\right)^N
  \prod\limits_{i=1}^{N}
  \left.\left(\frac{dR_n(J_1,\ldots,J_n)}{d{x}_1\ldots d{x}_r}(\omega)\;
      \frac{d\sigma^\NLO}{dR_n(J_1,\ldots,J_n)}(\omega)
    \right)\right|_{\vec{x}=\vec{x}_i}.
\end{eqnarray}
This NLO likelihood is a function of the model parameters $\omega$ entering the NLO cross section calculations. The factor $dR_n/d{x}_1\ldots d{x}_r$ denotes the Jacobian of the parameterisation of the $n$ jet momenta in terms of the $r$ event variables. 

To simulate the \MEM at NLO accuracy the event samples generated in section~\ref{sec:evgen} can be used as the outcome of toy experiments. Estimators $\widehat{\omega}$ for the model parameter $\omega$ can be extracted by maximising \Eq{eq:likeliNLO} for the event sample $\{\vec{x}_i\}$ (cf. \Eq{eq:likelimax}). Thereby reproducing the input value of the model parameter $\omega^{\true}$ used in the generation of the event sample $\{\vec{x}_i\}$ serves as a consistency check and validation of the \MEM at NLO and allows to explore the potential of the analysis. Instead of maximising the likelihood it is convenient to minimise the log-likelihood (previously defined in section~\ref{sec:MLvar} ) 
\begin{equation}
  -\log\Lik^{\NLO}\left(\widehat{\omega}\right)
  =\inf\limits_{\omega}\left(-\log\Lik^{\NLO}\left(\omega\right)\right)=-\log\Lik_{\tmin}.
\end{equation}
According to section~\ref{sec:MLvar} the log-likelihood can be approximated around its minimum by a parabola (see \Eq{eq:apprLikeli}) 
and the uncertainty of the extracted estimator $\Delta\widehat{\omega}$ can be assessed by an increase of the log-likelihood by $1/2$ with respect to its minimal value (see \Eq{eq:apprLikelisig}).

\subsection{Extended likelihood at NLO accuracy}\label{sec:extlikeli}
As indicated in \Eq{eq:likeliNLO}, the fiducial cross section $\sigma^\NLO$, thus the expected number of observed events, might in general also depend on the model parameters $\omega$. However, the likelihood defined in \Eq{eq:likeliNLO} depends on the normalised differential cross section only. To make use of the additional information contained in the total event number $N$ the extended likelihood can be applied in the analysis (see section~\ref{sec:MLext}). For collider experiments the expected number of observed events $\nu$ is given by the product of the (parameter dependent) fiducial cross section $\sigma^\NLO(\omega)$ and the integrated luminosity of the collider $L$
\begin{equation}
\nu(\omega)=\sigma^\NLO(\omega)L.
\end{equation}
The extended likelihood at NLO is given by the NLO likelihood in \Eq{eq:likeliNLO} weighted with the Poisson probability to observe an event sample of size $N$ under the parameter-dependent model assumption incorporated in $\sigma^\NLO(\omega)$ (cf. \Eq{eq:MLextlikeli})
\begin{eqnarray}\label{eq:extLike}
  \nonumber\mathcal{L}^\NLO_{\text{ext}}\left(\omega\right)
  &=& \frac{\nu(\omega)^N}{N!}e^{-\nu(\omega)}
 \mathcal{L}^\NLO
  \left(\omega\right)\\
 \nonumber&=&  \frac{\left(\sigma^\NLO(\omega)L\right)^N}{N!}
 e^{-\sigma^\NLO(\omega)L}\prod\limits_{i=1}^{N} 
 \frac{1}{\sigma^\NLO(\omega)}\frac{d\sigma^\NLO(\omega)}{d\vec{x}_i}\\
 &=&  \frac{L^N}{N!}e^{-\sigma^\NLO(\omega)L}
 \prod\limits_{i=1}^{N}\frac{d\sigma^\NLO(\omega)}{d\vec{x}_i}.
\end{eqnarray}
In the last line of \Eq{eq:extLike} the normalisation in front of the differential jet cross section is cancelled revealing that the \MEM employing extended likelihoods is also sensitive to the total number of observed events. Using this additional information may turn the \MEM into an even more efficient analysis tool. 

However, the integrated luminosity $L$ (as a measure of the total number of particle collisions in the detectors) needed in \Eq{eq:extLike} has to be determined by the experiments introducing an additional systematic uncertainty. In \Ref{Aaboud:2016hhf} the uncertainty of the luminosity determination is quoted to be of the order of $\frac{\Delta L}{L}\approx 2\%$ for the LHC. This additionally introduced systematic uncertainty has the potential to spoil the enhanced efficiency gained by utilising the extended likelihood. This issue is studied in more detail in the context of top-quark mass extraction from single top-quark events in section~\ref{sec:mtextsgt}.

\subsection{Application I: Top-quark mass extraction from top-quark pairs}\label{sec:mtexttt}
As a first concrete example, \Eq{eq:likeliNLO} is used to apply the \MEM at NLO to the sample of $N=73128$ exclusive $t\bar{t}$ events $\vec{x}_i=(\cos{\theta_{t}},\;\phi_{t})_i$ generated in section~\ref{sec:evgentt}. An NLO estimator $\widehat{m}^{\NLO}_t$ for the top-quark mass is extracted from these events by maximising the NLO likelihood (cf. \Eq{eq:likeliNLO} and \Eq{eq:jetwgttt} with $\omega=m_t$)
\begin{eqnarray}\label{eq:likeliNLOtt}
  \Lik^\NLO\left(m_t\right)
  &=& 
  \prod\limits_{i=1}^{N} \Lik^\NLO
  \left(m_t|\vec{x}_i\right)=\left(\frac{1}{\sigma^\NLO(m_t)}\right)^N
  \prod\limits_{i=1}^{N}\frac{d\sigma^\NLO(m_t)}{d\vec{x}_i}\\
  \nn&=& \left(\frac{ 1}{\sigma^\NLO(m_t)}\right)^N
  \prod\limits_{i=1}^{N}
  \left.\left(\frac{\beta_t}{32\pi^2}\;
      \frac{d\sigma^\NLO}{dR_n(J_t,J_{\bar{t}})}(m_t)
    \right)\right|_{\vec{x}=\vec{x}_i}
\end{eqnarray}
with respect to the top-quark mass. It is important that the event definition $\vec{x}_i$ does not allow a reconstruction of the top-quark mass since this would essentially lead to a kinematic fit instead of enabling a consistent variation of $m_t$ in the likelihood minimisation.
Note that the events are generated with an input value of the top-quark mass of $m^{\true}_t=174$~GeV in section~\ref{sec:evgen}.

\begin{figure}[htbp]
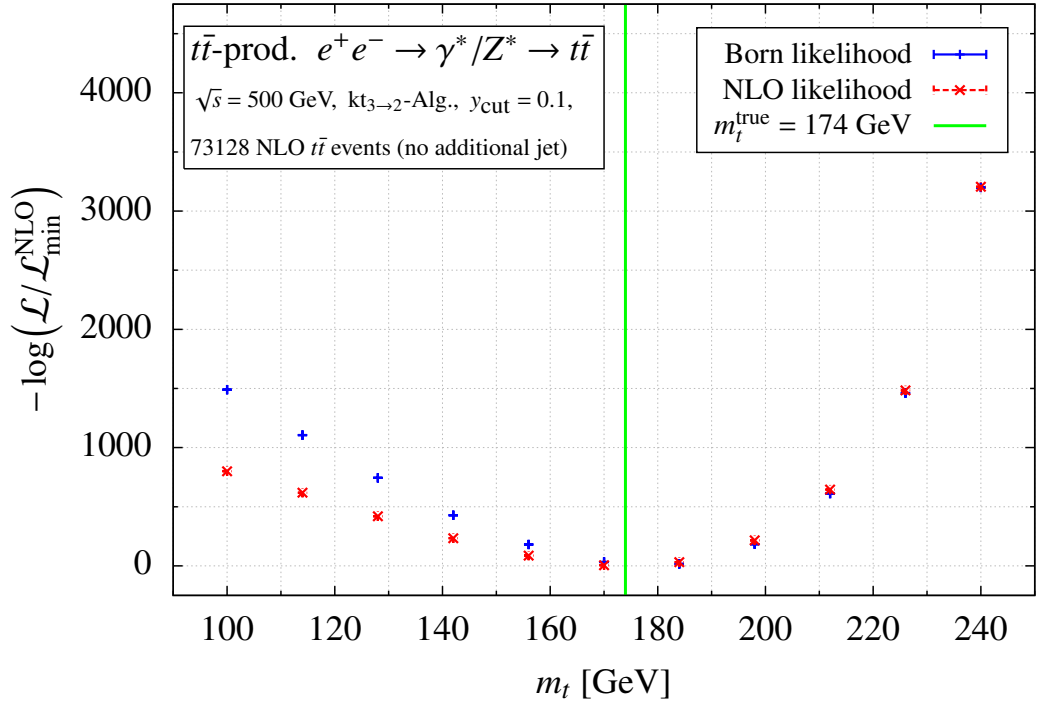
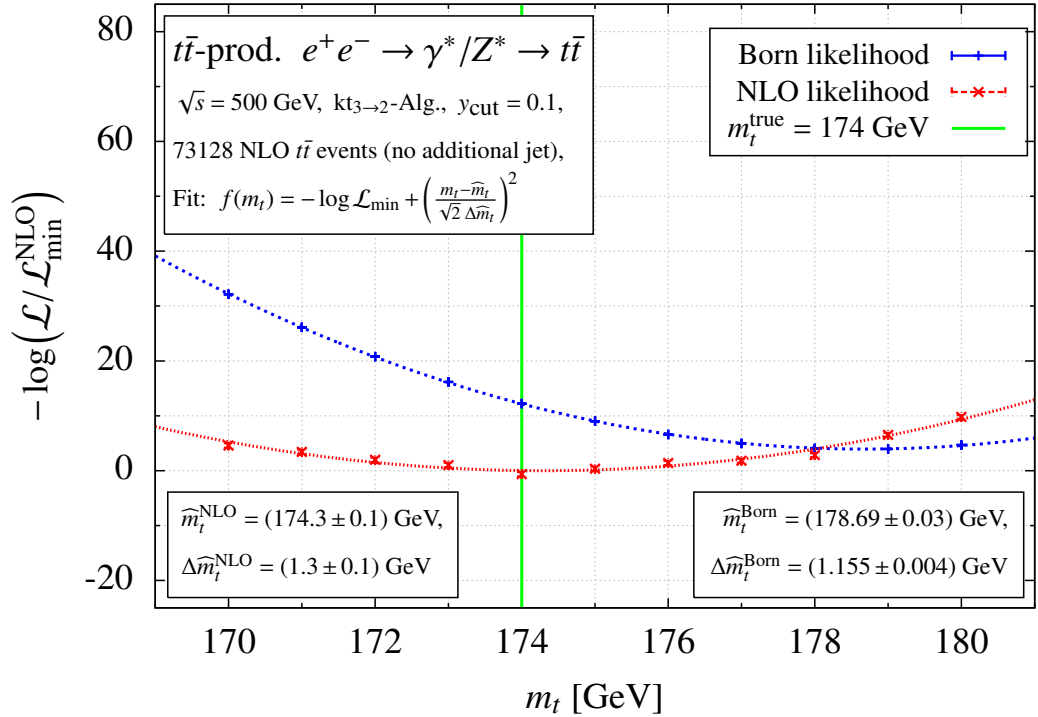

  \begin{center}
    \leavevmode
    \begin{subfigure}[b]{1\textwidth}
    \includegraphics[width=0.87\textwidth]{{{%
          dytt/eeQQKT3-2ycut01mem-150evts-crop}}}
    \caption{NLO and Born Log-likelihood for $73128$ NLO $t\bar{t}$ events between 
      $m_t=100$ GeV and $m_t=240$ GeV.}
 \label{fig:memttwdrng}
    \end{subfigure}\\    
    \begin{subfigure}[b]{1\textwidth}
    \includegraphics[width=0.87\textwidth]{{{%
          dytt/eeQQKT3-2ycut01memdet-150evts-crop}}}
    \caption{Same as in \Fig{fig:memttwdrng} but zoomed in between
      $m_t=170$~GeV and $m_t=180$~GeV to extract $\widehat{m}_t$ and
      $\Delta\widehat{m}_t$ by a parabola fit.}
 \label{fig:memttzoom}
    \end{subfigure}
\caption{Top-quark mass extraction from $t\bar{t}$ events with the \MEM.}
 \label{fig:memtt}
  \end{center}
\end{figure}
In \Fig{fig:memttwdrng} the log-likelihood is shown as a function of the top-quark mass $m_t$ in the range between $100$~GeV and $240$~GeV evaluated for the sample of $73128$ $t\bar{t}$ events to locate the minimum. The dashed red points show the log-likelihood based on NLO predictions (see \Eq{eq:likeliNLOtt}). To study the impact of the NLO corrections, the NLO events are also analysed with a likelihood based on the Born approximation (obtained from \Eq{eq:likeliNLOtt} but with $d\sigma^{\Born}$ and shown as solid blue points in the plots). In \Fig{fig:memttzoom} the focus is on the region between $m_t=170$~GeV and $m_t=180$~GeV. Around the minimum the estimator $\widehat{m}_t$ and its uncertainty $\Delta\widehat{m}_t$ can be extracted by a parabola fit (see section~\ref{sec:MLvar}).The extraction of the top-quark mass with the likelihood based on NLO predictions yields an estimator $\widehat{m}^\NLO_t$ which is consistent with the
input value of $m^\true_t=174$~GeV within the statistical uncertainty $\Delta\widehat{m}^\NLO_t$
\begin{equation}
    \widehat{m}^\NLO_t\pm\Delta\widehat{m}^\NLO_t = (174.3\pm1.3)\mbox{~GeV}.
\end{equation}
On the other hand, when extracting the top-quark mass from the same NLO events with a likelihood based on the Born approximation an estimator $\widehat{m}^\textrm{Born}_t$ is obtained
\begin{equation}
  \widehat{m}^\Born_t\pm\Delta\widehat{m}^\Born_t = (178.7\pm1.2)\mbox{~GeV}
\end{equation}
 which significantly deviates from the input value.
In this analysis $\widehat{m}^\Born_t$ is $4 \sigma$ away from the `true' top-quark mass used to generate the events. Hence $\widehat{m}^\Born_t$ cannot be viewed as an unbiased estimator. 

\begin{figure}[htbp]
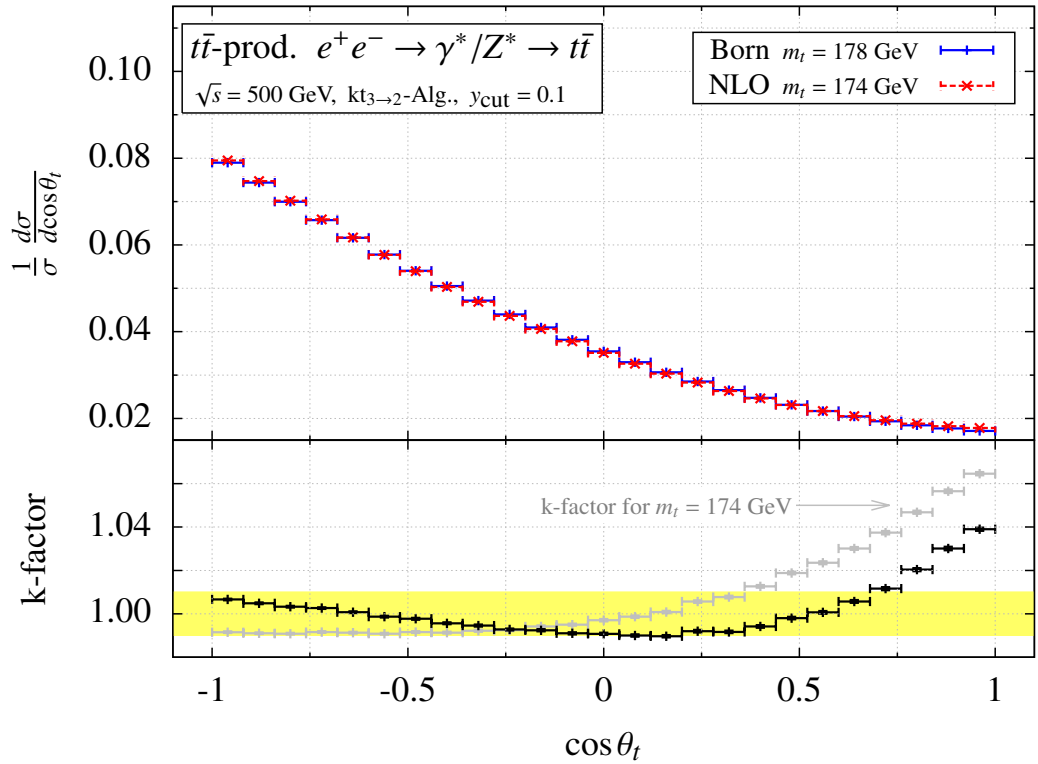

  \begin{center}
    \leavevmode
    \includegraphics[width=0.87\textwidth]{{{%
          dytt/eeQQKT3-2ycut0.1compbornCTH1-50x1e7mt178-crop}}}
    \caption{NLO predictions calculated with $\mt=174$~GeV (red dashed points) compared to the Born approximation calculated for
      $\mt=178$~GeV (blue solid points). The ratio of the NLO and the Born prediction (`k-factor') is shown in black at the bottom of the plot. Additionally, the k-factor with the Born prediction calculated with $m_t=174$~GeV is shown in grey for comparison.}
    \label{fig:born178vsNLO174}
  \end{center}
\end{figure}

In principle, it is not surprising that a biased estimator  $\widehat{m}^\Born_t$ is obtained when analysing events distributed according to the NLO cross section with a likelihood based on the Born approximation. However, the large size of the effect is remarkable considering that the NLO effects in the relevant $\cos\theta_t$ distribution are rather small (see grey points in the lower part of \Fig{fig:born178vsNLO174}). The bias in the extracted Born estimator suggests that the Born cross section calculated with $\mt=178$ GeV should give a better approximation of the NLO cross section calculated with $\mt=174$~GeV than the cross section in the Born approximation calculated with $\mt=174$~GeV. For illustration, \Fig{fig:born178vsNLO174} shows the comparison of the two predictions together with the k-factor distributions.  
Indeed, the comparison of the black and the grey k-factor distributions in \Fig{fig:born178vsNLO174} confirms that setting the top-quark mass to $\mt=178$ GeV in the Born cross section gives a slightly better description of the NLO result: The absolute value of the k-factor is below $1\%$ in the phase space region with $-1<\cos\theta_t\lesssim0.75$. The maximal deviation at $\cos\theta_t=1$ is $4\%$. On the other hand, when the top-quark mass is set to $\mt=174$ GeV in the Born approximation, the k-factor is in the $1\%$ band in the range $-1<\cos\theta_t\lesssim0.38$ with a maximal deviation at $\cos\theta_t=1$ of $6\%$. Obviously, the NLO effects cannot be absorbed into a shift of the mass parameter in the Born approximation. This can also be seen from the fact that the minimal value of the NLO likelihood at $\widehat{m}^\NLO_t$ in \Fig{fig:memttzoom} is below the minimum of the Born likelihood at $\widehat{m}^\Born_t$ indicating that the NLO cross section gives a better description of the analysed events.
Note that in the Born approximation top-quark pair production at a lepton collider $e^+e^-\rightarrow t\bar{t}$ is a purely electroweak process. Since the LO results do not depend on $\alpha_s$ it is not possible to estimate their theoretical uncertainty due to missing higher orders in QCD by a variation of  the renormalisation scale $\mur$.

\begin{figure}[htbp]
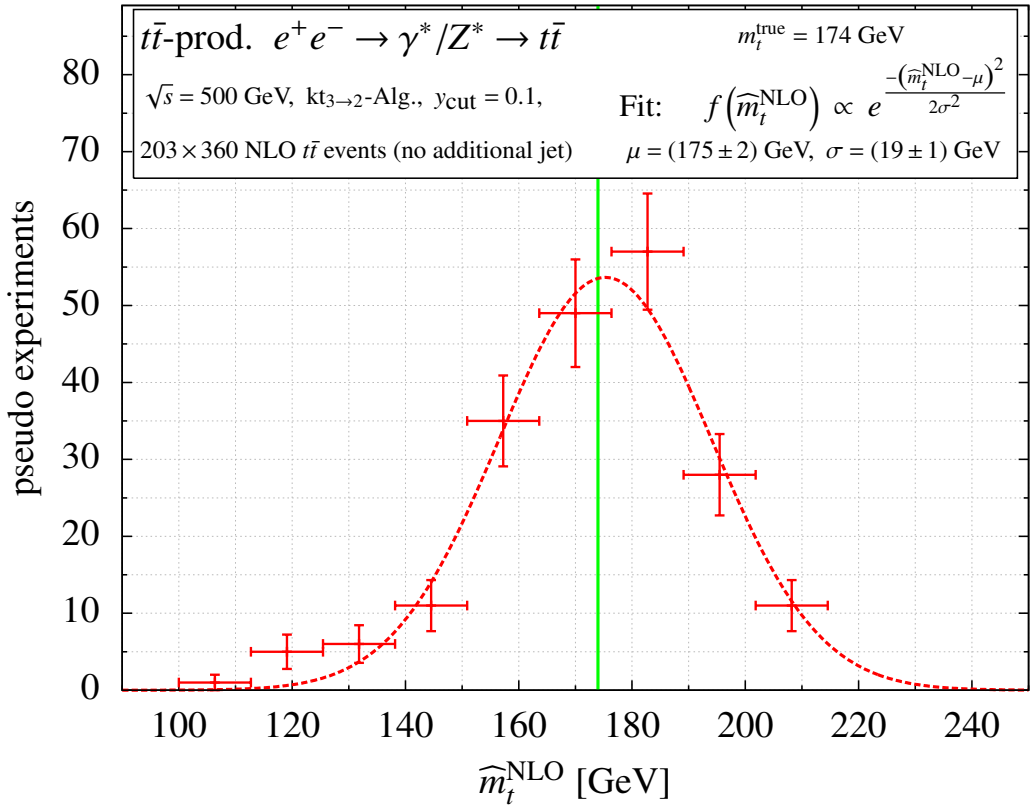

  \begin{center}
    \leavevmode 
    \includegraphics[width=0.87\textwidth]{{{%
           dytt/eeQQKT3-2ycut01mem-203x360-crop}}}
    \caption{Distribution of the estimator $\widehat{m}^\NLO_t$ around the 
      true value $m^\true_t=174$~GeV.}
 \label{fig:memttgauss}
  \end{center}
\end{figure}
\begin{figure}[htbp]
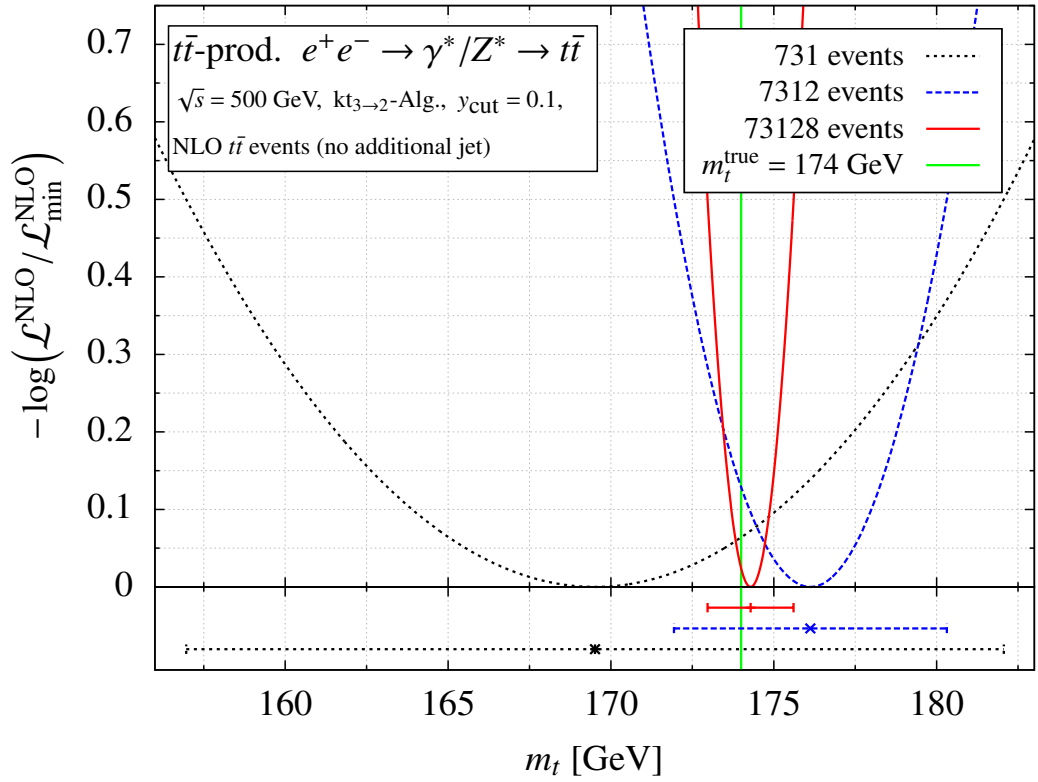

  \begin{center}
    \leavevmode 
    \includegraphics[width=0.87\textwidth]{{{%
           dytt/eeQQKT3-2ycut01mem-73128-7312-731evts-crop}}}    
    \caption{Log-likelihood depending on the number of events (estimators $\widehat{m}^\NLO_t\pm\Delta\widehat{m}^\NLO_t$ shown at the bottom).}
    \label{fig:memttnev}
  \end{center}
\end{figure}
The consistency of the implementation of the \MEM at NLO is illustrated in \Fig{fig:memttgauss} and \Fig{fig:memttnev} using the example of \ML properties (cf. section~\ref{sec:MLprop}). For the analysis presented in \Fig{fig:memttgauss} the event sample is split in $203$ sub-samples with $360$ events each. The distribution of the extracted estimators $\widehat{m}^\NLO_t$ from the sub-samples can be fitted by a Gaussian function (dashed line) with mean value $\mu=175$~GeV and standard deviation $\sigma=19$~GeV which is consistent with the input value of the top-quark mass. Thus, \Fig{fig:memttgauss} illustrates that $\widehat{m}^\NLO_t$ provides indeed an unbiased \ML estimator. 
Increasing the number of events $N$ in a \ML parameter extraction results in a tightening of the dip in the log-likelihood around the true value of the extracted parameter. In \Fig{fig:memttnev} the \MEM at NLO is used to extract estimators for the the top-quark mass from $N=731$, $N=7312$ and $N=73128$ top-quark pair events distributed according to the NLO cross section.  At the bottom of the plot the respective estimators $\widehat{m}^\NLO_t\pm\Delta\widehat{m}^\NLO_t$ are shown. The uncertainties of the estimators $\Delta\widehat{m}^\NLO_t$ show approximate scaling with $\sqrt{N}$ as
\begin{equation}
\Delta\widehat{m}^\NLO_t\propto N^{-\frac{1}{2}}.
\end{equation} 
This study, although presented here for validation and illustration purposes, has some relevance for a future linear collider. These top-quark mass measurements in the continuum pose a complementary alternative to threshold scans if competitive  precision can be achieved in the analysis (see e.g. \Ref{Seidel:2013sqa}). 
With regard to that the impact of the modified jet algorithm is briefly studied. 

\begin{figure}[htbp]
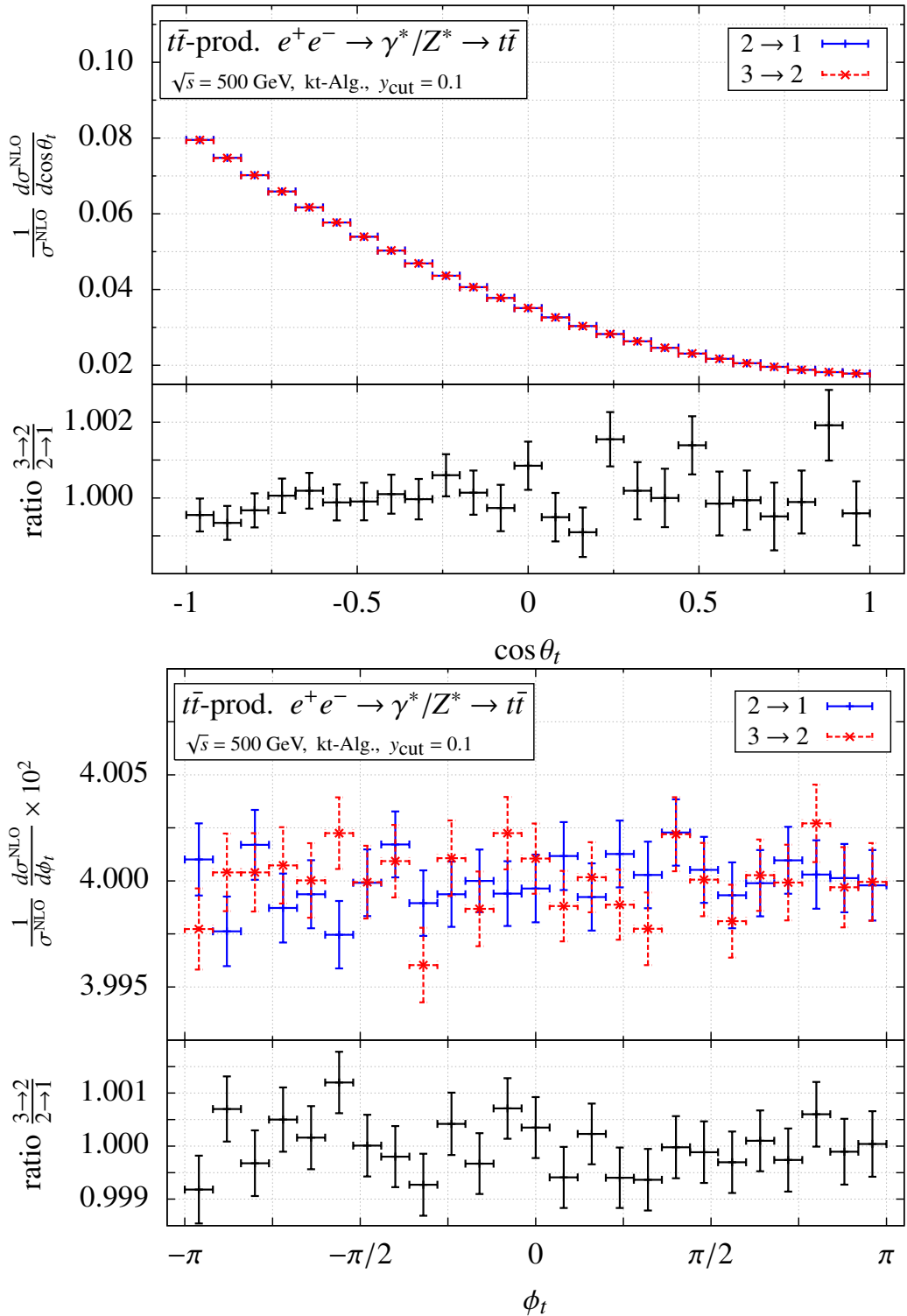

  \begin{center}
    \leavevmode
    \includegraphics[width=0.87\textwidth]{{{%
          dytt/eeQQKT3-2ycut0.1compnloCTH1-50x1e7-crop}}}\\
    \includegraphics[width=0.87\textwidth]{{{%
          dytt/eeQQKT3-2ycut0.1compnloPH1-50x1e7-crop}}}
    \caption{Impact of $3\to2$ clustering with respect to 
      $2\to1$ clustering on the angular distributions for
      top-quark pair production in $e^+e^-$ annihilation.}
    \label{fig:imp3221_tt}
  \end{center}
\end{figure}

\begin{figure}[htbp]
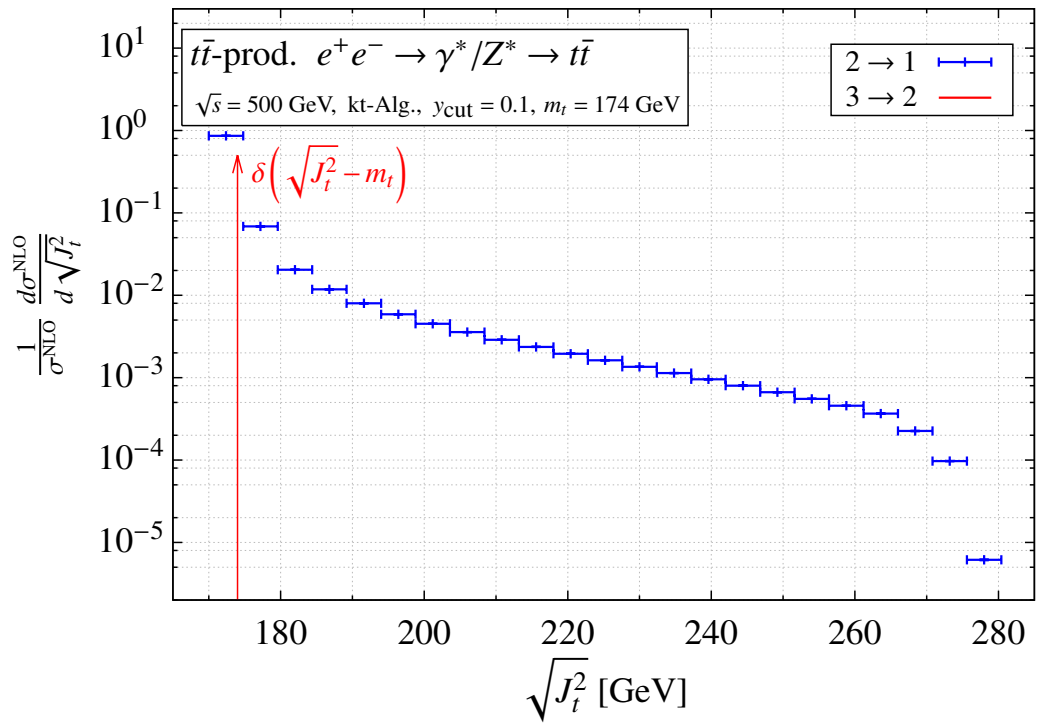

  \begin{center}
    \leavevmode
    \includegraphics[width=0.87\textwidth]{{{%
          dytt/eeQQKT3-2ycut0.1konvM1-50x1e7_21-crop}}}
    \caption{Top-quark jet mass distribution stemming from the $2\to1$ recombination.}
    \label{fig:massdistr_2to1_clustering}
  \end{center}
\end{figure}

As stated before, the kinematics of the top-quark pair produced in  $e^+e^-$ annihilation is highly constrained by symmetries of the interaction and by imposing momentum conservation and on-shell conditions. The $t\bar{t}$ events are completely determined by the polar and azimuthal angle of the top-quark jet (cf. \Eq{eq:jmomtt}). In \Fig{fig:imp3221_tt} the NLO distributions of these angular variables are shown as obtained with the modified $3\to2$ (red dashed) and the conventional $2\to1$ (solid blue) clustering prescriptions. For both angular variables the two clustering algorithms result in the same distributions at the permille level. This is not surprising since both clustering prescriptions conserve the overall momentum which dominantly affects the angular kinematics. No major differences in the \MEM analysis presented above are expected in case the observed events would
be obtained with the conventional $2\to1$ jet algorithm by the experiment.
However, this is not a general statement and similar studies would have to be conducted for other processes and even different event definitions. The latter is illustrated by \Fig{fig:massdistr_2to1_clustering} showing the distribution of the mass of the top-quark jet stemming from the $2\to1$ recombination in contrast to the fixed jet mass in the $3\to2$ clustering. Consequently, a corresponding impact on distributions which show sensitivity to the masses of the final-state objects (e.g. thresholds in energy distributions) is expected if the modified jet algorithm is used.
\clearpage
\subsection{Application II: Top-quark mass extraction from single top-quark events}\label{sec:mtextsgt}
\subsubsection{Using the exclusive event definition}\label{sec:mtextsgtexcl}
For the first application of the \MEM at NLO accuracy the hadronic single top-quark production is taken as an example and the sample of $N_\excl=12755$ ($N_\excl=24088$) exclusive $s$-channel ($t$-channel) single top-quark events generated in section~\ref{sec:event-definition-exclusive-case} is treated as the outcome of a toy experiment. Since the event definition $\vec{x}_i=(\eta_t,\;E_j,\;\eta_j,\;\phi_j)_i$ does not fix the top-quark mass, the \MEM can be consistently applied to extract estimators for the top-quark mass from the event samples.
Again, reproducing the input value of the event generation serves as a consistency check of the procedure and a useful check of the numerical implementation. Furthermore, this analysis allows to estimate the impact of higher-order corrections to top-quark mass measurements at a hadron collider with the \MEM. 

The NLO likelihood as a function of the top-quark mass for the exclusive jet events is obtained by using the NLO jet event weight from \Eq{eq:diffjxsex} in \Eq{eq:likeliNLO} with $\omega=m_t$
\begin{eqnarray}\label{eq:likeliex}
 \nn \mathcal{L}^\NLO\left(m_t\right)
  &=& 
  \prod\limits_{i=1}^{N_\excl} \mathcal{L}^\NLO
  \left(\vec{x}_i|m_t\right)=\left(\frac{1}{\sigma_\excl^\NLO(m_t)}\right)^{N_\excl}
  \prod\limits_{i=1}^{N_\excl}\frac{d\sigma_\excl^\NLO(m_t)}{d\vec{x}_i}\\
  &=& \left(\frac{ 1}{\sigma_\excl^\NLO(m_t)}\right)^{N_\excl}
  \prod\limits_{i=1}^{N_\excl}
  \left.\left(\frac{{J^{\perp}}^2\cosh{\eta_t}}{2\;s\;E_t\;\cosh{\eta_j}}\;
      \frac{d\sigma_\excl^\NLO}{dR_2(J_{t},J_{j})}(\mt)
    \right)\right|_{\vec{x}=\vec{x}_i}
\end{eqnarray}
where $\vec{x}_i$ denotes an event taken from the exclusive event sample $\{\vec{x}_1, \ldots, \vec{x}_{N_\excl}\}$. Maximising the likelihood with respect to the top-quark mass yields an NLO estimator $\widehat{m}^\NLO_t$ for the input value of the event generation of $m^\true_t=173.2\GeV$ (see section~\ref{sec:event-definition-exclusive-case}). 
As a start, $s$- and $t$-channel production of single top-quark events are studied separately to validate the \MEM at NLO accuracy and to investigate some of its features. At the end of this section also the production of single top-quark events from either the $s$- or the $t$-channel is considered in order to study the individual influence of both production channels on the analysis. 

\begin{figure}[htbp]
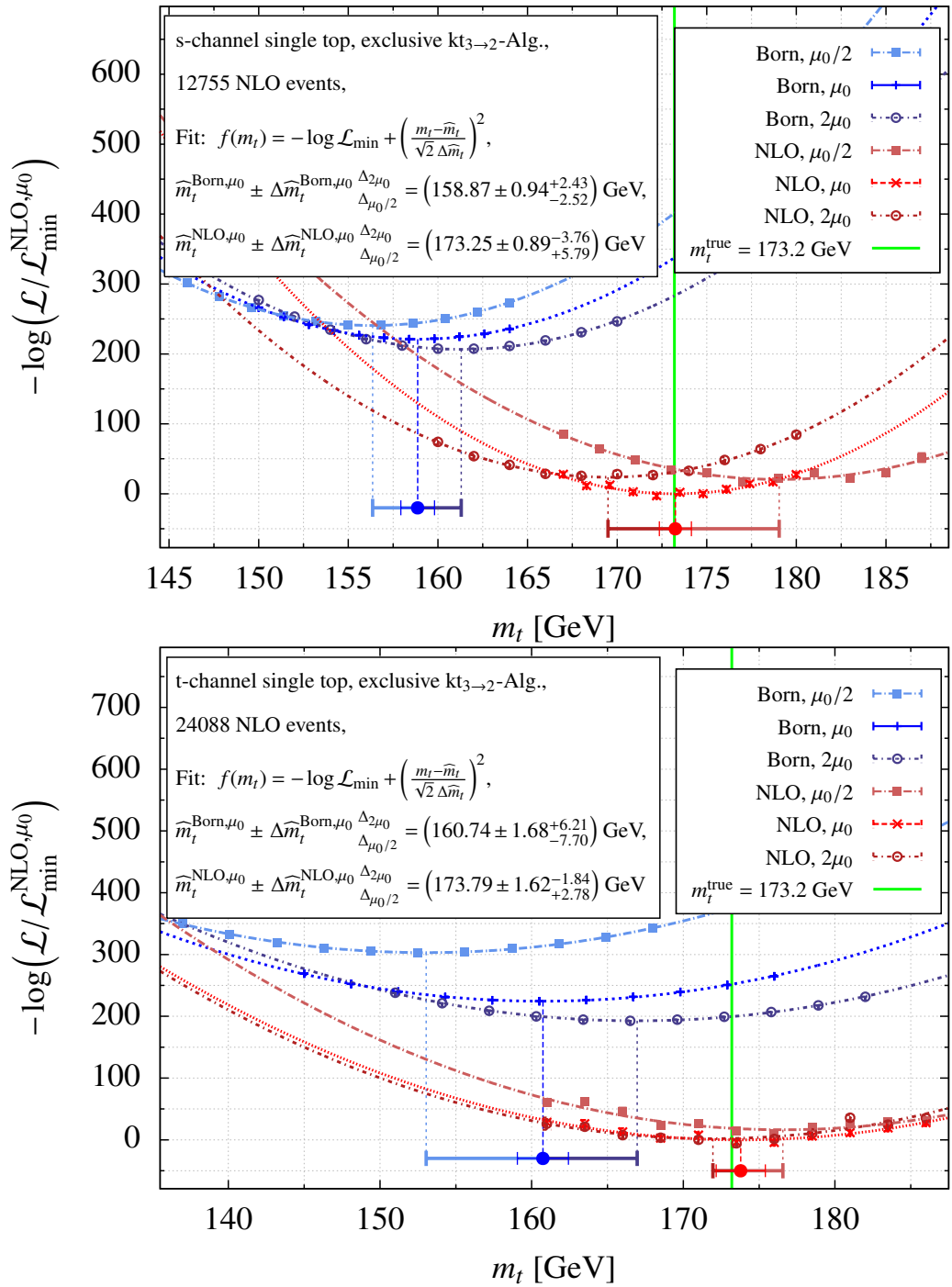

  \begin{center}
    \includegraphics[width=0.87\textwidth]{{{%
          sgt/mem/sgtsKT3-2ycut30membornnlomu-94evts-crop}}}\\
    \includegraphics[width=0.87\textwidth]{{{%
          sgt/mem/sgttKT3-2ycut30membornnlomu-90evts-crop}}}
     \caption{Top-quark mass extraction from exclusive NLO $s$-channel
       (top) and $t$-channel (bottom) single top-quark events employing NLO
       and Born likelihoods using a parabola fit around the minimum.}
    \label{fig:MEM_sgtex}
  \end{center}
\end{figure}
In \Fig{fig:MEM_sgtex} the log-likelihood (cf. \Eq{eq:TaylorLikeli}) evaluated with the exclusive $s$-channel and $t$-channel single top-quark jet events is shown as a function of the top-quark mass. To study the impact of the NLO corrections on the mass extraction, these events---which follow the NLO cross section---are analysed with likelihoods based on NLO predictions for the event weights (reddish lower points) as well as the Born approximation only (bluish upper points) in \Eq{eq:likeliex}. In order to estimate the uncertainty due to missing higher orders in the predictions, the renormalisation and factorisation scales are simultaneously varied by factors $1/2$ and $2$ in the calculation of the likelihood. Note that all events from section~\ref{sec:event-definition-exclusive-case} are generated for the central scale choice $\mur=\muf=\mu_0$ (see \Eq{eq:scale-definition} for the definition of $\mu_0$). The results for the three different scale settings are shown as open circles ($\mur=\muf=2\mu_0$), crosses ($\mur=\muf=\mu_0$) and solid squares ($\mur=\muf=\mu_0/2$). The results of parabola fits to the data points are shown by lines in the plots.
In \Fig{fig:MEM_sgtex} the values for the log-likelihood including the NLO corrections are significantly smaller than the values corresponding to the Born likelihoods. Overall, the NLO likelihood evaluated with the central scale choice $\mur=\muf=\mu_0$ yields the lowest values. As expected, the NLO predictions at the central scale give the best description of the unweighted events.  

\begin{table}[htbp]
\centering
\caption{Extracted top-quark mass estimators from exclusive jet events following the NLO cross section.}
\label{tab:exclmasses}
\def\arraystretch{1.8}
\begin{tabular}{|c|c|c|}
\hline
{exclusive events} & \multicolumn{1}{c|}{$12755$ $s$-channel events} & \multicolumn{1}{c|}{$24088$ $t$-channel events} \\ \hline
{$\widehat{m}_t^{\NLO,\;\mu_0}\pm \left.\Delta\widehat{m}_t^{\NLO,\;\mu_0}\right.^{\Delta_{2\mu_0}}_{\Delta_{\mu_0/2}}$ [GeV]} & {$173.25\pm0.89^{-3.76\;(-2.2\%)}_{+5.79\;(+3.3\%)}$} &{$173.79\pm1.62^{-1.84\;(-1.1\%)}_{+2.78\;(+1.6\%)}$} \\ \hline
{$\widehat{m}_t^{\Born,\;\mu_0}\pm \left.\Delta\widehat{m}_t^{\Born,\;\mu_0}\right.^{\Delta_{2\mu_0}}_{\Delta_{\mu_0/2}}$ [GeV]} & {$158.87\pm0.94^{+2.43\;(+1.5\%)}_{-2.52\;(-1.6\%)}$} & {$160.74\pm1.68^{+6.21\;(+3.9\%)}_{-7.70\;(-4.8\%)}$}\\
\hline
\end{tabular}
\end{table}
The estimators for the top-quark mass are extracted as the positions of the minima of the log-likelihood functions. The results of the top-quark mass extraction from the exclusive NLO $s$- and $t$-channel events are summarised in table~\ref{tab:exclmasses}. The superscripts `NLO' and `Born' indicate whether the estimator is obtained with likelihoods based on NLO predictions or the Born approximation.
The statistical uncertainties of the estimators $\Delta\widehat{m}_t$ are obtained according to \Eq{eq:apprLikelisig}.
The systematic uncertainties due to missing higher orders in the predictions are estimated as the differences of the estimators resulting from the scale variations in the likelihoods with respect to the result for the central scale. The shifts in the extracted values for the upwards
$\mur=\muf=2\mu_0$ (downwards $\mur=\muf=\mu_0/2$) variation are indicated by superscripts (subscripts). This systematic scale uncertainty constitutes a lower limit on the reachable accuracy. The results are also illustrated as data points in the lower part of the plots in \Fig{fig:MEM_sgtex} with errorbars indicating the statistical as well as the systematic uncertainties. 

For both the $s$- and the $t$-channel the top-quark mass values extracted with likelihoods based on NLO and Born predictions significantly differ. The NLO estimators $\widehat{m}^\NLO_t$ perfectly reproduce the input value $m_t^\true=173.2\GeV$ within their statistical uncertainties $\Delta\widehat{m}^\NLO_t$. However, the Born estimators $\widehat{m}^\Born_t$ deviate by  around $-8.3\%$ in case of the $s$-channel and by $-7.2\%$ in case of the $t$-channel. These differences are neither covered by the statistical nor the systematic uncertainties. Recalling the impact of the NLO corrections on the normalised distributions given in \Fig{fig:kfac_sgts_e2} to \Fig{fig:kfac_sgtt_eta1}, these deviations are not surprising: The normalised Born and NLO distributions differ up to $25\%$ while their theoretical uncertainty bands due to scale variation do not overlap in wide ranges of the phase space. As mentioned at the end of section~\ref{sec:SmatMEM}, if the data is not well described by the assumed probability distribution used in the likelihood construction, the resulting \ML estimators are prone to be systematically biased. 

\begin{figure}[htbp]
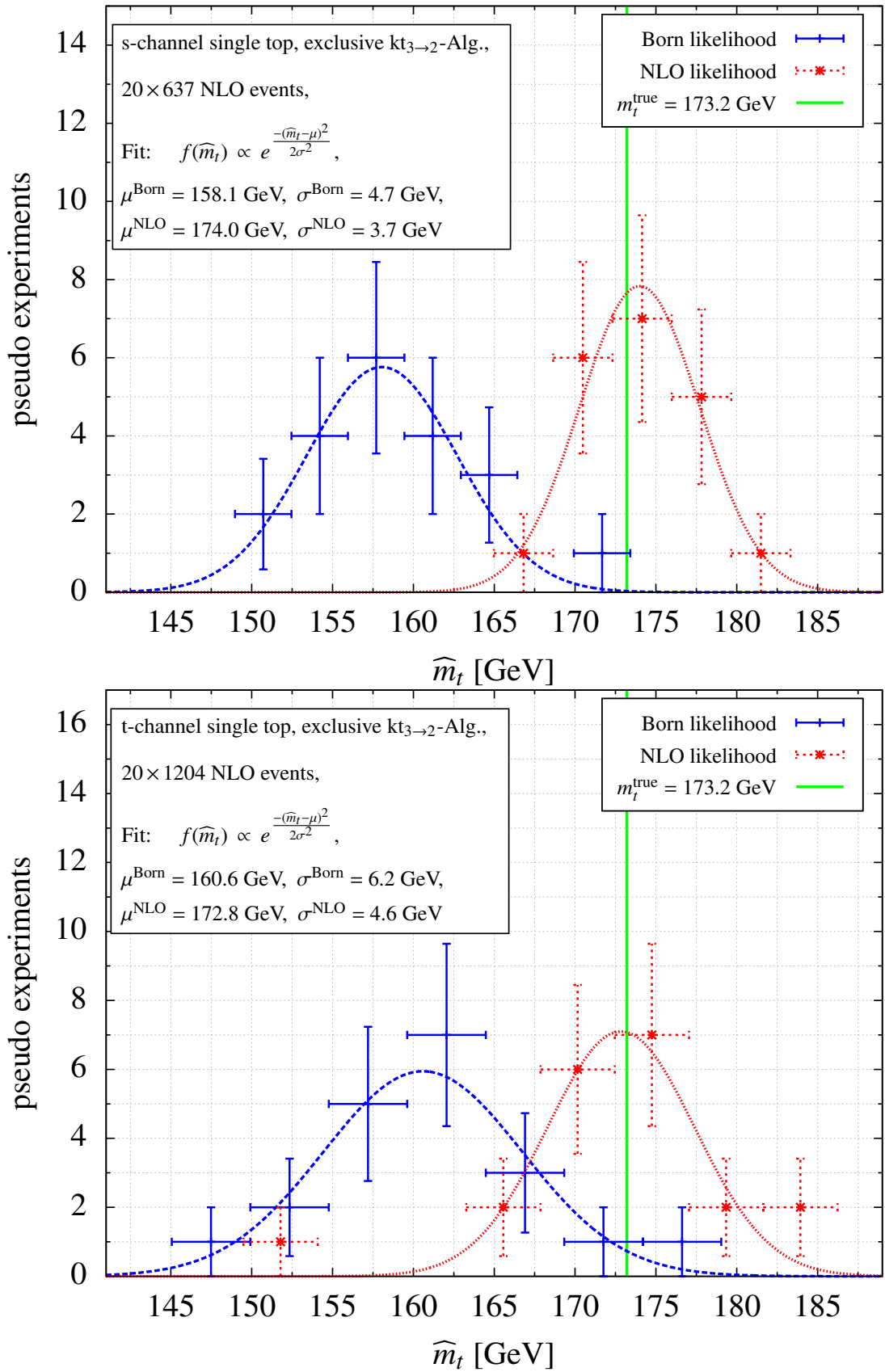

  \begin{center}
    \includegraphics[height=0.47\textheight]{{{%
          sgt/mem/sgtsKT3-2ycut30mem-20x637-crop}}}\\
   \includegraphics[height=0.47\textheight]{{{%
          sgt/mem/sgttKT3-2ycut30mem-20x1204-crop}}}
     \caption{Distribution of the extracted top-quark mass estimators from exclusive NLO
       $s$-channel (top) and $t$-channel (bottom) single top-quark events
       employing NLO and Born likelihoods.}
    \label{fig:MEM_sgtdistr}
  \end{center}
\end{figure}

To carve out the systematic nature of the observed deviations, the distribution of mutiple extracted estimators are investigated and the bias is identified as the deviation of their mean value from $m_t^\true$  according to \Eq{eq:bias}. The samples of pseudo data are split into $20$ subsamples each representing an individual experiment to be analysed.
The distributions of the extracted estimators for $s$-channel (with $637$ events per subsample) and $t$-channel (with $1204$ events per subsample) are shown in \Fig{fig:MEM_sgtdistr} and their mean and standard deviation are obtained by a Gaussian fit.
In accordance with the findings summarised in table~\ref{tab:exclmasses}, \Fig{fig:MEM_sgtdistr} shows that the distributions of the NLO estimators are in agreement with $\widehat{m}^\NLO_t$ being an unbiased estimator for both channels. In contrast, the distributions of the Born estimators show significant biases which are consistent with the ones seen in \Fig{fig:MEM_sgtex}. Thus, employing Born likelihoods in the \MEM for top-quark mass extraction does not yield an unbiased estimator. 

Per se, a bias in the estimator does not necessarily preclude the application of the \MEM based on Born predictions. With an additional calibration procedure any shift can be accounted for (cf. the end of section~\ref{sec:SmatMEM}). However, this calibration introduces additional uncertainties and might even mask important effects. It is therefore preferable to rely on the most accurate predictions available. Furthermore, when the \MEM is used for parameter extraction, only a full NLO calculation allows to uniquely define the renormalisation scheme enabling a clear interpretation of the extracted parameter. 

Last but not least, assessing the theory uncertainties by a common scale variation in the Born predictions does not necessarily provide a reliable estimate of missing higher-order corrections. While in the $t$-channel the uncertainty estimated by the (downwards/upwards) scale variation is reduced:
\[
 (\;-4.8\%\; / \;+3.9\%\;)_{\Born} \longrightarrow (\;+1.6\% \;/ \;-1.1\%\;)_\NLO 
\]
by including NLO corrections in the predictions (see table~\ref{tab:exclmasses} and \Fig{fig:MEM_sgtex}), this is not the case for the $s$-channel: 
\[
 (\;-1.6\%\; / \;+1.5\%\;)_{\Born} \longrightarrow (\;+3.3\% \;/ \;-2.2\%\;)_\NLO. 
\]
Scale variation by factors $1/2$ and $2$ in the calculation of the likelihood needed to extract an estimator for the top-quark mass results in larger shifts of the NLO estimator than in the Born estimator. In contrast to the naive expectation, the deduced uncertainties thus increase when including NLO corrections in the theoretical predictions. Concerning this matter, we stress again that in the Born approximation the process is purely mediated by electroweak interactions with only a factorisation scale dependence which enters through the PDFs. Since the probability densities in the likelihood are obtained from normalised differential cross sections, this dependence on the factorisation scale cancels in the ratio to some extent (cf. section~\ref{sec:kfacexcl}).
Bearing that in mind, the estimated theoretical uncertainties based on the scale variation in the Born predictions most likely underestimate missing higher-order effects. 

Altogether, we recapitulate that while biases apparent in the \MEM without NLO corrections may be accounted for by a calibration procedure, the extracted parameter values elude clear theoretical interpretation without a well-defined renormalisation scheme. Additionally, customary estimates of theoretical uncertainties due to missing higher orders inferred from scale variations might be unreliable. 

\begin{figure}[htbp]
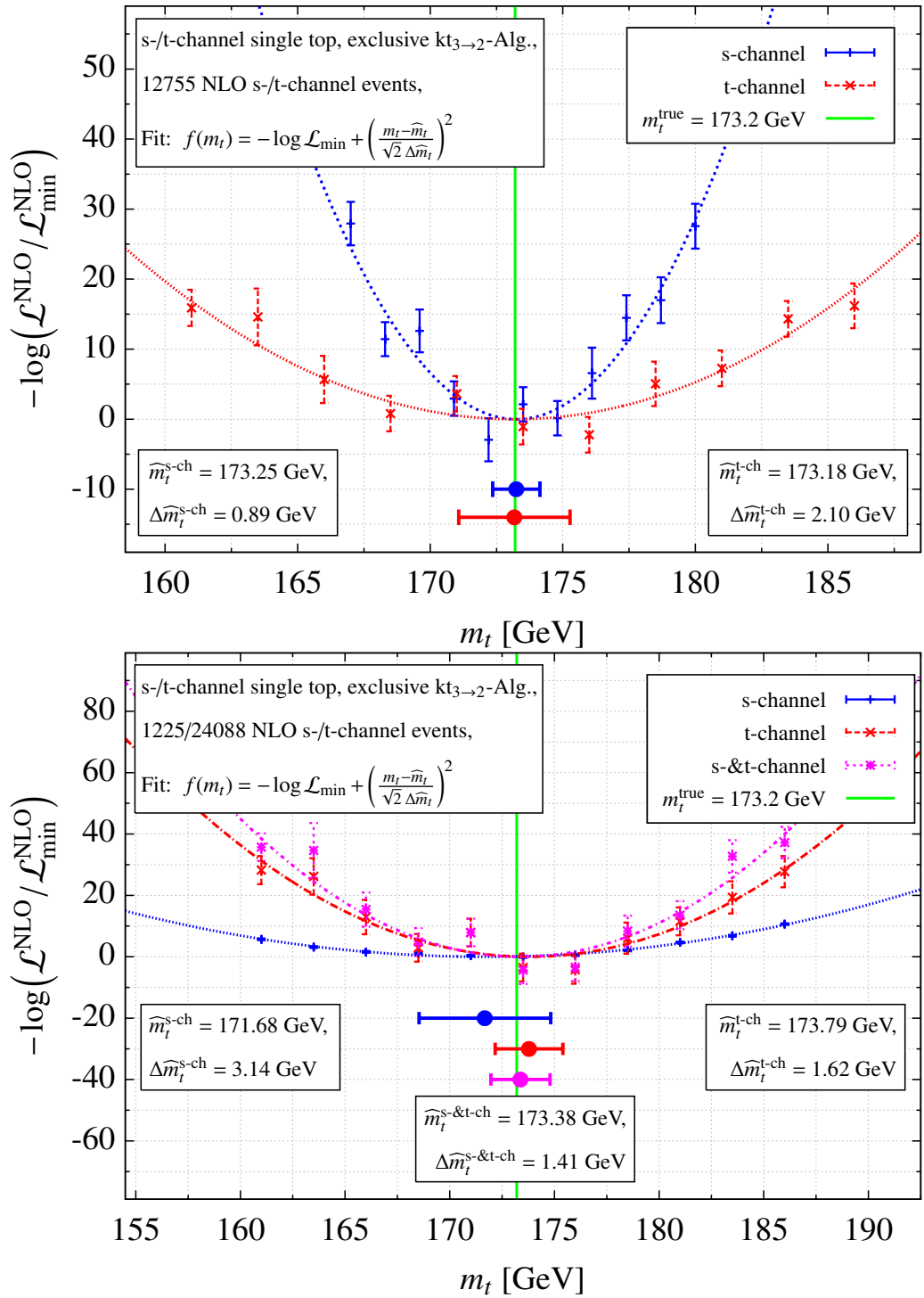

  \begin{center}
    \includegraphics[width=0.87\textwidth]{{{%
          sgt/mem/sgtKT3-2ycut30memdet_st-crop}}}\\
    \includegraphics[width=0.87\textwidth]{{{%
          sgt/mem/sgtKT3-2ycut30memdet_strel-crop}}}
     \caption{Extraction of the top-quark mass from a fixed 
       number of NLO $s$- and $t$-channel single top-quark events (upper plot) and for 
       fixed luminosity (lower plot).}
    \label{fig:MEM_sgt}
  \end{center}
\end{figure}
For $12755$ $s$-channel events and $24088$ $t$-channel events the statistical uncertainties amount to $\Delta\widehat{m}^\NLO_t=0.9\GeV$ for the $s$-channel analysis and to $\Delta\widehat{m}^\NLO_t=1.6\GeV$ in the $t$-channel. So already for samples of these sizes the uncertainties of the top-quark mass extractions are dominated by the systematic scale uncertainties quoted above.
The sensitivity to the top-quark mass of the NLO analyses can be explicitly studied in the $s$- and the $t$-channel. For that purpose, the top-quark mass extraction employing the \MEM at NLO accuracy in both channels is compared in \Fig{fig:MEM_sgt} for a fixed number of events (upper plot) and for fixed luminosity (lower plot). Given samples of equal sizes, the $s$-channel analysis shows a higher sensitivity to the top-quark mass allowing a more precise mass measurement as far as only statistical uncertainties are concerned. This can be seen from the upper plot in \Fig{fig:MEM_sgt} where the log-likelihood for the $s$-channel is significantly narrower than the log-likelihood for the $t$-channel. The parabola fit yields a value of the statistical uncertainty $\Delta\widehat{m}^\NLO_t$ in the $s$-channel analysis which is smaller by roughly a factor of $2.3$ compared to the respective value in the $t$-channel analysis. 
\begin{figure}[htbp]
  \begin{center}
    \includegraphics[width=0.87\textwidth]{{{%
          sgt/mem/sgttKT3-2ycut30sgt-excl-crop}}}
     \caption{Top-quark mass dependence of the fiducial exclusive $s$- and $t$-channel cross section calculated at NLO accuracy.}
    \label{fig:MEM_sgtmass}
  \end{center}
\end{figure}
This observation is in agreement with the top-quark mass sensitivity of the fiducial exclusive NLO cross sections: In \Fig{fig:MEM_sgtmass} the dependence of the fiducial exclusive cross sections calculated including NLO corrections is shown and the top-quark mass sensitivity is inferred by approximating the dependence of the cross section on the top-quark mass by a polynomial as presented in \Refs{Campbell:2009gj,Kant:2014oha}. The factor of $2.3$ is consistent with the top-quark mass sensitivities of the $s$- and $t$-channel fiducial exclusive NLO cross sections estimated from \Fig{fig:MEM_sgtmass} (and also consistent with the results in \Ref{Kant:2014oha} for the inclusive cross sections at $\sqrt{s}=14$~GeV)
\begin{equation}\label{eq:fidxsmassexcl}
  \frac{\Delta\sigma^\NLO_{\excl\;s\text{-ch}}}{\sigma^\NLO_{\excl\;s\text{-ch}}} = -3.50\times
  \frac{\Delta\mt}{\mt}\,
  ,\quad
  \frac{\Delta\sigma^\NLO_{\excl\;t\text{-ch}}}{\sigma^\NLO_{\excl\;t\text{-ch}}} = -1.44\times
  \frac{\Delta\mt}{\mt}\,. 
\end{equation}
However, because the $t$-channel cross section is about $20$ times larger than the $s$-channel cross section, the higher top-quark mass sensitivity in the $s$-channel analysis is compensated by the $20$ times larger event rate of the $t$-channel in practice. This is illustrated in the lower plot in \Fig{fig:MEM_sgt} where the top-quark mass measurements are performed from event samples of realistic proportions corresponding to a fixed luminosity. The statistical uncertainty of the top-quark mass extraction from $1225$ $s$-channel events is roughly twice as large as in the $t$-channel with respectively $24088$ events. Combining both channels to a proportionally mixed sample of $25313$ events leads to a minor improvement in the statistical uncertainty in the top-quark mass extraction of around $200~\text{MeV}$ compared to the $t$-channel only case.
This improvement is consistent with a weighted mean of the separate mass determinations from the samples of $24088$ $t$-channel events and $1225$ $s$-channel events which yields $\overline{m}_t\pm\Delta\overline{m}_t=173.35\pm1.44$ GeV.

\subsubsection{Extended likelihood for the exclusive event definition}
As illustrated in section~\ref{sec:extlikeli}, the information of the total number of observed events can be incorporated into the \MEM through the extended likelihood, potentially improving the sensitivity of the analysis. According to \Eq{eq:extLike}, the respective extended likelihood is given by weighting \Eq{eq:likeliex} with the top-quark mass dependent Poisson factor
\begin{equation}
\mathcal{L}^\NLO_{\text{ext}}\left(m_t\right)
  =  \frac{\left(\sigma^\NLO_\excl(m_t)L\right)^{N_\excl}}{N_\excl!}
 e^{-\sigma^\NLO_\excl(m_t)L}
 \mathcal{L}^\NLO
  \left(m_t\right).
\end{equation}
\begin{figure}[htbp]
  \begin{center}
    \includegraphics[width=0.87\textwidth]{{{%
          sgt/mem/sgtsKT3-2ycut30membornnlomu-94evts_ext-crop}}}\\
    \includegraphics[width=0.87\textwidth]{{{%
          sgt/mem/sgttKT3-2ycut30membornnlomu-90evts_ext-crop}}}
     \caption{Same as  \Fig{fig:MEM_sgtdistr} but with extended
       likelihoods.
       }
     \label{fig:MEM_sgt_ext}
  \end{center}
\end{figure}
The $s$-channel and $t$-channel analyses with extended likelihoods are shown in \Fig{fig:MEM_sgt_ext} using the exclusive event definition.
\begin{table}[htbp]
\centering
\caption{Same as in table~\ref{tab:exclmasses} but with extended likelihoods}
\label{tab:exclmassesext}
\def\arraystretch{1.8}
\begin{tabular}{|c|c|c|}
\hline
{excl. events, extended likelihoods} & \multicolumn{1}{c|}{$12755$ $s$-channel events} & \multicolumn{1}{c|}{$24088$ $t$-channel events} \\ \hline
{$\widehat{m}_t^{\NLO,\;\mu_0}\pm\left. \Delta\widehat{m}_t^{\NLO,\;\mu_0}\right.^{\Delta_{2\mu_0}}_{\Delta_{\mu_0/2}}$ [GeV]} & {$173.35\pm0.40^{-0.23\;(-0.1\%)}_{+0.65\;(+0.4\%)}$} &{$173.53\pm0.70^{+4.70\;(+2.7\%)}_{-3.39\;(-2.0\%)}$} \\ \hline
{$\widehat{m}_t^{\Born,\;\mu_0}\pm \left.\Delta\widehat{m}_t^{\Born,\;\mu_0}\right.^{\Delta_{2\mu_0}}_{\Delta_{\mu_0/2}}$ [GeV]} & {$171.22\pm0.41^{+1.62\;(+0.9\%)}_{-1.97\;(-1.2\%)}$} & {$201.43\pm0.73^{+8.89\;(+4.4\%)}_{-12.50\;(-6.2\%)}$}\\
\hline
\end{tabular}
\end{table}
The results of the top-quark mass extraction from the NLO events with extended likelihoods based on Born and NLO predictions are summarised in table~\ref{tab:exclmassesext}. The top-quark mass extraction with the NLO extended likelihood correctly reproduces the input value $m_t^\true$ for the top-quark mass used in the event generation in both channels. Again, including NLO corrections evaluated at the central scale $\mu_0$ yields the lowest values of the extended log-likelihoods. As expected, not only the distribution of the events in both samples but also their total number are best described by the NLO cross sections evaluated with a scale choice and a value of the top-quark mass consistent with the input parameters of the event generation in section~\ref{sec:event-definition-exclusive-case}. 

Utilising the extended likelihoods significantly reduces the statistical uncertainty of the top-quark mass extractions $\Delta\widehat{m}^\NLO_t$ by roughly a factor of two. As a cross check, the impact of the additional information contained in the number of observed events on the top-quark mass sensitivity can be estimated: When assuming that the uncertainty of a fiducial cross section measurement is completely dominated by the Poisson standard error $\sqrt{N}$ of counting $N$ events, the relative uncertainties of top-quark mass determinations from fiducial cross section measurements can be estimated as (according to \Eq{eq:fidxsmassexcl})
\begin{equation}\label{eq:hypmassmeas}
\left|{\Delta\mt^{\excl\;s\text{-ch}}\over m_t}\right|=\underbrace{{\Delta\sigma^\NLO_{\excl\;s\text{-ch}}\over\sigma^\NLO_{\excl\;s\text{-ch}}}}_{={12755}^{-1/2}}\,{1\over 3.50} = 0.25\%
  ,\quad
\left|{\Delta\mt^{\excl\;t\text{-ch}}\over m_t}\right|=\underbrace{{\Delta\sigma^\NLO_{\excl\;t\text{-ch}}\over\sigma^\NLO_{\excl\;t\text{-ch}}}}_{={24088}^{-1/2}}\,{1\over 1.44} = 0.45\%.
\end{equation}
Note that these hypothetical top-quark mass measurements are depending on the sample sizes only. On the other hand, the \MEM analysis without extended likelihoods presented in the previous section (cf. table~\ref{tab:exclmasses}) only depends on normalised cross sections. To estimate the impact of the information contained in the sample sizes on the top-quark mass sensitivity, the results from \Eq{eq:hypmassmeas} and table~\ref{tab:exclmasses} are combined. For the weighted averages relative uncertainties of $\pm0.23\%$ for the $s$-channel and $\pm0.40\%$ for the $t$-channel are obtained which are in fact compatible with the uncertainties in the extended likelihood analyses given in table~\ref{tab:exclmassesext}.

As indicated in section~\ref{sec:extlikeli} the imperfect knowledge of the integrated luminosity introduces an additional systematic uncertainty in the extended likelihood calculation. The impact of this systematic uncertainty can be estimated by repeating the analyses with luminosities varied by $\pm2\%$ (cf. \Ref{Aaboud:2016hhf}). In these analyses shifts in the extracted top-quark mass estimators of about $\pm 1\GeV$ in the $s$-channel and $\pm 2\GeV$ in the $t$-channel are observed. Hence, using the extended likelihood does not allow a more precise top-quark mass measurement unless the uncertainty of  the integrated luminosity determination can be significantly reduced. 

The estimators for the top-quark mass extracted from the NLO events with extended likelihoods based on the Born approximation $\widehat{m}^\Born_t$ deviate from the input value $m_t^\true$ by about $-1.1\%$ in the case of $s$-channel events and $+16.3\%$ for $t$-channel events. The large deviation from the input value $m_t^\true$ of the Born estimator $\widehat{m}^\Born_t$ extracted from the $t$-channel events of around $+28.2\GeV$ can be explained by the large negative NLO correction to the fiducial exclusive $t$-channel cross section (cf. table~\ref{tab:fidxsex}): Employing extended likelihoods in the \MEM yields a parameter value adjusted not only to reproduce the distribution of the events in the sample in the best way but also their total number. Increasing the value of the top-quark mass limits the available phase space volume because of the production threshold of the massive top quark. This compensates for the decreased number of events in the NLO event sample compared to the Born expectation. In contrast, the negative NLO corrections to the fiducial exclusive $s$-channel cross section given in table~\ref{tab:fidxsex} are comparatively small. Hence, the \MEM using extended likelihoods favours only a small deviation of the Born estimator $\widehat{m}^\Born_t$ from the input value in the $s$-channel analysis of around $-2.0\GeV$ as a compromise between the distribution of the events only (cf. table~\ref{tab:exclmasses}) and their total number. In both channels the shifts are not covered by the estimated theoretical uncertainties based on the scale variation.

Since the normalisation of the differential cross sections cancels in the construction of the extended likelihood (cf. \Eq{eq:extLike}), there is no cancellation of the (factorisation) scale dependence in the extended likelihood calculated in the Born approximation. Indeed, when using the extended likelihoods calculated at NLO accuracy to extract estimators for the top-quark mass, the estimated theoretical uncertainties with respect to the Born approximation are improved in both channels. The shifts in the estimators due to simultaneous (downwards/upwards) variation of the central scale $\mu_0$ by a factor of $2$ are reduced  in the $s$-channel:
\[(\;-1.2\%\; /\; +0.9\%\;)_\Born\longrightarrow  (\;+0.4\%\; / \;-0.1\%\;)_\NLO\]
and in the $t$-channel:
\[(\;-6.2\%\; /\; +4.4\%\;)_\Born\longrightarrow  (\;-2.0\%\; / \;+2.7\%\;)_\NLO\] when including NLO corrections. 

\subsubsection{Using the inclusive event definition without vetoing an additional jet}\label{sec:mtextsgtincl}
Treating the sample of $N_\incl=16964$ ($N_\incl=32278$) inclusive $s$-channel ($t$-channel) single top-quark events from section~\ref{sec:event-definition-inclusive-case} as the outcome of a toy experiment, the analyses of the exclusive case can be repeated for the inclusive event definition $\vec{x}_i=(\eta_t,\;E_j,\;\eta_j,\;\phi_j)_i$ where $j$ represents the hardest resolved light jet.
For the extraction of an estimator for the top-quark mass, the NLO likelihood is needed as a function of $m_t$. Plugging the NLO event weight for inclusive jet events from \Eq{eq:diffjxsin} into \Eq{eq:likeliNLO} yields (with $\omega=m_t$)
\begin{eqnarray}\label{eq:likeliin}
 \nn \mathcal{L}^\NLO\left(m_t\right)
  &=& 
  \prod\limits_{i=1}^{N_\incl} \mathcal{L}^\NLO
  \left(\vec{x}_i|m_t\right)=\left(\frac{1}{\sigma_\incl^\NLO(m_t)}\right)^{N_\incl}
  \prod\limits_{i=1}^{N_\incl}\frac{d\sigma_\incl^\NLO(m_t)}{d\vec{x}_i}\\
\nn  &=& \left(\frac{ 1}{\sigma_\incl^\NLO(m_t)}\right)^{N_\incl}
  \prod\limits_{i=1}^{N_\incl}
 \Bigg(\frac{{J^{\perp}}^2\cosh{\eta_t}}{2\;s\;E_t\;\cosh{\eta_j}}\;
      \frac{d\sigma_\excl^\NLO}{dR_2(J_{t},J_{j})}(\mt)\\
  &&\left.+\int\limits_{p^{\perp}_{\text{min}}}^{J^{\perp}_{j}}dJ^{\perp}_{X}
  \int\limits_{-\eta_{\text{max}}}^{\eta_{\text{max}}}d\eta_{X}
  \int\limits_{0}^{2\pi}d\phi_{X}\;
  \frac{{J^{\perp}_j}{J^{\perp}_X}{J^{\perp}_t}
    \cosh{\eta_t}}{4s\;E_t\;\cosh{\eta_j}} \;
  \frac{d\sigma_{3\text{-jet}}}{ dR_3(J_t,J_j,J_X)}
    \Bigg)\right|_{\vec{x}=\vec{x}_i}
\end{eqnarray}
where the event $\vec{x}_i$ is now taken from the inclusive event sample $\{\vec{x}_1, \ldots, \vec{x}_{N_\incl}\}$.

\begin{figure}[htbp]
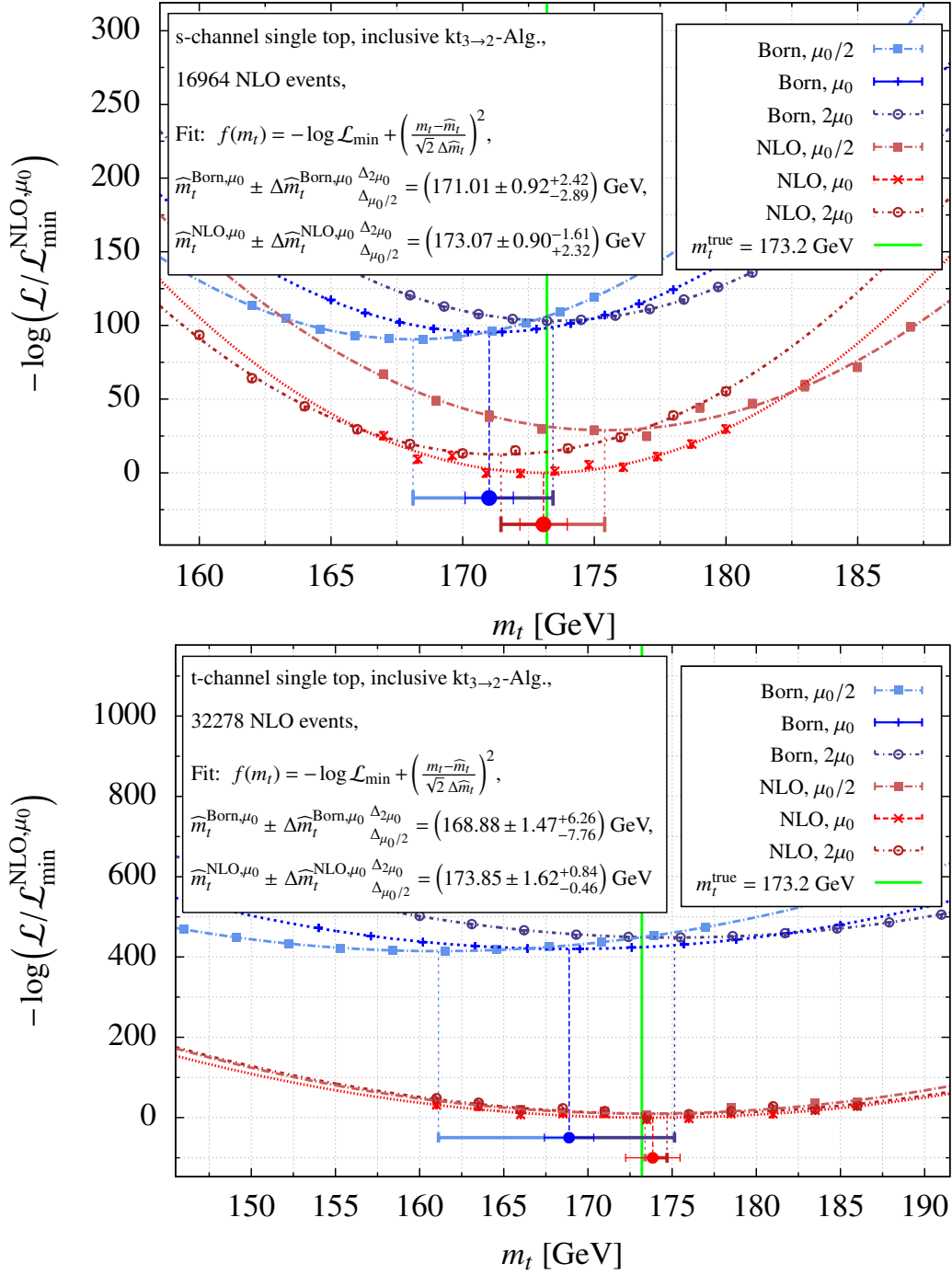

  \begin{center}
    \includegraphics[width=0.87\textwidth]{{{%
          sgt/mem/sgtsKT3-2ycut30membornnloincmu-100evts-crop}}}\\
    \includegraphics[width=0.87\textwidth]{{{%
          sgt/mem/sgttKT3-2ycut30membornnloincmu-114evts-crop}}}
     \caption{Top-quark mass extraction from inclusive NLO $s$-channel
       (top) and $t$-channel (bottom) single top-quark events employing NLO
       and Born likelihoods using a parabola fit around the minimum.}
    \label{fig:MEM_sgtin}
  \end{center}
\end{figure}
In \Fig{fig:MEM_sgtin} the extraction of the estimators for the top-quark mass by parabola fits to the log-likelihoods evaluated with the inclusive $s$- and $t$-channel events is shown. The likelihood is calculated including NLO corrections as in \Eq{eq:likeliin} (reddish lower points) as well as based on the Born approximation only (bluish upper points). While the inclusive events from section~\ref{sec:event-definition-inclusive-case} are generated for the central scale choice $\mur=\muf=\mu_0$ (see \Eq{eq:scale-definition}), the renormalisation and factorisation scales are simultaneously varied  by factors $1/2$ and $2$ in the calculation of the likelihood. The analyses corresponding to the three different scale choices are depicted as open circles ($\mur=\muf=2\mu_0$), crosses ($\mur=\muf=\mu_0$) and solid squares ($\mur=\muf=\mu_0/2$). The systematic uncertainty owed to missing higher orders on the top-quark mass extraction is estimated by the shifts in the extracted estimators due to the scale variation in the likelihoods. The minima of the log-likelihoods obtained by parabola fits are illustrated as data points with error bars indicating the statistical as well as the aforementioned systematic uncertainties in the lower part of the plots.

From \Fig{fig:MEM_sgtin}, it is verified that the NLO predictions evaluated at the central scale give the best description of the unweighted events in both inclusive event samples since the corresponding log-likelihood's minima have the lowest values.  
The results of the top-quark mass extraction from the inclusive NLO $s$- and $t$-channel events are summarised in table~\ref{tab:exclmasses} with shifts in the extracted values for the upwards (downwards) scale variation indicated by superscripts (subscripts).
\begin{table}[htbp]
\centering
\caption{Extracted top-quark mass estimators from inclusive NLO jet events.}
\label{tab:inclmasses}
\def\arraystretch{1.8}
\begin{tabular}{|c|c|c|}
\hline
{inclusive events} & {$16964$ $s$-channel events} &{$32278$ $t$-channel events} \\ \hline
{$\widehat{m}_t^{\NLO,\;\mu_0}\pm\left. \Delta\widehat{m}_t^{\NLO,\;\mu_0}\right.^{\Delta_{2\mu_0}}_{\Delta_{\mu_0/2}}$ [GeV]} & {$173.07\pm0.90^{-1.61\;(-0.9\%)}_{+2.32\;(+1.3\%)}$} &{$173.85\pm1.62^{+0.84\;(+0.5\%)}_{-0.46\;(-0.3\%)}$} \\ \hline
{$\widehat{m}_t^{\Born,\;\mu_0}\pm \left.\Delta\widehat{m}_t^{\Born,\;\mu_0}\right.^{\Delta_{2\mu_0}}_{\Delta_{\mu_0/2}}$ [GeV]} & {$171.01\pm0.92^{+2.42\;(+1.4\%)}_{-2.89\;(-1.7\%)}$} & {$168.88\pm1.47^{+6.26\;(+3.7\%)}_{-7.76\;(-4.6\%)}$}\\
\hline
\end{tabular}
\end{table}
In both the $s$- and the $t$-channel, the NLO estimators $\widehat{m}^\NLO_t$ perfectly reproduce the input value $m_t^\true=173.2\GeV$ within their statical uncertainties $\Delta\widehat{m}^\NLO_t$. Compared to the large deviations observed in the exclusive case the Born estimators $\widehat{m}^\Born_t$ extracted from the inclusive events show smaller deviations of about $-1.3\%$ in case of the $s$-channel and of $-2.4\%$ in case of the $t$-channel. Additionally, these differences are now covered by the scale uncertainties (see \Fig{fig:MEM_sgtin}). These observations match the findings regarding the impact of the NLO corrections on the normalised distributions in section~\ref{sec:kfacincl}: Born and NLO predictions show better agreement together with more overlap of their theoretical uncertainty bands due to scale variation in regions of the phase space containing the majority of events with respect to the exclusive case. Thus, smaller biases are expected when analysing the inclusive events following the NLO cross section with likelihoods calculated in the Born approximation than in the respective analysis of the exclusive events. 

\begin{figure}[htbp]
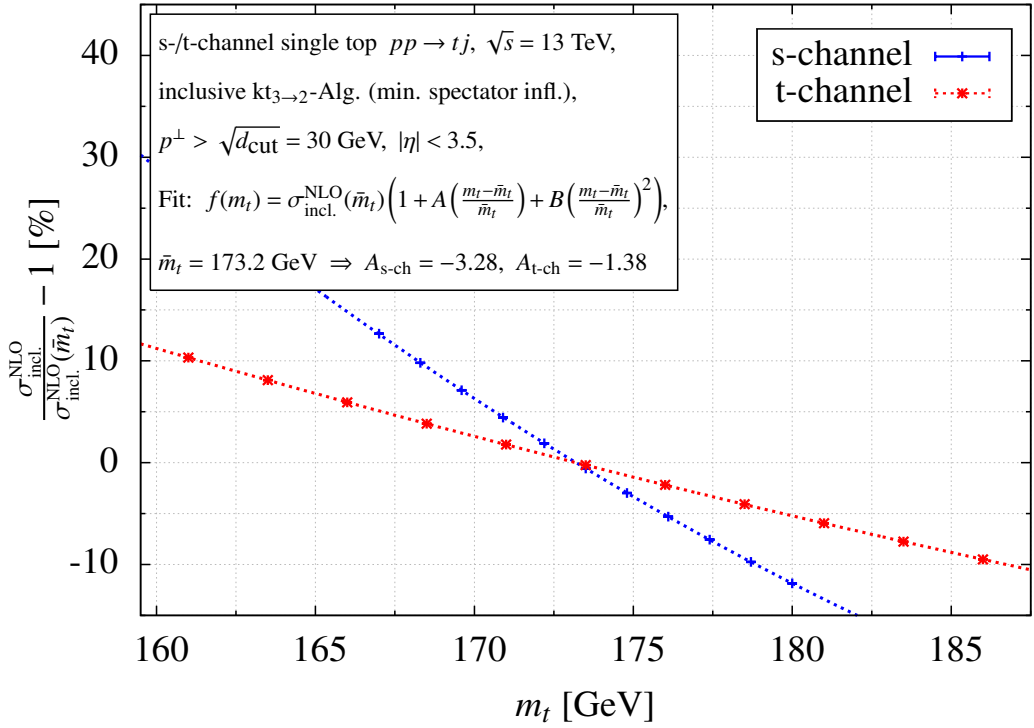

  \begin{center}
    \includegraphics[width=0.87\textwidth]{{{%
          sgt/mem/sgttKT3-2ycut30sgt-incl-crop}}}
     \caption{Top-quark mass dependence of the fiducial inclusive $s$- and $t$-channel cross section calculated at NLO accuracy.}
    \label{fig:MEM_sgtmassincl}
  \end{center}
\end{figure}

The increase in the sample sizes by also considering events with an additional resolved jet does not result in an improvement of the statistical uncertainties of the top-quark mass extractions still amounting to $\Delta\widehat{m}^\NLO_t=0.9\GeV$ in the inclusive $s$-channel analysis and $\Delta\widehat{m}^\NLO_t=1.6\GeV$ for the inclusive $t$-channel. The top-quark mass sensitivity of the fiducial inclusive cross sections is illustrated in \Fig{fig:MEM_sgtmassincl}. Indeed, the top-quark mass dependence of the fiducial cross sections is weakened by allowing the emission of an additional resolved light jet compared to the exclusive case (cf. \Fig{fig:MEM_sgtmass}). This effect compensates the gain in the sample sizes from more events passing the event selection.

While not vetoing events with an additional resolved jet does not have an impact on the statistical uncertainty of the top-quark mass extraction  it improves the behaviour of the systematic uncertainties estimated by scale variations when including NLO corrections in the likelihood calculation. The shifts in the estimators due to a simultaneous (downwards/upwards) variation of the renormalisation and factorisation scale are reduced in the $s$-channel:
\[ (\;-1.7\%\; / \;+1.4\%\;)_\Born \longrightarrow (\;+1.3\%\; / \;-0.9\%\;)_\NLO\]
and in the $t$-channel:
\[ (\;-4.6\%\; / \;+3.7\%\;)_\Born \longrightarrow  (\;-0.3\%\; /\; +0.5\%\;)_\Born\] 
by taking NLO corrections into account. This observation together with the fact that the smaller shifts in the estimators extracted from the NLO events with Born likelihoods are covered by the scale variations suggests a better convergence of the perturbative expansion for this analysis compared to the exclusive case. Regarding this, we stress again that vetoing additional resolved jets above a certain transverse-momentum threshold $p^\perp_\tmin$ introduces an additional scale in the NLO calculation. In particular, cutting the real phase space for the emission of additional radiation by the jet veto gives rise to potentially large logarithms of the additional scale $p^\perp_\tmin$ (cf. \PSS illustrated in section~\ref{sec:PSS}) . When exclusively requiring one resolved top-tagged jet together with one resolved light jet only, these logarithms in the real contribution are not cancelled and can possibly spoil the convergence of the perturbative expansion.

In comparison to the \MEM analyses of the exclusive events, the results obtained from a more inclusive event definition fit better to naive expectations of perturbation theory: Born estimators exhibit smaller biases which are covered by the scale variations underpinning the reliability of the deduced systematic uncertainties. However, the issue of ambiguous parameter interpretation due to the undefined renormalisation scheme persists when applying the \MEM in the Born approximation.

\subsubsection{Extended likelihood for the inclusive event definition}
To study the impact of the information of the total event number in the inclusive event samples on the top-quark mass determination, the extended likelihood (see \Eq{eq:extLike}) in the \MEM by weighting \Eq{eq:likeliin} with the top-quark mass dependent Poisson factor is considered:
\begin{equation}
\mathcal{L}^\NLO_{\text{ext}}\left(m_t\right)
  =  \frac{\left(\sigma^\NLO_\incl(m_t)L\right)^{N_\incl}}{N_\incl!}
 e^{-\sigma^\NLO_\incl(m_t)L}
 \mathcal{L}^\NLO
  \left(m_t\right).
\end{equation}
\begin{figure}[htbp]
  \begin{center}
    \includegraphics[width=0.87\textwidth]{{{%
          sgt/mem/sgtsKT3-2ycut30membornnloincmu-100evts_ext-crop}}}\\
    \includegraphics[width=0.87\textwidth]{{{%
          sgt/mem/sgttKT3-2ycut30membornnloincmu-114evts_ext-crop}}}
     \caption{Same as  \Fig{fig:MEM_sgtdistr} but with extended
       likelihoods.
       }
     \label{fig:MEM_sgtinc_ext}
  \end{center}
\end{figure}
In analogy to the exclusive case, the results of the top-quark mass extraction from the inclusive $s$- and $t$-channel events with extended Born and NLO likelihoods shown in \Fig{fig:MEM_sgtinc_ext} are summarised in  table~\ref{tab:inclmassesext}.
\begin{table}[htbp]
\centering
\caption{Same as in table~\ref{tab:inclmasses} but with extended likelihoods}
\label{tab:inclmassesext}
\def\arraystretch{1.8}
\begin{tabular}{|c|c|c|}
\hline
{incl. events, extended likelihoods} &{$16964$ $s$-channel events} & {$32278$ $t$-channel events} \\ \hline
{$\widehat{m}_t^{\NLO,\;\mu_0}\pm \left.\Delta\widehat{m}_t^{\NLO,\;\mu_0}\right.^{\Delta_{2\mu_0}}_{\Delta_{\mu_0/2}}$ [GeV]} & {$173.22\pm0.37^{-0.80\;(-0.5\%)}_{+1.39\;(+0.8\%)}$} & {$173.53\pm0.65^{+2.03\;(+1.2\%)}_{-0.51\;(-0.3\%)}$} \\ \hline
{$\widehat{m}_t^{\Born,\;\mu_0}\pm \left.\Delta\widehat{m}_t^{\Born,\;\mu_0}\right.^{\Delta_{2\mu_0}}_{\Delta_{\mu_0/2}}$ [GeV]} & {$160.88\pm0.35^{+1.53\;(+0.9\%)}_{-2.06\;(-1.3\%)}$} &{$172.37\pm0.58^{+8.65\;(+5.0\%)}_{-12.31\;(-7.1\%)}$}\\
\hline
\end{tabular}
\end{table}
The results confirm the consistency of the implementation of the \MEM at NLO accuracy: Extracting estimators for the top-quark mass with NLO extended likelihoods correctly reproduces the input value $m_t^\true$ used in the inclusive event generation (see section~\ref{sec:event-definition-inclusive-case}) with the lowest values of the log-likelihoods for the central scale choice $\mu_0$ in both channels. 

Taking the additional information of the inclusive event sample sizes into account results in improvements of the statistical uncertainties of the top-quark mass extractions $\Delta\widehat{m}^\NLO_t$ by almost a factor of $2.5$. Again, the impact of using the extended likelihoods on the top-quark mass sensitivities is estimated by combining hypothetical top-quark mass determinations based on the number of inclusive events alone with results from the \MEM using normal likelihoods given in table~\ref{tab:inclmasses}. According to the fit values given in \Fig{fig:MEM_sgtmassincl}, the relative uncertainties of top-quark mass determinations from idealised fiducial cross section measurements can be estimated as
\begin{equation}\label{eq:hypmassmeasincl}
\left|{\Delta\mt^{\incl\;s\text{-ch}}\over m_t}\right|=\underbrace{{\Delta\sigma^\NLO_{\incl\;s\text{-ch}}\over\sigma^\NLO_{\incl\;s\text{-ch}}}}_{={16964}^{-1/2}}\,{1\over 3.28} = 0.23\%
  ,\quad
\left|{\Delta\mt^{\incl\;t\text{-ch}}\over m_t}\right|=\underbrace{{\Delta\sigma^\NLO_{\incl\;t\text{-ch}}\over\sigma^\NLO_{\incl\;t\text{-ch}}}}_{={32278}^{-1/2}}\,{1\over 1.38} = 0.40\%.
\end{equation}
By forming the weighted averages of the hypothetical results given in \Eq{eq:hypmassmeasincl} and the results given in table~\ref{tab:inclmasses}, the relative uncertainties of the combinations amount to $\pm0.21\%$ for the $s$-channel and $\pm0.37\%$ for the $t$-channel. These estimated top-quark mass sensitivities are consistent with the results of the extended likelihood analyses in table~\ref{tab:inclmassesext}. The observed significant reductions of the statistical uncertainties from using extended likelihoods in the analyses are again supported by these simple estimates.

Concerning the additional systematic uncertainty introduced by the uncertainty from the integrated luminosity determination (see section~\ref{sec:extlikeli}), the extracted top-quark mass estimators shift about $\pm 1\GeV$ in the $s$-channel and $\pm 2\GeV$ in the $t$-channel when varying the luminosity by $\pm2\%$ in the extended likelihoods. Again, the improvement regarding statistical uncertainties acquired through the extended likelihoods is nullified by bartering for systematic uncertainties owing to the integrated luminosity. 

Estimators for the top-quark mass extracted with extended likelihoods calculated in the Born approximation ($\widehat{m}^\Born_t$) from the inclusive event samples---which follow the NLO cross section---show discrepancies with the input value $m_t^\true$ of about $-7.1\%$ in the case of $s$-channel events and about $+0.5\%$ for $t$-channel events. To explain this deviation of $-12.3\GeV$ of the $s$-channel Born estimator, recall the impact of NLO corrections on the fiducial inclusive $s$-channel cross section given in table~\ref{tab:fidxsin}: The NLO calculation predicts the number of observed $s$-channel events to be roughly four thirds of the Born prediction. To account for this increase in the fiducial inclusive cross section, the \MEM using extended likelihoods calculated in the Born approximation favours lower top-quark mass values thereby enlarging the available phase space. On the contrary, the number of observed $t$-channel events predicted by the NLO calculation is roughly the same as predicted in the Born approximation. Hence, there is not a lot of margin for the Born estimator. 

In addition, the large shift observed in the $s$-channel analysis is not covered by the estimated theoretical uncertainties inferred from the scale variation.
Using extended likelihoods in the \MEM results in reductions of the estimated theoretical uncertainties in both channels when including NLO corrections in the likelihood calculations. When simultaneously varying the renormalisation and factorisation scales downwards and upwards by a factor of $2$, the Born estimators vary between $-1.3\%$ and  $+0.9\%$ in the $s$-channel and between $-7.1\%$ and $+5.0\%$ in the $t$-channel. Conversely, the NLO estimators vary only between $-0.5\%$ and $+0.8\%$ in the $s$-channel and between $-0.3\%$ and $+1.2\%$ in the $t$-channel. These observations regarding the shifts and scale variations of the Born estimators with respect to the NLO estimators are consistent with the qualitative picture of the unnormalised k-factor distributions given in \Fig{fig:kfac_sgtt_eta2nonorm} in section~\ref{sec:kfacincl} and \Fig{fig:kfac_sgts_e2nonorm} to \Fig{fig:kfac_sgtt_eta1nonorm} in appendix~\ref{app:impNLO_sgtexcl}.

\subsection{Application III: Top-quark mass extraction from exclusive single top-quark events obtained with a $2\to1$ jet algorithm}\label{sec:mtextsgt21}
The exclusive event definition $\vec{\tilde{x}}_i=(\eta_t,\;E_j,\;\eta_j,\;\phi_j)_i$ given in section~\ref{sec:21event-definition} allows to define event weights including NLO corrections with the top-quark mass as a free parameter for jet events obtained with a $2\to1$ jet algorithm. Hence, the NLO weight from \Eq{eq:2to1evwgtexcl} can be used in the \MEM at NLO accuracy to extract an estimator for the input value for the top-quark mass used in the generation of the sample of $N_\excl=12755$ ($N_\excl=24088$) exclusive $s$-channel ($t$-channel) single top-quark events in section~\ref{sec:21event-definition}. Reproducing the input value for the top-quark mass in this way validates the possibility to formulate the \MEM at NLO accuracy also for jet events defined by a $2\to1$ jet algorithm in situations where the events definition does not fix the masses of the final-state objects. 

The NLO likelihood corresponding to the $2\to1$ clusterings for the exclusive jet events is given as a function of the top-quark mass by using the respective NLO jet event weight from \Eq{eq:2to1evwgtexcl} in \Eq{eq:likeliNLO} with $\omega=m_t$:
\begin{eqnarray}\label{eq:likeliex21}
 \nn \mathcal{L}^\NLO\left(m_t\right)
  &=\quad
  \displaystyle\prod\limits_{i=1}^{N_\excl} \mathcal{L}^\NLO
  \left(\vec{x}_i|m_t\right)&=\quad\left(\frac{1}{\sigma_\excl^\NLO(m_t)}\right)^{N_\excl}
  \prod\limits_{i=1}^{N_\excl}\frac{d\sigma_\excl^\NLO(m_t)}{d\vec{\tilde{x}}_i}\\
\nn&\;\;\;=\;\displaystyle\left(\frac{ 1}{\sigma_\excl^\NLO(m_t)}\right)^{N_\excl}
  \prod\limits_{i=1}^{N_\excl}
  &\Bigg(\frac{d\sigma^{\text{BV}}}{d\vec{\tilde{x}}}
  + \left.\frac{d\sigma^{\text{R}}}{d\vec{\tilde{x}}}\right|_{\text{light}}+\left.\frac{d\sigma^{\text{R}}}{d\vec{\tilde{x}}}\right|_{\text{top}}+\left.\frac{d\sigma^{\text{R}}}{d\vec{\tilde{x}}}\right|_{\text{beam}}\Bigg)\Bigg|_{\vec{\tilde{x}}=\vec{\tilde{x}}_i}\\
\end{eqnarray}
where the event $\vec{\tilde{x}}_i$ is taken from the exclusive event sample $\{\vec{\tilde{x}}_1, \ldots, \vec{\tilde{x}}_{N_\excl}\}$ obtained with the $2\to1$ jet algorithm. 
\begin{figure}[htbp]
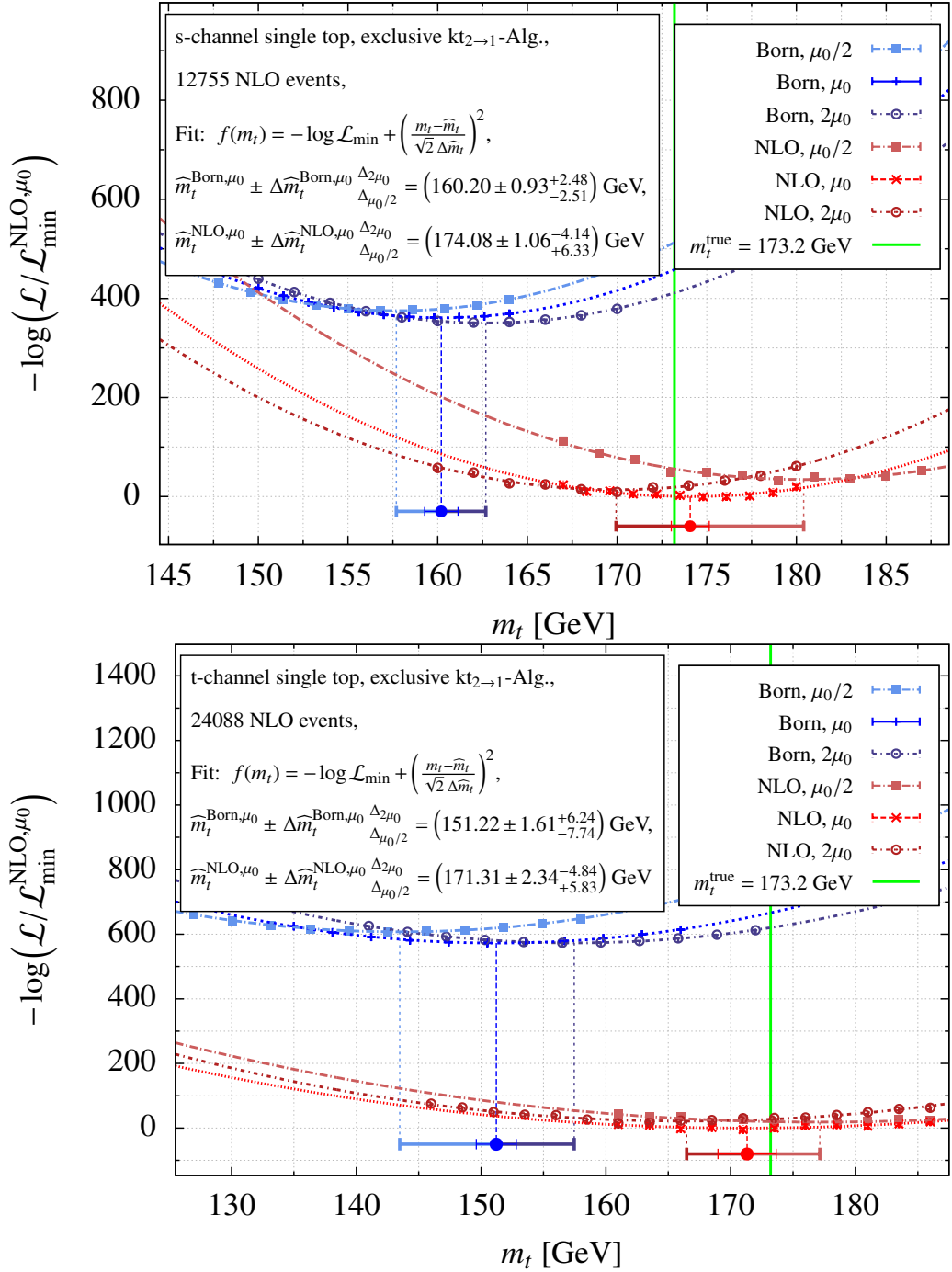

  \begin{center}
    \includegraphics[width=0.87\textwidth]{{{%
          sgt/2to1/mem/sgtsKT3-2ycut30memtwobornnlomu-90evts-crop}}}
\\
    \includegraphics[width=0.87\textwidth]{{{%
          sgt/2to1/mem/sgttKT3-2ycut30memtwobornnlomu-100evts-crop}}}
     \caption{Top-quark mass extraction  employing NLO and Born likelihoods from exclusive NLO $s$-channel (top) and $t$-channel (bottom) single top-quark events obtained with a $2\to1$ jet algorithm.}
    \label{fig:MEM_sgtex21}
  \end{center}
\end{figure}
\begin{table}[htbp]
\centering
\caption{Extracted top-quark mass estimators from exclusive NLO jet events (obtained with a $2\to1$ jet algorithm).}
\label{tab:exclmasses21}
\def\arraystretch{1.8}
\begin{tabular}{|c|c|c|}
\hline
{excl. events, $2\to1$ clustering} & \multicolumn{1}{c|}{$12755$ $s$-channel events} & \multicolumn{1}{c|}{$24088$ $t$-channel events} \\ \hline
{$\widehat{m}_t^{\NLO,\;\mu_0}\pm \left.\Delta\widehat{m}_t^{\NLO,\;\mu_0}\right.^{\Delta_{2\mu_0}}_{\Delta_{\mu_0/2}}$ [GeV]} & {$174.08\pm1.06^{-4.14\;(-2.4\%)}_{+6.33\;(+3.6\%)}$} &{$171.31\pm2.34^{-4.84\;(-2.8\%)}_{+5.83\;(+3,4\%)}$} \\ \hline
{$\widehat{m}_t^{\Born,\;\mu_0}\pm \left.\Delta\widehat{m}_t^{\Born,\;\mu_0}\right.^{\Delta_{2\mu_0}}_{\Delta_{\mu_0/2}}$ [GeV]} & {$160.20\pm0.93^{+2.48\;(+1.5\%)}_{-2.51\;(-1.6\%)}$} & {$151.22\pm1.61^{+6.24\;(+4,1\%)}_{-7.74\;(-5,1\%)}$}\\
\hline
\end{tabular}
\end{table}

The log-likelihoods for the extraction of estimators for the top-quark mass from the respective exclusive $s$- and $t$-channel events through the \MEM employing likelihoods based on NLO predictions or the Born approximation are shown in \Fig{fig:MEM_sgtex21}. The results are summarised in table~\ref{tab:exclmasses21}. The plots in \Fig{fig:MEM_sgtex21} confirm that analysing the event samples with the NLO likelihoods yields the overall lowest values for the minima of the log-likelihoods at top-quark mass values and scale choices which are consistent with the input values $m^\true_t=173.2\GeV$ and $\mur=\muf=\mu_0$. Relying only on the Born approximation in the analyses yields top-quark mass estimators $\widehat{m}_t^\Born$ which deviate from the input value by about $-7.5\%$ in the $s$-channel and by about $-12.7\%$ in the $t$-channel. These deviations of the Born estimators are not accounted for by varying the renormalisation and factorisation scales in the calculation of the likelihoods.

\begin{figure}[htbp]
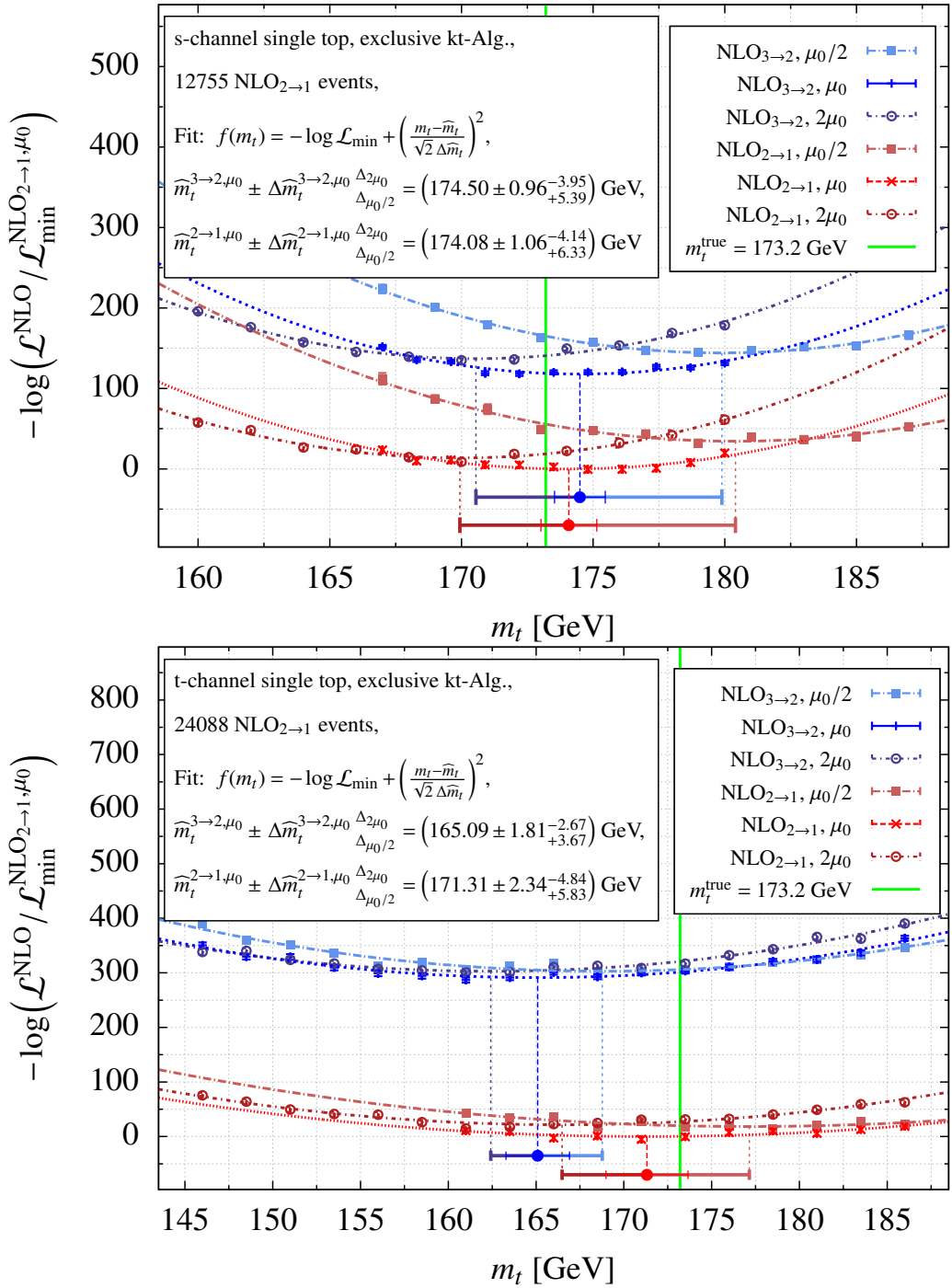

  \begin{center}
    \includegraphics[width=0.87\textwidth]{{{%
          sgt/2to1/mem/sgtsKT3-2ycut30memtwobornnlomu-90evts32lik-crop}}}
\\
    \includegraphics[width=0.87\textwidth]{{{%
          sgt/2to1/mem/sgttKT3-2ycut30memtwobornnlomu-100evts32lik-crop}}}
     \caption{Top-quark mass extraction from exclusive NLO $s$-channel
       (top) and $t$-channel (bottom) single top-quark events obtained with the $2\to1$ jet algorithm employing NLO
       likelihoods based on the $2\to1$ or $3\to2$ clustering prescriptions.}
    \label{fig:MEM_sgtex2132}
  \end{center}
\end{figure}
The results can be compared to the respective ones relying on the $3\to2$ clustering prescriptions presented in section~\ref{sec:mtextsgtexcl}. Analysing NLO $s$-channel jet events obtained with the $2\to1$ jet algorithm with the corresponding NLO likelihood based on the $2\to1$ clustering prescriptions shows a slightly larger relative statistical uncertainty of $\Delta \widehat{m}^\NLO_t/ \widehat{m}^\NLO_t= 0.6\%$ (see \Fig{fig:MEM_sgtex21}) than the respective $3\to2$ analysis with $\Delta \widehat{m}^\NLO_t/ \widehat{m}^\NLO_t= 0.5\%$ (see \Fig{fig:MEM_sgtex}). The $t$-channel analysis exhibits a lower top-quark mass sensitivity in the first place (cf. \Fig{fig:MEM_sgt}). The associated statistical uncertainty is also further enlarged to $\Delta \widehat{m}^\NLO_t/ \widehat{m}^\NLO_t= 1.4\%$ when extracting an estimator for the top-quark mass from $t$-channel events obtained with the $2\to1$ jet algorithm (see \Fig{fig:MEM_sgtex21}) compared to the respective $3\to2$ analysis with $\Delta \widehat{m}^\NLO_t/ \widehat{m}^\NLO_t= 0.9\%$ (see \Fig{fig:MEM_sgtex}). This can be understood by recalling that requiring the on-shell condition for the clustered jets introduces an additional direct kinematic dependence on the top-quark mass for the clustered final states. In contrast, the dependence on the top-quark mass is therefore weaker with the $2\to1$ clustering. 

The theoretical uncertainties inferred from scale variations in the NLO likelihood calculations are also enlarged in the $2\to1$ analyses with respect to the corresponding $3\to2$ analyses. Using likelihoods based on the Born approximation to extract estimators for the top-quark mass yields similar shifts with respect to the input value and similar theoretical as well as statistical uncertainties for samples of $s$-channel jet events either obtained with the $2\to1$ or the $3\to2$ jet algorithm. In case of the $t$-channel however, larger shift in the Born estimator and overall larger statistical uncertainties are apparent in the analyses of the events obtained with the $2\to1$ jet algorithm compared to the results of the respective $3\to2$ analyses. These observations are in agreement with the conclusions drawn from the comparison of the k-factor distributions for both clustering prescriptions in both channels (cf. section~\ref{sec:21event-definition}). 

To further accentuate the dependence of the NLO analysis on the clustering prescription, estimators for the top-quark mass from the jet events obtained with the $2\to1$ jet algorithm are also extracted with NLO likelihoods based on the $3\to2$ clustering (see \Fig{fig:MEM_sgtex2132}). As expected, the respective minima of the log-likelihoods in \Fig{fig:MEM_sgtex2132} confirm that the jet events obtained with the $2\to1$ jet algorithm are best described by the theoretical predictions based on the $2\to1$ clustering evaluated at the central scale $\mur=\muf=\mu_0$ for values of the top-quark mass which are consistent with the input value of the event generation of $m^\true_t=173.2\GeV$. Nevertheless, analysing these $s$-channel events with an NLO likelihood corresponding to the $3\to2$ clustering prescription introduces a shift in the top-quark mass estimator of only $1.3\sigma$ with respect to $m^\true_t$. This weak dependence of the $s$-channel analysis on the clustering prescription is in accordance with the impact study of the $3\to2$ jet algorithm presented in section~\ref{sec:sgtimpac}: The differences in the distributions of the variables in the exclusive $s$-channel event definition are of the order of $1\%$ in wide ranges of the phase space and in particular in regions where most of the events are expected. However, using NLO likelihoods based on the the $3\to2$ clustering prescriptions to analyse the $t$-channel events obtained with the $2\to1$ jet algorithm yields a top-quark mass estimator which deviates from the input value by $-4.5\sigma$. This result is also consistent with the impact study in section~\ref{sec:sgtimpac} which shows that the $t$-channel is more sensitive to the clustering prescription. Using different clustering prescriptions in the theoretical predictions than to obtain the jets in the experiment introduces systematic biases into the analysis whose significance depends on the process under study. However, \Fig{fig:MEM_sgtex2132} also shows that for the $s$- as well as the $t$-channel---despite the shifts in the estimators---the theoretical uncertainty bands inferred from the scale variation overlap for both NLO analyses. Thus, the difference in the clustering prescriptions employed in the analyses---which is an NLO effect---is covered by the estimated theoretical uncertainties. This observation strengthens the confidence in the reliability of the estimated theoretical uncertainties from the NLO analyses.

\chapter{Conclusion}\label{sec:concl}
In this work an extension of the \MEM to consistently incorporate full NLO QCD corrections into the theoretical predictions entering the likelihood calculation has been presented.  As a proof of concept, the first application of the \MEM at NLO to the hadronic production of jets using the example of single top-quark production at the LHC has been demonstrated.

It has been shown that modifying the recombination prescription of conventional $2\to 1$ jet algorithms by introducing a $3\to 2$ clustering yields an algorithm allowing to factorise the real phase space in terms of a phase space corresponding to the resolved jets and a phase space of the unresolved partonic configurations contributing to the resolved jets. These unresolved phase space regions can be integrated numerically in an efficient way resulting in a real contribution to the cross section which is differential in the momenta of the resolved jets. $3\to 2$ clusterings inspired by the Catani--Seymour dipole formalism are employed which keep the clustered jet momenta on shell while overall
$4$-momentum conservation is maintained. That way, the virtual matrix element can be evaluated for the momenta of the resolved jets which ensures a point-wise cancellation of the (appropriately regularised) \IR divergences in the sum of the real and virtual contributions.
The definition of a fully differential event weight accurate to NLO QCD for jets instead of partons is thereby facilitated.  

The first important application of this NLO event weight is the possibility to generate unweighted jet events which are distributed according to the NLO cross section. In turn, NLO weights can be attributed to jet events at hand (e.g. which are observed in experiments). By using these weights to construct a cumulative likelihood for the event sample, the powerful \MEM is elevated to a sound theoretical footing at NLO accuracy.

The phase space factorisations corresponding to the $3\to2$ clusterings have been separately validated for initial-state radiation (exemplified by Drell-Yan production) and radiation from a massive final state (exemplified by leptonic top-quark pair production). Subsequently, all aspects of the presented phase space factorisations have been tested altogether for the hadronic production of single top quarks. 

As proof of concept, the NLO weights have been used to generate unweighted top-quark pair events produced at a lepton collider as well as $s$- and $t$-channel single top-quark events produced at the LHC. By defining events in terms of sets of jet variables instead of the jets' $4$-momenta, it is possible to specify not only exclusive but also inclusive event definitions which allow the presence of additional jet activity in the final state. For specific event variables it is even possible to define event weights at NLO accuracy for jet events obtained with a jet algorithm employing common $2\to1$ clustering prescriptions. The distribution of the unweighted events according to the NLO cross section has been checked by reproducing respective differential distributions obtained using a conventional parton level Monte Carlo integration for all cases.  

To explore the potential of the \MEM, the generated unweighted events are treated as the outcome of toy experiments to be analysed with the \MEM at NLO accuracy. The successful reproduction of the input value for the top-quark mass used in the respective event generation serves both as a further consistency check and a first example of the application of the \MEM at NLO accuracy. The impact of the NLO corrections on the analysis has been studied by repeating the particular top-quark mass determinations with likelihoods based on the Born approximation only. Using Born likelihoods to analyse events following the NLO cross section can introduce significant biases in the extracted estimators depending on the process and event definition. The theoretical uncertainties of the parameter extraction have been estimated by repeating the analyses with varied renormalisation and factorisation scales in the calculation of the likelihood. While theoretical uncertainty estimates are reduced when incorporating NLO corrections in the likelihood in most instances, the shifts in the Born estimators are in general not accounted for by the associated theoretical uncertainties. Even though it might be possible to remove the bias in the extracted Born estimator through a calibration procedure of the \MEM, assigning theoretical uncertainties to the Born results inferred from scale variation would most likely be unreliable. The inclusion of NLO corrections in the \MEM thus not only reduces the required amount of calibration (and the associated uncertainty) but is also needed to provide a more reliable estimate of the theoretical uncertainty of the analysis. Regarding the interpretation of the extracted parameters, the NLO calculation in the likelihood is needed to uniquely define the renormalisation scheme of the extracted model parameters. Jet events obtained with a $2\to1$ jet algorithm have also been analysed with likelihoods based on the NLO predictions corresponding to $3\to2$ clusterings to investigate the dependence of the analysis on the employed clustering prescriptions for exclusive $s$- and $t$-channel production.

This work also presents top-quark mass determinations from single top-quark events produced at the LHC as a promising, complementary alternative to prevailing top-quark mass measurements from top-quark pair events. Additionally, the presented study of the leptonic $t\bar{t}$-production illustrates an interesting possibility to increase the precision of top-quark mass measurements at energies beyond the production threshold.

As an outlook, it should be emphasised that the work presented here and in \Refs{Martini:2015fsa,Martini:2017ydu} should be seen as a starting
point towards the actual experimental application of the \MEM at NLO accuracy. The presented algorithm based on the $3\to2$ clusterings is process independent and can be automated (a library providing implementations of the $3\to2$ clusterings and the corresponding phase space factorisations is under development). Additionally, modifications of the presented parameterisations of the real phase space in favour of invariants tailored to the application of the \PSS method have been introduced. In general, numerical NLO calculations relying on the \PSS method may benefit from these phase space factorisations.
Further studies are required to describe more realistic final states like the ones for example presented in \Ref{Frederix:2016rdc}: First of all, the inclusion of the top-quark decay should be straightforward and poses no difficulties within the introduced method. However, more work is needed to combine the fixed-order calculation in the likelihoods with parton showers to allow further modelling of the QCD radiation cascades and hadronisation. In order to realistically match the experimental  conditions, the effects due to non-trivial transfer functions should be investigated as well.

\appendix
\chapter{Supplementary figures}
\section{Validation of the phase space parameterisation}
\subsection{Drell-Yan production}
\label{app:phspval_DY}
\begin{figure}[htbp]
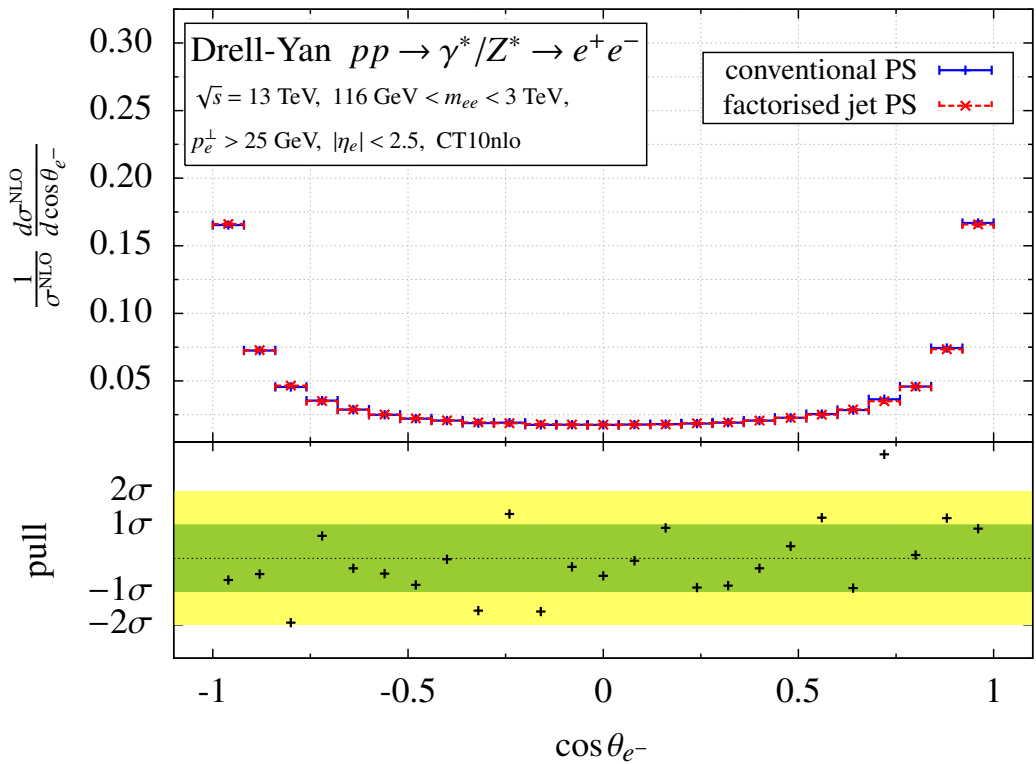

  \begin{center}
    \leavevmode
    \makebox[\textwidth]{\includegraphics[width=\ftw\textwidth]{{{%
          dytt/qqee3-2ptmn25etamx2.5compCTH2-50x1e7-crop}}}}
    \caption{Polar angle distribution of the electron from Drell-Yan production
      calculated at NLO accuracy using a conventional parton level MC (solid blue) compared to 
      results from the factorised jet phase space as described in
      section~\ref{sec:ii} (red dashed).}
    \label{fig:partonMCvsjetMC_DrellYan_cth2}
  \end{center}
\end{figure}
\begin{figure}[htbp]
  \begin{center}
    \leavevmode
    \makebox[\textwidth]{\includegraphics[width=\ftw\textwidth]{{{%
          dytt/qqee3-2ptmn25etamx2.5compMEE-50x1e7-crop}}}}
    \caption{Same as \Fig{fig:partonMCvsjetMC_DrellYan_cth2} but for the invariant mass of the lepton pair.}
    \label{fig:partonMCvsjetMC_DrellYan_mee}
  \end{center}
\end{figure}
\begin{figure}[htbp]
  \begin{center}
    \leavevmode
    \makebox[\textwidth]{\includegraphics[width=\ftw\textwidth]{{{%
          dytt/qqee3-2ptmn25etamx2.5compYEE-50x1e7-crop}}}}
    \caption{Same as \Fig{fig:partonMCvsjetMC_DrellYan_cth2} but for the rapidity of the lepton pair.}
    \label{fig:partonMCvsjetMC_DrellYan_yee}
  \end{center}
\end{figure}
\clearpage
\subsection{Leptonic top-quark pair production}
\label{app:phspval_tt}
\begin{figure}[htbp]
  \begin{center}
    \leavevmode
    \makebox[\textwidth]{\includegraphics[width=\ftw\textwidth]{{{%
          dytt/eeQQKT3-2ycut0.1compPH1-50x1e7-crop}}}}
    \caption{Azimuthal angle distribution of the top-tagged jet from top-quark pair production 
      in $e^+e^-$ annihilation calculated at NLO accuracy using a conventional parton
      level MC (solid blue) compared to results from the factorised jet
      phase space as described in section~\ref{sec:ffm} (red dashed).}
    \label{fig:partonMCvsjetMC_ttprod_ph1}
  \end{center}
\end{figure}
\begin{figure}[htbp]
  \begin{center}
    \leavevmode
    \makebox[\textwidth]{\includegraphics[width=\ftw\textwidth]{{{%
          dytt/eeQQKT3-2ycut0.1compPTR1-50x1e7-crop}}}}
    \caption{Same as \Fig{fig:partonMCvsjetMC_ttprod_ph1} but for the transverse momentum of the top-tagged jet.}
    \label{fig:partonMCvsjetMC_ttprod_ptr1}
  \end{center}
\end{figure}
\begin{figure}[htbp]
  \begin{center}
    \leavevmode
    \makebox[\textwidth]{\includegraphics[width=\ftw\textwidth]{{{%
          dytt/eeQQKT3-2ycut0.1compY1-50x1e7-crop}}}}
    \caption{Same as \Fig{fig:partonMCvsjetMC_ttprod_ph1} but for the rapidity of the top-tagged jet.}
    \label{fig:partonMCvsjetMC_ttprod_y1}
  \end{center}
\end{figure}
\clearpage

\subsection{Hadronic single top-quark production}
\label{app:phspval_sgt}
\begin{figure}[htbp]
  \begin{center}
    \leavevmode
    \makebox[\textwidth]{\includegraphics[width=\ftw\textwidth]{{{%
          sgt/phsp/sgtsKT3-2ycut30compmnspecETA1-50x1e7-crop}}}}
   \caption{Pseudo rapidity distribution of the top-tagged jet from exclusive $t$-channel single top-quark production
      calculated at NLO accuracy using a conventional parton level MC (solid blue) and the factorised jet phase space (dashed red) as described in section~\ref{sec:PhaseSpaceparameterisation}.}
    \label{fig:partonMCvsjetMC_sgts_eta1}
  \end{center}
\end{figure}

\begin{figure}[htbp]
  \begin{center}
    \leavevmode
    \makebox[\textwidth]{\includegraphics[width=\ftw\textwidth]{{{%
          sgt/phsp/sgtsKT3-2ycut30compmnspecE2-50x1e7-crop}}}}
    \caption{Same as \Fig{fig:partonMCvsjetMC_sgts_eta1} but for the energy of the light jet from the $s$-channel.}
    \label{fig:partonMCvsjetMC_sgts_e2}
  \end{center}
\end{figure}
\begin{figure}[htbp]
  \begin{center}
    \leavevmode
    \makebox[\textwidth]{\includegraphics[width=\ftw\textwidth]{{{%
          sgt/phsp/sgtsKT3-2ycut30compmnspecETA2-50x1e7-crop}}}}
    \caption{Same as \Fig{fig:partonMCvsjetMC_sgts_eta1} but for the pseudo rapidity of the light jet from the $s$-channel.}
    \label{fig:partonMCvsjetMC_sgts_eta2}
  \end{center}
\end{figure}
\begin{figure}[htbp]
  \begin{center}
    \leavevmode
    \makebox[\textwidth]{\includegraphics[width=\ftw\textwidth]{{{%
          sgt/phsp/sgtsKT3-2ycut30compmnspecMEE-50x1e7-crop}}}}
     \caption{Same as \Fig{fig:partonMCvsjetMC_sgts_eta1} but for the invariant mass of the top-light jet system from the $s$-channel.}
    \label{fig:partonMCvsjetMC_sgts_mee}
  \end{center}
\end{figure}

\begin{figure}[htbp]
  \begin{center}
    \leavevmode
    \makebox[\textwidth]{\includegraphics[width=\ftw\textwidth]{{{%
          sgt/phsp/sgttKT3-2ycut30compmnspecE2-50x1e7-crop}}}}
    \caption{Same as \Fig{fig:partonMCvsjetMC_sgts_eta1} but for the energy of the light jet from the $t$-channel.}
    \label{fig:partonMCvsjetMC_sgtt_e2}
  \end{center}
\end{figure}
\begin{figure}[htbp]
  \begin{center}
    \leavevmode
    \makebox[\textwidth]{\includegraphics[width=\ftw\textwidth]{{{%
          sgt/phsp/sgttKT3-2ycut30compmnspecET2-50x1e7-crop}}}}
    \caption{Same as \Fig{fig:partonMCvsjetMC_sgts_eta1} but for the transverse energy of the light jet from the $t$-channel.}
    \label{fig:partonMCvsjetMC_sgtt_e2}
  \end{center}
\end{figure}
\begin{figure}[htbp]
  \begin{center}
    \leavevmode
    \makebox[\textwidth]{\includegraphics[width=\ftw\textwidth]{{{%
          sgt/phsp/sgttKT3-2ycut30compmnspecETA1-50x1e7-crop}}}}
    \caption{Same as \Fig{fig:partonMCvsjetMC_sgts_eta1} but for the pseudo rapidity of the top-tagged jet from the $t$-channel.}
    \label{fig:partonMCvsjetMC_sgtt_e2}
  \end{center}
\end{figure}
\begin{figure}[htbp]
  \begin{center}
    \leavevmode
    \makebox[\textwidth]{\includegraphics[width=\ftw\textwidth]{{{%
          sgt/phsp/sgttKT3-2ycut30compmnspecYEE-50x1e7-crop}}}}
     \caption{Same as \Fig{fig:partonMCvsjetMC_sgts_eta1} but for the rapidity of the top-light jet system from the $t$-channel.}
    \label{fig:partonMCvsjetMC_sgtt_yee}
  \end{center}
\end{figure}
\clearpage

\subsection{Slicing parameter (in)dependence}
\label{app:smindep_sgt}
\begin{figure}[htbp]
  \begin{center}
    \leavevmode
     \makebox[\textwidth]{\includegraphics[width=\ftw\textwidth]{{{%
          sgt/smin/sgtsKT3-2ycut30compmnspecET1-50x1e7sminc-crop}}}}
    \caption{Transverse energy distribution of the top-tagged jet from exclusive $s$-channel single top-quark production
      calculated at NLO accuracy with three different values of the slicing parameter $s_\tmin$.}
    \label{fig:sminc_sgts_et1}
  \end{center}
\end{figure}
\begin{figure}[htbp]
  \begin{center}
    \leavevmode
     \makebox[\textwidth]{\includegraphics[width=\ftw\textwidth]{{{%
          sgt/smin/sgtsKT3-2ycut30compmnspecETA1-50x1e7sminc-crop}}}}
    \caption{Same as \Fig{fig:sminc_sgts_et1} but for the pseudo rapidity of the top-tagged jet from the $s$-channel.}
    \label{fig:sminc_sgts_eta1}
  \end{center}
\end{figure}
\begin{figure}[htbp]
  \begin{center}
    \leavevmode
     \makebox[\textwidth]{\includegraphics[width=\ftw\textwidth]{{{%
          sgt/smin/sgtsKT3-2ycut30compmnspecETA2-50x1e7sminc-crop}}}}
    \caption{Same as \Fig{fig:sminc_sgts_et1} but for the pseudo rapidity of the light jet from the $s$-channel.}
    \label{fig:sminc_sgts_eta2}
  \end{center}
\end{figure}
\begin{figure}[htbp]
  \begin{center}
    \leavevmode
     \makebox[\textwidth]{\includegraphics[width=\ftw\textwidth]{{{%
          sgt/smin/sgtsKT3-2ycut30compmnspecYEE-50x1e7sminc-crop}}}}
    \caption{Same as \Fig{fig:sminc_sgts_et1} but for the rapidity of the top-light jet system from the $s$-channel.}
    \label{fig:sminc_sgts_yee}
  \end{center}
\end{figure}
\begin{figure}[htbp]
  \begin{center}
    \leavevmode
     \makebox[\textwidth]{\includegraphics[width=\ftw\textwidth]{{{%
          sgt/smin/sgttKT3-2ycut30compmnspecE2-50x1e7sminc-crop}}}}
    \caption{Same as \Fig{fig:sminc_sgts_et1} but for the energy of the light jet from the $t$-channel.}
    \label{fig:sminc_sgtt_e2}
  \end{center}
\end{figure}
\begin{figure}[htbp]
  \begin{center}
    \leavevmode
     \makebox[\textwidth]{\includegraphics[width=\ftw\textwidth]{{{%
          sgt/smin/sgttKT3-2ycut30compmnspecETA2-50x1e7sminc-crop}}}}
    \caption{Same as \Fig{fig:sminc_sgts_et1} but for the pseudo rapidity of the light jet from the $t$-channel.}
    \label{fig:sminc_sgtt_eta2}
  \end{center}
\end{figure}
\begin{figure}[htbp]
  \begin{center}
    \leavevmode
     \makebox[\textwidth]{\includegraphics[width=\ftw\textwidth]{{{%
          sgt/smin/sgttKT3-2ycut30compmnspecE1-50x1e7sminc-crop}}}}
    \caption{Same as \Fig{fig:sminc_sgts_et1} but for the energy of the top-tagged jet from the $t$-channel.}
    \label{fig:sminc_sgtt_e1}
  \end{center}
\end{figure}
\begin{figure}[htbp]
  \begin{center}
    \leavevmode
     \makebox[\textwidth]{\includegraphics[width=\ftw\textwidth]{{{%
          sgt/smin/sgttKT3-2ycut30compmnspecMEE-50x1e7sminc-crop}}}}
    \caption{Same as \Fig{fig:sminc_sgts_et1} but for the invariant mass of the top-light jet system from the $t$-channel.}
    \label{fig:sminc_sgtt_mee}
  \end{center}
\end{figure}
\clearpage

\section{Distribution of the unweighted events}
\subsection{Top-quark pair events from $e^+e^-$ annihilation}
\label{app:unwgtev_tt}
\begin{figure}[htbp]
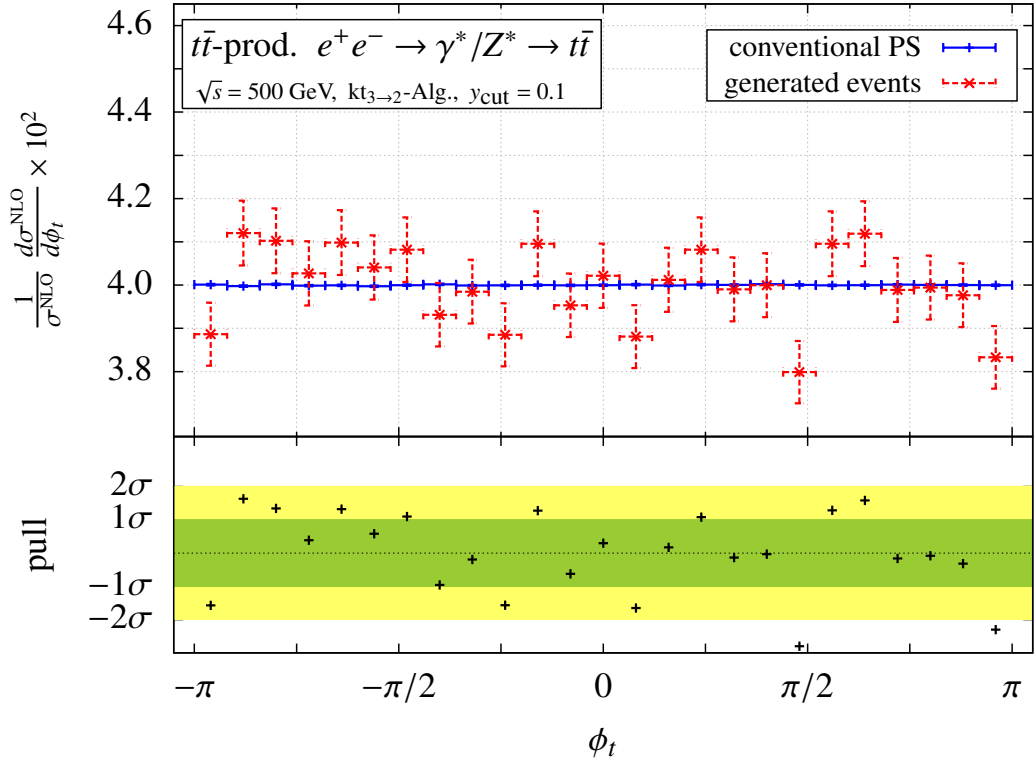

  \begin{center}
    \leavevmode
    \makebox[\textwidth]{\includegraphics[width=\ftw\textwidth]{{{%
          dytt/eeQQKT3-2ycut0.1compPH1-150evts-crop}}}}
    \caption{Validation of the generation of unweighted NLO 
      top-quark pair events by comparing the azimuthal angle distribution of the top-tagged jet calculated with a conventional parton level MC (solid blue) to histograms filled with the unweighted events (dashed red).}
    \label{fig:partonMCvsunwgtEv_ttprod_ph1}
  \end{center}
\end{figure}
\begin{figure}[htbp]
  \begin{center}
    \leavevmode
    \makebox[\textwidth]{\includegraphics[width=\ftw\textwidth]{{{%
          dytt/eeQQKT3-2ycut0.1compPTR1-150evts-crop}}}}
    \caption{Same as \Fig{fig:partonMCvsunwgtEv_ttprod_ph1} but for the transverse momentum of the top-tagged jet.}
    \label{fig:partonMCvsunwgtEv_ttprod_ptr1}
  \end{center}
\end{figure}
\begin{figure}[htbp]
  \begin{center}
    \leavevmode
    \makebox[\textwidth]{\includegraphics[width=\ftw\textwidth]{{{%
          dytt/eeQQKT3-2ycut0.1compY1-150evts-crop}}}}
    \caption{Same as \Fig{fig:partonMCvsunwgtEv_ttprod_ph1} but for the rapidity of the top-tagged jet.}
    \label{fig:partonMCvsunwgtEv_ttprod_y1}
  \end{center}
\end{figure}
\clearpage

\subsection{Single top-quark events at the LHC (exclusive event definition)}
\label{app:unwgtev_sgt}
\begin{figure}[htbp]
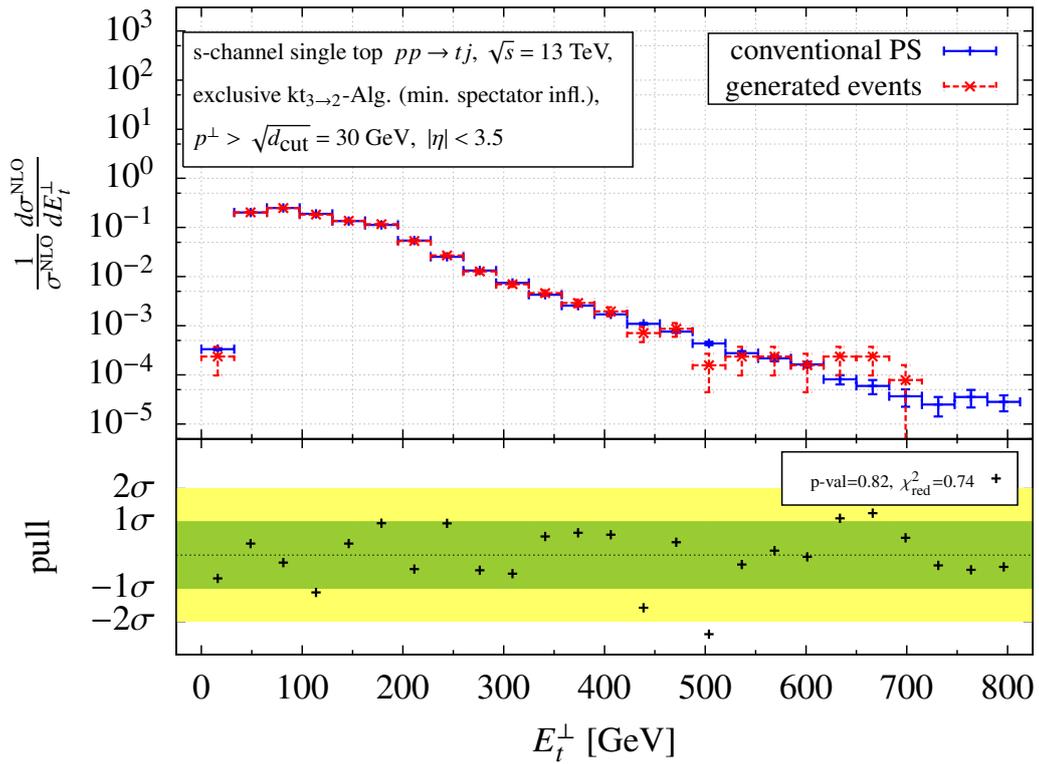

  \begin{center}
    \leavevmode
    \makebox[\textwidth]{\includegraphics[width=\ftw\textwidth]{{{%
          sgt/evgen/sgtsKT3-2ycut30compmnspecET1-94evts-crop}}}}
    \caption{Transverse energy distribution of the top-tagged jet from exclusive $s$-channel single top-quark production
      calculated at NLO accuracy using a conventional parton level MC (solid blue) compared to histograms filled with generated NLO events (dashed red).}
       \label{fig:partonMCvsgenev_sgts_et1}
  \end{center}
\end{figure}
\begin{figure}[htbp]
  \begin{center}
    \leavevmode
    \makebox[\textwidth]{\includegraphics[width=\ftw\textwidth]{{{%
          sgt/evgen/sgtsKT3-2ycut30compmnspecE2-94evts-crop}}}}
    \caption{Same as \Fig{fig:partonMCvsgenev_sgts_et1} but for the energy of the light jet from the $s$-channel.}
       \label{fig:partonMCvsgenev_sgts_e2}
  \end{center}
\end{figure}
\begin{figure}[htbp]
  \begin{center}
    \leavevmode
    \makebox[\textwidth]{\includegraphics[width=\ftw\textwidth]{{{%
          sgt/evgen/sgtsKT3-2ycut30compmnspecETA2-94evts-crop}}}}
    \caption{Same as \Fig{fig:partonMCvsgenev_sgts_et1} but for the pseudo rapidity of the light jet from the $s$-channel.}
       \label{fig:partonMCvsgenev_sgts_eta2}
  \end{center}
\end{figure}
\begin{figure}[htbp]
  \begin{center}
    \leavevmode
    \makebox[\textwidth]{\includegraphics[width=\ftw\textwidth]{{{%
          sgt/evgen/sgtsKT3-2ycut30compmnspecYEE-94evts-crop}}}}
    \caption{Same as \Fig{fig:partonMCvsgenev_sgts_et1} but for the rapidity of the top-light jet system from the $s$-channel.}
       \label{fig:partonMCvsgenev_sgts_yee}
  \end{center}
\end{figure}
\begin{figure}[htbp]
  \begin{center}
    \leavevmode
    \makebox[\textwidth]{\includegraphics[width=\ftw\textwidth]{{{%
          sgt/evgen/sgttKT3-2ycut30compmnspecE1-90evts-crop}}}}
    \caption{Same as \Fig{fig:partonMCvsgenev_sgts_et1} but for the energy of the top-tagged jet from the $t$-channel.}
       \label{fig:partonMCvsgenev_sgtt_et1}
  \end{center}
\end{figure}
\begin{figure}[htbp]
  \begin{center}
    \leavevmode
    \makebox[\textwidth]{\includegraphics[width=\ftw\textwidth]{{{%
          sgt/evgen/sgttKT3-2ycut30compmnspecETA1-90evts-crop}}}}
    \caption{Same as \Fig{fig:partonMCvsgenev_sgts_et1} but for the pseudo rapidity of the top-tagged jet from the $t$-channel.}
       \label{fig:partonMCvsgenev_sgtt_eta1}
  \end{center}
\end{figure}
\begin{figure}[htbp]
  \begin{center}
    \leavevmode
    \makebox[\textwidth]{\includegraphics[width=\ftw\textwidth]{{{%
          sgt/evgen/sgttKT3-2ycut30compmnspecETA2-90evts-crop}}}}
    \caption{Same as \Fig{fig:partonMCvsgenev_sgts_et1} but for the pseudo rapidity of the light jet from the $t$-channel.}
       \label{fig:partonMCvsgenev_sgtt_eta2}
  \end{center}
\end{figure}
\begin{figure}[htbp]
  \begin{center}
    \leavevmode
    \makebox[\textwidth]{\includegraphics[width=\ftw\textwidth]{{{%
          sgt/evgen/sgttKT3-2ycut30compmnspecMEE-90evts-crop}}}}
    \caption{Same as \Fig{fig:partonMCvsgenev_sgts_et1} but for the invariant mass of the top-light jet system from the $t$-channel.}
       \label{fig:partonMCvsgenev_sgtt_mee}
  \end{center}
\end{figure}
\clearpage

\subsection{Single top-quark events at the LHC (inclusive event definition)}
\label{app:unwgtev_sgtinc}
\begin{figure}[htbp]
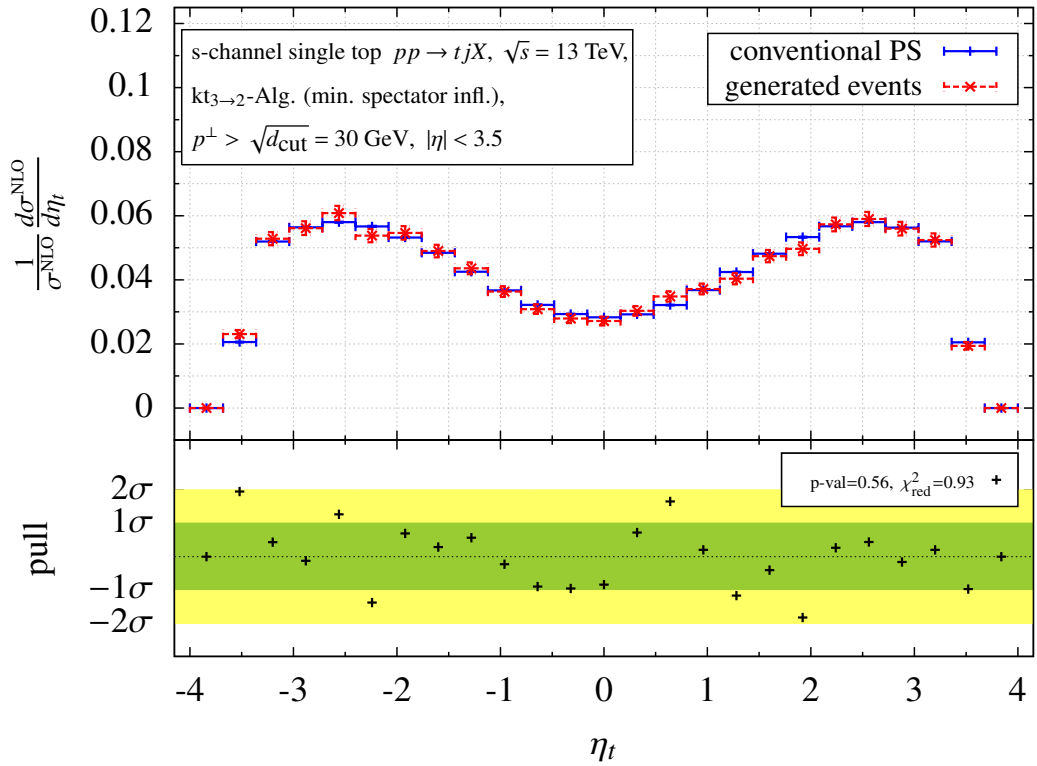

  \begin{center}
    \leavevmode
     \makebox[\textwidth]{\includegraphics[width=\ftw\textwidth]{{{%
          sgt/evgen/sgtsKT3-2ycut30compmnspecincETA1-100evts-crop}}}}
    \caption{Pseudo rapidity distribution of the top-tagged jet from inclusive $s$-channel single top-quark production calculated at NLO accuracy using a conventional parton level MC (solid blue) compared to histograms filled with generated NLO events (dashed red).}
    \label{fig:partonMCvsgenev_sgtsinc_eta1}
  \end{center}
\end{figure}
\begin{figure}[htbp]
  \begin{center}
    \leavevmode
     \makebox[\textwidth]{\includegraphics[width=\ftw\textwidth]{{{%
          sgt/evgen/sgtsKT3-2ycut30compmnspecincETA2-100evts-crop}}}}
    \caption{Same as \Fig{fig:partonMCvsgenev_sgtsinc_eta1} but for the pseudo rapidity of the light jet from the $s$-channel.}
    \label{fig:partonMCvsgenev_sgtsinc_eta2}
  \end{center}
\end{figure}
\begin{figure}[htbp]
  \begin{center}
    \leavevmode
     \makebox[\textwidth]{\includegraphics[width=\ftw\textwidth]{{{%
          sgt/evgen/sgttKT3-2ycut30compmnspecincE2-114evts-crop}}}}
    \caption{Same as \Fig{fig:partonMCvsgenev_sgtsinc_eta1} but for the energy of the light jet from the $t$-channel.}
    \label{fig:partonMCvsgenev_sgttinc_e2}
  \end{center}
\end{figure}
\begin{figure}[htbp]
  \begin{center}
    \leavevmode
     \makebox[\textwidth]{\includegraphics[width=\ftw\textwidth]{{{%
          sgt/evgen/sgttKT3-2ycut30compmnspecincETA2-114evts-crop}}}}
    \caption{Same as \Fig{fig:partonMCvsgenev_sgtsinc_eta1} but for the pseudo rapidity of the light jet from the $t$-channel.}
    \label{fig:partonMCvsgenev_sgttinc_eta2}
  \end{center}
\end{figure}
\clearpage

\subsection{Single top-quark events at the LHC ($2\to1$ clustering prescription)}
\label{app:unwgtev_sgt21}
\begin{figure}[htbp]
  \begin{center}
    \leavevmode
     \makebox[\textwidth]{\includegraphics[width=\ftw\textwidth]{{{%
          sgt/2to1/evgen/sgtsKT3-2ycut30comptwoE2-90evts-crop}}}}
    \caption{Energy distribution of the light jet from exclusive $s$-channel single top-quark production calculated at NLO accuracy using a conventional parton level MC (solid blue) with $2\to1$ jet clustering compared to histograms filled with generated NLO events (dashed red).}
    \label{fig:partonMCvsgenev_sgts21_e2}
  \end{center}
\end{figure}
\begin{figure}[htbp]
  \begin{center}
    \leavevmode
     \makebox[\textwidth]{\includegraphics[width=\ftw\textwidth]{{{%
          sgt/2to1/evgen/sgtsKT3-2ycut30comptwoETA2-90evts-crop}}}}
    \caption{Same as \Fig{fig:partonMCvsgenev_sgts21_e2} but for the pseudo rapidity distribution of the light jet from the $s$-channel.}
    \label{fig:partonMCvsgenev_sgts21_eta2}
  \end{center}
\end{figure}
\begin{figure}[htbp]
  \begin{center}
    \leavevmode
     \makebox[\textwidth]{\includegraphics[width=\ftw\textwidth]{{{%
          sgt/2to1/evgen/sgttKT3-2ycut30comptwoETA1-100evts-crop}}}}
    \caption{Same as \Fig{fig:partonMCvsgenev_sgts21_e2} but for the pseudo rapidity distribution of the top-tagged jet from the $t$-channel.}
    \label{fig:partonMCvsgenev_sgtt21_eta1}
  \end{center}
\end{figure}
\begin{figure}[htbp]
  \begin{center}
    \leavevmode
     \makebox[\textwidth]{
    \includegraphics[width=\ftw\textwidth]{{{%
          sgt/2to1/evgen/sgttKT3-2ycut30comptwoETA2-100evts-crop}}}}
    \caption{Same as \Fig{fig:partonMCvsgenev_sgts21_e2} but for the pseudo rapidity distribution of the light jet from the $t$-channel.}
    \label{fig:partonMCvsgenev_sgtt21_eta2}
  \end{center}
\end{figure}
\clearpage

\section{Impact of the $3\to2$ clustering on single top-quark distributions}
\label{app:impact3to2_sgt}
\begin{figure}[htbp]
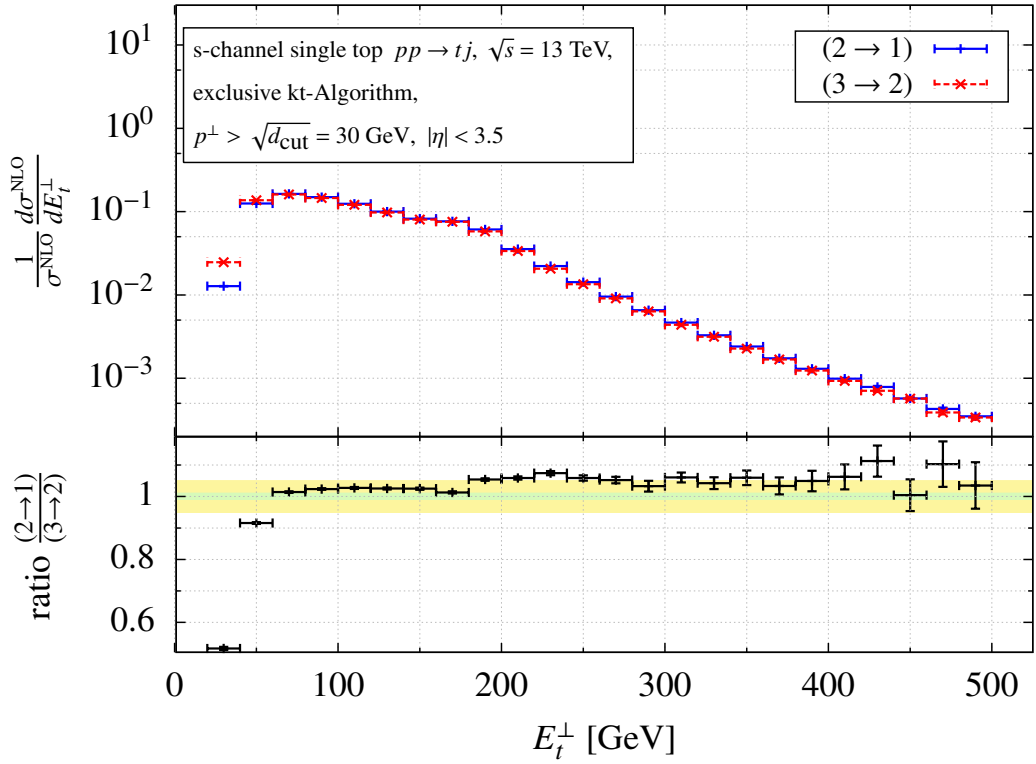

  \begin{center}
    \leavevmode
    \makebox[\textwidth]{\includegraphics[width=\ftw\textwidth]{{{%
          sgt/jetcomp/sgtsKT3-2ycut30compjetET1-200x1e7-crop}}}}
    \caption{Transverse energy distribution of the top-tagged jet from exclusive $s$-channel single top-quark production calculated at NLO accuracy with $3\rightarrow 2$ and $2\rightarrow 1$ jet clusterings.}
    \label{fig:jetcom_sgts_et1}
  \end{center}
\end{figure}
\begin{figure}[htbp]
  \begin{center}
    \leavevmode
    \makebox[\textwidth]{\includegraphics[width=\ftw\textwidth]{{{%
          sgt/jetcomp/sgtsKT3-2ycut30compjetETA1-200x1e7-crop}}}}
    \caption{Same as \Fig{fig:jetcom_sgts_et1} but for the pseudo rapidity of the top-tagged jet from the $s$-channel.}
    \label{fig:jetcom_sgts_eta1}
  \end{center}
\end{figure}
\begin{figure}[htbp]
  \begin{center}
    \leavevmode
    \makebox[\textwidth]{\includegraphics[width=\ftw\textwidth]{{{%
          sgt/jetcomp/sgttKT3-2ycut30compjetE2-200x1e7-crop}}}}
    \caption{Same as \Fig{fig:jetcom_sgts_et1} but for the energy of the light jet from the $t$-channel.}
    \label{fig:jetcom_sgtt_e2}
  \end{center}
\end{figure}
\begin{figure}[htbp]
  \begin{center}
    \leavevmode
    \makebox[\textwidth]{\includegraphics[width=\ftw\textwidth]{{{%
          sgt/jetcomp/sgttKT3-2ycut30compjetETA2-200x1e7-crop}}}}
    \caption{Same as \Fig{fig:jetcom_sgts_et1} but for the pseudo rapidity of the light jet from the $t$-channel.}
    \label{fig:jetcom_sgtt_eta2}
  \end{center}
\end{figure}
\clearpage

\section{Impact of NLO corrections to unnormalised single top-quark distributions}
\subsection{Exclusive event definition}
\label{app:impNLO_sgtexcl}
\begin{figure}[htbp]
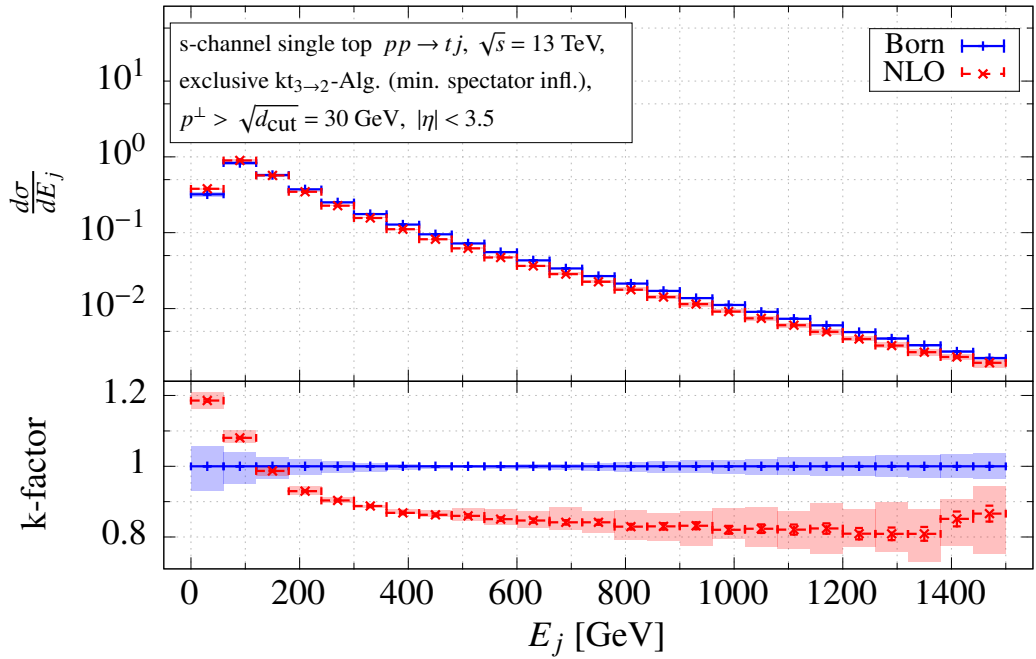

  \begin{center}
    \leavevmode
    \makebox[\textwidth]{\includegraphics[width=\ftw\textwidth]{{{%
          sgt/kfmu/sgtsKT3-2ycut30compmumnspecE2-200x1e7nonorm-crop}}}}
    \caption{Energy distribution and k-factors of the energy of the light jet from $s$-channel
      single top-quark production using the exclusive event
      definition.}  
    \label{fig:kfac_sgts_e2nonorm}
  \end{center}
\end{figure}
\begin{figure}[htbp]
  \begin{center}
    \leavevmode
    \makebox[\textwidth]{\includegraphics[width=\ftw\textwidth]{{{%
          sgt/kfmu/sgtsKT3-2ycut30compmumnspecETA2-200x1e7nonorm-crop}}}}
    \caption{Same as \Fig{fig:kfac_sgts_e2nonorm} but for the pseudo rapidity of the light jet from the $s$-channel.}  
    \label{fig:kfac_sgts_eta2nonorm}
  \end{center}
\end{figure}
\begin{figure}[htbp]
  \begin{center}
    \leavevmode
    \makebox[\textwidth]{\includegraphics[width=\ftw\textwidth]{{{%
          sgt/kfmu/sgtsKT3-2ycut30compmumnspecETA1-200x1e7nonorm-crop}}}}
    \caption{Same as \Fig{fig:kfac_sgts_e2nonorm} but for the pseudo rapidity of the top-tagged jet from the $s$-channel.}  
    \label{fig:kfac_sgts_eta1nonorm}
  \end{center}
\end{figure}
\begin{figure}[htbp]
  \begin{center}
    \leavevmode
    \makebox[\textwidth]{\includegraphics[width=\ftw\textwidth]{{{%
          sgt/kfmu/sgttKT3-2ycut30compmumnspecE2-200x1e7nonorm-crop}}}}
    \caption{Same as \Fig{fig:kfac_sgts_e2nonorm} but for the energy of the light jet from the $t$-channel.}  
    \label{fig:kfac_sgtt_e2nonorm}
  \end{center}
\end{figure}
\begin{figure}[htbp]
  \begin{center}
    \leavevmode
    \makebox[\textwidth]{\includegraphics[width=\ftw\textwidth]{{{%
          sgt/kfmu/sgttKT3-2ycut30compmumnspecETA1-200x1e7nonorm-crop}}}}
    \caption{Same as \Fig{fig:kfac_sgts_e2nonorm} but for the pseudo rapidity of the top-tagged jet from the $t$-channel.}  
    \label{fig:kfac_sgtt_eta1nonorm}
  \end{center}
\end{figure}
\clearpage

\subsection{Inclusive event definition}
\label{app:impNLO_sgtincl}
\begin{figure}[htbp]
  \begin{center}
    \leavevmode
    \makebox[\textwidth]{\includegraphics[width=\ftw\textwidth]{{{%
          sgt/kfmu/sgtsKT3-2ycut30compmumnspecincE2-200x1e7nonorm-crop}}}}
    \caption{Energy distribution and k-factors of the energy of the light jet from $s$-channel
      single top-quark production using the inclusive event
      definition.}  
    \label{fig:kfac_sgtsinc_e2nonorm}
  \end{center}
\end{figure}
\begin{figure}[htbp]
  \begin{center}
    \leavevmode
    \makebox[\textwidth]{\includegraphics[width=\ftw\textwidth]{{{%
          sgt/kfmu/sgtsKT3-2ycut30compmumnspecincETA2-200x1e7nonorm-crop}}}}
    \caption{Same as \Fig{fig:kfac_sgtsinc_e2nonorm} but for the pseudo rapidity of the light jet from the $s$-channel.}  
    \label{fig:kfac_sgtsinc_eta2nonorm}
  \end{center}
\end{figure}
\begin{figure}[htbp]
  \begin{center}
    \leavevmode
    \makebox[\textwidth]{\includegraphics[width=\ftw\textwidth]{{{%
          sgt/kfmu/sgtsKT3-2ycut30compmumnspecincETA1-200x1e7nonorm-crop}}}}
    \caption{Same as \Fig{fig:kfac_sgtsinc_e2nonorm} but for the pseudo rapidity of the top-tagged jet from the $s$-channel.}  
    \label{fig:kfac_sgtsinc_eta1nonorm}
  \end{center}
\end{figure}
\begin{figure}[htbp]
  \begin{center}
    \leavevmode
    \makebox[\textwidth]{\includegraphics[width=\ftw\textwidth]{{{%
          sgt/kfmu/sgttKT3-2ycut30compmumnspecincE2-200x1e7nonorm-crop}}}}
    \caption{Same as \Fig{fig:kfac_sgtsinc_e2nonorm} but for the energy of the light jet from the $t$-channel.}  
    \label{fig:kfac_sgttinc_e2nonorm}
  \end{center}
\end{figure}
\begin{figure}[htbp]
  \begin{center}
    \leavevmode
    \makebox[\textwidth]{\includegraphics[width=\ftw\textwidth]{{{%
          sgt/kfmu/sgttKT3-2ycut30compmumnspecincETA1-200x1e7nonorm-crop}}}}
    \caption{Same as \Fig{fig:kfac_sgtsinc_e2nonorm} but for the pseudo rapidity of the top-tagged jet from the $t$-channel.}  
    \label{fig:kfac_sgttinc_eta1nonorm}
  \end{center}
\end{figure}
\clearpage

\providecommand{\href}[2]{#2}\begingroup\raggedright\endgroup
\begingroup
\selectlanguage{ngerman}
\chapter*{Danksagung}
\begin{itemize}
\item An erster Stelle gilt mein Dank meinem Doktorvater Herr Prof. Peter Uwer f\"ur seine wissenschaftliche
sowie kollegiale Unterst\"utzung. Mir wurde sowohl jede Freiheit zur eigenst\"andigen Erforschung m\"oglicher L\"osungswege geboten als auch die Gewissheit gegeben, dass meine (Irr-)Wege stets mit ernsthaften Interesse verfolgt werden, um im Zweifelsfall zu jeder Zeit geduldig mit konstruktivem Rat zur Seite stehen zu k\"onnen. Dies und die tatkr\"aftige Unterst\"utzung bei der Vereinbarkeit von (wissenschaftlichem) Beruf und Familie, sei es durch pers\"onlichen Erfahrungsaustausch, besondere R\"ucksichtnahme oder die M\"oglichkeit zur Heimarbeit, hat eine f\"ur mich ideale Arbeitsatmosph\"are geschaffen.
\item Unserer Arbeitsgruppe danke ich f\"ur die zahlreichen fachlichen Gespr\"ache und das ernsthafte Interesse an meiner Arbeit. Die damit verbundenen Ratschl\"age und Anmerkungen lie{\ss}en mich oft neue Aspekte und Ans\"atze entdecken. Auch die vielen angenehmen, nicht-wissenschaftlichen Gespr\"ache haben entscheidend zur o.g. idealen Arbeitsatmosph\"are beigetragen.
\item Den Kollegen aus der experimentellen Teilchenphysik danke ich f\"ur geduldige Erl\"auterungen, den regen Austausch und die vielen erhellenden Einblicke. 
\item Ich bedanke mich bei den anderen Promotionsstudenten aus unserer Arbeitsgruppe und auch aus dem Graduiertenkolleg f\"ur den kollegialen Austausch und die gegenseitige Motivation auf dem Weg zum Doktortitel. Besonders bedanke ich mich bei Christoph Meyer und Felix Stollenwerk, die stets bereit waren ein der Situation angebrachtes Getr\"ank mit mir zu teilen.
\item Ich m\"ochte mich bei meiner Familie und meinen Freunden bedanken, die viel Verst\"andnis daf\"ur aufbringen mussten, dass ich einerseits einen nicht unerheblichen Teil meiner Aufmerksamkeit und Energie in die Promotion gesteckt habe und anderseits nicht immer eine klare Linie zwischen Arbeit und Freizeit ziehen konnte.  Besonders dankbar bin ich Esther und meinen beiden Kindern daf\"ur, dass sie mir die wirklich wichtigen Dinge im Leben immer wieder vor Augen f\"uhren konnten.
\end{itemize}
\endgroup
\end{document}